\documentclass[%
 reprint,
 amsmath,amssymb,aps
]{revtex4-2}

\usepackage{graphicx}
\usepackage{dcolumn}
\usepackage{comment}
\usepackage{upgreek}

\usepackage{bm}

\usepackage{xcolor}
\usepackage{subcaption}
\usepackage{float}
\usepackage{siunitx}
\usepackage{hyperref}
\usepackage{tcolorbox}

\begin{document}

\preprint{APS/123-QED}

\title{Fluorescence Microscopy: a statistics-optics perspective}

\author{$^{1,2}$Mohamadreza Fazel, $^3$Kristin S. Grussmayer, $^4$Boris Ferdman, $^5$Aleksandra Radenovic, $^4$Yoav Shechtman, $^6$J{\"o}rg Enderlein, $^{1,2}$Steve Press{\'e}} 
 \affiliation{$^1$Department of Physics, Arizona State University, Tempe, Arizona, USA\\
 $^2$Center for Biological Physics, Arizona State University, Tempe, Arizona, USA\\
 $^3$Department of Bionanoscience, Faculty of Applied Science and Kavli Institute for Nanoscience, Delft University of Technology, Delft, Netherlands\\
 $^4$Russel Berrie Nanotechnology Institute and  Department of Biomedical Engineering, Technion - Israel Institute of Technology, Haifa, Israel\\
 $^5$Laboratory of Nanoscale Biology, Institute of Bioengineering, Ecole Polytechnique Federale de Lausanne (EPFL), Lausanne, Switzerland \\
 $^6$III. Institute of Physics - Biophysics, Georg August University, G{\"o}ttingen, Germany
 }%

\date{\today}

\begin{abstract}
Fundamental properties of light unavoidably impose features on images collected using fluorescence microscopes. Modeling these features is ever more important in quantitatively interpreting microscopy images collected at scales on par or smaller than light's wavelength. Here we review the optics responsible for generating fluorescent images, fluorophore properties, microscopy modalities leveraging properties of both light and fluorophores, in addition to the necessarily probabilistic modeling tools imposed by the stochastic nature of light and measurement.
\end{abstract}

\maketitle

\tableofcontents
\section{\label{Intro}Introduction }

\subsection{\label{Intro_History} A brief history of optics and statistics}

The ancient Greeks were divided over whether vision arose from rays entering or leaving the eyes~\cite{darrigol2012history,thibodeau2016ancient}. For instance, atomists believed that perception arose from an atom flux traveling through space to the eyes. Aristotle (384--322 BCE) later proposed the notion of ether serving as a medium for transmission of intrinsic qualities of objects to the eye rather than fluxes of atoms.  An alternative formulation, advocated by Pythagoras (570--495 BCE) and Euclid (325--270 BCE), proposed the notion of ocular fire whose rays impassively scanned their surroundings. Following this logic, Euclid established a geometric optics explaining the perception of size and angles from the geometry of these ocular rays. Along these same lines, the Chinese philosopher Mo Di (470--391 BCE) established a geometric optics similar to Euclid's explaining the formation of shadows and images in mirrors~\cite{Wang2008}.

An amalgam of these ideas--with fire originating from the eyes coalescing with 
another fire derived from objects enabling vision--was perhaps now demanded on philosophical grounds and promoted by Plato (427--347 BCE). In Ptolemy's optics (100-170 CE), sunlight activated objects whose emitted rays now interacted with visual rays to give rise to perception. In Ptolemy's theory, perception relied on the angular distribution, length, refraction and reflection of rays from the eye ~\cite{darrigol2012history,smith1996ptolemy}.

Although these early Greek theories appear manifestly naive, emerging notions of geometric optics served as a clear starting point to Medieval Arabs who took a decidedly more phenomenological approach. For example, inspired by Euclid's geometric optics, Al-Kindi (801--873 CE) demonstrated that visual rays travel in straight lines by simple experiments on shadows~\cite{darrigol2012history}. This early progress was followed by insights from Ibn al-Heytham--latinized as Alhazen (965--1040 CE)--who showed that eyesight is derived from light rays received by the eyes from objects~\cite{darrigol2012history,nasr1968Science}. Further, he consistently devised experiments to test his optical theories including theories on refractive and reflective properties of light rays on boundaries, lenses and spherical mirrors among others~\cite{nasr1968Science,kriss1998history,tbakhi2007ibn,darrigol2012history}.

The distribution of Latin translations of Alhazen's \textit{Book of Optics}~\cite{al2015retrospect} amongst other ancient works, ultimately sparked a Renaissance that presages the onset of modern optics in Europe. From the democratization of knowledge driven by the indefatigable Gutenberg presses followed refractive telescopes attributed to the Dutch spectacle-makers Zacharias Janssen (1585--1638 CE) and Hans Lippershey (1570--1619 CE) and reflecting telescopes attributed to Issac Newton (1643--1727 CE)~\cite{bardell2004BioOne}. In contrast to telescopes, there is uncertainty regarding the original inventor of microscope, though often credited to Zacharias Janssen~\cite{kriss1998history,chung2017pioneers}.

From the very start, the world of microscopy and biology were intertwined: the Dutch businessman and scientist Antonie van Leeuwenhoek (1632--1732 CE) exploited his microscope to single-handedly discover bacteria, sperm cells, and red blood cells amongst other actors dominating the microscopic realm~\cite{chung2017pioneers}. Little, in this regard, has changed throughout history with sizes, features, and other optical properties of the Natural world motivating the design of modern microscopes. Subsequent \emph{compound} microscopes~\cite{gest2004Discovery}, also credited to Janssen and foreshadowing our multi-lens microscopes, provided improved magnification and were widely used by Robert Hooke (1635--1703 CE)~\cite{gest2004Discovery}, author of the first book on microscopes \textit{Micrographia}. 

Now taken for granted, successive properties of light--including diffraction, refraction, reflection as well as light's particulate nature--were each individually leveraged in microscope development with diffraction through an aperture first reported by the Italian Jesuit Francesco Maria Grimaldi (1618--1663 CE), followed by a number of discoveries culminating in Maxwell's (1831--1879 CE) electromagnetic theory, and theories on light's quantization~\cite{planck1901law,einstein1905molekularkinetischen} due to Planck (1858--1947 CE) and Einstein (1879--1955 CE).

Setting aside remarkable later microscopy advances--including phase imaging~\cite{popescu2011quantitative,park2018quantitative}--we interrupt history to pause at fluorescence microscopy which has dominated the scene of the last half century as smaller scales demanded increased contrast between background and object of interest~\cite{lichtman2005fluorescence}. At such scales, the stochastic properties of light, intrinsic to quantum mechanics, dictate our ability to interpret fluorescence microscopy data and bring us back to the primary focus of this review: fluorescence microscopy from a statistics-optics perspective.  

Modeling light's stochastic properties isn't an exercise in mitigating the recurring nuisance of shot noise. It is, instead, fundamental to how we draw insights at the scales fluorescence microscopy has unraveled. In fact, a fluorescent photon's emission time, its absorption time, emission wavelength, and detection location, \textit{i.e.}, where a photon is detected on an image plane, are all random variables. These random variables themselves are drawn from probability distributions. In the classical limit, the probability density for locating photons is proportional to the time-averaged energy flux given by Poynting's theorem~\cite{poynting1884xv}, introduced by John Henry Poynting (1852--1914 CE). For point-like sources of light, \textit{e.g.}, fluorophores, the normalized spatial distribution, coinciding with a slice orthogonal to the propagation direction, is termed the Point Spread Function (PSF). This inherent randomness in a photon's location, imperfectly detected and reporting only probabilistically on a fluorescent object of interest, now introduces multiple levels of stochasticity between the object whose properties we care to characterize and measurement output. This, unavoidably, introduces statistical concepts--including notions of latent variables and hierarchical probabilistic models--in the quantitative modeling of imaging systems. 

The manipulation of hierarchical dependencies between random variables then requires what is known today as Bayes' theorem. The theorem, attributed to its namesake Thomas Bayes (1702--1761 CE), was popularized by Pierre-Simon de Laplace (1749--1827 CE) who introduced and codified, through seminal texts on probability~\cite{de1820theorie,marquis1840essai}, probabilistic modeling to the Sciences~\cite{dale1982bayes}. 

Before we return to microscopy, we now take a brief detour to discuss statistical modeling relevant to our future applications. 

\subsection{\label{prob-stat}Introduction to statistical modeling}

The electromagnetic force carrying particle, the photon, is intrinsically both wave-like and particulate. While the continuous spatial distributions over a photon's location
are dictated by the photon's wave properties, photon detections themselves are necessarily pointillistic and probabilistic. As such, even before considering other sources of stochasticity like detection, a quantitative picture of microscopy demands, at its most fundamental level, an exposition of the theory of statistical sampling.

Here, we first lay out the main concepts for probabilistic modeling. We then discuss the concept of likelihoods and Bayesian inference key to the statistical frameworks introduced throughout this review.

\vspace{-5mm}
\subsubsection{Basic concepts and notation}

Stochasticity in a system arises either from the inherent random nature of the physical system or measurement noise or both. Both are relevant in quantitative microscopy and thus we minimally require two layers of stochasticity: at the level of photon shot noise and at the level of detection; see Appendix~\ref{Se:Detector}. Soon, we will also see that additional levels of stochasticity may arise from the behavior of fluorescent labels. 

For this reason, we begin by defining the requisite notions of a random variable. A random variable, $\mathfrak{R}$, represents a collection of possible options, either numeric or non-numeric, following the statistics of a probability distribution $\mathbb{P}$.

As such, we often write $\mathfrak{R} \sim \mathbb{P}$,
where the above reads ``the random variable $\mathfrak{R}$ is sampled from the probability distribution $\mathbb{P}$". We then denote $\mathfrak{r}$ a particular realization of $\mathfrak{R}$ and $p(\mathfrak{r})$ the probability density associated with the probability distribution $\mathbb{P}$. 

Generally, the probability distribution itself depends on parameters, $\vartheta$. To make such dependency explicit, we may write $p(\mathfrak{r}; \vartheta)$ and $\mathbb{P}(\vartheta)$~\cite{presse2023data}. For example, the location at which the photon is detected is itself a random variable, $R$, sampled from a distribution centered at the emitting molecule's location, $\mathbf{r}_0$. As such, we write
\begin{align}
     R & \sim U(\mathbf{r}_0), \\
     p(\mathbf{r};\mathbf{r}_0) & = U(\mathbf{r};\mathbf{r}_0),
\end{align}
where $\vartheta \equiv \mathbf{r}_0$, and $p(\mathbf{r};\vartheta)$ is the probability density, {\it i.e.}, the PSF, from which $\mathbf{r}$ is drawn.

It is often of interest to compute the probability of obtaining a value from a subset $\eta$ of the possible values ($\mathfrak{r}\in \eta$), given by
\begin{equation}
    \mathcal{P}_{\eta} = \int_{\eta} d\mathfrak{r}\,p(\mathfrak{r};\vartheta).
    \label{eq:subset}
\end{equation}
By definition if $\eta$ is the entire set of options then $\mathcal{P}_{\eta} = 1$. For instance, the probability of a photon reaching a pixel is given by the integral of the PSF over the pixel area $\mathcal{A}$
\begin{equation}
    \mathcal{P}_{\mathrm{pix}} = \int_{\mathcal{A}} U(\mathbf{r};\mathbf{r}_0) d\mathbf{r}.
    \label{eq:PSFintPix}
\end{equation}

In probabilistic modeling, we often work with many random variables, $\mathfrak{R}_1,\,\mathfrak{R}_2,\, ...,\, \mathfrak{R}_N$, at once. For this reason, we define the joint density
\begin{equation}
    p(\mathfrak{r}_{1:N};\vartheta) = p(\mathfrak{r}_1,\mathfrak{r}_2, ..., \mathfrak{r}_N;\vartheta). 
    \label{eq:jointprob}
\end{equation}
The density of any individual $\mathfrak{r}_n$ is then obtained by integrating the joint density with respect to all values of $\mathfrak{r}_{1:n-1}$ and $\mathfrak{r}_{n+1:N}$
\begin{equation}
    p(\mathfrak{r}_n;\vartheta) = \int d\mathfrak{r}_{1:n-1}\,d\mathfrak{r}_{n+1:N} \,p(\mathfrak{r}_{1:N};\vartheta).
    \label{eq:mrinalization}
\end{equation}
 This integration, termed a marginalization, results in a marginal density, $p(\mathfrak{r}_n;\vartheta)$. Marginalization is often useful in computing, say, the probability over the diffusion coefficient of an emitter (a fluorescently labeled molecule or dye) irrespective (and thus integrating over) its exact location in space. This is later explored in, \emph{e.g.}, Fig.~\ref{fig:Diff-FCS}.

If random variables $\mathfrak{R}_{1:N}$ are independent and identically distributed, \textit{iid}, then Eq.~\ref{eq:jointprob} assumes the simpler form
\begin{equation}
\mathfrak{R}_1,\mathfrak{R}_2,\,...,\mathfrak{R}_N \overset{iid}{\sim} \mathbb{P}(\vartheta)
\end{equation} 
with the understanding that the joint density decomposes into the product of independent densities $p(\mathfrak{r}_1;\vartheta), p(\mathfrak{r}_2;\vartheta), ..., p(\mathfrak{r}_N;\vartheta)$. For example, \textit{iid} random variables include photon arrival times following pulsed excitation for a static distribution of molecules; \textit{e.g.}, as later explored in the Box~\ref{box:SingleFLIM}.

In general, random variables are not independent, \emph{e.g.}, the position of a molecule in time where the system's state depends on its state at a previous time point either exactly or by approximation. This dependency, explored in the context of fluorophore dynamics in Sec.~\ref{Non_Markov} is termed the Markov assumption. In this case, we say that values that can be ascribed to $\mathfrak{R}_2$ depend on the realization of a preceding random variable $\mathfrak{r}_1$. This dependency is often expressed as 
\begin{equation}   \mathfrak{R}_2|\mathfrak{r}_1,\vartheta \sim \mathbb{P}(\mathfrak{r}_1,\vartheta),
    \label{eq:conditionalsample}
\end{equation}
which reads ``the random variable $\mathfrak{R}_2$ given the parameters $\vartheta$ and realization (or ``conditioned on") $\mathfrak{r}_1$ of $\mathfrak{R}_1$ is sampled from the probability distribution $\mathbb{P}(\mathfrak{r}_1,\vartheta)$". The  density we associate to this probability distribution then reads $p(\mathfrak{r}_2|\mathfrak{r}_1,\vartheta)$ and is referred to as a conditional density. In general, a random variable $\mathfrak{R}_N$ can depend on many other random variables $\mathfrak{R}_{1:N-1}$ with associated conditional density $p(\mathfrak{r}_N|\mathfrak{r}_{1:N-1},\vartheta)$. Such conditionals will become useful as we build hierarchical models relating random variables across our boxed environments.

Bayes' theorem, of central importance in expressing hierarchical random variable dependencies, then follows from the observation that conditional densities, such as $p(\mathfrak{r}_{1:2})=p(\mathfrak{r}_{2}|\mathfrak{r}_{1})p(\mathfrak{r}_{1})$, 
satisfy $p(\mathfrak{r}_{1:2})=p(\mathfrak{r}_{2:1})$ and thus
\begin{equation} 
p(\mathfrak{r}_{1}|\mathfrak{r}_{2})p(\mathfrak{r}_{2})=
p(\mathfrak{r}_{2}|\mathfrak{r}_{1})p(\mathfrak{r}_{1}).
    \label{eq:conditionalsampleprob2}
\end{equation}

As is customary in physics, we will now denote both random variables and their realizations with lower case letters. The distinction between both notions will be implied by the context.

\subsubsection{Likelihood} 
We can now introduce the object at the heart of quantitative analysis of microscopy data: the likelihood. The likelihood is a probability distribution over those random variables coinciding with $K$ experimental observations, $w_{1:K}$, conditioned on $\vartheta$. The likelihood's density is thus written as $p(w_{1:K}|\vartheta)$ where $w_{1:K} = \{w_1, w_2, ..., w_K\}$. It is  also convenient to denote this set with an overbar, $\overline{w}$.

The term likelihood follows from the notion that $p(w_{1:K}|\vartheta)$ is a likelihood of observing the sequence of observations $w_{1:K}$ under the assumptions of the model (\textit{i.e.}, calibrated values for parameters $\vartheta$ of a particular model). Indeed, all box environments will contain likelihoods for each statistical framework presented.

Often as the parameters are themselves unknown, we ask what values for these parameters maximize the likelihood of the observed sequence $w_{1:N}$. These parameter values are called estimators and are denoted by $\hat{\vartheta}$. For example, we can ask what values of the excited state lifetime (assuming one fluorophore species) make the photon arrival times observed most probable; \emph{e.g.}, Box~\ref{box:SingleFLIM}.

For practical reasons, it is common to work with, and maximize, the likelihood's logarithm $\mathcal{L}(w_{1:K}|\vartheta)=\log\left(p(w_{1:K}|\vartheta)\right)$, sometimes termed log-likelihood, rather than the likelihood itself, \textit{e.g.}, see Sec.~\ref{PSFeng_sec}. This is because the logarithm is both monotonic with the original function and avoids numerical underflow typical of small probability densities arising as $K$ grows.

Within a Maximum Likelihood Estimation (MLE) framework, $\vartheta$ are treated as fixed (deterministic) parameters and the data, $w_{1:K}$, are understood as realized random variables. While the MLE yields a single value (estimator) for the parameters, the uncertainty around the parameter estimate is captured by computing the likelihood's breadth around its maximum. The breadth is often estimated as
\begin{equation}
\sigma^2_{\vartheta_l}=\left[\mathcal{Q}(\vartheta)^{-1} \right]_{ll},
\label{eq:CRLB}
\end{equation}
where $l$ counts the elements of the model parameter set, $\vartheta$. Here, $\mathcal{Q}(\vartheta)$ is the Fisher information matrix defined as~\cite{cramer1999mathematical,rao1992information}
\begin{equation}
\mathcal{Q}_{ll'}(\vartheta) =-\mathbb{E}\left[\frac{\partial^2 \mathcal{L}(w_{1:K}|\vartheta)}{\partial \vartheta_l \vartheta_{l'}} \Bigg|_{\hat{\vartheta}}\right],
\label{eq:FisherInfo}
\end{equation}
where $\mathbb{E}$ denotes the expected value of the expression within the parenthesis. As Eq.~\ref{eq:CRLB} sets the variance, an uncertainty bound, around the MLE, it is sometimes termed the Cram\'er-Rao Lower Bound (CRLB).

As may be evident, MLE-based approaches present challenges for likelihoods with multiple degenerate maxima or, more importantly, when the model is unknown. What is more, even assuming a model form, the MLE only provides a point estimate not a full distribution over the putative parameter values. 

It is for all these reasons that we often turn to a more general Bayesian paradigm. In this setting, we use the likelihood to construct the distribution over the parameters of interest given the observed data,  $p(\vartheta|w_{1:K})$. The latter object is termed the posterior and is central to Bayesian inference.

\vspace{-4mm}
\subsubsection{Posterior}
In working with likelihoods, the data is understood as random variables and parameters, $\vartheta$, as fixed but to be determined. In contrast, in a Bayesian setting both data and parameters are treated as random variables. In particular, the data are random variables already realized and whose values are used to construct the probability, $p(\vartheta|w_{1:K})$, over the unknown random variables, $\vartheta$. The Bayesian paradigm allows us to properly propagate uncertainty over $\vartheta$ from all sources including detector noise, camera intensity pixelation, motion aliasing, photon shot noise, and many more.

The posterior is constructed from the likelihood by invoking Bayes' theorem, Eq.~\ref{eq:conditionalsampleprob2}, 
\begin{equation}
    p(\vartheta|w_{1:K}) = \frac{p(w_{1:K}|\vartheta)p(\vartheta)}{p(w_{1:K})},
    \label{eq:postdef}
\end{equation}
where, by normalization, 
\begin{equation}
    p(w_{1:K}) = \int d\vartheta p(w_{1:K}|\vartheta)p(\vartheta).
\end{equation}
Here $p(\vartheta)$, termed the prior, provides a mean to regulate the parameters. For instance, determining a range over which non-zero values of the density arise, \textit{e.g.}, positive or integer values, prior to considering the data. 

Thus, from Bayes' theorem, we obtain a clear recipe by which the prior distribution is updated based on data, $w_{1:K}$, encoded in the likelihood, to arrive at the posterior $p(\vartheta|w_{1:K})$. It is thus clear that to avoid the prior biasing the posterior, $K$ must be sufficiently large ~\cite{mcneish2016using,smid2020bayesian,van2020small,zitzmann2021performance}. To mitigate the size of $K$ needed, roughly ``flat" or featureless prior distributions between $\vartheta$'s upper and lower bounds are preferred.

As we will see in all applications, likelihoods can generally be constructed from knowledge of the microscopy technique and the physics of the problem while priors are normally motivated by computational convenience. The broad question then arises: Can we determine whether the posterior is peaked at some value of $\vartheta$? More concretely, what does our posterior look like? 

Unfortunately, posteriors rarely attain a simple, analytic form, on account of the measurement and physics informing the likelihood. As such, values of $\vartheta$ are typically numerically sampled from posteriors using Monte Carlo methods. For example, as we later discuss in the context of confocal microscopy, \textit{e.g.}, Sec.~\ref{Im_Detect}, we will see that $\vartheta$ includes quantities such as diffusion coefficients, emission rates, and emitter locations. As posteriors are thus often multi-variate, a common Monte Carlo strategy, loosely speaking, involves sampling one random variable at a time in a scheme termed Gibbs sampling~\cite{geman1984stochastic}.

Whether sampling a posterior exactly or numerically, {\it e.g.}, via Gibbs sampling, it is often computationally convenient to judiciously select the functional form of priors. Indeed, some prior forms play a special role in Bayesian modeling by having the unique mathematical property that, when multiplied by the likelihood, result in a posterior of the same form as the original prior (albeit with updated, ``re-normalized", parameters). As such, we often speak of conjugate prior-likelihood pairs or, for succinctness, conjugate priors when such priors can be identified. While we will not dwell on specialized notions of Bayesian inference, we make the reader aware that computational efficiency is what makes it possible to include measurement noise details at marginal added computational cost whilst improving the spatiotemporal resolution of any fluorescence analysis method. Indeed, whenever possible, specialized Monte Carlo schemes (from Gibbs sampling~\cite{geman1984stochastic}, to Metropolis-Hastings~\cite{metropolis1953equation,hastings1970monte}, to slice sampling~\cite{murray2010elliptical}, and beyond~\cite{bishop2006pattern,brooks2011handbook,presse2023data}) used across all applications discussed herein benefit from any computational advantage thrown their way.
\vspace{-5mm}
\subsubsection{Bayesian non-parametrics}

From Eq.~\ref{eq:postdef}, we see that constructing a posterior demands a mathematical, \emph{i.e.} ``parametric", form of the likelihood. However, for most practical cases, we often do not know what competing models describe a given data set. We also know, and can demonstrate by way of examples, that the more complicated we make a model, the larger its likelihood, \emph{i.e.}, we over-fit the data. 

Compromising between data under- and over-fitting is at the heart of the fundamental model selection problem. From the onset, progress in model selection has been critical, for instance, in clustering problems where the number of clusters (\textit{i.e.}, the model) are unknown~\cite{richardson1997bayesian,neal2000markov,gelfand2005bayesian,sgouralis2017introduction}. Indeed, the model selection problem manifests itself across microscopy applications. For example: determining the number of molecules within a diffraction-limited spot (\textit{i.e.}, the model) explored in Box~\ref{box:Counting}; or determining the number of fluorophore species in lifetime imaging explored in Box~\ref{box:SingleFLIM}.

While heuristically comparing a fixed set of models to resolve model selection---for example by relying on information criteria~\cite{quan2011high}, and other tools introduced as post-processing steps---is computationally advantageous, such an approach presents theoretical problems. For example, it is often limited to cases where we can  exhaustively enumerate models. For example, how many emitters in each frame across a stack of frames can we consider in any wide-field tracking application? Even if enumerable, how do we assign probabilities to these competing models given the data?  

Answers to these questions, outside the realm of the Natural Sciences, have led to the formal development of Bayesian Non-Parametrics (BNPs)~\cite{ferguson1973bayesian,presse2023data} alongside Monte Carlo tools to sample from the resulting non-parametric posteriors, including Reversible Jump Markov Chain Monte Carlo (RJMCMC)~\cite{green1995reversible}. In short, BNPs treat model and parameter estimation on the same footing ~\cite{orieux2011bayesian,hines2015analyzing, sgouralis2017introduction,gabitto2021bayesian} and construct non-parametric posteriors over both models and their associated parameters. 

In particular, within a non-parametric treatment, we consider \textit{a priori} an infinite number of competing models. We place priors on these models alongside their associated parameters just as we place priors on parameters alone within the regular (parametric) Bayesian paradigm. 

One catch is that BNPs are limited to a particular class of models termed nested models. Fortuitously, many models considered across microscopy applications belong to this class. Briefly, nested models include all models that can be generated from a more general model by setting parameters to different values (including zero) with the most general model itself being infinite dimensional. For example, a two state model used in analyzing a F\"orster Resonance Energy Transfer (FRET) time trace, later explored in greater depth in Sec.~\ref{Non_Chem}, follows from a three state model where transitions to the third state are all set to zero. Other examples of nested models we will explore include: 1) the number of molecules in a diffraction-limited spot (see Boxes~\ref{box:Counting} and \ref{box:SMLM}); 2) the number of fluorophore species in lifetime imaging (see Box~\ref{box:SingleFLIM}); and perhaps less intuitively 3) all competing two-dimensional lifetime maps obtained from scanning confocal lifetime imaging; see Box~\ref{box:MultiFLIM}.

These examples were intentionally numbered. They allow us to introduce three commonly used non-parametric priors used in constructing non-parametric posteriors. In the order in which these examples are listed, we have: the Beta-Bernoulli process prior~\cite{hjort1990nonparametric,paisley2009nonparametric,broderick2012beta,shah2015empirical,al2018approximations,presse2023data}; the Dirichlet process prior~\cite{neal2000markov,gelfand2005bayesian,sgouralis2017introduction,sgouralis2018single,gabitto2021bayesian,presse2023data}; and the Gaussian Process (GP) prior~\cite{rasmussen2003gaussian,quinonero2005unifying,presse2023data}. 

The Beta-Bernoulli process prior is used when we try to estimate the number of discrete elements contributing to the data. These could be, for example, the number of clusters or, equivalently, the number of emitters contributing photons generating an image frame or producing a stream of photons within a confocal spot, \emph{e.g.}, Box~\ref{box:FCS}. Within a BNP paradigm, we assign a Bernoulli variable (binary random variable), $b_m$ called a load, to each discrete element (molecule). Considering as many as $M$ loads (and letting $M$ eventually tend to infinity), the unknowns appearing in $\vartheta$ are augmented to include $b_{1:M}$. Thus $\vartheta$ for the single spot confocal would now include the diffusion coefficient, emission rate, molecular locations, as  well as loads, $b_{1:M}$. 

When multiplying the likelihood by the appropriate Beta-Bernoulli prior process, we may then construct a posterior, whose parameters we wish to sample, include the loads. The resulting posterior is, in turn, often sampled using Monte Carlo techniques to determine which loads are sampled mostly as 0's (and thus coincide with molecules not warranted by the data) or coincide with 1's (and thus coincide with molecules warranted by the data). The number of molecules in each draw from the posterior are then determined by summing all loads. 

We now turn, much more briefly, to the subsequent two non-parametric priors. For instance, the Dirichlet process prior is used when we wish to assign probabilities to an infinite number of components. For example, when we wish to determine to what degree each unique chemical species contributes photons in a lifetime experiment; see Box~\ref{box:SingleFLIM}. Ideally, based on Monte Carlo sampling of the non-parametric posterior (obtained from the product of the likelihood and the Dirichlet process prior), we would find which of the infinite species introduced in modeling contribute non-negligibly to the data. 

Finally, GP priors are used in estimating smooth functions. Smooth functions of interest in microscopy include, for example, fluorophore density maps explored in Sec.~\ref{Im_Detect} or even smooth background for large numbers of emitters. 
Each of these maps consists of an infinite set of correlated random variables, \textit{i.e.}, values of the map at every point in space. Draws from the (non-parametric) posterior then assign values to each point on the map. In practice, the number of map points whose value we wish to deduce is kept finite and limited to a fixed number of points typically over a uniform mesh grid termed inducing points~\cite{quinonero2005unifying,bryan2020inferring,fazel2022high}. The value of the map on a finer spatial grid can then be interpolated from the spatial correlation function already informing the GP prior.

Having now introduced key notions from statistics, we turn to microscopy.

\vspace{-5mm}
\subsection{\label{Intro_Limit}Basic characteristics of fluorescence microscopy}

\begin{figure}[H]
    \centering
  \includegraphics[width=0.9\linewidth]{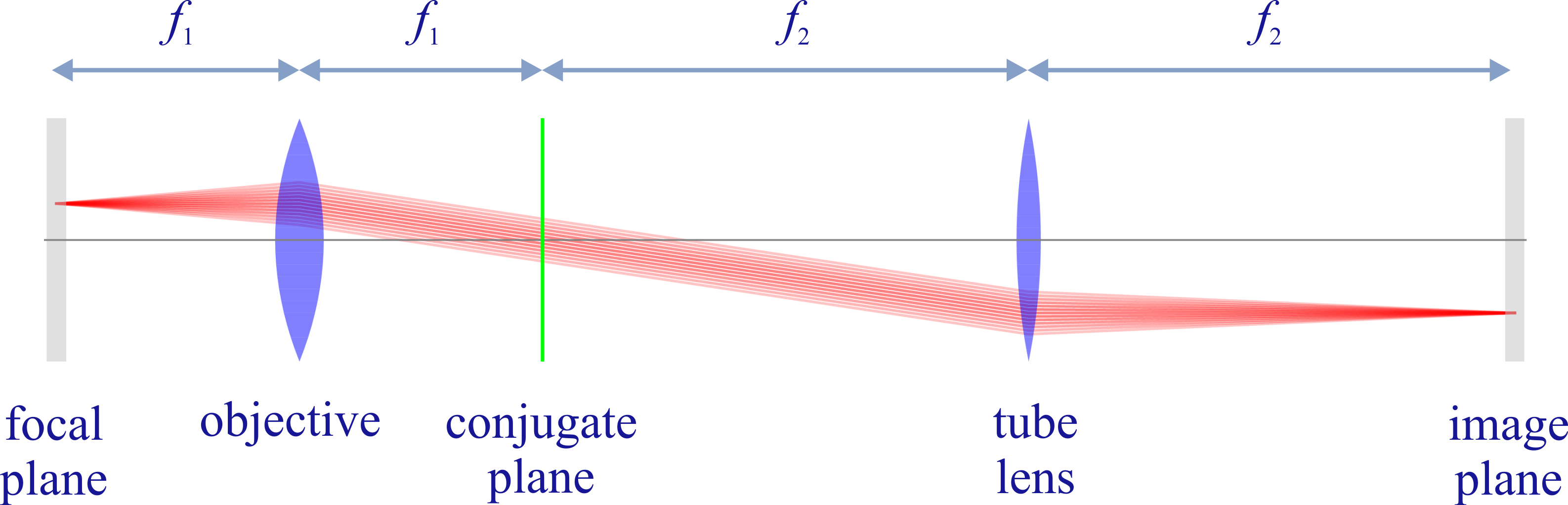}
    \caption{Schematic of an infinity-corrected wide-field microscope consisting of an ideal objective lens with focal length $f_1$ and an ideal tube lens with focal length $f_2$. We show light propagation from a point source in the focal plane (sample space) to the image point in image space. The plane between the lenses, a distance $f_1$ away from the objective lens and $f_2$ from the tube lens, is called the conjugate plane (green vertical line). The conjugate plane is also sometimes termed the back focal plane, Fourier plane or pupil plane. Here the light from any point source on the focal plane crosses through the same lateral position. By considerations of geometric proportion, it can be seen that the ratio of lateral displacement of the image point to lateral displacement of the source point is equal to the ratio of the focal lengths, $f_2/f_1$. This ratio is the microscope's magnification $\mathcal{M}$.}
    \label{fig:magnification}
    \vspace{-4mm}
\end{figure}

All optical microscopes use light, one way or another, to interact with the sample under observation. Indeed, bright-field, dark-field, or even phase contrast imaging differ from each other in details pertaining to which part of the excitation or detection arms are altered or blocked to create images at the detector.

However, these microscopes are limited in their ability to discern contrast at molecular and even supra-molecular length scales at which life unravels. At such scales, we exploit fluorescence microscopy, involving fluorophore-labeled samples, later detailed in Sec.~\ref{Nonlinear}. When excited, fluorophores emit light that can be selectively filtered from the excitation beam to form an image. In its simplest form, a fluorescence microscope is a two-lens system: an objective lens with small focal length $f_1$ and a tube lens with long focal length $f_2$; see Fig.~\ref{fig:magnification}.  

In modern infinity-corrected research microscopes, the objective converts the diverging spherical wavefront emitted by a point emitter in the focal plane in sample space into a planar wavefront. The planar wavefront is then reconverted by the tube lens into a spherical wavefront converging into a point on the image plane.

\begin{figure}[H]
    \centering
    \includegraphics[width=0.9\linewidth]{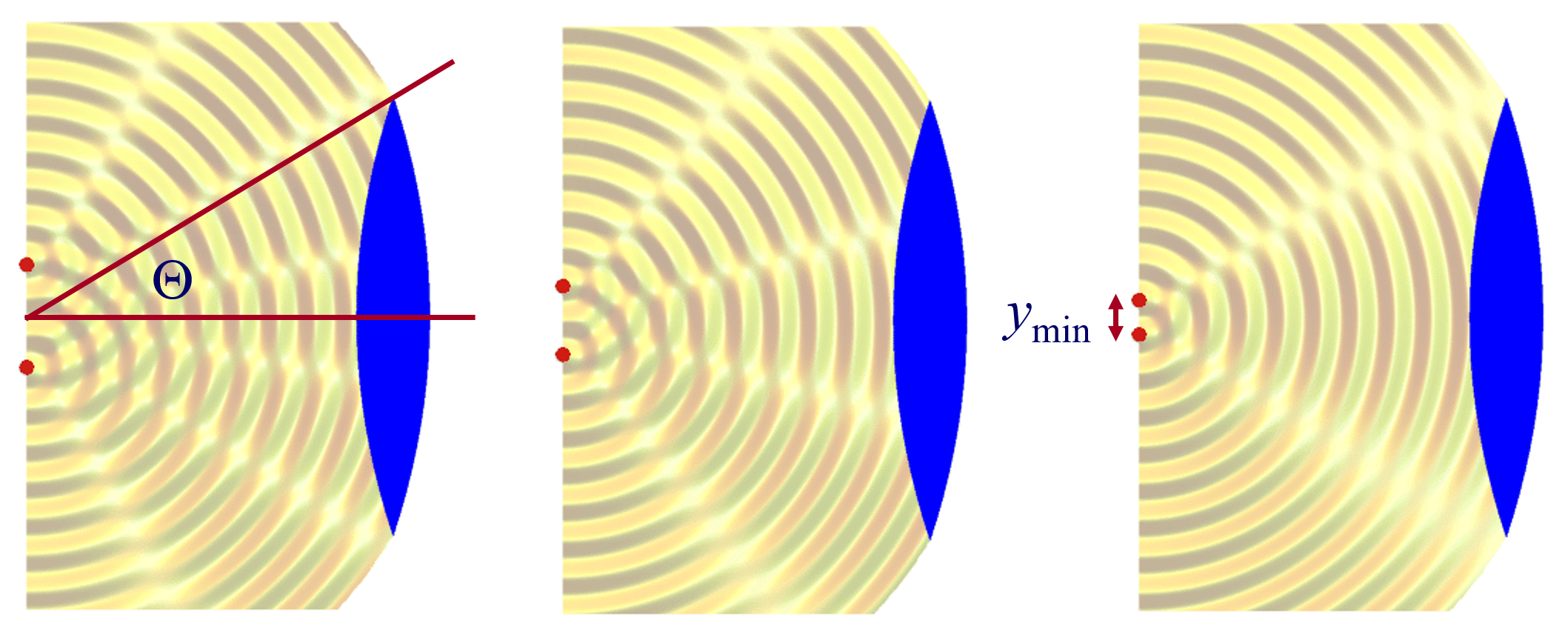}
    \caption{Visualization of the diffraction limit of resolution. Here, we show interference patterns of two coherently emitting point emitters, shown by red dots, for three different distances between emitters across panels. The closer the emitters are positioned with respect to each other, the larger the angular positions of the destructive interference lanes (directions of zero light intensity). At a critical distance, shown in the right panel, the first lane of destructive interference is positioned at the half angle $\Theta$ of light collection of the objective, and the objective lens receives a continuous wavefront absent intensity minima appearing as a single emitter wavefront.}
    \label{fig:resolution}
    \vspace{-4mm}
\end{figure}

The two most important characteristics of a microscope are its magnification and its resolution, {\it i.e.}, how well sample features are resolved. From Fig.~\ref{fig:magnification}, the system's magnification is given by the ratio $f_2/f_1$ (from the proportionality of vertical to horizontal distances). However, the magnification of an optical microscope today is of secondary importance as images are recorded with array detectors, such as Complementary Metal-Oxide Semiconductor (CMOS) or Electron Multiplying Charge-Coupled Device (EMCCD) cameras with varying pixel size; see Appendix~\ref{Se:Detector}. This is in contrast to visual inspection of a sample where our rod and cone cell sizes are fixed. For such wide-field microscopes equipped with a camera, the detector's physical pixel size divided by the microscope's magnification set an upper bound on the image quality. This effective pixel size should be at least two times smaller than the microscope's optical resolution (Nyquist criterion).

This leads us to the second important microscope characteristic: its resolution. The microscope resolution is limited by a number of factors including the diffraction of light and light collection by objective lenses. These two effects lead to a fundamental resolution limit of approximately half of the wavelength. As such, if the emitted light's wavelength were to be far smaller than typical dimensions of the molecular species of interest, then our review article would stop here and textbooks would be replenished with real life images reminiscent of David Goodsell's artistic renderings of life inside the cell~\cite{goodsell2009machinery}. However, this is not the case.

We will discuss more thoroughly resolution of different microscope modalities shortly though we start with a heuristic albeit useful visualization of a fundamental microscope's optical resolution limit; see Fig.~\ref{fig:resolution}. Here we show the (far-field) electric-field distribution of light from two coherent point sources, designated by red dots, before an objective lens. As both point sources are assumed to emit light coherently, the resulting intensity distribution shows characteristic lanes of constructive and destructive interference. When the distance between the two point emitters, $y$, is gradually reduced (from left to right panels in Fig.~\ref{fig:resolution}), the two symmetric lanes of destructive interference (directions of zero light intensity) closest to the optical axis migrate towards higher emission angles, until they reach the objective lens' edge. At that point, the objective detects only light of a continuous spherical wavefront absent any zero-intensity minima within its light detection cone (with half angle $\Theta$), similar to what the objective would see from a single emitter. 

\begin{figure}[H]
\vspace{-2mm}
    \centering
    \includegraphics[width=0.65\linewidth]{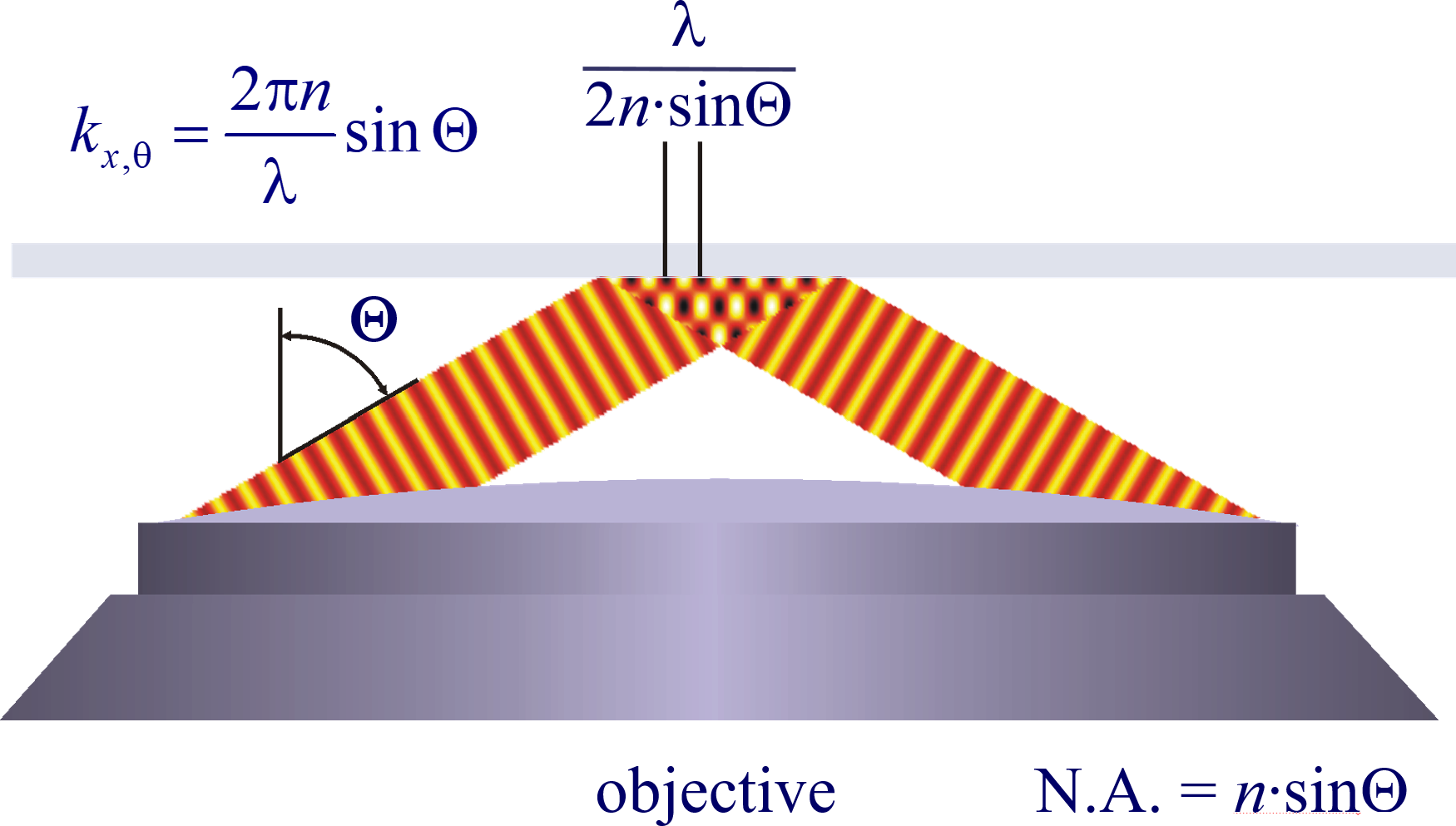}
    \caption{Lateral resolution limit of a CLSM. The resolution is determined by the highest lateral spatial frequency contained in a focused bright spot. This is generated by the interference of two rays traveling from the edges of the objective to the focal point with the highest possible incidence angle $\Theta$ with respect to the optical axis as shown. The associated wave vectors are of equal magnitude, $2\pi n/\lambda$, where $\lambda$ is the vacuum wavelength. The corresponding lateral components, $k_{x,\theta}$, of these wave vectors are of equal magnitude given by $k_{x,\theta}=2\pi n \sin\Theta/\lambda$, and opposite directions resulting in a difference of $4\pi n\sin\Theta/\lambda$. As such, the interference of the two beams leads to a periodic interference pattern in the lateral direction with periodicity $\lambda/2n\sin\Theta$, equal to the lateral resolution limit of a CLSM.}
\label{fig:lateralresolution}
\vspace{-5mm}
\end{figure}

Simple trigonometry dictates that the path difference between 1) the first emitter, and the edge of the lens, and 2) the second emitter, and the same edge of the lens is $y\sin\Theta$. In doing so, we assumed that the separation of the lens, and emitters is much larger than $y$ in the far-field limit. The first destructive interference lane therefore occurs at angle $\Theta$ if the path difference is half the wavelength, {\it i.e.}, $y_\mathrm{min} \sin\Theta = \lambda/2n$, where $\lambda$ is the vacuum emission wavelength, and $n$ is the refractive index of medium in which the emitters are embedded. As such, the wavelength in this medium is $\lambda/n$. From this result follows Abbe's famous resolution limit, first formulated by Ernst Abbe (1840-1905 CE) in 1873 ~\cite{Abbe_1873}, as
\begin{equation}
y_\mathrm{min} = \frac{\lambda}{2 n \sin\Theta} = \frac{\lambda}{2 \mathrm{NA}}, 
\label{eq:abberesolutionlimit}
\end{equation}
where NA is the objective's numerical aperture. 

A similar simplified consideration can also be applied toward understanding the spatial resolution of a Confocal Laser Scanning Microscope (CLSM). In a CLSM, the sample is scanned with a tightly focused laser beam, and the excited fluorescence light is collected by the microscope optics, focused through a confocal pinhole to suppress out-of-focus light, and finally detected with a point detector (usually silicon-based photo-diode, or photo-electron multiplier tube); see Sec.~\ref{CLSM-Confocal}. The recorded fluorescence light intensity as a function of scan position is then used to reconstruct an image. The fundamental advantage of a CLSM as compared to a wide-field imaging microscope is its optical sectioning capability, {\it i.e.}, its capability to record true three-dimensional sample images, later detailed when considering the Optical Transfer Functions (OTF) of both microscope types. Neglecting momentarily a CLSM's confocal detection volume, then its lateral resolution is determined by how tightly a laser beam can be focused into an excitation spot. In a mathematically more precise manner, one asks about the tightest spatial intensity modulation still present in a diffraction-limited focus. The answer is given by Fig.~\ref{fig:lateralresolution}, which shows that the tightest modulation is achieved by the interference of the two light rays exiting the objective at the highest possible angle, which is exactly the half angle of light detection $\Theta$ of the objective. As can be seen, the spatial periodicity of this intensity modulation is again given by Abbe's formula, Eq.~\ref{eq:abberesolutionlimit}, only with the emission wavelength now replaced by the excitation wavelength (usually shorter than the emission wavelength due to the spectral Stokes shift of fluorescence emission with respect to excitation; see Sec.~\ref{Nonlinear}). 

\begin{figure}[H]
    \centering
    \includegraphics[width=0.75\linewidth]{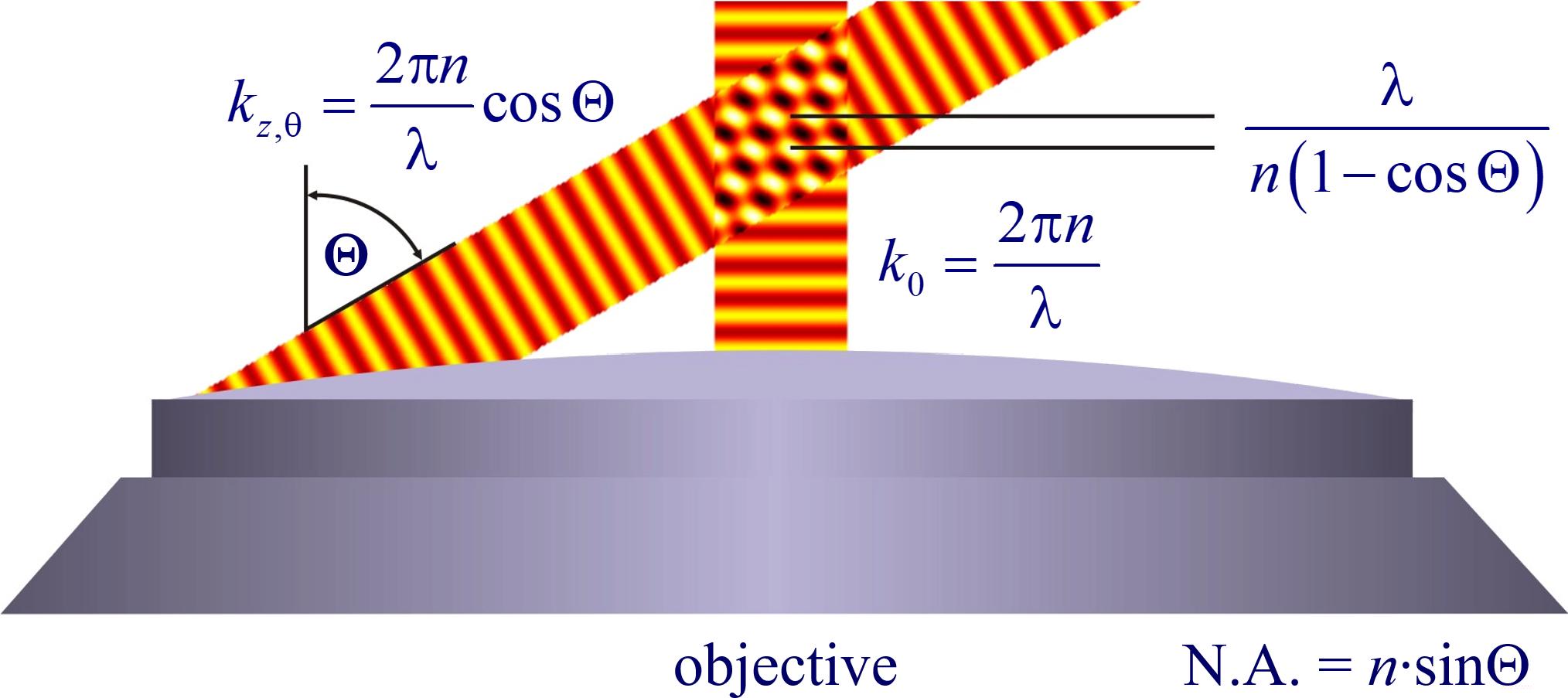}
    \caption{Axial resolution of a CLSM: Similar to the lateral resolution, the axial resolution is determined by the tightest spatial modulation of light that can be generated along the optical axis. This is achieved by interfering an axially propagating beam with one traveling at the highest possible incidence angle. The axial component of the wave vector of the former is equal to the full wave vector length $k_0=2\pi n/\lambda$, and the axial component for the latter is $k_{z,\Theta}=2\pi n\cos\Theta/\lambda$. The resulting interference therefore leads to a spatial intensity modulation along the optical axis with periodicity $\lambda/n(1-\cos\Theta)$ setting a CLSM's axial resolution limit.}
    \label{fig:axialresolution}
    \vspace{-4mm}
\end{figure}

In a similar vein, we can also obtain the axial resolution limit of a (confocal laser scanning) microscope, by asking about the tightest spatial intensity modulation achievable when focusing light through the objective. The answer is presented in Fig.~\ref{fig:axialresolution}, where the tightest modulation is now generated by the interference of an axial light ray with a light ray traveling at the highest possible incidence angle $\Theta$. This directly yields the axial resolution limit of an optical microscope, complementary to Abbe's lateral resolution limit, and is given by
\begin{equation}
z_\mathrm{min} = \frac{\lambda}{n (1-\cos\Theta)} \approx \frac{2n\lambda}{(\mathrm{NA})^2},
\label{eq:axialresolutionlimit}
\end{equation}
where the approximation on the right hand side is valid only for small numerical aperture values.

\begin{figure}[H]
    \centering
    \includegraphics[width=0.9\linewidth]{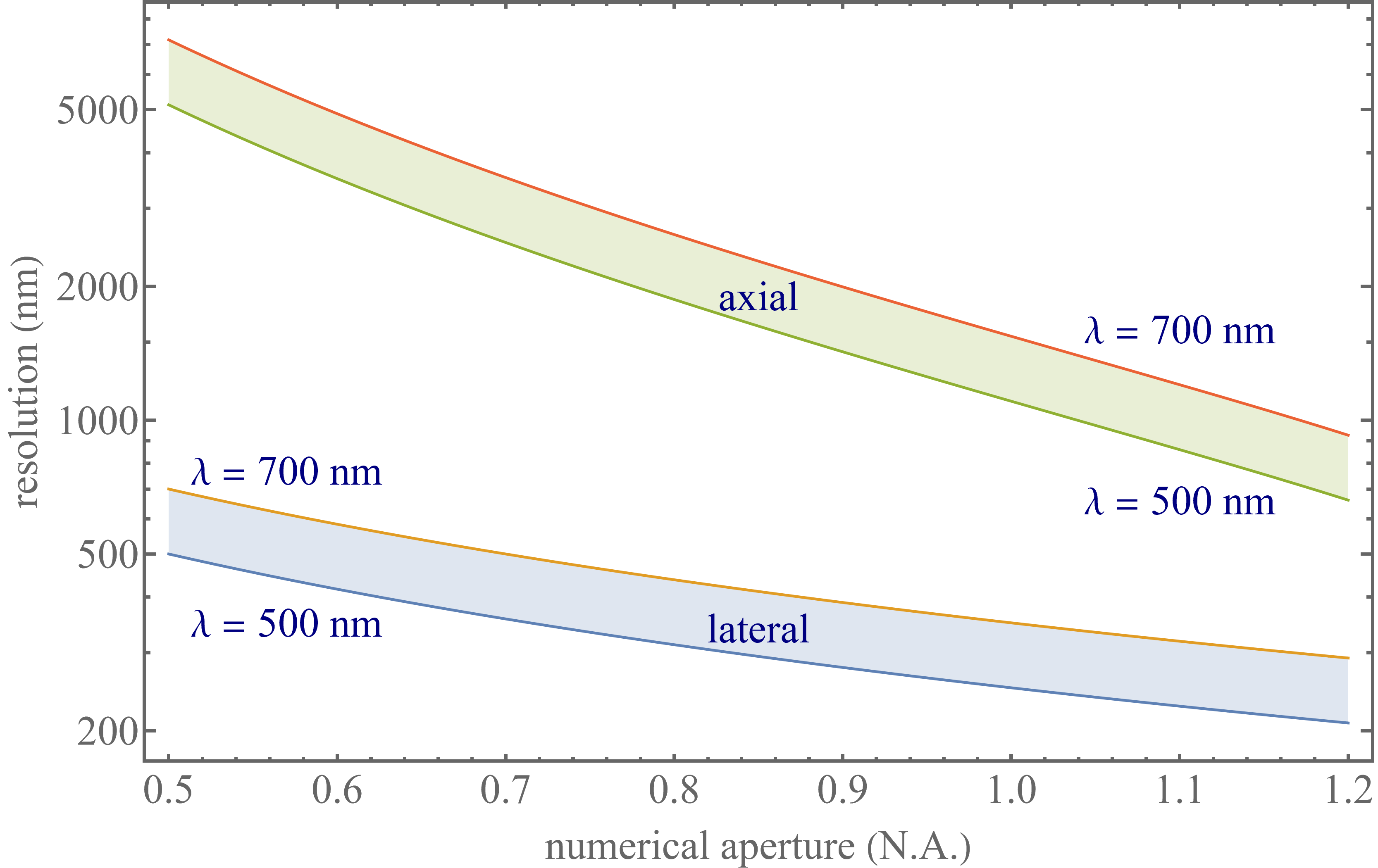}
    \caption{Lateral and axial resolution of diffraction-limited optical microscopy using a water immersion objective (designed for imaging in water with refractive index 1.33) as a function of numerical aperture NA and wavelength.}
    \label{fig:resolutioninnumbers}
\end{figure}

We summarize physical scales associated with lateral and axial resolution of diffraction-limited optical microscopes in Fig.~\ref{fig:resolutioninnumbers}. Here we show lateral and axial resolutions as functions of the numerical aperture, NA, for optical wavelengths across the visual spectrum using for concreteness a water immersion objective (\textit{i.e.}, designed for imaging in water with refractive index 1.33). 

While providing qualitative guidance on optical system design, axial and lateral spatial resolution expressions provided in Eq.~\ref{eq:abberesolutionlimit}-\ref{eq:axialresolutionlimit} remain theoretical. In particular, such expressions provide an upper bound on the resolution otherwise limited by factors including crucial notions of stochastic nature of photons, and undesired out-of-focus light among others. 

A final important note is warranted on light (information) collection efficiency and suppression of out-of-focus light from regions outside the focal plane, \textit{i.e.}, limiting light collection to a certain axial range termed optical sectioning. For this purpose, specialized sample illumination and fluorescent light detection techniques have been developed including Total Internal Reflection Fluorescence (TIRF) microscopy~\cite{axelrod1981JCB}, Super-critical Angle Fluorescence (SAF) microscopy ~\cite{enderlein1999highly,ruckstuhl2004supercritical}, Metal-Induced Energy Transfer (MIET) microscopy~\cite{chizhik2014metal}, confocal microscopy~\cite{marvin1961USPatent}, Image Scanning Microscopy (ISM)~\cite{Sheppard1988,Muller2010}, two-photon microscopy~\cite{denk1990Science}, 4pi microscope ~\cite{hell1992OpticsLetters}, Structured Illumination Microscopy (SIM) ~\cite{bailey1993Nature}, light-sheet microscopy ~\cite{voie1993Microscopy,huisken2004optical}, and multi-plane microscopy~\cite{Blanchard1999,Prabhat2004}.

All methods mentioned accomplish optical sectioning and enhance photon collection efficiency in improving image resolution and contrast. These techniques pushed the optical resolution to its very limits as dictated by Abbe's diffraction barrier. However, it was not until the end of 20th century that this barrier was overcome to achieve spatial resolutions in far-field light microscopy far beyond the diffraction limit~\cite{hell1994OpticsLetter}. Research in this front is still ongoing leveraging advances in four main components of fluorescence microscopes: fluorescent emitters; optical setups; detectors; and analysis. In what follows, we first discuss fluorescent light sources and then proceed to review optics of different microscope modalities while presenting statistical analysis frameworks throughout. 

\vspace{-10mm}
\section{\label{Nonlinear}Fluorophores}

Point fluorescent emitters or light sources, often molecules termed fluorophores, are key to fluorescence imaging of labeled samples. Both conventional fluorescence imaging, as well as microscopy techniques achieving resolution beyond light's diffraction limit, rely on tunable properties of fluorophores including emission rates, brightness, absorption and emission spectra, excited state lifetimes, and other photo-physical properties such as blinking and photo-bleaching~\cite{moerner2015single}. Here, we will discuss quantum fluorophore properties, alongside their statistical modeling, and relegate classical models to Sec.~\ref{sec:classicfluorophore}, where we derive their emission fields. 

\vspace{-4mm}
\subsection{\label{Non_Chem}Fluorophore properties}

Most molecules do not naturally fluoresce in regimes detectable by modern detectors and cannot easily be excited without inducing photo-damage. Thus, one must often resort to specific fluorescence labeling~\cite{Specht2017} of biological samples, \textit{e.g.}, to identify and investigate structures against the vast cellular background of proteins, nucleic acids, lipids, and small molecules. 

While the addition of fluorescent labels introduces challenges, their intrinsic properties as well as non-linear response to light in themselves open windows of opportunity, {\it e.g.}, to study molecular interactions~\cite{Ciruela2008, Luo2020}, determine molecular copy numbers ~\cite{Grumayer2019,bryan2022diffraction,hummert2021photobleaching}, and improve optical resolution ~\cite{huang2009super,Schermelleh2019} as later detailed in Sec.~\ref{SR}.

The most common labels include: fluorescent proteins~\cite{tsien1998green,zhang2002creating,Dedecker2013}; organic dyes~\cite{dempsey2011evaluation,Lavis2017}, generally small organic molecules containing conjugated $\pi$-electron systems; and semiconductor quantum dots, inorganic nano-crystals with especially broad excitation and correspondingly narrow emission spectra~\cite{Resch-Genger2008}. 

Fluorescent labels include a large variety of fluorophores with excitation and emission wavelength maxima spanning the near-infrared, visible and UV spectrum~\cite{Giepmans2006, Li2018}. Less common, ``exotic", fluorescent labels providing an even larger color palette and increasingly tunable photo-physical properties include carbon nanorods, carbon dots, polymer dots, fluorocubes, and fluorescent defects in diamond or 2D materials~\cite{aharonovich2014diamond,Jin2018,saurabh2022modeling}.

\begin{figure}[H] \centering
\includegraphics[width=0.8\linewidth]{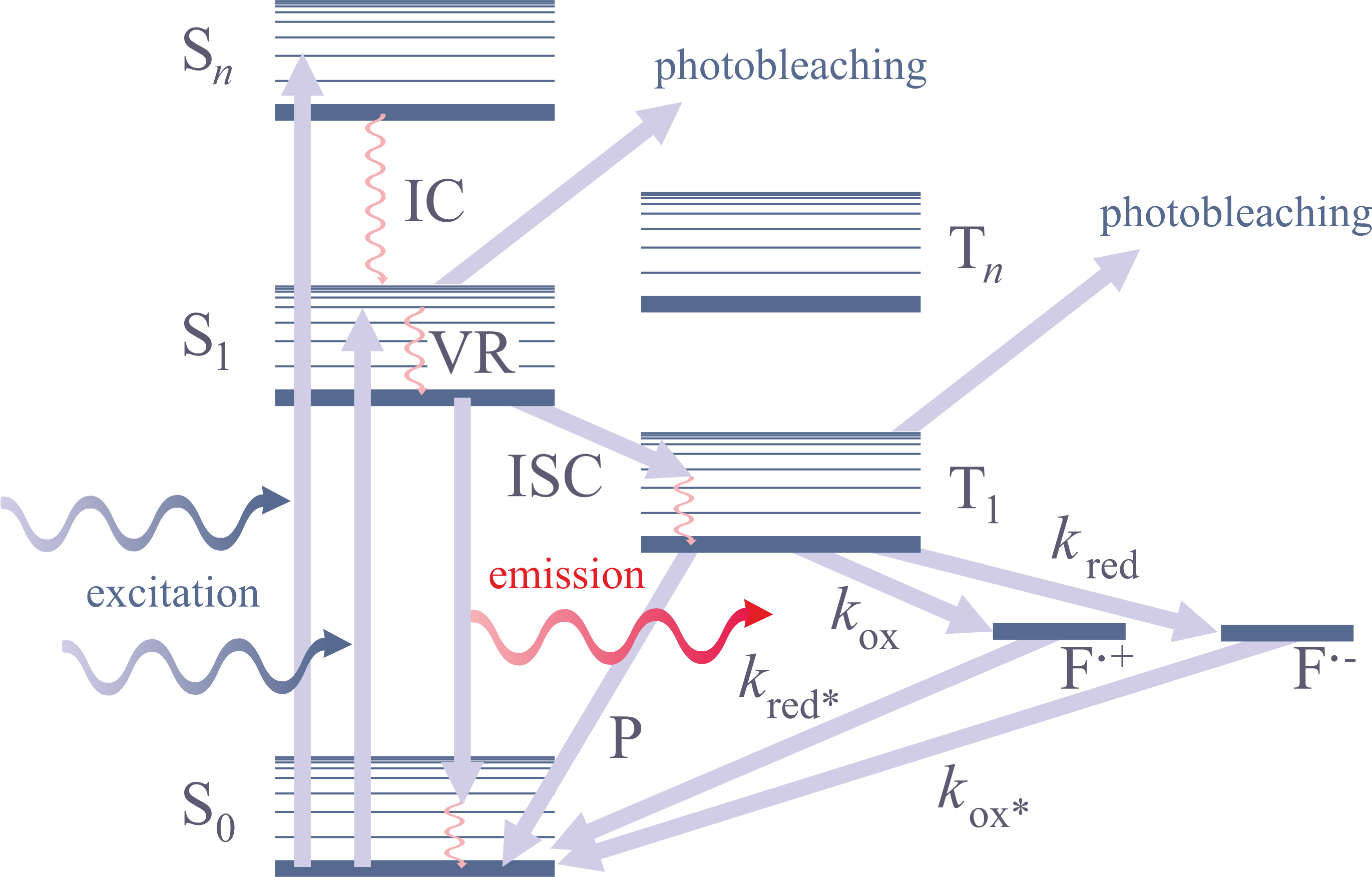} 
\caption{Simplified Jablonski diagram. The electronic ground state S$_0$, the singlet excited states S$_n$, the triplet excited states T$_n$, and radical cation F$^{\cdot+}$
or anion states F$^{\cdot-}$. Thick lines represent electronic energy levels, thin lines vibrational energy levels, while rotational energy states are left unmarked. Here we denote: Phosphorescence by P; Vibrational Relaxation by VR; Internal Conversion by IC; Inter System Crossing by ISC; and rates of oxidation and reduction are $k_{\mathrm{ox}}$ and $k_{\mathrm{red}}$, respectively. Arrows represent a subsample from all possible transitions between different states.}
\label{fig:Jablonski-extended} 
\vspace{-4mm}
\end{figure}

Basic fluorophore photo-physics are captured by Jablonski diagrams such as Fig.~\ref{fig:Jablonski-extended} for an organic dye illustrating select transitions between different molecular energy and spin states. A more rigorous treatment of transition rules, molecular spectra, and interactions of light and matter, can be found in the books of Lakowicz~\cite{Lakowicz2006} and Valeur {\it et al}~\cite{Valeur2012}.

A molecule in the (typically singlet) ground state is excited to a singlet excited state by absorbing a photon with a probability depending on the excitation light intensity and the molecule's absorption cross-section (linearly related to the molar extinction coefficient~\cite{Lakowicz2006}). The molar extinction coefficient $\epsilon_{\lambda}$ is a measure of how strongly a solution containing one mole of a fluorophore absorbs (attenuates) light at wavelength $\lambda$ expressed using the Lambert-Beer law~\cite{Lakowicz2006}
\begin{align}
    \epsilon_{\lambda} &= \frac{A_{\lambda}}{c_Ml}=\frac{\log_{10}\left(I_{0\lambda}/I_{\lambda}\right)}{c_Ml}, \label{eq:lambertBeer}
\end{align}
where $A_{\lambda}$ is the absorbance measured, $I_{0\lambda}$ is the initial light intensity of wavelength $\lambda$, and $I_{\lambda}$ is the light intensity after traveling the path length $l$ through the solution with molar concentration $c_M$. From Eq.~\ref{eq:lambertBeer}, it is clear that the SI unit of molar extinction coefficient is $\mathrm{m}^2/\mathrm{mol}$, but the commonly used unit is lit/cm/mol. 

From $\epsilon_{\lambda}$, we immediately arrive at another important fluorophore property, namely molecular brightness $B_\lambda$. To achieve high Signal to Noise Ratio (SNR), fluorescent labels with high molecular brightness, $B_{\lambda} = Q_f\epsilon_{\lambda}$, are desired. Here $Q_f$ is a unitless quantity called fluorescence quantum yield describing how many fluorescence photons are emitted relative to the number absorbed. This is given by the ratio of the sum of radiative transitions to the total transitions, \textit{i.e.}, the sum of transition rates corresponding to all transition paths out of the excited state,
\begin{align}
    Q_f = \frac{\sum k_f}{\sum k_f+\sum k_{\text{non}}} \label{eq:Q}, 
\end{align}
where $k_{f}$ and $k_{\mathrm{non}}$, are, respectively the rate of fluorescence or radiative decay, and rate of non-radiative decay. 

Another important fluorophore property is the average time, $\tau$, a fluorophore remains excited prior to emitting a photon
\begin{align}
    \tau &= \frac{1}{\sum k_f+\sum k_{\text{non}}}. 
    \label{eq:lifetime}
\end{align}
Here $\tau$, termed fluorescence lifetime, typically lasts on the order of nanoseconds for organic dyes. The fluorescence lifetime is a characteristic property of fluorophores in their unique environment tuned by pH, ion or oxygen concentration, molecular binding, or proximity dependent inter-molecular energy transfers primarily influencing the rate of non-radiative decay~\cite{Lakowicz2006, Valeur2012}. As such, differences in fluorophore lifetimes can be employed to distinguish fluorophore species thereby broadening the appeal of Fluorescence Lifetime Imaging Microscopy (FLIM)~\cite{digman2008phasor,datta2020fluorescence} in functional and multiplexed imaging of disparate fluorophores with otherwise overlapping spectra~\cite{fereidouni2012spectral,valm2017applying,scipioni2021phasor}; see Sec.~\ref{CLSM-Confocal}. 

As described above, the quantum yield is tied to the number of possible transitions out of the excited state either non-radiatively or radiatively. Upon fluorophore excitation, one such radiative transition occurs via rapid vibrational relaxation to the lowest energy level of the $S_1$ excited state followed by radiative decay to a vibrational ground state level with spontaneous fluorescence emission; see Fig.~\ref{fig:Jablonski-extended}. The fluorescence emission is shifted towards longer wavelengths (Stokes shift) as compared to excitation, due to fast internal conversion and vibrational relaxation to the lowest level of the $S_1$ excited state (Kasha's rule ~\cite{Kasha1950}). Another radiative transition out of the excited state, of later interest, is stimulated emission. Typically, stimulated emission does not play a role at room temperature so long as the excitation intensity is low. However, this non-linear process is exploited in STimulated Emission Depletion (STED) super-resolution imaging~\cite{hell1994OpticsLetter} later described in Sec.~\ref{sec:STED}. 

In addition to radiative transitions, several alternative non-radiative pathways are available for transition from the first singlet excited state, S$_1$, to the ground state. For instance, the molecule can return to the ground state dissipating the energy to the environment as heat. For example, the non-radiative transition to the triplet state, T$_1$, via inter-system crossing is often employed in Single Molecule Localization Microscopy (SMLM); see Sec.~\ref{sec:SMLM}. Return from  T$_1$ to the ground singlet state (phosphorescence) is typically delayed on account of a forbidden spin flip transition; see Fig.~\ref{fig:Jablonski-extended}. As such, transitions to and from triplet, or further reduced/oxidized off-states (also referred to as bright and dark states, respectively)
occur on longer timescales (\SIrange{0.1}{100}{\milli\second}). 

To control fluorophore switching between triplet dark and bright states, \textit{i.e.}, to control blinking, oxygen concentration may be adjusted. Upon reaction with dissolved molecular oxygen, fluorophores may transition from the triplet dark (off-state) to singlet ground (on-state) states by interacting with molecular oxygen's ground triplet state. Molecular oxygen can also accept an electron from a triplet fluorophore inducing typically undesirable phototoxic effects, {\it i.e.}, irreversible photo-bleaching~\cite{zheng2014contribution} occurring from many states as shown in Fig.~\ref{fig:Jablonski-extended}. 

Though in some applications photo-bleaching is desirable, in others, such as particle tracking~\cite{Shen2017,sgouralis2023dynamic,xu2023bnp} and protein-protein interactions via FRET~\cite{forster1948zwischenmolekulare,lerner2018toward}, photo-bleaching and blinking are problematic and suppressed by removal of dissolved oxygen 
via oxygen scavenging systems, such as glucose oxidase coupled with catalase~\cite{aitken2008oxygen}, or by depopulating dark states leveraging both reducing and oxidizing agents~\cite{Vogelsang2008}. 

In many cases, such as in STochastic Optical Reconstruction Microscopy~\cite{rust_sub-diffraction-limit_2006}, blinking of fluorophores is desirable to achieve spatial resolution below the diffraction limit; see Sec.~\ref{sec:SMLM}.
Here, many cyanine and rhodamine dyes are used as they can be reversibly photo-switched from a bright state to a dark state (blink) in a buffer containing enzymatic oxygen scavengers and a primary thiol such as $\beta$-mercaptoethylamine or $\beta$-mercaptoethanol~\cite{Li2018,Jradi2019}. Alexa Fluor 647 is the organic dye of choice for state-of-the-art direct STochastic Optical Reconstruction Microscopy (dSTORM) imaging due to its high brightness and efficient switching behavior~\cite{Diekmann2020}. For several cyanines, \textit{e.g.}, Cy5, it has been shown that thiolate anions covalently bind to the fluorophore~\cite{Dempsey2009}, thereby disrupting the conjugated system resulting in dark state. The dyes can also be chemically reduced by $\text{NaBH}_4$ to a non-fluorescent form or synthesized in a caged form that can later be photo-activated, which has been used in different SMLM techniques~\cite{Vaughan2012, Lehmann2015}. Rhodamine dyes can as well reversibly switch from a fluorescent to a non-fluorescent form by intra-molecular spirocyclization either spontaneously or driven by UV light. This has been exploited to generate sensors and switches or can be used across  SMLM applications~\cite{Li2018, Zheng2019}. 

Examples of SMLM include (fluorescence) Photo-Activated Localization Microscopy ((f)PALM)~\cite{betzig_imaging_2006,hess_ultra-high_2006}, as well as derivatives such as single particle tracking PALM (sptPALM)~\cite{Manley2008a}. In these applications, advanced fluorescent proteins are used. These switch between fluorescent states through at the chromophore either reversibly ({\it e.g.}, on and off for Dronpa by cis-trans isomerization) or through photo-activation ({\it e.g.}, PA-GFP by decarboxylation) or photo-conversion ({\it e.g.}, green to red wavelength for mEos by $\beta$-elimination)~\cite{Dedecker2013, Jradi2019}. 

More recently, studies of protein activity and SMLM have benefited from the discovery of a new class of ligand-activated fluorescent proteins~\cite{Kumagai2013}. The prototype UnaG binds the small molecule bilirubin via multiple noncovalent interactions and forms a fluorescent complex. The oxidized (and photo-bleached) ligand can detach from the protein, allowing a fresh bilirubin molecule to bind and act as a sensor for small molecules thereby reporting on protein activity~\cite{Kwon2020}.

In general, fluorescent proteins have the advantage of being genetically encodeable, allowing fluorescent labeling of nearly arbitrary target proteins in living cells and organisms by creating fusion constructs. However, this also means that proteins must undergo appropriate folding followed by chromophore maturation, {\it i.e.}, formation of a fluorescent molecule typically starting from three amino acids~\cite{Dedecker2013}. This process can take minutes to hours, may be incomplete, and can impair the temporal accuracy of measurements of rapid processes such as gene expression dynamics~\cite{Balleza2018}.  
While organic dyes circumvent some of these difficulties, both organic dyes and fluorescent proteins often exhibit complex photo-physical and photo-chemical behaviors complicating quantitative analysis. For instance, organic dyes can exhibit spectral blue shifts upon high laser radiation~\cite{Helmerich2021, Cho2021} or spectral shifts from substrate (green) to product state (orange) like in the epoxidation of a double bond in conjugation to a BODIPY dye~\cite{Rybina2013} useful in mechanistic studies of chemical reactions at the single molecule level~\cite{Cordes2013}. However, such spectral shifts may affect multi-color applications, \textit{e.g.}, in super-resolution imaging or Single Particle Tracking (SPT), and are problematic to FRET experiments. Moreover, many proteins have additional dark states, \textit{e.g.}, mEos cis-trans isomerization~\cite{Annibale2011a,DeZitter2019}, and organic dyes may have several conformations with different intensity levels, \textit{e.g.}, Atto647N, with at least three  states differing in fluorescent lifetimes~\cite{Wang2013} complicating quantitative single molecule read-outs. 
\vspace{-6mm}
\subsection{F\"orster resonance energy transfer \label{Sec:FRET}}

In the previous section, we discussed fluorophore properties involving radiative transitions or non-radiative transitions. Here, we continue by considering non-radiative transitions through inter-molecular energy transfer~\cite{Valeur2012}. A few example of these transitions include: Photo-induced Electron Transfer (PET) ~\cite{escudero2016revising}, collisional quenching or FRET, Bioluminescence Resonance Energy Transfer (BRET) \cite{Kobayashi2019,syed2021applications}, Protein Induced Fluorescence Enhancement (PIFE)~\cite{myong2009cytosolic,hwang2011protein}, or the recently discovered Proximity-Assisted Photo-Activation (PAPA)~\cite{Graham2021}. Such transitions are distance dependent and thus have been leveraged to probe binding interactions or conformational changes. 

In what follows, we focus on FRET, an inter-molecular energy transfer process widely used to measure molecular interactions serving as a distance ruler for structural biology~\cite{Lakowicz2006,wu2020forster,agam2023reliability}. In FRET, non-radiative energy transfer from a donor to an acceptor fluorophore occurs through dipole-dipole coupling with rate constant $k_{\text{FRET}}$ when the donor's emission spectrum overlaps with the acceptor's absorption spectrum~\cite{Forster1948}. Under the dipolar approximation, the probability for energy transfer to occur, termed FRET efficiency ($E_{\text{FRET}}$), scales with the donor-acceptor distance to the inverse $6^{th}$ power~\cite{jones2019resonance} and is 50\% at the F\"orster radius $R_0$ 
\begin{align}
    E_{\text{FRET}} &= \frac{1}{1+\left(r/R_0\right)^6} = \frac{k_{\text{FRET}}}{\sum k_f+\sum k_{\text{non}}} \nonumber \\
    &= 1-\frac{\tau_{DA}}{\tau_D},
    \label{eq:FRET} 
\end{align}
where $\tau_{DA}$ and $\tau_D$ are, respectively, the donor fluorescence lifetime in the acceptor's presence and absence. For typical donor-acceptor pairs, $R_0$ is a few nanometers~\cite{Lakowicz2006} and depends on the donor emission-acceptor absorption spectral overlap and the relative orientation of donor-acceptor dipole moments. It is explicitly given by 
\begin{align}
    R^6_0 &= \frac{9000 \ln 10}{128 \pi^5 N_{\mathrm{A}} n^4}\kappa^2Q_{f,D} \int{I_D\left(\lambda\right)\epsilon_A\left(\lambda\right)\lambda^4 d\lambda},
    \label{eq:FRETdistance}
\end{align}
where $\kappa$ is the so-called orientation factor, 
\begin{align}
    \kappa &= 3\cos{\theta_D}\cos{\theta_A} - \cos{\theta_{DA}},
\end{align}
$Q_{f,D}$ is the donor's quantum yield in the absence of the acceptor, $n$ is the solution's refractive index, $N_{\mathrm{A}}$ is the Avogadro constant, $I_D$ is the donor's normalized fluorescence emission spectrum, $\epsilon_A$ is the acceptor's molar extinction coefficient, $\theta_{DA}$ is the angle between donor and acceptor transition moments, and $\theta_{D}$ and $\theta_{A}$ are the angles between these moments
and the vector connecting donor to acceptor, respectively. For $\epsilon_A$ and $\lambda$, respectively, given in lit/cm/mol and cm units, $R_0$ is in cm.

Ignoring the angular dependence of the energy transfer, as described in Eq.~\ref{eq:FRETdistance}, for fixed dipoles can yield significant biases in FRET distance assessments~\cite{Hellenkamp2018}. Fortunately,
in practice, the dipoles are often freely and rapidly rotating (rapid compared to the donor de-excitation rate) leading to an average value of $\kappa^2=2/3$.  

FRET can also occur between spectrally identical molecules (homo-FRET), and is observed by measuring its effect on fluorescence polarization anisotropy~\cite{Gradinaru2010} 
\begin{align}
    r &= \frac{I_{\parallel}-G I_{\perp}}{I_{\parallel}+2G I_{\perp}}. \label{eq:anisotropy}
\end{align}
Here, $I_{\parallel/\perp}$ is the intensity measured when the polarizers in the detection path are aligned parallel/perpendicular to those in the excitation, and $G$ is a correction factor for the difference in the instrument’s sensitivity to the two orthogonal polarization orientations.

Upon exposure to linearly polarized light, the excitation probability is highest for molecules whose absorption dipole moments are aligned parallel to the polarization vector of the exciting light. In most cases, the absorption and emission dipoles of a molecule are co-linear, such that fluorescence emission remains polarized immediately after excitation. Fluorescence remains anisotropic unless the molecule rotates over the fluorescence lifetime or the excitation energy is transferred to a different molecule. Thus anisotropy or polarization measurements inform us on molecular parameters such as orientation, oligomerization or size, and environmental conditions like viscosity~\cite{Bader2011, Gradinaru2010}. Polarization can also be read out in super-resolution imaging, \textit{e.g.}, using polarized light in illumination or detection and capturing polarized emission by implementing specifically engineered PSFs sensitive to polarization~\cite{hulleman2021}; see Sec.~\ref{PSFeng_sec}. 

Polarization, lifetime, FRET efficiency, or other photo-physical markers we have discussed herein are only interesting in so far as their changes report back on the kinetics of the underlying labeled molecules. We now turn to Markov models describing discrete molecular events to extract molecular kinetics from photo-physical changes.
\vspace{-4mm}
\subsection{\label{Non_Markov} Markov models for fluorophores}

To help motivate the use of Markov models, we consider them in the analysis of FRET data and the enumeration of fluorophores within a diffraction-limited Region Of Interest (ROI). 

For example, observations from FRET experiments with photons individually recorded (at avalanche photo-diodes abbreviated as APDs) include a set of photon arrival times along with a set of corresponding colors (wavelengths), designated by $c=1,2$, attributing photons to either donor or acceptor channels, respectively.

The set of photon arrival times (data) are either measured with respect to the start of the experiment, for continuous illumination \cite{saurabh2022single}, or with respect to the pulse immediately preceding a photon detection, such as in pulsed illumination ~\cite{safar2022single}. Here, for sake of illustration, we assume continuous illumination where data consists of intervals between photon arrivals. We let $K+1$ coincide with the total number of photons and denote the data with $\Delta t_{1:K}=\{\Delta t_1, ...,\Delta t_{K}\}$. The sets of inter-arrival times are then used to learn transition kinetics between system states comprised of molecular and label photo-physical states. For concreteness, we assume that molecular states coincide with conformational states of a typically large biomolecule.

To collect such typical FRET data sets, the donor is excited using an illumination laser and we assume, only for simplicity here though performed more generally in Ref.~\cite{saurabh2022single_I}, that acceptors become excited exclusively via FRET. The rate of donor and acceptor emission then depends on their separation characterizing a conformational state and its corresponding FRET efficiency; see Sec.~\ref{Sec:FRET}. As the number of conformational states associated with different FRET efficiencies ($E_{\mathrm{FRET}}$, Eq.~\ref{eq:FRET}) may be unknown, these may be learned non-parametrically~\cite{saurabh2022single_I, saurabh2022single}. However, for simplicity here again, we presume two states termed high and low FRET designated by $\xi_m,\, m=1,2$. 
Further, given that both donors and acceptors are rarely simultaneously excited, we only consider three possible photo-physical states: $f_1=(\mathrm{Ground},\mathrm{Ground}),\,f_2=(\mathrm{Excited},\mathrm{Ground})$, and $f_3=(\mathrm{Ground},\mathrm{Excited})$ where the first elements represent the donor's state. The entire problem's state space is then spanned by a set of states obtained from the tensor product of photo-physical and conformational states termed composite states. To facilitate the notation, we designate composite states by $s_m\in\{(\xi_1,f_1),(\xi_1,f_2),(\xi_1,f_3),(\xi_2,f_1),(\xi_2,f_2),(\xi_2,f_3)\}$ with $m=1:6$. 

We can now write a generative model required in constructing the likelihood used in the analysis of FRET experiments. To do so, we start from the rate matrix 
\begin{equation}
    \mathbb{K} = 
    \begin{bmatrix} 
        0 & k_{s_1\rightarrow s_2} & ... & k_{s_1\rightarrow s_6} \\
        k_{s_2 \rightarrow s_1} & 0 & ... & k_{s_2\rightarrow s_6} \\
        \vdots & \vdots & \ddots & \vdots \\
        k_{s_6\rightarrow s_1} & k_{s_6\rightarrow s_2} & ... & 0
    \end{bmatrix},
\end{equation}
where self-transitions are, by definition, disallowed and $k_{s_m\rightarrow s_{m'}}$ is the transition rate from state $s_m$ to $s_{m'}$. Furthermore, elements of the rate matrix coinciding with simultaneous conformational and photo-physical transitions are set to zero owing to their rarity. Non-zero matrix elements of the rate matrix thus coincide with: 1) transitions between the two FRET conformational states ($k_{\xi_1\rightarrow\xi_2},k_{\xi_2\rightarrow\xi_1}$) while the photo-physical states remain fixed; or 2) transitions between different photo-physical states while conformational states remain fixed. To be more precise, photo-physical transitions include donor excitation ($k_{s_1\to s_2}=k_{\mathrm{ex}}$), donor radiative relaxation ($k_{s_2\to s_1}=k_d$), acceptor relaxation ($k_{s_3\to s_1}=k_a$), FRET transition when in $\xi_1$ ($k_{s_2\to s_3}=k_{\mathrm{FRET}}^{(1)}$), and FRET transition when in $\xi_2$ ($k_{s_5\to s_6}=k_{\mathrm{FRET}}^{(2)}$). As such, written explicitly, the rate matrix for this simple case reads
\begin{equation}
    \mathbb{K} = 
    \begin{bmatrix} 
        0 &  k_{\mathrm{ex}} & 0 & k_{\xi_1\rightarrow\xi_2} & 0 & 0\\
        k_d & 0 & k_{\mathrm{FRET}}^{(1)} & 0 & k_{\xi_1\rightarrow\xi_2} & 0 \\
        k_a & 0 & 0 & 0 & 0 & k_{\xi_1\rightarrow\xi_2} \\
        k_{\xi_2\rightarrow\xi_1} & 0 & 0 & 0 & k_{\mathrm{ex}} & 0\\
        0 & k_{\xi_2\rightarrow\xi_1} & 0 & k_d & 0 & k_{\mathrm{FRET}}^{(2)} \\
        0 & 0 & k_{\xi_2\rightarrow\xi_1}  & k_a & 0 & 0
    \end{bmatrix}.
    \label{eq:FRET_rate}
\end{equation}

Observations only occur when either the donor or acceptor emit radiatively. As such, the system may visit intermediate states between photon emissions such as undergo conformational transitions. For a perfect detector, \textit{e.g.,} ignoring detector dead time~\cite{saurabh2022single_I} and assuming complete detection efficiency (otherwise $k_{\mathrm{ex}}$ is understood as an effective excitation rate), the photon inter-arrival time coincides with the total time the system spends avoiding radiative transitions. 

Now to construct the likelihood for a FRET data set (inter-photon arrival times and detection channels), we begin by illustrating how such data set can be obtained from a generative model.  
To do so, we first designate the state of the composite system at time $t_n$ as $s(t_n)$. Next, following the notation introduced in Sec.~\ref{prob-stat} (see Eq.~\ref{eq:conditionalsample}), a state trajectory is constructed following the Gillespie algorithm~\cite{gillespie1976general} by first selecting the state to which we transition and then deciding when this transition occurs
\begin{align}
    s(t_{n+1})|s(t_n) \sim & \, \mathrm{Categorical}\left(\frac{k_{s(t)\rightarrow s_1}}{k_{s(t_n)}},...,\frac{k_{s(t)\rightarrow s_6}}{k_{s(t_n)}}\right) \label{eq:Cat_Gillespi}, \\
    \delta t_n  \sim & \, \mathrm{Exponential} (k_{s(t_n)}).
    \label{eq:HoldingTime}
\end{align}
Here, $\delta t_n = t_{n+1}-t_n$ is the time the system spends in state $s(t_n)$, and $k_{s(t_n)}$ is the escape rate out of $s(t_n)$, \textit{i.e.}, sum of rates pointing out of $s(t_n)$. The Categorical distribution introduced herein is treated here as the generalization of the Bernoulli albeit with more than two outcomes. 

Taken together, Eqs.~\ref{eq:Cat_Gillespi}-~\ref{eq:HoldingTime} constitute what is called a generative model, \emph{i.e.}, a model both helpful in generating the data but also in constructing the likelihood. This generative model can indeed be further generalized to include imperfect detectors, dead time, and other artifacts such as direct acceptor excitation and cross-talk ~\cite{saurabh2022single_I,roy2008practical,bacia2012correcting,sgouralis2018bayesian}.

We are now presented with a modeling choice. That is, we may learn the trajectory in composite state space (states occupied across time points) and kinetic rates populating the rate matrix~\cite{kilic2021generalizing,safar2022single}. Alternatively, as is more commonly done, we may marginalize (see Eq.~\ref{eq:mrinalization}) over all trajectories and learn only kinetic rates~\cite{gopich2005theory,saurabh2022single_I}.

As it is most common, we select the latter path and marginalize over all possible (non-radiative) paths between observations. To achieve this, we use the master equation ~\cite{van1992stochastic,gopich2005theory,lee2012derivation,saurabh2022single_I,presse2023data}
\begin{equation}
    \frac{d}{dt}\mathbf{P}(t) = \mathbf{P}(t)\mathbf{G}
    \label{eq:master}
\end{equation}
describing the evolution of the probability vector $\mathbf{P}(t)$ collecting the probabilities of occupying different states at time $t$. Here, $\mathbf{G}$, the generator matrix, is related to the rate matrix as follows 
\begin{equation}
    \mathbf{G} = \mathbb{K} - 
    \left[
    \begin{matrix}
    k_{s_1} & 0 & \hdots & 0 \\
    0 & k_{s_2} & \hdots & 0\\
    \vdots & \vdots & \ddots & \vdots\\
    0 & 0 & ... & k_{s_6}
    \end{matrix}
    \right],
\end{equation}
where the diagonal matrix has the same size as $\mathbb{K}$ and its non-zero elements coincide with the escape rates. From the generator matrix, we obtain a propagator matrix $\mathbb{Q}$ collecting transition probabilities over an infinitesimal period $\varepsilon$
\begin{equation}
    \mathbb{Q} = \exp\left[\mathbf{G}\varepsilon\right].
    \label{eq:generator}
\end{equation}
Therefore, given the probability vector at time $t-\varepsilon$, $\mathbf{P}(t-\varepsilon)$, the probability vector at time $t$ reads $\mathbf{P}(t) = \mathbf{P}(t-\varepsilon) \mathbb{Q}$. As such, given the initial probability vector $\mathbf{P}_{\mathrm{in}}$, we find the probability at any time by dividing the time interval into $N$ small periods of $\varepsilon$
\begin{equation}
    \mathbf{P} = \mathbf{P}_{\mathrm{in}}\mathbb{Q}_1 ... \mathbb{Q}_N,
    \label{eq:finalProp}
\end{equation}
where $\mathbb{Q}_1 = ... = \mathbb{Q}_N = \mathbb{Q}$ in the absence of observations. However, in the presence of observations, the propagators in Eq.~\ref{eq:finalProp} are modified according to the  monitored transitions~\cite{saurabh2022single_I}. For example, observation of no photon over the $n$th period $\varepsilon$ signifies no radiative transitions allowing us to set $k_a=k_d=0$ for this period, which in turn results in a modified propagator, designated by $\mathbb{Q}^{\mathrm{non}}_n$. Furthermore, a photon arrival, indicating a radiative transition, forces non-radiative transition rates to be zero leading to a modified propagator $\mathbb{Q}^{\mathrm{rad}}_k$ for the $k$th photon over an infinitesimal period $\varepsilon$.

The likelihood over a set of observations is now expressed in terms of these modified propagators~\cite{gopich2006theory,saurabh2022single_I} 
\begin{equation}
    P(\Delta t_{1:K}|\mathbb{K},\mathbf{P}_{\mathrm{in}}) \propto \mathbf{P}_{\mathrm{in}}\mathbb{Q}_1^{\mathrm{non}} ... \mathbb{Q}_k^{\mathrm{rad}} ... {\mathbb{Q}}_N^{\mathrm{non}} \,\mathbf{P}_{\mathrm{norm}}^{T},
    \label{eq:parametricFRETLike}
\end{equation}
where $\mathbf{P}_{\mathrm{norm}}$ is a row vector of ones.  

\begin{tcolorbox}[colback=brown!5!white,colframe=brown!75!black,title=Statistical Framework \ref{box:FRET}: FRET]

Data: Photon inter-arrival times
\begin{eqnarray}
    \Delta t_{1:K} = \left\{\Delta t_1, ..., \Delta t_{K}\right\}. \nonumber 
\end{eqnarray}
Parameters: loads, transition rates, initial probability vector
\begin{equation}
    \vartheta = \left\{ \overline{b},\mathbb{K}, \mathbf{P}_{\mathrm{in}}\right\}. \nonumber
\end{equation}
Likelihood:
\begin{equation}
    P(\overline{\Delta t}|\vartheta) \propto \mathbf{P}_{\mathrm{in}}\mathbb{Q}_1^{\mathrm{non}} ... \mathbb{Q}_k^{\mathrm{rad}} ... {\mathbb{Q}}_N^{\mathrm{non}} \, \mathbf{P}_{\mathrm{norm}}^{T}. \nonumber
\end{equation}
Priors:
\begin{align}
    q_m \sim & \mathrm{Beta}(A_q,B_q), m = 1:\infty, \nonumber \\
    b_m \sim & \mathrm{Bernoulli}(q_m), \nonumber \\
    \mathbb{K} \sim & \, \mathrm{Gamma}(\alpha_{\mathbb{K}},\beta_{\mathbb{K}}), \nonumber\\
    \mathbf{P}_{\mathrm{in}} \sim & \, \mathrm{Dirichlet}(\alpha_{\Pi}). \nonumber
\end{align}
Posterior:
\begin{equation}
    P(\vartheta|\overline{\Delta t}) \propto P(\overline{\Delta t}|\vartheta)P(\vartheta). \nonumber
\end{equation}
\label{box:FRET}
\end{tcolorbox}

Until now, we have assumed a parametric framework with a fixed number of conformational states, often set to two, low and high FRET~\cite{mckinney2003structural}, in the literature. Now we lift this constraint and treat the number of conformational states as unknown and extend the formulation above to the non-parametric regime. To do so, we assume an infinite number of conformational states with a load $b_m$ (see Sec.~\ref{prob-stat}) associated to each $m$th state resulting in an infinite dimensional generator matrix; see Refs.~\cite{saurabh2022single_I,saurabh2022single}. From the non-parametric generator matrix, we compute the corresponding propagator matrices and use them to build a likelihood similar to Eqs.~\ref{eq:generator}-\ref{eq:parametricFRETLike}. The non-parametric posterior over the set of unknowns $\vartheta=\{\overline{b},\mathbb{K},\mathbf{P}_{\mathrm{in}}\}$ is then constructed by including a Beta-Bernoulli process prior (see Sec.~\ref{prob-stat}) over the loads and appropriate priors over the remaining unknowns (ideally conditionally conjugate priors if available~\cite{presse2023data}); see Box ~\ref{box:FRET}. Strictly speaking, in computational applications, we often use large albeit finite load numbers, $M$, and verify that for large enough $M$ the conclusions drawn are independent of $M$. Finally, the FRET posterior obtained is sampled using Monte Carlo methods to deduce the set of unknowns~\cite{saurabh2022single_I,saurabh2022single,safar2022single}. 

\vspace{-4mm}
\begin{figure}[H]
    \centering
    \includegraphics[width=1\linewidth]{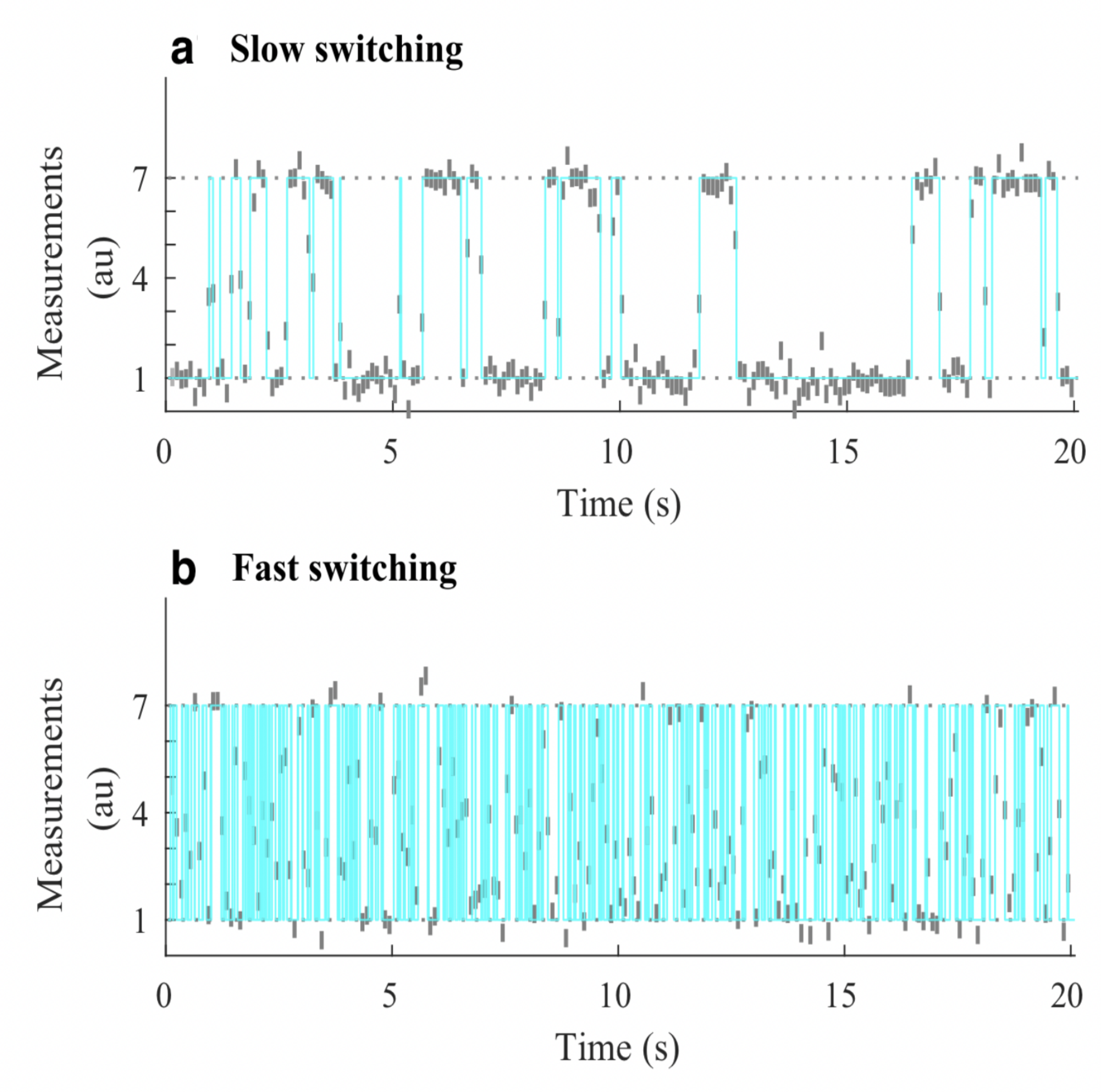}
    \vspace{-6mm}
    \caption{Data simulated for discrete measurements of two state systems with fast and slow transitions depicted in panels a and b, respectively. The system trajectories in the state space, measurements at different times intervals ($\delta T$), \textit{i.e.}, bins, and the state signal levels in the absence of noise are, respectively, denoted by cyan, gray, and dotted lines. The measurements between the state signals level coincide with time intervals where the system has switched to a different state at some point during those intervals. In the simulations, data acquisitions take place at every $\delta T = 0.1$~s where the average time spent in each state is, respectively, $0.8$~s and $0.066$~s for slow and fast kinetics. The figure is adapted from Ref.~\cite{kilic2021generalizing}.}
    \label{fig:HMJP}
    \vspace{-4mm}
\end{figure}

An alternative statistical FRET framework makes use of photon counts over equal time windows, \textit{i.e.}, bins, during the experiment rather than single photons~\cite{gopich2005theory,saurabh2022single_I}. In this case, the likelihood is derived using the fact that photon counts over fixed periods are Poisson distributed (ignoring detector noise convoluted with Poisson shot noise required of quantitative analyses)
~\cite{saurabh2022single_I}. 
The derivation of such likelihoods is more straightforward than the single photon case ~\cite{sgouralis2018bayesian,patel2019hidden} and learning rates (or, more accurately, transition probabilities) is achieved using Hidden Markov Models (HMMs) ~\cite{roy2008practical}. While traditional HMM frameworks require the number of FRET states as input, more recent iterations have leveraged variational tools to determine states \emph{e.g.}, vbFRET~\cite{bronson2009learning}, with recent developments in non-parametric infinite HMMs (iHMMs) now allowing posterior probabilities over states warranted by the data to be sampled simultaneously alongside kinetics~\cite{sgouralis2017introduction,sgouralis2018bayesian}.

However, by virtue of binning photon arrivals, whether by choice or due to the detector used, HMM frameworks naturally compromise our ability to resolve fast kinetics, occurring on timescales at or below the bin size. For this reason, other than the potential for computational speed-up, there is no reason to bin single photon data. On the other hand if using detectors that unavoidably bin counts across pixels commonly used in wide-field applications (see Appendix~\ref{Se:Detector}), then fast transitions may be deduced on timescales exceeding data acquisition. This is achieved by leveraging the fact that the signal amounts to an average of the properties over the state visited~\cite{pirchi2016photon,kilic2021generalizing,kilic2021extraction}; see Fig.~\ref{fig:HMJP}.

Such strategies used to deduce dynamics on timescales at or exceeding data acquisition rely on the Markov jump process (MJP)~\cite{hobolth2009simulation,kilic2021generalizing} which assumes that the system evolves in continuous time. This is by contrast to the HMM paradigm which approximates dynamics as occurring discretely and only at the measurement time. Put differently, the MJP accurately pre-supposes a continuous time trajectory $\mathcal{S}(t)$ in the discrete state space of the composite system generated using the same procedure as described by Eqs.~\ref{eq:Cat_Gillespi}-\ref{eq:HoldingTime}. The observation for the $k$th data acquisition period (bin) is therefore~\cite{kilic2021generalizing,kilic2021extraction}
\begin{equation}
    w_k \sim \mathrm{Poisson}\left(\int^{t_k+\delta T}_{t_k} \mu_{_{\mathcal{S}(t)}} dt\right),
    \vspace{-3mm}
\end{equation}
where $\mu_{_{\mathcal{S}(t)}}$ represents the photon emission rate for the instantaneous state occupied at time $t$, $\mu_{_{\mathcal{S}(t)}}$.

Having briefly highlighted Markov model applications for FRET, here we describe how Markov models are used when enumerating fluorophores ~\cite{ulbrich2007subunit,rollins2015stochastic,tsekouras2016novel,lee2017unraveling,bryan2022diffraction} typically with the intent of determining the stoichiometry of a labeled protein complex within a diffraction-limited spot.

For a single fluorophore we assume, for simplicity of demonstration alone, a state space spanned by 3 photo-physical states, though this treatment is generalized elsewhere ~\cite{bryan2022diffraction,patel2019hidden}. These include the: 1) bright state, $f_A$; 2) dark state, $f_D$; 3) photo-bleached state, $f_{B}$. Transitions between these states include: $f_A\rightarrow f_A$, $f_A\rightarrow f_D$, $f_A\rightarrow f_B$, $f_D\rightarrow f_D$, $f_D\rightarrow f_A$, $f_B\rightarrow f_B$. Here, the photo-bleached state is an absorbing state from which escape is impossible; see Sec.~\ref{Non_Chem}.   

\begin{figure}[H] \centering
\includegraphics[width=\linewidth]{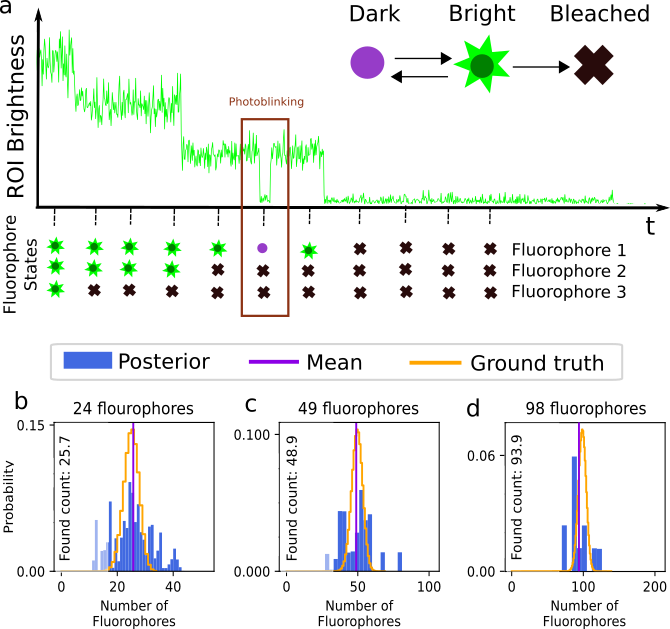} 
\caption{Fluorophore enumeration. (a) Cartoon representation of the enumeration problem where the ROI intensity varies as fluorophores switch between the dark, bright, and photo-bleached states. (b-d) Histogram of the sampled posterior over the number of fluorophores, \textit{i.e.}, sum of sampled loads, for experimental data with, respectively, 24, 49 and 98 fluorophores using the statistical framework appearing in Box~\ref{box:Counting}. The figure is adapted from Ref.~\cite{bryan2022diffraction}. }
\label{fig:Counting}
\vspace{-4.5mm}
\end{figure}

Typically, in such applications, a wide-field detector (see Appendix~\ref{Se:Detector}) is used to record data from ROIs containing one or multiple putative complexes. The ROIs may contain one or more pixels. The input to the analysis then consists of the sum of the intensity or brightness in each ROI typically obtained by summing the pixel values (Analogue-to-Digital Units or ADUs) in each pixel involved. The sum of ADUs for each ROI is then recorded over $K$ successive frames and designated by $\overline{w}_{1:K} = \left\{w_{1:K}^1, ..., w_{1:K}^R\right\}$, where the overbar represents the set of $R$ ROIs. Typically, the last frame is taken after all fluorophores within the ROI have photo-bleached; see Fig.~\ref{fig:Counting}. Assuming only photo-bleaching and ignoring transitions from bright to dark states, the number of discrete intensity drops in the time trace, if all fluorophores are initially bright, should coincide with the number of photo-bleaching events and thus the complex stoichiometry. However, not all fluorophores may initially be active such as in the case of PALM~\cite{rollins2015stochastic}. What is more, fluorophores blink; see Sec.~\ref{Non_Chem} and Fig.~\ref{fig:Counting}.

If our goal is to enumerate the fluorophores, assuming identical complexes across ROIs, then for independent ROIs (\textit{iid} variables), the likelihood reads (see Sec.~\ref{prob-stat})
\begin{equation}
P(\overline{w}_{1:K}|\overline{\Lambda}_{1:K},\Xi) = \prod_r \prod_k P(w_k^r|\Lambda_k^r,\Xi),
\vspace{-2.5mm}
\end{equation}
where $\Xi$ denotes the camera parameters (see Appendix~\ref{Se:Detector}) and the elements of $\overline{\Lambda}_{1:K}$, namely $\Lambda_k^r$, coincide with the expected photon count, \textit{i.e.}, brightness obtained from the emission rate multiplied by the camera exposure time, of the $r$th ROI at frame $k$. 

Decomposed in terms of emission due to background and fluorophores, $\Lambda_k^r$ reads \vspace{-2mm}
\begin{equation}
    \Lambda^r_k = \mathcal{B}^r + I_A \sum_{m=1}^{M_r} \delta_{A,s_k^{rm}},
    \label{eq:ROIbright}
    \vspace{-2mm}
\end{equation}
where $m$ counts $M_r$ fluorophores within the $r$th ROI. Here $I_A, \, \mathcal{B}^r$, and $s_k^{rm}$ respectively, denote the fluorophore's brightness, background brightness of the $r$th ROI per frame, and the state of the $m$th fluorophore within the $r$th ROI at frame $k$. The Kronecker delta, $\delta_{A,s_k^{rm}}$, assumes fluorophores only emit in the bright state. This decomposition assumes, perhaps erroneously in some cases, that the fluorphores do not interact~\cite{saurabh2022modeling}. 

Next, approximating the fluorophore state as remaining the same over each frame and the state at frame $k$ only depending on its (potentially different) state at frame $k-1$, \emph{i.e.}, the Markov assumption, we may formulate the problem using transition probabilities between different states and avoid transition rates altogether. The transition probabilities associated to a single fluorophore can be collected as elements of a matrix, designated by $\mathbf{\Pi}$, analogous to the propagator, $\mathbb{Q}$ in Eq.~\ref{eq:generator}, for finite time windows
\begin{equation}
    \mathbf{\Pi} = \exp \left[ \mathbf{G}\delta T\right] = 
    \begin{bmatrix}
    \pi_{A\rightarrow A} & \pi_{A\rightarrow D} & \pi_{A\rightarrow B}\\
    \pi_{D\rightarrow A} & \pi_{D\rightarrow D} & 0 \\
    0 & 0 & 1
    \end{bmatrix}.
    \label{eq:ProbMatrix}
\end{equation}
Here, $\delta T$ is the fixed period of time between measurements (frame exposure time) and each line of the transition matrix contains transition probabilities out of a certain state. For instance, we have $\pi_A=\left[\pi_{A\rightarrow A},\, \pi_{A\rightarrow D},\, \pi_{A\rightarrow B} \right]$ for the bright state. The structure of the last row in $\mathbf{\Pi}$ reflects the absorbing nature of the bleached state. 

The state of a single fluorophore at frame $k$ given its state at $k-1$ is sampled as follows
\begin{equation}
    s_k^{mr}|s_{k-1}^{mr} \sim \mathrm{Categorical}(\pi_{s_{k-1}^{mr}}),
    \label{eq:TrajCatCount}
\end{equation}
where $\pi_{s_{k-1}^{mr}}$ collects the set of possible transitions' probabilities out of $s_{k-1}^{mr}$. Finally, as fluorophore transitions are assumed independent, transitions of the full system are obtained from the product of the individual fluorophore transition probabilities. 

While the photo-physics of individual fluorophores may be known, the number of fluorophores are themselves unknown. This presents a model selection challenge warranting a non-parametric formulation. Conceptually, this is achieved by assuming an infinite number of fluorophores with associated loads; see Sec.~\ref{prob-stat}. Concretely, we modify Eq.~\ref{eq:ROIbright} as follows
\begin{equation}
\vspace{-2mm}
    \Lambda_{k}^r = \mathcal{B}^r + I_A\sum_{m=1}^{\infty} b_m^r\delta_{A,s_k^{mr}},
\end{equation}
where $b_m^r$ is the load associated to the $m$th fluorophore in the $r$th ROI. In this case, the number of fluorophores is replaced by loads for each ROI. We collect the set of unknowns in $\vartheta=\left\{\overline{\overline{b}},I_A,\overline{\mathcal{B}},\mathbf{\Pi},\overline{\overline{\mathcal{S}}}\right\}$. Here, double overbars represent the set of all possible values for the two indices associated to each of parameters $b$ and $\mathcal{S}$.

\begin{tcolorbox}
[colback=brown!5!white,colframe=brown!75!black,title=Statistical Framework \ref{box:Counting}: Counting]

Data: Sum of ROIs' pixel values in ADUs
\begin{equation}
    \overline{w}_{1:K} = \left\{w_{1:K}^1, ..., w_{1:K}^R\right\}. \nonumber
\end{equation}
Parameters: loads, fluorophore intensity, background, transition probabilities, fluorophores' trajectories in the state space
\begin{equation}
    \vartheta = \left\{\overline{\overline{b}},I_A,\overline{\mathcal{B}},\mathbf{\Pi},\overline{\overline{\mathcal{S}}}\right\}. \nonumber
\end{equation}
Likelihood: 
\begin{equation}
    P\left(\overline{w}_{1:K}|\vartheta\right) = \prod_{k,r} P(w_k^r|\Lambda_k^r,\Xi). \nonumber
\end{equation}
Priors:
\begin{align}
    q_m \sim & \, \mathrm{Beta}(A_q,B_q), \,\, m = 1:\infty, \nonumber \\
    b_m \sim & \, \mathrm{Bernoulli}(q_m), \nonumber \\
    I_A \sim & \, \mathrm{Gamma}(\alpha_A,\beta_A), \nonumber\\
    \mathcal{B} \sim & \, \mathrm{Gamma}\left(\alpha_{\mathcal{B}},\beta_{\mathcal{B}}\right), \nonumber \\
    \mathbf{\Pi} \sim & \, \mathrm{Dirichlet}(\alpha_{\Pi}), \nonumber\\
    s_k^{mr}|s_{k-1}^{mr},\mathbf{\Pi} \sim & \, \mathrm{Categorical}(\pi_{s_{k-1}^{mr}}). \nonumber
\end{align}
Posterior:
\begin{equation}
    P(\vartheta|\overline{w}_{1:K}) \propto P(\overline{w}_{1:K}|\vartheta)P(\vartheta). \nonumber
\end{equation}
\label{box:Counting}
\end{tcolorbox}

Finally, to construct the posterior for the set of  parameters in $\vartheta$, we introduce priors. The most notable priors are the Beta-Bernoulli process priors on loads and the prior on the transition probabilities, the Dirichlet prior, due to its conjugacy to the Categorical distribution Eq.~\ref{eq:TrajCatCount}. For the remaining priors in Box~\ref{box:Counting}, we opt for computationally efficient priors when possible leveraging the mathematical structure for the likelihood (see Sec.~\ref{prob-stat})~\cite{bryan2022diffraction}. In particular, we invoke multiple Monte Carlo to draw samples of $\vartheta$ from the posterior with forward filter backward sampling specifically used to sample fluorophore trajectories~\cite{scott2002bayesian,bishop2006pattern,bryan2022diffraction}. 

Having discussed how to decode temporal data, we now turn to spatiotemporal data and, for this, we discuss the optics of different microscope modalities and derive their corresponding PSFs.
\vspace{-5mm}

\section{Fluorescence microscopy: point spread function}
\label{fluorescence}

In this section, we develop in a brief but otherwise self-contained manner the physical theory of optical imaging within a wide-field fluorescence microscope. We start by deriving the Abbe sine condition subsequently used to describe fundamental properties of electromagnetic wave propagation through optical systems. We then continue by deriving the basic principles of how to compute the OTF and PSF of a microscope, discuss the lack of optical sectioning of wide-field microscopes, and illustrate the effect of optical aberrations on PSFs.

\vspace{-4mm}
\subsection{\label{Abbe_sine}Fundamental property of microscopic imaging: Abbe's sine condition}
\vspace{-2mm}

\begin{figure}[H]
    \centering
    \includegraphics[width=0.85\linewidth]{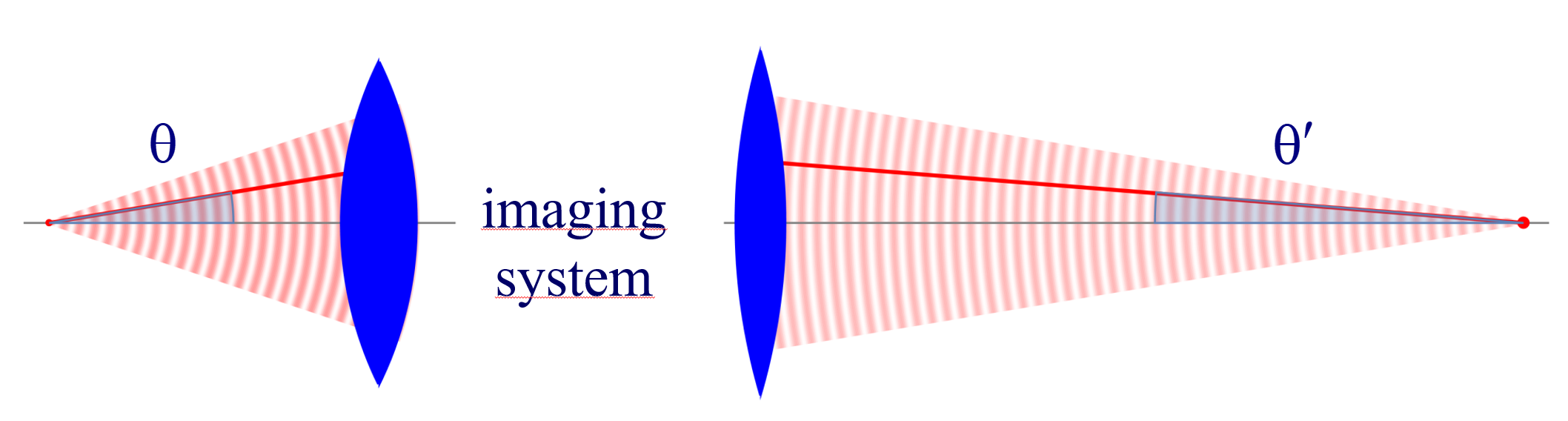}
    \caption{The optical microscope, \textit{i.e.}, imaging system, is a wavefront transforming system converting the outgoing spherical wavefront of a point emitter in sample space (left) into a concentric spherical wavefront in image space (right) converging into an image point in the image space.  
    }
    \label{fig:wavefrontconverter}
    \vspace{-4mm}
\end{figure}

To gain a deeper understanding of how a microscope forms an image alongside fundamental principles governing image formation, we start by considering the imaging of a generic point source in sample space into an image point in image space; see Fig.~\ref{fig:wavefrontconverter}. To do so, we denote parameters associated to the image and sample spaces with and without prime, respectively, hereafter. A point source in the focal plane on the optical axis (symmetry axis designated by blue lines) emits concentric (electromagnetic) waves. The segment of the spherical wavefront collected by the objective is then converted by the microscope into a segment of a spherical wavefront converging onto the corresponding image point. To facilitate subsequent derivations, we assume that the distance between the sample point and the objective lens is large enough such that the spherical wavefront incident on the objective can be considered as a super-position of planar wavefront segments traveling at different propagation angles $\theta$ with respect to the optical axis (Fraunhofer diffraction limit). Correspondingly, the transformed spherical wavefront in image space is also considered to be a super-position of planar wavefront segments traveling at angles $\theta'$ with respect to the optical axis. 

\begin{figure}[H]
    \centering
    \includegraphics[width=0.85\linewidth]{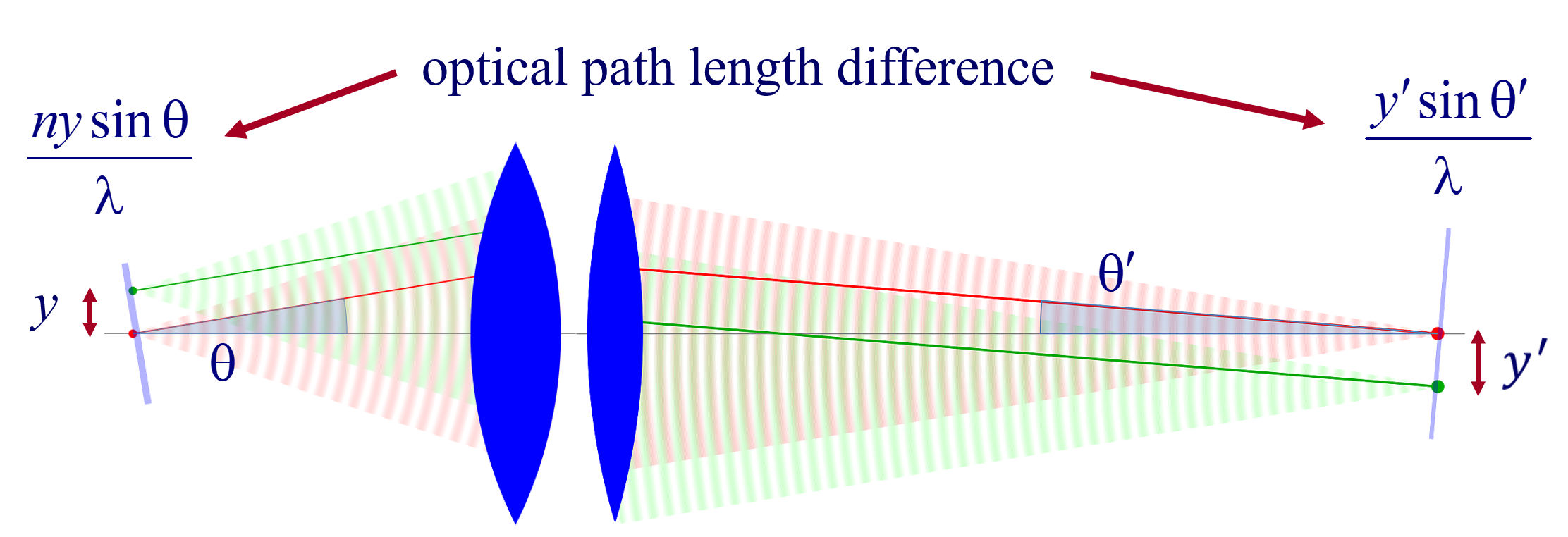}
    \caption{The phase relation between planar wavefront segments propagating along the same angle $\theta$ but emanating from two different point sources, where one point source is on the optical axis (red) and the other is laterally shifted by a distance $y$ (green). The image point (point of convergence of the spherical wavefront segment) corresponding to the shifted point source is translated by a distance $y'$ away from the optical axis. The ratio between $y'$ and $y$ is the magnification $\mathcal{M}$. Optical path length differences between wavefront segments traveling along angles $\theta$ or $\theta'$, respectively, are shown as thin bluish lines at the emitters' positions and oriented perpendicular to the propagation directions $\theta$ and $\theta'$.}
    \label{fig:abbesine}
    \vspace{-5mm}
\end{figure}

We can now obtain a relation between the angle $\theta$ and the corresponding angle $\theta'$ of a planar wavefront segment within the sample and image spaces, respectively; see Fig.~\ref{fig:magnification}. We begin by assuming that the point source is shifted laterally away from the optical axis by a distance $y$; see Fig.~\ref{fig:abbesine}. Considering a perfect imaging system, the spherical wavefronts from the shifted point source, shown in green, will be converted into spherical wavefronts converging onto a point shifted a distance $y'$ away from the optical axis in the image space where the relation between $y'$ and $y$ is given by $y'=\mathcal{M}y$. Here, $\mathcal{M}$ denotes the microscope's magnification. 

Now, consider two planar wavefront segments traveling at angle $\theta$ from a source located at $y$ and on the optical axis. There is a phase difference between these two planar wavefront segments proportional to $n y\sin\theta$. The microscope transforms these planar wavefront patches into two planar wavefront patches traveling along angle $\theta'$ in the image space with a phase difference of $y'\sin\theta'$ between the patches (assuming both here and later that the refractive index of the image space is always that of air, {\it i.e.}, $\approx 1.0$). Now, to attain perfect focus, all planar wavefront patches originating from a point source and converging at a corresponding focal point in the image space must have the same phase at the focal point (maximum constructive interference). In other words, the phases of all planar wave components constituting the spherical wavefront must be the same at the image point where the spherical wavefront converges. We thus find $ny\sin\theta = y'\sin\theta'$. When considering that the ratio between $y'$ and $y$ is the image magnification, this yields
\begin{equation}
    n\sin\theta = \mathcal{M}\sin\theta',
    \label{eq:AbbeSin}
    \vspace{-2mm}
\end{equation}
which is the so-called Abbe sine condition~\cite{mansuripur1998abbe,mansuripur2002classical} for a perfect aplanatic imaging system (\textit{i.e.}, emission from a point at lateral distance $y$ in the focal plane in sample space is converted into a perfect spherical wavefront segment converging into an image point at position $y'=\mathcal{M}y$ in the image plane). 

\begin{figure}[H]
    \centering
    \includegraphics[width=0.85\linewidth]{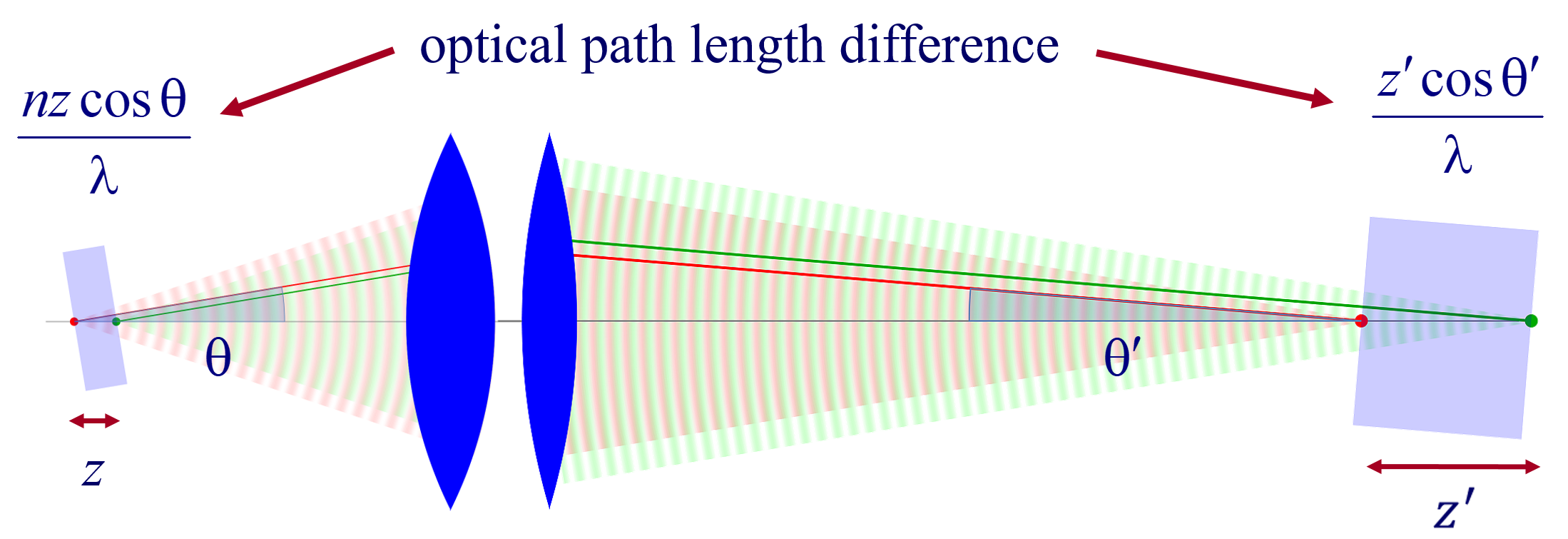}
    \caption{Phase relation between planar wavefront segments propagating along the same angle $\theta$ but emanating from two different point sources along the optical axis. Similar to Fig.~\ref{fig:abbesine}, optical path differences (phase differences) between wavefront segments traveling along angles $\theta$ or $\theta'$, respectively, are shown as blue rectangles.}
    \label{fig:herschelcondition}
\end{figure}

Invoking similar arguments, we can derive the relation between $\theta$ and $\theta'$ required for the perfect imaging of point sources along the optical axis into corresponding image points in image space. This situation is illustrated in Fig.~\ref{fig:herschelcondition} where we again compare the phase differences between: 1) wavefronts from the point source in the focal plane with the shifted point source; and 2) corresponding wavefronts converging in the image points. As such, we now find the following relation between $\theta$ and $\theta'$ 
\begin{equation}
    \label{eq:herschel0}
    n (\cos\theta-1) = \mathcal{M}_z (\cos\theta'-1),
\end{equation}
where $\mathcal{M}_z$ denotes the \emph{axial} magnification~\cite{hopkins1946herschel,born2013principles,braat1997abbe}. As can be seen, it is \emph{impossible} for both the Abbe sine condition and Eq.~\ref{eq:herschel0} to be simultaneously satisfied. This shows that an optical system which perfectly images points from the focal plane onto the conjugate image plane can do so only on these two specific planes and exhibit aberrations, \textit{i.e.}, deviations of wavefronts from spherical shape, away from the focal plane. Interestingly, for small values of $\theta$, we can expand Eq.~\ref{eq:herschel0} into a first order Taylor series, \textit{i.e.}, $n\theta^2/2\approx \mathcal{M}_z \theta'^2/2$, which can simultaneously be satisfied with Abbe's sine condition if 
\begin{equation}
    \label{eq:herschel1}
    n \sin^2\theta/2 \approx \mathcal{M}_z \sin^2 \theta'/2 
\end{equation}
and $\mathcal{M}_z \approx \mathcal{M}^2/n$.
Eq.~\ref{eq:herschel1} is called Herschel's condition \cite{gross2005handbook,born2013principles,steward1927herschel,braat1997abbe,botcherby2008optical}. This shows that a system satisfying Abbe's sine condition (aplanatic imaging system) has an axial magnification of roughly the square of the lateral magnification divided by the sample medium's refractive index. 

\vspace{-10mm}
\subsection{\label{EMField}Electromagnetic field of image formation}
\vspace{-2mm}

\begin{figure}[H]
    \centering
    \includegraphics[width=1\linewidth]{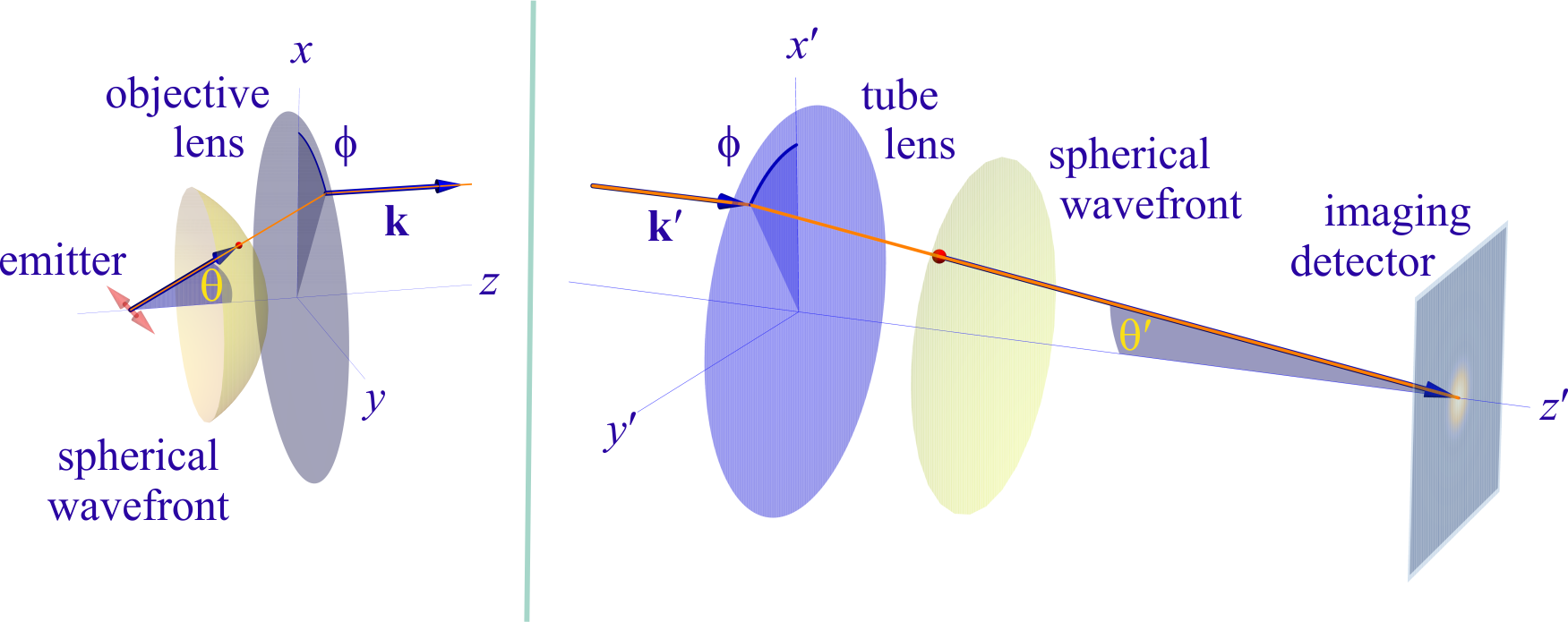}
    \caption{Geometry of propagation of a narrow section of the wavefront from the emitter to the image plane.}
    \label{fig:psfderivation1}
    \vspace{-4mm}
\end{figure}

In this section, we consider a point emitter with incoherent emission in sample space and proceed to derive a relation between the corresponding electromagnetic fields in the sample and image spaces. Specifically, we operate in the Fourier domain to derive electric and magnetic field components in image space in terms of the emissive electric fields in sample space. To begin, we write the emitter's electric field plane wave (Fourier) representation in sample space
\begin{equation}
    \mathbf{E}(\mathbf{r}) = \int_0^\Theta d\theta \sin\theta \int_0^{2\pi} d\phi \mathbf{E}_0(\theta,\phi) \exp\left(i \mathbf{k}\cdot\mathbf{r}\right),
    \label{eq:EfieldSampleSpace}
\end{equation}
where $\mathbf{r}$ is the position vector in sample space with respect to the objective focal point in sample space; see Fig.~\ref{fig:magnification}. Moreover, $\mathbf{E}_0(\theta,\phi)$ is the electric field amplitude for the plane wave traveling along wave vector $\mathbf{k}$ with length $\left\vert\mathbf{k}\right\vert=2\pi n/\lambda$ and direction $\hat{\mathbf{k}} = \left(\cos\phi\sin\theta,\sin\phi\sin\theta,\cos\theta\right)$ (a hat above a vector always designates a unit vector with components $\left(x,y,z\right)$ in Cartesian coordinates); see Fig.~\ref{fig:psfderivation1}. Furthermore, the angular integration extends over the whole cone of light with angle $\Theta$ detected by the objective (recalling that $n\sin\Theta$ is the objective's numerical aperture; see Fig.~\ref{fig:resolution}). 

In Fig.~\ref{fig:psfderivation1}, considering the plane on which both the optical axis (z-axis in Fig.~\ref{fig:psfderivation1}) and $\mathbf{k}$ lie, then it is convenient to split the electric field amplitude $\mathbf{E}_0(\theta,\phi)$ into two orthogonal polarization components, namely parallel and perpendicular to this plane $\mathbf{E}_0 = E_{0,\parallel}(\theta,\phi) \hat{\mathbf{e}}_\parallel + E_{0,\perp}(\theta,\phi) \hat{\mathbf{e}}_\perp$, where $E_{0,\parallel}$ and $E_{0,\perp}$ are the corresponding electric field amplitudes along the two polarization orientations, and the corresponding unit vectors are denoted by $\hat{\mathbf{e}}_\parallel$ and $\hat{\mathbf{e}}_\perp$. These two unit vectors with the unit vector $\hat{\mathbf{k}}$ form an orthonormal set of unit vectors, given as follows in Cartesian coordinates
\begin{equation}
    \begin{split}
        \hat{\mathbf{k}} & = \left(\cos \phi\sin\theta,\sin\phi \sin \theta,\cos \theta \right),\\
        \hat{\mathbf{e}}_\parallel &= \left(-\sin\phi,\cos\phi,0\right), \\
        \hat{\mathbf{e}}_\perp &= \hat{\mathbf{e}}_\parallel\times\hat{\mathbf{k}} = \left(\cos\phi\cos\theta,\sin\phi\cos\theta,-\sin\theta\right).
    \end{split}
    \label{eq:unitvecs}
\end{equation}
This representation immediately allows us to write down the magnetic field in sample space. We do so by recalling that for a plane wave with wave vector $\mathbf{k}$ and electric field amplitude $\mathbf{E}_0$, the magnetic field amplitude is $\mathbf{B}_0=n\hat{\mathbf{k}}\times\mathbf{E}_0$ ~\cite{jackson1999classical}. Thus, the magnetic field amplitude in sample space reads $\mathbf{B}_0 = n \left[ -E_{0,\parallel}(\theta,\phi) \hat{\mathbf{e}}_\perp + E_{0,\perp}(\theta,\phi)  \hat{\mathbf{e}}_\parallel \right]$.

The microscope's optics now converts each plane wave component of Eq.~\ref{eq:EfieldSampleSpace} into a corresponding plane wave component $\mathbf{E}'_0(\theta',\phi) \exp(i\mathbf{k}'\cdot\mathbf{r'})$ in the image space; see right panel of Fig.~\ref{fig:psfderivation1}. Here, $\mathbf{r}'$ is centered at the focus of the tube lens (see Fig.~\ref{fig:magnification}), the angle $\phi$ remains the same, and the propagation angles $\theta$ and $\theta'$ are connected via Abbe's sine condition given by Eq.~\ref{eq:AbbeSin}. As before, we conveniently split the electric field amplitude into two principal polarization directions $\mathbf{E}'_0 = E'_{0,\parallel}(\theta,\phi) \hat{\mathbf{e}}_\parallel + E'_{0,\perp}(\theta,\phi) \hat{\mathbf{e}}'_\perp$, where the set of unit vectors in the image space is obtained by substituting $\theta$ by $\theta'$ in Eq.~\ref{eq:unitvecs}. Moreover, we note that $\hat{\mathbf{e}}'_\parallel = \hat{\mathbf{e}}_\parallel$ due to its independence of $\theta$. 
Now, the corresponding magnetic field amplitude can be obtained as $\mathbf{B}'_0 = -E'_{0,\parallel}(\theta',\phi) \hat{\mathbf{e}}'_\perp + E'_{0,\perp}(\theta',\phi) \hat{\mathbf{e}}_\parallel$,
assuming a refractive index in image space of unity. 

We now relate the electric field amplitudes in sample and image spaces by considering the conservation of energy flux density along the optical axis for every plane wave component absent attenuation (attenuation can be considered as a form of aberration discussed in Sec.~\ref{Aberrations}). This flux density is given by the $z$-component of the time-averaged Poynting vector $\mathbf{P}$ ~\cite{jackson1999classical} which reads
\begin{equation}
    P_z = \frac{c}{8\pi} \hat{\mathbf{e}}_z\cdot(\mathbf{E}_0\times\mathbf{B}^{*}_0) = \frac{c}{8\pi} \hat{\mathbf{e}}_z\cdot(\mathbf{E}'_0\times\mathbf{B}^{\prime *}_0),
    \label{eq:PoyntingZ}
\end{equation}
where a star denotes complex conjugation. For $\mathbf{B}_0=n\hat{\mathbf{k}}\times\mathbf{E}_0$ in sample space and $\mathbf{B}'_0=\hat{\mathbf{k}}'\times\mathbf{E}'_0$ in image space, we obtain $n \left\vert\mathbf{E}_0\right\vert^2 \cos\theta = \left\vert\mathbf{E}'_0\right\vert^2 \cos\theta'$ from which the electric field amplitudes in image and sample spaces are related
\begin{equation}
    \left\vert\mathbf{E}_0'\right\vert = \sqrt{n \frac{\cos\theta}{\cos\theta'}} \left\vert\mathbf{E}_0\right\vert.
    \label{eq:Efieldconversion}
\end{equation}
Furthermore, by combining Abbe's sine condition $n \sin\theta = \mathcal{M}\sin\theta'$, Eq.~\ref{eq:AbbeSin}, and its differential $n \cos\theta d\theta = \mathcal{M}\cos\theta' d\theta'$, we have 
\begin{equation}
    \sin\theta d\theta = \left(\frac{\mathcal{M}}{n}\right)^2 \frac{\cos\theta'}{\cos\theta} \sin\theta' d\theta'.
    \label{eq:AbbeDiff}
\end{equation}
Substituting the above into the electric field's plane wave representation, Eq.~\ref{eq:EfieldSampleSpace}, and leveraging Eq.~\ref{eq:Efieldconversion}, we arrive at the following expression for the image space electric field plane wave representation 
\begin{equation}
    \begin{split}
        \mathbf{E}'(\mathbf{r}') = \frac{\mathcal{M}^2}{n^{3/2}} &\int_0^{\Theta'} d\theta' \sin\theta' \sqrt{\frac{\cos\theta'}{\cos\theta}} \int_0^{2\pi} d\phi\\
        & \left[E_{0,\parallel} \hat{\mathbf{e}}_\parallel + E_{0,\perp} \hat{\mathbf{e}}'_\perp\right] \exp\left(i\mathbf{k}'\cdot\mathbf{r}'\right),
    \end{split}
    \label{eq:Efield}
\end{equation}
where the maximum integration angle, derived from Abbe's sine condition for $\Theta$ and $\Theta'$, is now $\Theta' = \arcsin \left(n\sin\Theta/\mathcal{M}\right) = \arcsin \left(\mathrm{NA}/\mathcal{M}\right)$. Similarly, for the magnetic field, we find 
\begin{equation}
    \begin{split}
        \mathbf{B}'(\mathbf{r}') = \frac{\mathcal{M}^2}{\sqrt{n}} &\int_0^{\Theta'} d\theta' \sin\theta' \sqrt{\frac{\cos\theta'}{\cos\theta}} \int_0^{2\pi} d\phi\\
        & \left[-E_{0,\parallel} \hat{\mathbf{e}}'_\perp + E_{0,\perp} \hat{\mathbf{e}}_\parallel\right] \exp\left(i\mathbf{k}'\cdot\mathbf{r}'\right).
    \end{split}
    \label{eq:Bfield}
\end{equation}

Recognizing that the above equations for both electric and magnetic field components are nothing other than Fourier representations (expansion into plane waves $\exp(i\mathbf{k}'\cdot\mathbf{r}')$), we comment briefly on the frequency support restricted to wave vectors with $k' = \vert \mathbf{k}' \vert = (k_x^{'2}+k_y^{'2}+k_z^{'2})^{1/2}=2\pi/\lambda$, $0\leq\theta'\leq\Theta'$, and $0<\phi\leq2\pi$. This restriction is illustrated as a spherical cap of radius $k'=2\pi/\lambda$ in the frequency domain; see left panels in Figs.~\ref{fig:psfderivation}-\ref{fig:lateralaxialconvolution}. In other words, the Fourier amplitudes of the electric and magnetic fields are only non-zero on this spherical cap in Fourier space. To better see this, we rewrite Eq.~\ref{eq:Efield} as
\begin{equation}
    \begin{split}
        \mathbf{E}'(\mathbf{r}') = &\int \frac{d^3\mathbf{k'}}{(2\pi)^3} \tilde{\mathbf{E}}'(\mathbf{k}') \exp\left(i\mathbf{k}'\cdot\mathbf{r}'\right),
    \end{split}
    \label{eq:EfieldOTF}
\end{equation}
where a variable with tilde denotes Fourier representation of the variable hereafter. Now, assuming that the three-dimensional integration extends over the whole $\mathbf{k}$-space (Fourier space), the integration measure in spherical coordinates is $d^3\mathbf{k}'=k'^2 \sin\theta' dk' d\theta' d\phi$, and the electric field Fourier amplitude (integrand in Eq.~\ref{eq:EfieldOTF}) for angles $0\leq\theta'\leq\Theta'$ is given by (all constant pre-factors omitted) 
\begin{equation}
    \begin{split}
        \tilde{\mathbf{E}}'(\mathbf{k}') \propto \delta\left(k'-\frac{2\pi}{\lambda}\right) \sqrt{\frac{\cos\theta'}{\cos\theta}} \left(E_{0,\parallel} \hat{\mathbf{e}}_\parallel + E_{0,\perp} \hat{\mathbf{e}}'_\perp\right) 
    \end{split}
    \label{eq:EFourierAmplitude}
\end{equation}
while it is zero for angles $\theta'>\Theta'$. Here, $\delta$ denotes Dirac's delta function and guarantees that $k'=2\pi/\lambda$. The absolute value of the electric field in Eq.~\ref{eq:EFourierAmplitude} is obtained as (see left panels in Figs.~\ref{fig:psfderivation}-\ref{fig:lateralaxialconvolution})
\begin{equation}
    |\tilde{\mathbf{E}}'|\!\propto\!
    \begin{cases}
        \sqrt{\frac{\cos\theta'}{\cos\theta}(E_{0,\parallel}^2+E_{0,\perp}^2)}\,, \,\,\, k'\!=\!\frac{2\pi}{\lambda}\;\; \&\;\; 0 \leq \theta' \!\leq \! \Theta' \\
        \mathrm{\hspace{1.55cm}}0, \mathrm{\hspace{3cm}otherwise.}
    \end{cases}
    \label{eq:EFourierAmplitude2}
\end{equation}
A similar expression holds for the Fourier representation of the magnetic field, when replacing $E_{0,\perp}$ by $-n E_{0,\parallel}$ and $E_{0,\parallel}$ by $n E_{0,\perp}$. 
\vspace{-6mm}
\subsection{\label{Fluo_PSF}Point spread function}
\vspace{-2mm}

Now, we are in a position to calculate the PSF, denoted by $U(\mathbf{r}')$. The PSF is, by its very nature, a probability density over a photon reaching the point $\mathbf{r'}$ on the image plane, \textit{i.e.}, detector, where $\mathbf{r'}$ is a random variable. That is, the PSF plays the role of a normalized spatial distribution of light intensity recorded by a detector at the image plane for a point-like emitter located in the sample space. From this fundamental probabilistic property of light follows most statistical concepts inherent to the modeling of fluorescence microscopy. 

\begin{figure}[H]
    \centering
    \includegraphics[width=1\linewidth]{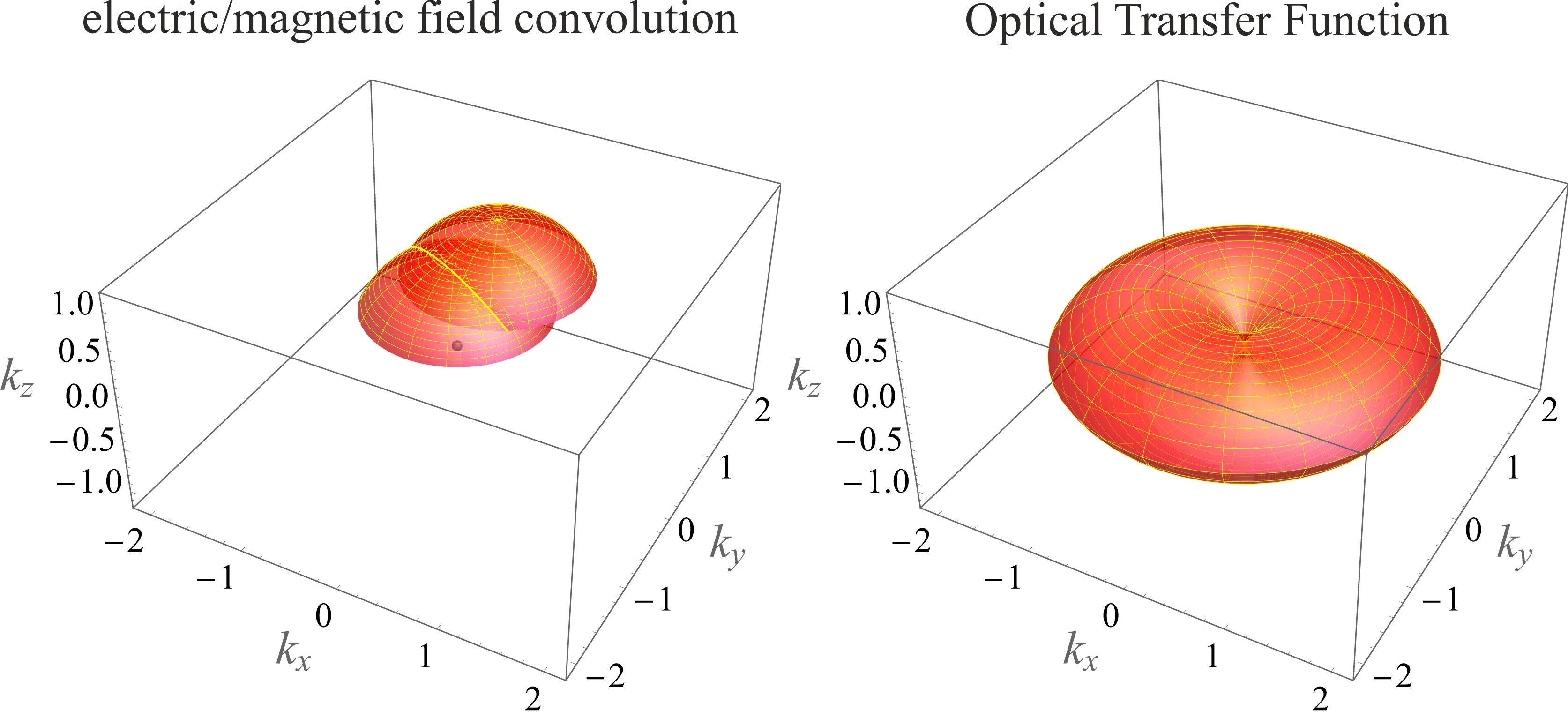}
    \caption{From electric/magnetic field to intensity. The two spherical caps in the left panel show the support of the Fourier representations of electric and magnetic fields given by Eq.~\ref{eq:EFourierAmplitude}. The right panel represents the extent of frequency support of the imaging OTF obtained by the convolution of the two caps on the left panel; see Eq.~\ref{eq:OTFconvolution}. The shape in the right panel is termed butterfly-shape and its missing cone in the middle highlights a wide-field microscope's inability to collect sufficient axial frequencies and thus lack of optical sectioning.
    }
    \label{fig:psfderivation}
    \vspace{-3mm}
\end{figure}

The PSF itself, once more, follows from the Poynting vector's $z$-component (see Eq.~\ref{eq:PoyntingZ})
\begin{equation}
    \begin{split}
        U(\mathbf{r}') &= \frac{c}{8\pi} \hat{\mathbf{e}}_z\cdot\left[\mathbf{E}'(\mathbf{r}')\times\mathbf{B}^{\prime *}(\mathbf{r}')\right]\\
        &= \frac{c}{8\pi} \left[E_x'(\mathbf{r}')B_y^{\prime *}(\mathbf{r}')-E_y'(\mathbf{r}')B_x^{\prime *}(\mathbf{r}')\right].
    \end{split}
    \label{eq:PSFdefinition}
\end{equation}
Knowing the PSF, the image model $\Lambda(\mathbf{r}')$, \emph{i.e.}, the spatial distribution of expected photon intensity or photon count, in image space, for an arbitrary sample follows from the convolution
\begin{align}
        \Lambda(\mathbf{r}') = I \int d^3\mathbf{r}_0 U\left(\mathbf{r}'-\mathcal{M}\mathbf{r}_0\right) S(\mathbf{r}_0),
    \label{eq:ImageFormation}
\end{align}
where $S(\mathbf{r_0})$ is the so-called sample function describing the fluorophore distribution. We assume the PSF, $U$, to be normalized to unity and $I$ to reflect the total photon emission per fluorophore.

For an aplanatic imaging system, which is shift-invariant (see Sec.~\ref{Aberrations}), Eq.~\ref{eq:ImageFormation} is exact for all emitters on the focal plane, \textit{i.e.}, for $z_0=0$. However, it is an approximate for emitters outside the focal plane, as follows from the discussion of the Abbe and Herschel conditions of Sec.~\ref{Abbe_sine}.

Using the electric field of Eq.~\ref{eq:EFourierAmplitude}, the lateral components of the electric and magnetic fields in the Fourier domain are explicitly given by (for $\theta'\leq\Theta'$) 
\begin{equation}
    \begin{split}
        \left(\begin{matrix}
        \tilde{E}_x'\\
        \tilde{E}_y'
        \end{matrix}\right)
         \propto\, &\delta\left(k'-\frac{2\pi}{\lambda}\right) \sqrt{\frac{\cos\theta'}{\cos\theta}} \\
        &\left(\begin{matrix}
        -E_{0,\parallel}\sin\phi + E_{0,\perp}\cos\theta'\cos\phi\\
        E_{0,\parallel}\cos\phi + E_{0,\perp}\cos\theta'\sin\phi
        \end{matrix}\right),
    \end{split}
    \label{eq:ElectricOTF}
\end{equation}
and
\begin{equation}
    \begin{split}
        \left(\begin{matrix}
        \tilde{B}_x'\\
        \tilde{B}_y'
        \end{matrix}\right)
         \propto\, &\delta\left(k'-\frac{2\pi}{\lambda}\right) \sqrt{\frac{\cos\theta'}{\cos\theta}}\\
        &\left(\begin{matrix}
        -E_{0,\parallel}\cos\theta'\cos\phi - E_{0,\perp}\sin\phi\\
        -E_{0,\parallel}\cos\theta'\sin\phi + E_{0,\perp}\cos\phi
        \end{matrix}\right),
    \end{split}
\end{equation}
where we also used the Cartesian representation of $\hat{\mathbf{e}}_{\parallel}$ and $\hat{\mathbf{e}}'_{\perp}$ similar to Eq.~\ref{eq:unitvecs}. Moreover, we remember that the refractive index in image space is assumed to be 1 (air). Thus, no additional prefactor appears in the coinciding magnetic field expression. 

Now, with the Fourier representations of the electric and magnetic fields at hand, we derive the imaging OTF then the PSF. To start, we note that the PSF is given by the product of the electric and magnetic field components in the spatial domain; see Eq.~\ref{eq:PSFdefinition}. However, within the Fourier domain, we use the well-known convolution theorem: the Fourier representation of the product of two functions is proportional to the convolution of their Fourier representations. As such, the imaging OTF is given by
\begin{align}
        \tilde{U}(\mathbf{k}') & \propto \tilde{E}_x'(\mathbf{k}') \otimes \tilde{B}_y^{\prime *}(\mathbf{k}') - \tilde{E}_y'(\mathbf{k}') \otimes \tilde{B}_x^{\prime *}(\mathbf{k}') \nonumber \\
        & = \int d^3\mathbf{k}'' \big[\tilde{E}_x'(\mathbf{k}'-\mathbf{k}'')\tilde{B}_y^{\prime *}(\mathbf{k}'') \nonumber\\
        &\hspace{13mm}-\tilde{E}_y'(\mathbf{k}'-\mathbf{k}'')\tilde{B}_x^{\prime *}(\mathbf{k}'')\big],
    \label{eq:OTFconvolution}
\end{align}
where $\otimes$ denotes convolution. The resulting OTF is then related to the PSF by Fourier transform
\begin{equation}
    \begin{split}
        U(\mathbf{r}') = &\int \frac{d^3\mathbf{k}'}{(2\pi)^3} \tilde{U}(\mathbf{k}') \exp\left(i\mathbf{k}'\cdot\mathbf{r}'\right).
    \end{split}
    \label{eq:PSFviaOTF}
\end{equation}

The convolution of Eq.~\ref{eq:OTFconvolution} is visualized in Fig.~\ref{fig:psfderivation}. The two spherical caps (note that it is only the area on the surface) shown in the left panel represent regions where the Fourier amplitudes of the electric and magnetic fields are non-zero (see Eq.~\ref{eq:EFourierAmplitude2}). The convolution of these caps results in the butterfly-shaped three-dimensional figure shown in the right panel, where the surface shown represents the maximum extent of frequency support of the OTF. That is, the OTF amplitude vanishes for all frequencies outside this region and takes non-zero values only for frequencies within the three-dimensional shape also termed microscope's band-pass. 

\begin{figure}[H]
    \centering
    \includegraphics[width=0.98\linewidth]{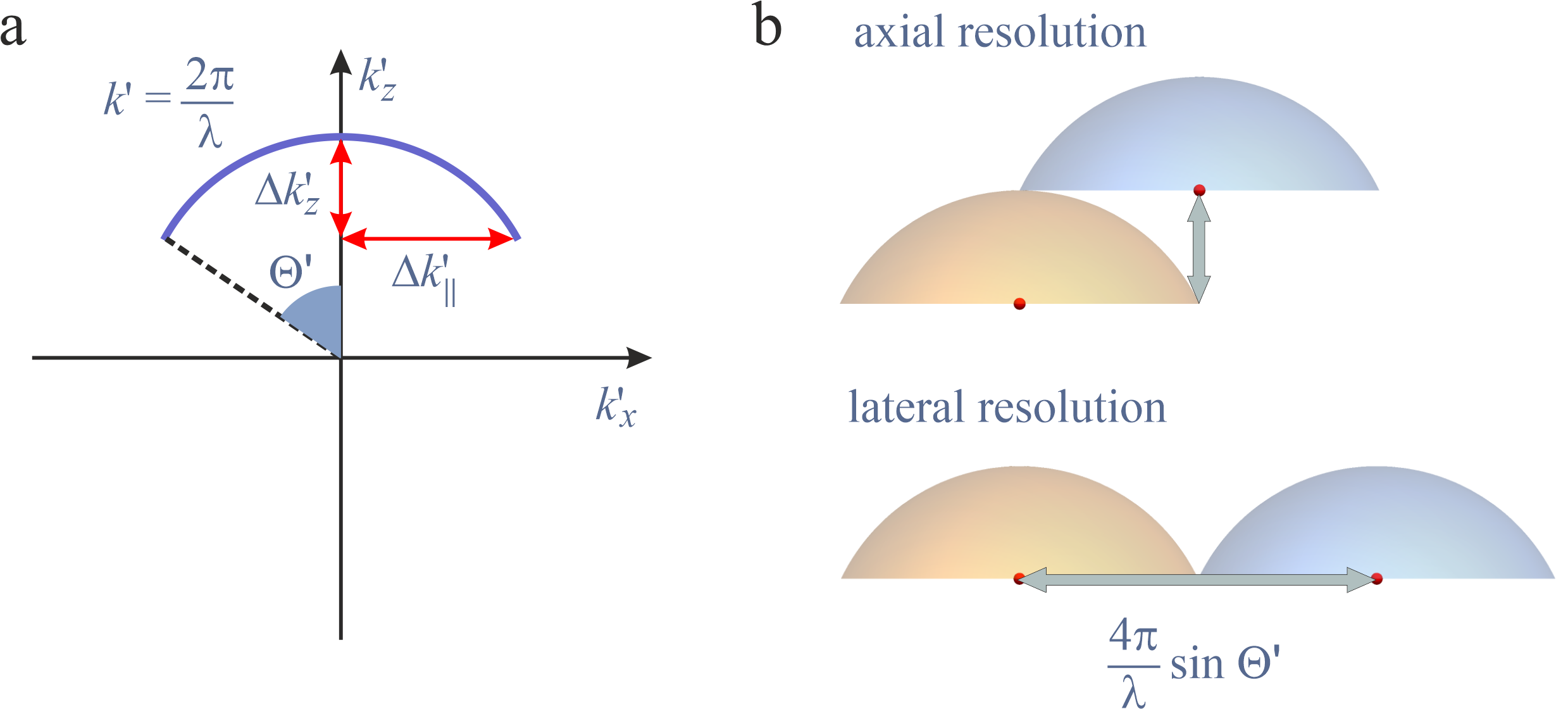}
    \vspace{-1mm}
    \caption{
    Visualization of the maximum axial and lateral extents of the Fourier representation of the electric field and the imaging OTF. (a) A cross-section of the Fourier representation of the electric field (cap) at $k'_y=0$. The cross-section is an arc with radius $k'=2\pi/\lambda$ and $0\leq\theta'<\Theta'$ (see Eq.~\ref{eq:EFourierAmplitude2}). The maximum extents of the cap along the lateral and axial directions are, respectively, given by $\Delta k'_{\parallel}=\frac{2\pi}{\lambda}\sin\Theta'$ and $\Delta k'_{z}=\frac{2\pi}{\lambda}(1-\cos\Theta')$. (b) Here we show the convolution of the caps associated to the electric and magnetic fields along the largest axial and lateral extents beyond which the convolution is zero. 
    }
    \label{fig:lateralaxialconvolution}
    \vspace{-4mm}
\end{figure}

From Fig.~\ref{fig:lateralaxialconvolution}a, one finds that the lateral and axial extents of the Fourier representations of the electric/magnetic fields are $\Delta k'_{\parallel} = 2\pi\sin\Theta'/\lambda$ and $\Delta k'_z = 2\pi(1-\cos\Theta')/\lambda$, respectively. As the OTF is computed from the auto-convolution of the cap associated to the electric/magnetic fields, the lateral and axial size of the OTF, respectively, is then found to be $4\pi\sin\Theta'/\lambda$ and $2\pi(1-\cos\Theta')/\lambda$, see Fig.~\ref{fig:lateralaxialconvolution}b.

Put differently, the microscope does not transmit lateral spatial frequencies beyond $k_\parallel' > 4\pi\sin\Theta'/\lambda$ or any axial spatial frequencies beyond $k_z' > 2\pi(1-\cos\Theta')/\lambda$, where $k_\parallel'=\sqrt{k_x^{'2}+k_y^{'2}}$ is the amplitude $\mathbf{k}'$'s projection in the $k_xk_y$-plane. Thus, the three-dimensional intensity distribution in image space does not transmit lateral spatial modulations smaller than $2\pi/\max k_\parallel' = \lambda/(2\sin\Theta')$. This leads to spatial modulations of $\mathcal{M}\lambda/(2 n \sin\Theta)$ in image space using Abbe's sine condition, and translates into the smallest discernible spatial variation $\lambda/(2 n \sin\Theta)$ in the sample space when taking into account that the lateral magnification is $\mathcal{M}$. Therefore, we recover Abbe's resolution limit, Eq.~\ref{eq:abberesolutionlimit}, as $2\pi$ over the largest lateral spatial frequency transmitted by the microscope from sample to image space 
\begin{equation}    
r^l_{\mathrm{min}} = \frac{2\pi}{k^l_{\mathrm{max}}},   \label{eq:resolutionMonotonic}
\end{equation}
where $r^l_{\mathrm{min}}$ and $k^l_{\mathrm{max}}$, respectively, denote the resolution and maximum extent of the OTF along the $l$th direction.

While Eq.~\ref{eq:resolutionMonotonic} provides a measure of resolution for lens-based imaging systems with OTF magnitudes consisting of a single lobe monotonically decaying to zero, \textit{e.g.}, lateral magnitude of wide-field microscope's OTF, it should be used with care for more complicated OTFs, \textit{e.g.}, axial resolution for wide-field microscope (see Figs.~\ref{fig:psfderivation} and Sec.~\ref{ScalarApprox}), SIM (see Sec.~\ref{sec:SIM}), some types of light-sheet microscopes with multiple gaps in their OTF magnitudes (see Sec.~\ref{Sec:LSFM}), and others. 

As such, regarding the wide-field microscope's axial resolution, the situation is more complicated due to the OTF's shape in the axial direction. To be more concrete, in the right panel of Fig.~\ref{fig:psfderivation}, one can see that the butterfly-shape imaging OTF does not support axial frequencies within a cone defined by $k'_z/k'_\parallel>\tan\Theta'$. This is often called the OTF's missing cone. One effect of this missing cone is that a wide-field microscope does not provide \emph{optical sectioning} ($z$-sectioning). That is, for $k'_{\parallel}\approx 0$ a wide-field microscope collects limited axial spatial frequencies. Put differently, the PSF pattern formed by light collected from a fluorophore using a wide-field setup varies slowly with the fluorophore's axial position. 

Yet, as can also be seen from Fig.~\ref{fig:psfderivation} axial frequencies $k'_z$ have non-zero amplitudes for $0<k'_\parallel<\max k'_\parallel = 4\pi\sin\Theta'/\lambda$. The maximal value $k'_z=2\pi(1-\cos\Theta')/\lambda$ contained in the OTF shows that the smallest possible spatial modulation of the PSF along the optical axis is approximately $\lambda/(1-\cos\Theta')$. For paraxial optics, \textit{i.e.}, for small values of $\Theta$ and $\Theta'$, where we have approximately an axial magnification $\mathcal{M}_z= \mathcal{M}^2/n$ (see Sec.~\ref{Abbe_sine}), and with the approximation
$1-\cos\Theta'\approx\Theta'^2/2\approx n^2\Theta^2/2\mathcal{M}^2$, this translates into a small axial modulation of $2 \lambda /n \Theta^2\approx 2 n \lambda/(\mathrm{NA})^2$ of the sample function  transmitted through the microscope. This is in accordance with our previous estimate of the axial resolution limit in Eq.~\ref{eq:axialresolutionlimit}. Problems associated with the OTF's missing cone, \textit{i.e.}, missing $z$-sectioning, is considered in Sec.~\ref{sec:modality} where discuss confocal microscopy alongside other modalities. 

\vspace{-5mm}
\subsection{\label{Dipole}Electromagnetic field emission of an oscillating electric dipole}
\label{sec:classicfluorophore}
\vspace{-2mm}

In the previous section, we derived integral expressions for the OTF and PSF of a wide-field microscope; \textit{e.g.}, see Eqs.~\ref{eq:PSFdefinition} and \ref{eq:OTFconvolution}-\ref{eq:PSFviaOTF}. Here, we evaluate these integrals and obtain a wide-field microscope's exact OTF and PSF using $\mathbf{E}_0(\theta,\phi)=E_{0,\parallel}\hat{\mathbf{e}}_\parallel + E_{0,\perp}\hat{\mathbf{e}}_\perp$ for a fluorescent  (point) emitter; see Eq.~\ref{eq:EFourierAmplitude}. We do so by noting that the electromagnetic emission of fluorescence emitters ({\it e.g.}, organic dyes, proteins, quantum dots), used in fluorescence microscopy are often well approximated as an oscillating electric dipole. Important exceptions, to which we can geenralize, include some emission bands of rare earth emitters, {\it e.g.}, europium complexes, exhibiting magnetic dipole or electric quadrupole properties ~\cite{moskovits1982intense,binnemans2015interpretation}.  

To compute the electric dipole's oscillating electromagnetic field, we start from a dipole moment with amplitude $\mathbf{p}$ and oscillation frequency $\omega$ located at $\mathbf{r}_d=\left(x_d,y_d,z_d\right)$ in the sample medium with refractive index $n_d$. Moreover, considering all fields oscillate as $\exp(-i\omega t)$ as the dipole moment oscillations, we focus on the amplitudes of the electric and magnetic fields. In this case, the Maxwell's equations read
\vspace{-2mm}
\begin{equation}
    \begin{split}
        \nabla\times\mathbf{E} &= \frac{i\omega}{c} \mathbf{B},\\
        \nabla\times\mathbf{B} &= -\frac{i\omega\epsilon_d}{c} \mathbf{E} + \frac{4\pi}{c} \mathbf{j},
    \end{split}
    \label{eq:MaxwellDipole}
\end{equation}
\vspace{-2.5mm}

\noindent
where $\epsilon_d=n_d^2$ is the dielectric constant of the sample solution in which the dipole is embedded and $\mathbf{j}=-i\omega\mathbf{p} \delta(\mathbf{r}-\mathbf{r}_d)$ is the electric current generated by the oscillating dipole. Thus, we find $\nabla\times\nabla\times\mathbf{E}_d - k_d^2 \mathbf{E}_d = 4\pi k_0^2 \mathbf{p} \delta\left(\mathbf{r}-\mathbf{r}_d\right)$ for the electric field $\mathbf{E}_d$ of the dipole emitter, where $k_0=\omega/c$ and $k_d=n_d k_0$. Using $\nabla\times\nabla\times\mathbf{E}_d = \nabla\nabla\cdot\mathbf{E}_d - \nabla^2 \mathbf{E}_d$ ~\cite{jackson1999classical} and passing to Fourier space yields for the Fourier amplitude $\tilde{\mathbf{E}}_d$ 
\begin{equation}
        (k^{\prime 2}-k_d^2)\tilde{\mathbf{E}}_d - \mathbf{k}'(\mathbf{k}'\cdot\tilde{\mathbf{E}}_d) = 4\pi k_0^2 \mathbf{p} \exp\left(-i\mathbf{k}'\cdot\mathbf{r}_d\right),
    \label{eq:FourierE}
\end{equation}
where $\mathbf{k}'$ is the Fourier space coordinate. Multiplying Eq.~\ref{eq:FourierE} by $\mathbf{k}'$ yields $\mathbf{k}'\cdot\tilde{\mathbf{E}}_d = -\frac{4\pi}{\epsilon_d}(\mathbf{k}'\cdot\mathbf{p})\exp\left(-i\mathbf{k}'\cdot\mathbf{r}_d\right)$ which we substitute into Eq.~\ref{eq:FourierE} to arrive at 
\begin{equation}
    \begin{split}
        \tilde{\mathbf{E}}_d = \frac{4\pi\exp(-i\mathbf{k}'\cdot\mathbf{r}_d)}{\epsilon_d(k'^2-k_d^2)}\left[k_d^2\mathbf{p}-\mathbf{k}'(\mathbf{k}'\cdot\mathbf{p})\right].
    \end{split}
\end{equation}
In real space, the above reads
\begin{equation}
        \mathbf{E}_d = \int \frac{d^3\mathbf{k}'}{2\pi^2\epsilon_d} \left[k_d^2\mathbf{p}-\mathbf{k}'(\mathbf{k}'\cdot\mathbf{p})\right] \frac{\exp\left[i\mathbf{k}'\cdot\left(\mathbf{r}-\mathbf{r}_d\right)\right]}{k'^2-k_d^2},
    \label{eq:MaxwellDipole2}
\end{equation}
where $|\mathbf{r}-\mathbf{r}_d|$ is the distance between electric dipole's location, $\mathbf{r}_d$, and the observation point, $\mathbf{r}$.

\begin{figure}[H]
    \centering
    \includegraphics[width=0.7\linewidth]{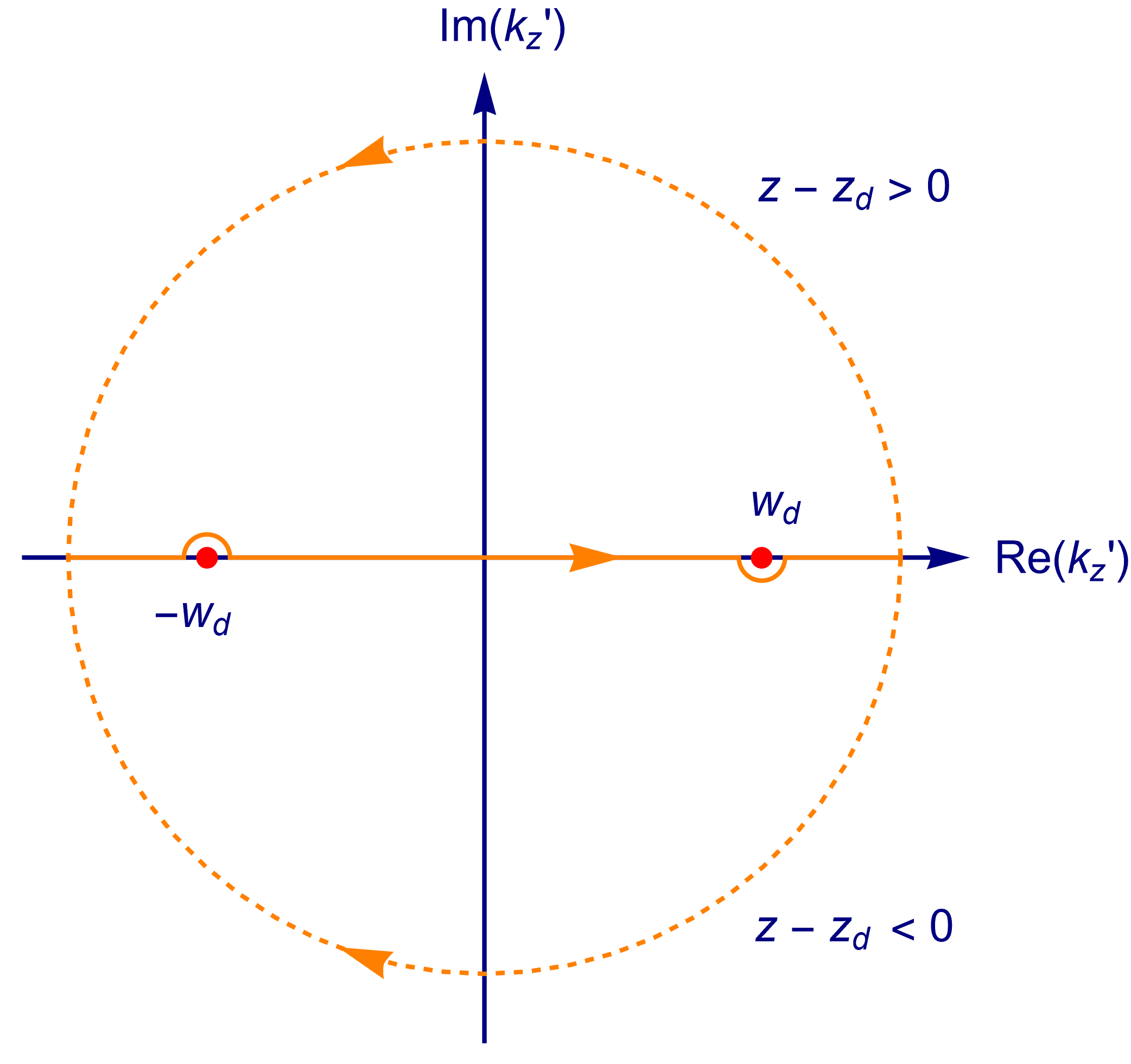}
    \vspace{-2mm}
    \caption{Contour for the integration over $k_z'$ of Eq.~\ref{eq:MaxwellDipole2} in the complex $k_z'$-plane. For positive values of $z-z_d$, the contour has to be closed, at infinity, over the positive $\mathrm{Im}(k_z')$ half-space, while for negative values of $z-z_d$ it is over the negative half-space. Along the real axis, the integrand has two poles at $\pm w_d = \pm\sqrt{k_d^2-q^2}$.}
    \label{fig:WeylIntegrationContour}
    \vspace{-5mm}
\end{figure}

To obtain an expression  well suited in modeling the emission of a dipole in a planar system (\textit{e.g.}, above a flat coverslide), we perform the integration along the $k_z'$-coordinate in the above expression, using Cauchy's residue theorem. To do so, we close the integration path along the real axis and complete a semi-circle at infinity over the complex $k_z'$-plane, as shown in Fig.~\ref{fig:WeylIntegrationContour}. To make sure that the exponent vanishes when extending the contour into the complex plane, one has to close the contour over the positive imaginary half plane when $z-z_d>0$ and over the negative imaginary half plane when $z-z_d<0$. Along the real axis, the integrand has two poles at positions $\pm w_d = \pm \sqrt{k_d^2-q^2}$, where $q^2=k_x^{'2}+k_y^{'2}$. However, the integration's result must contain only \emph{outgoing} plane waves (Sommerfeld radiation condition ~\cite{sommerfeld}), achieved by deforming the integration contour around the two poles as shown in Fig.~\ref{fig:WeylIntegrationContour}. Subsequently applying Cauchy's residue theorem yields 
\begin{equation}
    \begin{split}
        \mathbf{E}_d = &\frac{i}{2\pi\epsilon_d}\int \frac{d^2\mathbf{q}}{w_d} \left[k_d^2\mathbf{p}-\mathbf{k}_d(\mathbf{k}_d\cdot\mathbf{p})\right]\\ &\exp\left[i\mathbf{q}\cdot\left(\boldsymbol{\rho}-\boldsymbol{\rho}_d\right)+i w_d \left\vert z-z_d \right\vert \right],
    \end{split}
    \label{eq:DipoleE0}
\end{equation}
where we used the abbreviations $\mathbf{k}_d=\left(\mathbf{q},w_d\right)$, with $w_d=\sqrt{k_d^2-q^2}$ as the pole location. Here, $\boldsymbol{\rho}$ and $\mathbf{q}$, respectively, collect lateral coordinates in real and Fourier spaces. Further, ($\boldsymbol{\rho}_d$, $z_d$) denotes the dipole spatial coordinates. The two-dimensional integration over $\mathbf{q}$ extends over an infinite (Fourier) plane oriented perpendicular to the optical axis. Eq.~\ref{eq:DipoleE0} is the plane wave representation of the electric field of a free oscillating dipole, also called the Weyl representation~\cite{weyl1919ausbreitung,banos_1966,mertz2019introduction}. As we will see, the Weyl representation is particularly suited to modeling the imaging of an emitter through a microscope.  

\begin{figure}[H]
\vspace{-1.5mm}
    \centering
    \includegraphics[width=0.8\linewidth]{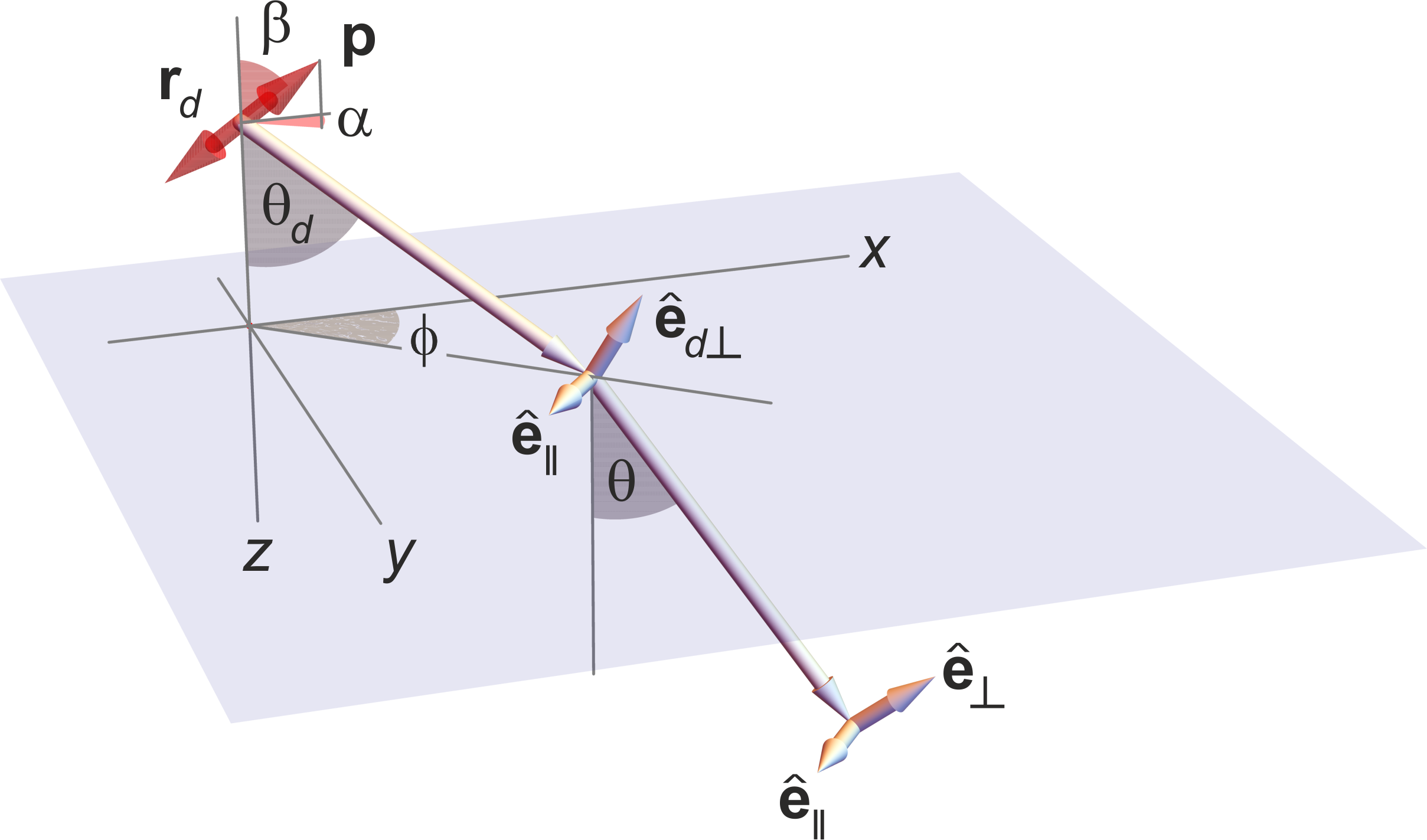}
    \vspace{-2.5mm}
    \caption{Angular distribution of the electric field generated by a single dipole emitter. Here, the gray rectangle represents the coverslide (commonly assumed to coincide with $z=0$ plane) which is the interface between the electric dipole's embedding medium (above the coverslide) and the immersion medium below the coverslide. The red two-headed arrow depicts the dipole; $\alpha$ and $\beta$ are, respectively, polar and inclination (azimuthal) angles describing the orientation of the dipole; $\phi$ is the polar angle of the wave vector; $\theta_d$ and $\theta$ are the azimuthal angles of the wave vector above and below the interface.  }
    \label{fig:dipoleabovesurface}
    \vspace{-4mm}
\end{figure}

Next, we consider the situation where the refractive index $n_d$ of the medium in which the emitting dipole is embedded and the refractive index $n$ of the immersion medium of the microscope's objective differ (\emph{e.g.}, imaging with an oil immersion objective with an emitter in water). This situation is schematically shown in Fig.~\ref{fig:dipoleabovesurface}. We use Eq.~\ref{eq:DipoleE0} to model the propagation of the electric field through an interface dividing the sample (dipole) and immersion medium, \textit{i.e.}, coverslide surface. To do so, it is convenient to recast the integrand in Eq.~\ref{eq:DipoleE0} as
\vspace{-1mm}
\begin{equation}
        k_d^2\mathbf{p}-\mathbf{k}_d(\mathbf{k}_d\cdot\mathbf{p}) = k_d^2 \left[\left(\mathbf{p}\cdot\hat{\mathbf{e}}_\parallel\right)\hat{\mathbf{e}}_\parallel + \left(\mathbf{p}\cdot\hat{\mathbf{e}}_{d\perp}\right)\hat{\mathbf{e}}_{d\perp}\right],
    \label{eq:DipoleE0Split}
\end{equation}
where we used $\mathbf{p}=\left(\mathbf{p}\cdot\hat{\mathbf{e}}_\parallel\right)\hat{\mathbf{e}}_\parallel + \left(\mathbf{p}\cdot\hat{\mathbf{e}}_{d\perp}\right)\hat{\mathbf{e}}_{d\perp}+(\mathbf{p}.\hat{\mathbf{k}}_d)\hat{\mathbf{k}}_d$ since the unit vectors $\hat{\mathbf{e}}_\parallel$, $\hat{\mathbf{e}}_{d\perp}$ and $\hat{\mathbf{k}}_d$ form an orthonormal set similar to Eq.~\ref{eq:unitvecs}. As such, the problem reduces to considering the propagation of $s$- and $p$-polarized plane waves through a planar interface. 

We now use Eqs.~\ref{eq:DipoleE0}-\ref{eq:DipoleE0Split} to write the electric field after it crosses the interface between both media and travels a distance through the immersion medium (with refractive index $n$) before arriving in front of the objective lens in term of the $p-$ and $s$-polarized components 
\begin{equation}
    \begin{split}
        \mathbf{E}_d = &\frac{i k_0^2}{2\pi}\int \frac{d^2\mathbf{q}}{w} \left[t_\parallel \left(\mathbf{p}\cdot\hat{\mathbf{e}}_\parallel\right)\hat{\mathbf{e}}_\parallel + t_\perp  \left(\mathbf{p}\cdot\hat{\mathbf{e}}_{\perp}\right)\hat{\mathbf{e}}_{\perp}\right]\\ &\exp\left[i\mathbf{q}\cdot\left(\boldsymbol{\rho}-\boldsymbol{\rho}_d\right) - i w_d z_d + i w \left(z-f\right)\right],
    \end{split}
    \label{eq:DipoleE}
\end{equation}
where the $t_{\parallel,\perp}$ are the Fresnel transmission coefficients, $(\boldsymbol{\rho},z)$ represent the observation point coordinates within the immersion medium, and the focal distance, $f$, is the location of the focal plane with respect to the interface $z=0$ coinciding with the coverslide surface separating the sample from the immersion medium; see Fig.~\ref{fig:dipoleabovesurface}. Here, the axial component $w$ of the wave vector $\mathbf{k}$ in the immersion medium is given by $w=\sqrt{k^2-q^2}=\sqrt{n^2k_0^2-q^2}$. Moreover, the unit vector $\hat{\mathbf{e}}_{\perp}$ is similar to $\hat{\mathbf{e}}_{d\perp}$ but formed from the wave vector $\left(\mathbf{q},\sqrt{n^2k_0^2-q^2}\right)$ instead of $\left(\mathbf{q},\sqrt{n_d^2k_0^2-q^2}\right)$. 

\begin{figure}[H]
\vspace{-2mm}
    \centering
    \includegraphics[width=0.95\linewidth]{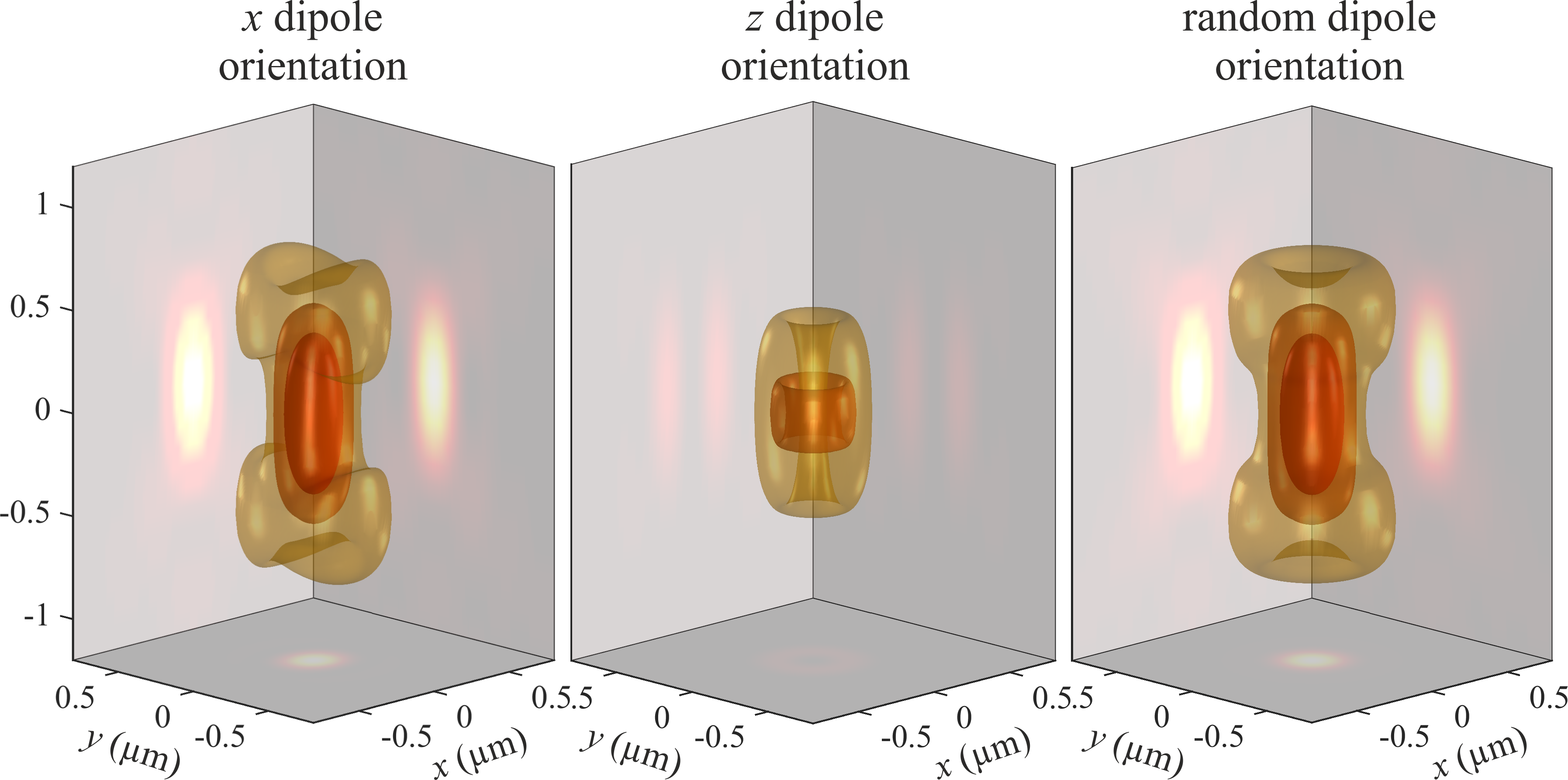}
    \caption{The PSF of a wide-field microscope, projected into sample space. Shown are plots of the $1/e$, $1/e^2$ and $1/e^3$ iso-surfaces of the maximum PSF value. The lateral coordinates refer to back-projected sample space coordinates $(x,y)=(x',y')/\mathcal{M}$, whereas the axial coordinate refers to an emitter's axial position $z_d$. We retain this PSF representation throughout the review. The individual panels are described in the main body. Calculations were performed for a NA~=~1.2 water immersion objective with $n=1.33$ and emission wavelength $\lambda=550$~nm.}
    \label{fig:psfwidefield}
    \vspace{-4mm}
\end{figure} 

The formulation above can be readily generalized to arbitrary number of interfaces. For instance, if an emitter is imaged through a stack of several layers characterized by different refractive indices, then the single interface's Fresnel transmission coefficients in Eq.~\ref{eq:DipoleE} must simply be replaced by those for the stacked structure. 

Finally, considering Fig.~\ref{fig:dipoleabovesurface}, we have $w= n k_0 \cos\theta$ and $\mathbf{q} = n k_0 (\sin\theta \cos\phi,\sin\theta \sin\phi,0)$ leading to $d^2\mathbf{q}/w = dq_x dq_y/w = n k_0 \sin\theta d\theta d\phi$ in spherical coordinates. Substituting this result into Eq.~\ref{eq:DipoleE} and comparing with Eq.~\ref{eq:EfieldSampleSpace} yields the following electric field amplitude $\mathbf{E}_0(\theta,\phi)$ for a dipole emitter (up to some constant factor) 
\begin{equation}
    \begin{split}
        \mathbf{E}_0 \propto &\left[t_\parallel \left(\mathbf{p}\cdot\hat{\mathbf{e}}_\parallel\right)\hat{\mathbf{e}}_\parallel + t_\perp  \left(\mathbf{p}\cdot\hat{\mathbf{e}}_{\perp}\right)\hat{\mathbf{e}}_{\perp}\right]\\ &\exp\left[-i\mathbf{q}\cdot\boldsymbol{\rho}_d - i w_d z_d - i w f\right],
    \end{split}
    \label{eq:DipoleEAmplitude}
\end{equation}
or more explicitly
\begin{equation}
    \begin{split}
        &\left(\begin{matrix}
        E_{0,\parallel}\\
        E_{0,\perp}
        \end{matrix}\right)
        =
        \left(\begin{matrix}
        \mathbf{E}_0\cdot\mathbf{e}_\parallel\\
        \mathbf{E}_0\cdot\mathbf{e}_\perp
        \end{matrix}\right)\\
        &\qquad\qquad \propto |\mathbf{p}| \exp\left(-i\mathbf{q}\cdot\boldsymbol{\rho}_d - i w_d z_d - i w f\right)\\
        &\qquad
        \left(\begin{matrix}
        -t_\parallel \sin\beta \sin(\phi-\alpha) \\
        t_\perp \left[\sin\beta \cos\theta \cos(\phi-\alpha) - \cos\beta \sin\theta\right]
        \end{matrix}\right),
            \label{eq:DipoleEAmplitude_2}
    \end{split}
\end{equation}
where $\alpha$ and $\beta$ are the dipole orientation angles as described in Fig.~\ref{fig:dipoleabovesurface}. By inserting these expressions into Eqs.~\ref{eq:Efield}, \ref{eq:Bfield} and \ref{eq:PSFdefinition}, one can compute the wide-field image PSF of the dipole emitter with arbitrary position and orientation. When doing so, it is convenient to present the results in terms of the lateral sample coordinates $\boldsymbol{\rho} = \boldsymbol{\rho}'/\mathcal{M}$ instead of the image space coordinates $\boldsymbol{\rho}'$, and as a function of the axial position $z_d$ (with respect to the coverslide) of the emitter. This notation will be applied to all PSF visualizations throughout this review. Thus, in what follows, when writing the PSF, $U(\mathbf{r})$, as a function of $\mathbf{r}$, it is silently assumed that the lateral coordinates $x$ and $y$ are the coordinates conjugate to $x'$ and $y'$, \textit{i.e.}, $x=x'/\mathcal{M}$ and $y=y'/\mathcal{M}$, and $z$ refers to the axial position $z_d$ of the emitter. 

As a first example of a PSF visualization, Fig.~\ref{fig:psfwidefield} shows three-dimensional representations of a dipole emitter's PSF along the optical axis
for a dipole oriented along the $x$-axis (left panel), $z$-axis (middle panel), and 
for a rapidly rotating emitter (right panel), where the isotropic PSF, $U_\mathrm{iso}(\mathbf{r})$, is given by an average of PSFs calculated for dipole orientations along the $x$, $y$ and $z$ axes~\cite{richards1959electromagnetic} 
\vspace{-2mm}
\begin{equation}
        U_\mathrm{iso}(\mathbf{r}) = \frac{1}{3} \left[ U_x(\mathbf{r}) + U_y(\mathbf{r}) + U_z(\mathbf{r}) \right].
    \label{eq:PSFarbitraryorientation}
    \vspace{-2mm}
\end{equation}
Accounting for effects of emitter orientation is of key interest in SMLM (Sec.~\ref{sec:SMLM}) as fixed orientations can lead to systematic mislocalization of emitters in space ~\cite{enderlein2006polarization,backlund2014role,deschout2014precisely,fazel2022analysis}. That being said, fluorescent labels are often coupled to structures with a sufficiently flexible linker allowing us to approximate labels as nearly freely rotating.

\begin{figure}[H]
    \centering
    \includegraphics[width=0.9\linewidth]{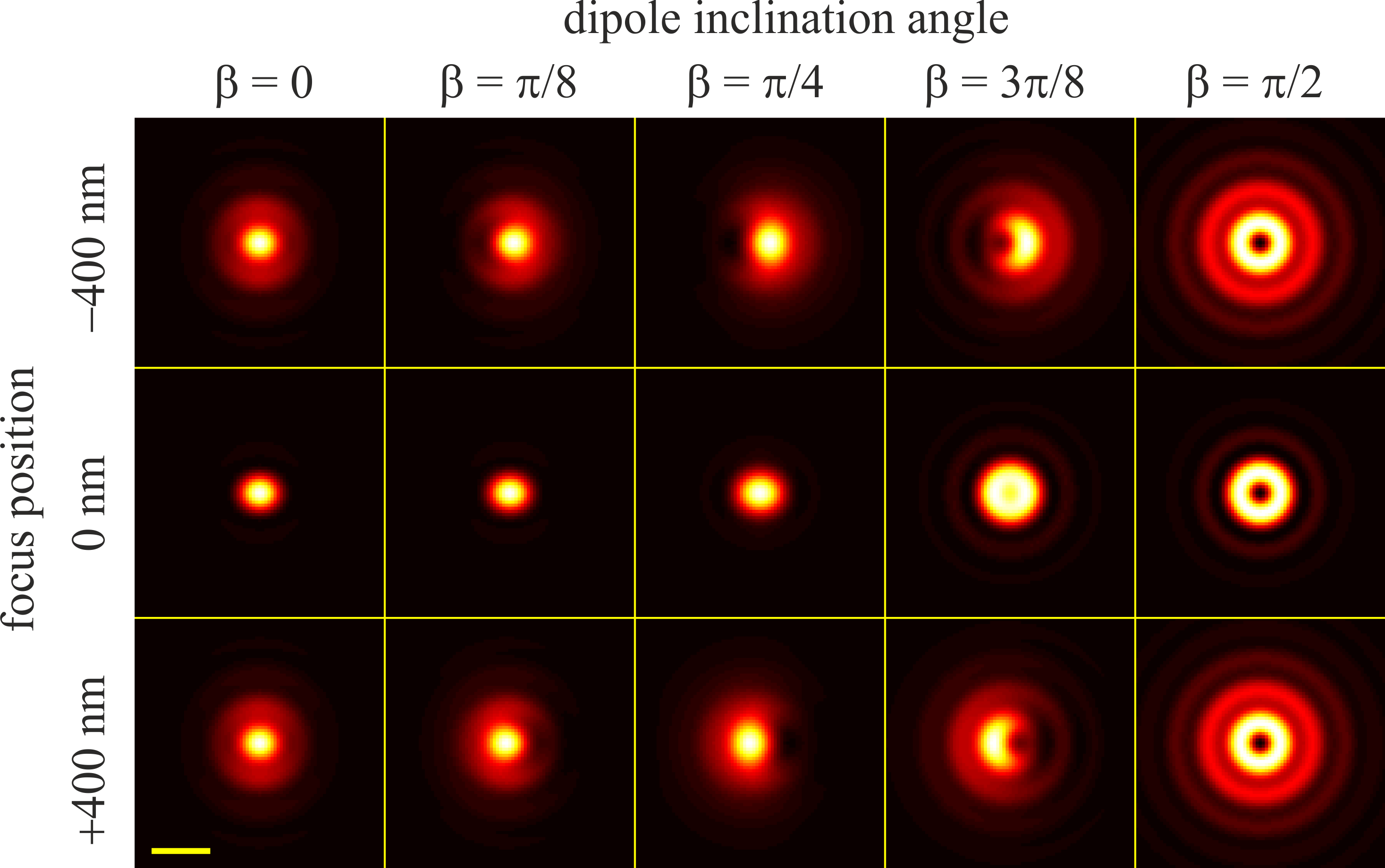}
    \caption{Effect of orientation on the emitter's image. Top row: images of electric dipole emitters of fixed strength but different orientations in the $xz$-plane, where $\beta$ is the inclination angle; see Fig.~\ref{fig:dipoleabovesurface}. The emitter is situated $400$~nm below the focal plane (NA~=~1.2, $n=1.33$). Middle row: same as top row, but for the emitter situated in the focal plane. Bottom row: same again but for an emitter situated 400~nm above the focal plane. The scale bar is 0.5~$\mu$m.}
    \label{fig:psfwidefieldorientation}
    \vspace{-7mm}
\end{figure}

As an example, Fig.~\ref{fig:psfwidefieldorientation} shows images of single emitters with different axial positions and inclination angles towards the optical axis. As can be seen, for out-of-focus emitters intermediate values of the inclination angle $\beta$ (see Fig.~\ref{fig:dipoleabovesurface}) can lead to considerable shifts in an emitter's image apparent center of mass especially significant for emitters away from the focal plane. The situation worsens when working with oil-immersion objectives with a larger Total Internal Reflection (TIR) critical angle than water immersion objective, which allows collection of fluorescent light with larger incident angles. In this case, even in-focus positions depend on emitter orientation. While this effect hinders the localizations of rigid single molecules under the assumption of a symmetric PSF, it can be exploited to learn three-dimensional orientations of molecules ~\cite{backer2019single,hulleman2021,wu2022dipole,rimoli20224polar}.

\begin{figure}[H]
    \centering
    \includegraphics[width=0.92\linewidth]{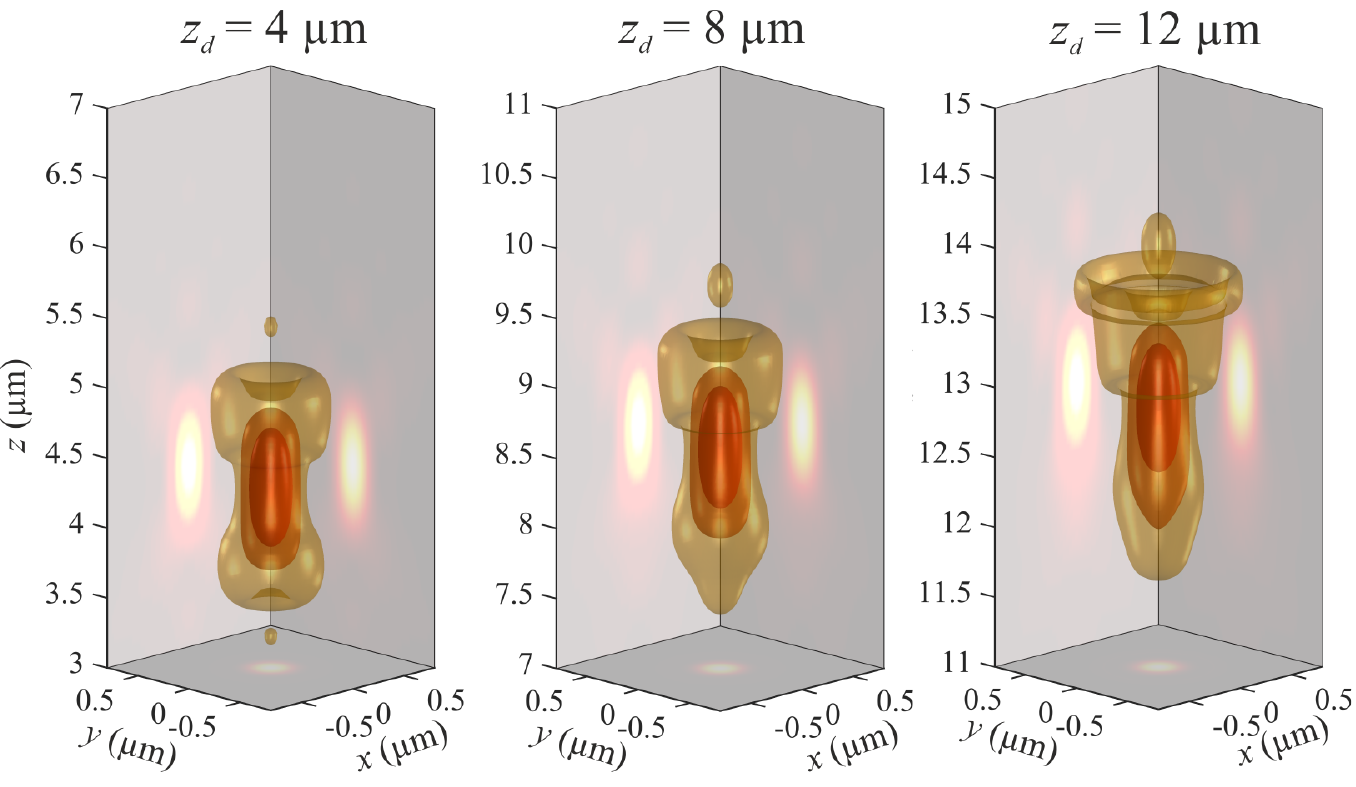}
    \caption{Effect of refractive index mismatch on the PSF. PSF of a rapidly rotating electric dipole emitter (isotropic emitter) positioned at various distances from a coverslide surface ($z=0$). Calculations were done for an NA~=~1.2 objective corrected for an immersion/medium with $n=1.33$, while the solution above the coverslide has $n=1.38$ ({\it i.e.}, refractive index mismatch $\Delta n=0.05$). The bottom of each box shows a density plot of the PSF's cross-section through its maximum value.}
    \label{fig:psfwidefieldrefindex}
    \vspace{-5mm}
\end{figure}

Finally, we briefly consider refractive index mismatch resulting in PSF distortion; see Sec.~\ref{Aberrations}. As an example, Fig.~\ref{fig:psfwidefieldrefindex} shows this effect for a slight refractive index mismatch of $\Delta n = 0.05$, again for a water immersion objective with NA~=~1.2. We further assumed that the objective lens is corrected for the light refraction introduced by the coverslide. As can be seen, this mismatch primarily results in PSF axial stretching and an axial shift between its center position towards larger $z$-values with respect to the actual position of the emitter. However, the lateral PSF cross-section at the axial location of its maximum does not change significantly, meaning that the refractive index mismatch does not affect the lateral position of the focused image of an emitter, but does result in its mislocalization along the optical axis.

\vspace{-5mm}
\subsection{\label{ScalarApprox}Scalar approximation of the PSF}
\vspace{-2mm}

In the previous section, we derived the exact electric field of an emitter, \textit{i.e.}, oscillating dipole, (see Eq.~\ref{eq:DipoleEAmplitude_2}) and used it to compute the PSF. However, these exact expressions are difficult to computationally manipulate. As such, here we provide a simple approximation to the emitter's electric field and the resulting PSF.

Along these lines, for many practical applications, we assume an isotropic emitter, {\it i.e.}, one with uniform emission amplitude in all directions. In such case, we can ignore the vectorial nature of the electric (and magnetic) fields resulting in an approximate scalar model. To derive such scalar approximations, we start from Eq.~\ref{eq:Efield} and replace the amplitude vector $E_{0,\parallel} \hat{\mathbf{e}}_\parallel + E_{0,\perp} \hat{\mathbf{e}}'_\perp$ by a scalar constant. Therefore the expression for the now ``scalar'' electric (magnetic) field in the image plane generated by an isotropic emitter on the optical axis at position $z=z_d$ simplifies to (up to a constant factor)
\begin{align}
        E(\mathbf{r}) &\propto \int_0^{\Theta'} d\theta' \sin\theta' \sqrt{\frac{\cos\theta'}{\cos\theta}} \int_0^{2\pi} d\phi e^{i \mathbf{q}'\cdot\boldsymbol{\rho}'-i k \cos\theta z} \nonumber \\
        &\propto \int_0^{\Theta'} d\theta' \sin\theta' \sqrt{\frac{\cos\theta'}{\cos\theta}} \int_0^{2\pi} d\phi e^{i |\mathbf{q}|\rho\cos\phi-i k \cos\theta z}\nonumber \\ 
        &\propto\int_0^{\Theta} d(\sin\theta) \sin\theta \frac{J_0(k\sin\theta\rho)}{\sqrt{\cos\theta'\cos\theta}} e^{-i k \cos\theta z}, 
    \label{eq:Escalar}
\end{align}
where we have used $\mathbf{q}'\cdot\boldsymbol{\rho}'=\mathbf{q}\cdot\boldsymbol{\rho}=|\mathbf{q}|\rho\cos\phi$ due to $\rho' = \mathcal{M}\rho$ and Abbe's sine condition $\sin\theta'=(n/\mathcal{M})\sin\theta$, while remembering $\left\vert \mathbf{q}' \right\vert = k_0 \sin\theta'$ and $\left\vert \mathbf{q} \right\vert = k \sin\theta = n k_0 \sin\theta$. 
In the second step, we performed the integral with respect to $\phi$ and used the Abbe's sine condition and its differential form (see Eq.~\ref{eq:AbbeDiff}), and ignored all the prefactors of $n$ and $\mathcal{M}$. Here, $J_m$ is the Bessel function of the first kind of order $m$~\cite{olver2010nist}. 

Further simplification is possible by replacing the square root factor for unity valid for small values of $\theta'$ and $\theta$ (far-field limit). Eq.~\ref{eq:Escalar} therefore simplifies to 
\begin{equation}
    \begin{split}
        E(\rho,z) \approx \left(\frac{n}{\mathcal{M}}\right)^2 \int_0^{\sin\Theta} d\eta\, \eta J_0\left(k \eta \rho\right) e^{-i k \sqrt{1-\eta^2} z},
    \end{split}
    \label{eq:Escalarsimple}
\end{equation}
where $\eta=\sin\theta$.
For the special case of $z=0$ (emitter in the focal plane), analytic integration then yields
\begin{equation}
    \begin{split}
        E(\rho) \approx \frac{\textrm{NA}}{\mathcal{M}^2 k_0 \rho} J_1\left(\textrm{NA} \,k_0 \rho \right),
    \end{split}
    \label{eq:Escalarfocal}
\end{equation}
where we have used $k\sin\Theta=\mathrm{N\hspace{-1pt}A}\,k_0$. Here, $J_1$ is the Bessel function of the first kind of order one ~\cite{olver2010nist}. The PSF is then given by the absolute square of the ``scalar'' electric field. Therefore, for the 2D PSF of an in-focus isotropic emitter in the far-field limit, we find the well-known Airy pattern \vspace{-3mm}
\begin{equation}
        U(\rho) \propto \left[\frac{J_1\left(\mathrm{NA}\, k_0 \rho \right)}{k_0 \rho}\right]^2,
    \label{eq:PSFscalar}
\end{equation}
where we have omitted a constant factor and, where we recall that $k_0 n \sin\Theta = \mathrm{NA}k_0$ is the maximum lateral wave vector component transmitted by the microscope from the sample to the image plane; see Sec.~\ref{Intro_Limit}. 

\begin{figure}[H]
    \centering
    \includegraphics[width=0.95\linewidth]{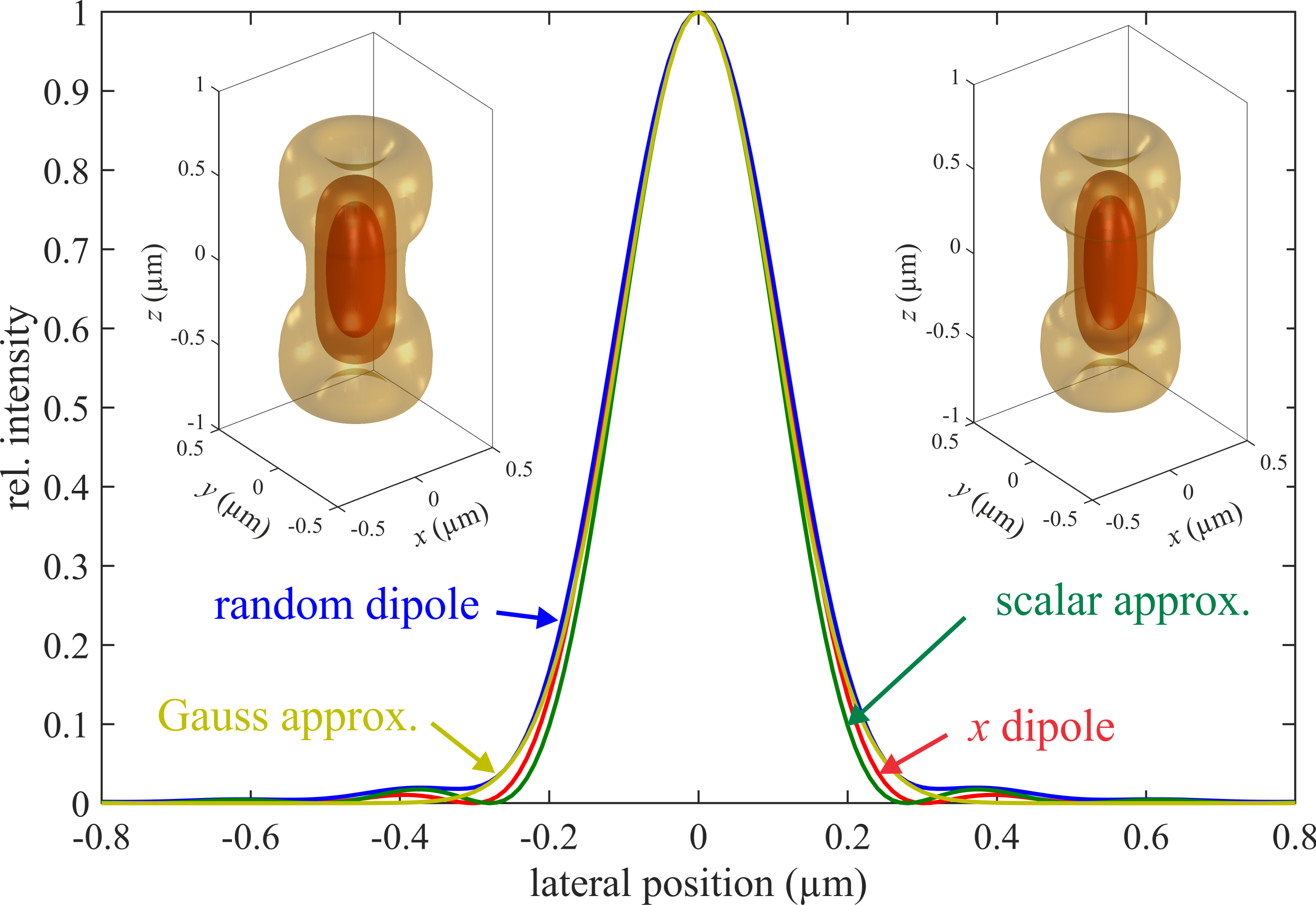}
    \caption{Comparison between scalar and vector PSF calculations. Shown are cross-sections of the PSF across the $x$-axis in the focal plane. The red curve shows results of the full wave-vector PSF calculation for an electric dipole emitter with fixed $x$-axis orientation, the blue curve the same calculation for a rapidly rotating (isotropic or random) emitter, the green curve presents the result of Eq.~\ref{eq:PSFscalar}, and the ochre curve shows the Gaussian approximation of Eq.~\ref{eq:PSFGaussian}. Insets show two three-dimensional iso-surface PSF plots, left using the exact vector field calculation for an isotropic emitter, right for the scalar approximation. All calculations were performed for a water immersion objective with NA~=~1.2.}
    \label{fig:psfwidefieldscalar}
    \vspace{-4mm}
\end{figure}

In situations where the scalar approximation is suitable (\textit{e.g.}, 3D imaging with molecules more than a wavelength away from the coverslide), this approximate PSF facilitates a computationally lighter model, as calculating Eq.~\ref{eq:PSFscalar} requires a single integration (Fourier transform) while evaluating Eq.~\ref{eq:PSFarbitraryorientation} requires three integrations. To check the accuracy of this approximation, Fig.~\ref{fig:psfwidefieldscalar} shows a comparison of the PSF's line cross-section through its center, calculated using the full vectorial model of Sec.~\ref{Fluo_PSF}-\ref{sec:classicfluorophore}, and the scalar approximation of Eq.~\ref{eq:PSFscalar}. As can be seen, the scalar approximation shows negligible deviations from the accurate model for the system considered (water immersion objective with NA~=~1.2, emission wavelength 500~nm). In most cases, this approximation is sufficient for quantitative analysis of fluorescence microscopy data, \textit{e.g.}, fitting single molecule images (see Sec.~\ref{sec:SMLM}) provided rapidly rotating molecules.

\begin{figure}[H]
    \centering
    \includegraphics[width=0.95\linewidth]{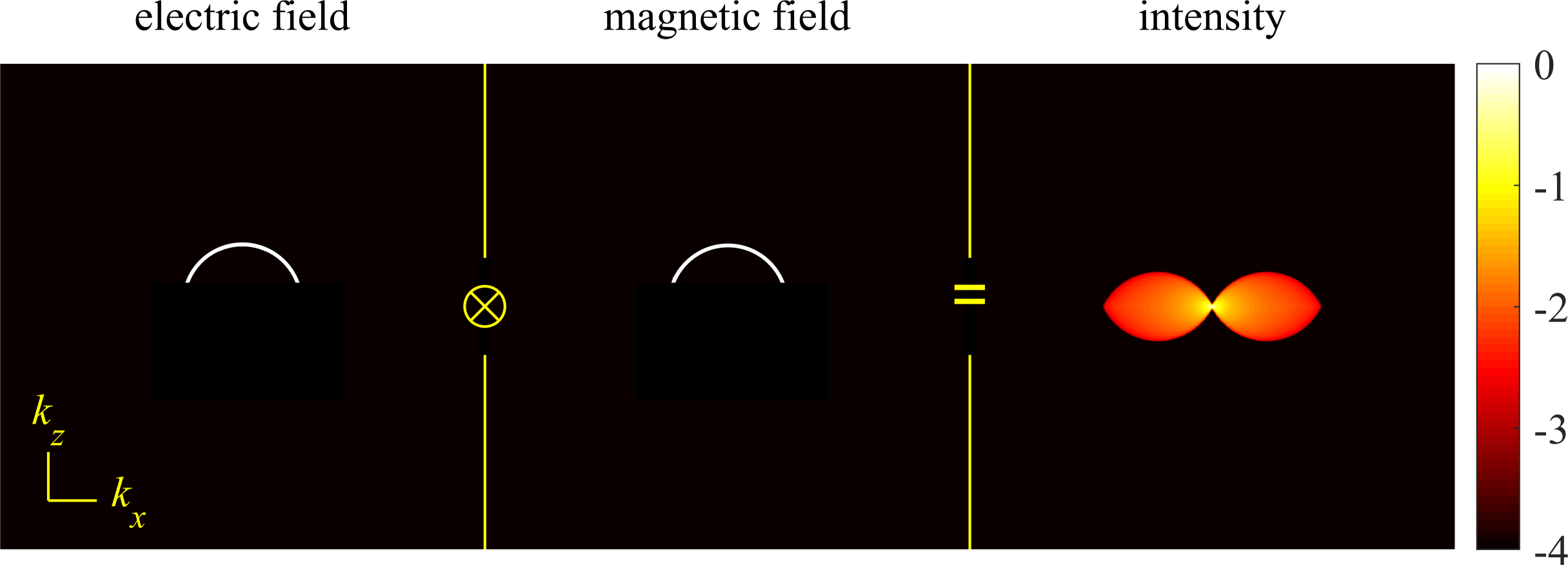}
    \caption{Scalar approximation of the OTF of a wide-field microscope. Calculations were done for NA~=~1.2 water immersion objective and an emission wavelength of 550~nm. The left panel shows the $k_x k_z$ cross-section of the electric field amplitudes in \emph{sample space}, having a frequency support (frequencies with non-zero amplitude) in the shape of a spherical cap with radius $k=2\pi n/\lambda$ and an opening half angle equal to the objective's maximum half angle $\Theta$. The middle panel shows the same distribution for the magnetic field. The right panel is the three-dimensional convolution of the left two panels, yielding the scalar approximation of the OTF amplitude. All panels show density plots of the \emph{decadic logarithm} of the Fourier amplitude's absolute value (see color bar on the right hand side) normalized by the maximum absolute value of the corresponding amplitudes. For all panels, the coordinate origin ($k_x=0,k_z=0$) is at the center. Throughout this review, we use the same representation for all OTFs shown.
    }
    \label{fig:otfwidefieldscalar}
    \vspace{-6mm}
\end{figure}

However, the usefulness of the scalar approximation is further evident in considering a microscope's OTF. When comparing Eqs.~\ref{eq:Efield} (also see Eq.~\ref{eq:EFourierAmplitude2}) and \ref{eq:Escalar}, the \emph{frequency support} of the Fourier transforms for the vector and scalar representations of the electric field are identical, given by a spherical cap centered at $\mathbf{k}'=0$ with radius $2\pi/\lambda$ and half opening angle $\Theta'$; see Figs.~\ref{fig:psfderivation}-\ref{fig:lateralaxialconvolution}. Similar to the PSF visualization, it is convenient to show the OTF back-projected to sample space, easily done using Abbe's sine condition as $(k'_x,k'_y) = n/\mathcal{M} (k_x,k_y)$ and the relation $k'=k/n$. Cross-sections of the corresponding electric (magnetic) field Fourier representation amplitude is shown in the left two panels of Fig.~\ref{fig:otfwidefieldscalar} at $k_y=0$. In the case of vectorial model, for each of the vector fields $\mathbf{E}$ and $\mathbf{B}$, one will have two  such cross-sections, one for the $E_\parallel$ ($B_\parallel$) and one for the $E_\perp$ ($B_\perp$) components. Here, Fig.~\ref{fig:otfwidefieldscalar} represents the scalar approximation with a uniform field amplitude over the whole spherical cap, \emph{cf}. with Eq.~\ref{eq:Escalar}. 
In both the exact vector field description as well as the scalar approximation, the PSF is found by products of the electric and magnetic fields, which translates in Fourier space to a convolution of the corresponding Fourier representations of these fields. 

A cross-section of the OTF amplitude at $k_y=0$ is visualized in the right panel of Fig.~\ref{fig:otfwidefieldscalar}, showing the (auto)convolution of the two Fourier amplitude distributions on the left. We note that, in general, the OTF is a complex quantity and all figures show the OTF amplitudes, sometimes termed modulation transfer function (MTF), but for brevity are simply termed OTFs for all subsequent figures. Although the exact amplitude distribution over the butterfly-shaped frequency support of the OTF will be slightly different for the full vector field (see Fig.~\ref{fig:psfderivation} and Eq.~\ref{eq:EFourierAmplitude2}) and the scalar approximation (see Eq.~\ref{eq:Escalarfocal}), the frequency support of the OTF remains identical. This is particularly important to emphasize, because the limits of this frequency support determines the microscope's optical resolution. Here, again, we emphasize that the resolution, along a given direction, is determined by the maximum frequency $k_{max}$ of this support along the chosen direction by Eq.~\ref{eq:resolutionMonotonic}. For the wide-field microscope in Fig.~\ref{fig:otfwidefieldscalar}, the lateral and axial extents of the OTF's frequency support are $k_{max,y} = 2 n k_0 \sin\Theta $ and $k_{max,z}= n k_0(1-\cos\Theta)$, respectively; also see Fig.~\ref{fig:lateralaxialconvolution}. This leads to the lateral and axial resolutions earlier derived (see Sec.~\ref{Intro_Limit}, \ref{Fluo_PSF} and Eq.~\ref{eq:resolutionMonotonic}) \vspace{-1mm}
\begin{align}
        &y_{\mathrm{min}} = \frac{2\pi}{k_{max,y}}=\frac{\lambda}{2 n \sin\Theta} = \frac{\lambda}{2 \mathrm{NA}}, 
    \label{eq:widefieldlateralresolution} \\
    \text{and} \hspace{6mm} & \nonumber\\
       &z_{\mathrm{min}} = \frac{2\pi}{k_{max,z}} = \frac{\lambda}{n \left(1-\cos\Theta\right)} \approx \frac{2 n \lambda}{\mathrm{NA}^2}. \hspace{4.75mm}
    \label{eq:widefieldaxialresolution}
\end{align}
The first equation is Abbe's famous lateral resolution limit for a wide-field microscope, while the approximate axial resolution in the second equation obtained is only valid for small numerical apertures. 

We can further simplify the PSF by approximating Eq.~\ref{eq:PSFscalar} with a 2D Gaussian function 
\begin{equation}
    U_\text{gauss}\left(\boldsymbol{\rho}-\boldsymbol{\rho}_0\right) \propto \\
    \exp\left(-\frac{\left|\boldsymbol{\rho}-\boldsymbol{\rho}_0\right|^2}{2\sigma_\mathrm{PSF}^2}\right),
    \label{eq:PSFGaussian}
\end{equation} 
where $\sigma_\mathrm{PSF} = \sqrt{2}/(\mathrm{NA}k_0)=\lambda/\sqrt{2}\pi n\sin\Theta$, as can be found by requiring the same curvature values at the maximum for both Eq.~\ref{eq:PSFGaussian} and Eq.~\ref{eq:PSFscalar}; see also Fig.~\ref{fig:psfwidefieldscalar}. This approximation is useful in creating a simple model, allowing straightforward fitting algorithms for many localization applications ~\cite{Stallinga2010Accuracy,fazel2022analysis}. This model fits the PSF's main lobe and thus is a good approximation when imaging within the depth of focus of an aberration-free microscope. The width $\sigma_{\mathrm{PSF}}$ is usually experimentally fit from a calibration sample or model ~\cite{santos2000model}.

\vspace{-4mm}
\subsection{\label{Aberrations}Optical aberrations}
\vspace{-3mm}

Finally, we discuss the impact of optical aberrations on the PSF. Optical aberrations refer to any deviation from idealized imaging models earlier presented and can be classified into various groups. The first distinction revolves around the wavelength, \emph{i.e.}, monochromatic aberrations occurring for a single wavelength, by contrast to chromatic aberrations, originating from the chromatic dispersion of the components in the optical system. The second distinction is characterized by shift-invariance, \emph{i.e.}, aberrations similar at every point in the Field Of View (FOV) \emph{versus} off-axis aberrations. In the presence of optical aberrations, modeling the PSF as a two-dimensional Fourier transform, $\mathcal{F}_{2D}$, operation is common as then the aberrations can be treated as part of the system's OTF. Here, we will focus on the scalar model, \emph{i.e.}, Eq.~\ref{eq:Escalar}. This approach can however be generalized to the vectorial case~\cite{ferdman2020vipr,siemons2018high,backer2014extending}.

As, generally, optical aberrations can be a function of $\left(\phi,\theta'\right)$, 
we return to Eq.~\ref{eq:Escalar} and extend it to include an additional amplitude/phase function that takes into account aberrations. Then, we can conveniently recast it as a $\mathcal{F}_{2D}$ operation prior to the integration over $\phi$ 
\begin{align}
        E(\rho,z;\mathbf{r}_0) \propto\mathcal{F}_{2D}\left(\mathcal{A}\left(\theta',\phi\right)e^{i\left[\Psi\left(\theta';z,f\right)+\Phi\left(\theta',\phi\right)\right]}\right), 
        \label{eq:PSFscalar_FFT}
\end{align}
where we ignored the term $\sqrt{\cos\theta'/\cos\theta}$ due to its negligible contribution. Here, $\mathcal{A}e^{i(\Psi+\Phi)}$ is the so-called pupil function, where $\mathcal{A}\left(\theta',\phi\right)$ is the pupil function's amplitude, which, neglecting all constant factors, simplifies to the Fourier plane support, limited by either the NA or $n_d$ as follows \vspace{-2mm}
\begin{equation}
    \mathcal{A}\left(\theta',\phi\right) =
    \begin{cases}
    1,& \text{if } \sin{\theta'} \leq  \min\left(\frac{n_d}{n},\frac{\mathrm{NA}}{n}\right)\\
    0,& \text{otherwise}
    \end{cases},
    \label{eq:circ_function}
\end{equation}
where $n$ and $n_d$ are, respectively, the refractive index of the objective immersion and the dipole (emitter) medium. In full generality, $\mathcal{A}$ can be a function of $\theta'$ and $\phi$, for instance, in the presence of aberrations in the form of attenuation of the transmitted electric and magnetic fields. However, these types of aberrations are rare and often induce negligible changes to the PSF compared to the phase terms~\cite{oppenheim1981importance}. Therefore, it is safe to neglect the effect of amplitude and focus on the phase. 

The first term in the phase, $\Psi\left(\theta';z,f\right)$, is induced by the molecule's shift off-axis and out-of-focus, \emph{i.e.}, the term $-\mathbf{q}\cdot\boldsymbol{\rho}_d -  w_d z_d -  w f z$ in Eq.~\ref{eq:DipoleEAmplitude_2}, 
\begin{multline}
    \Psi\left(\theta';z,f\right) =  k_{0} z  n_{d}   \sqrt{1-\sin{\theta'}^2} \\ - k_{0} f  n  \sqrt{1-\left(\frac{n_{d}}{n}\sin{\theta'}\right)^2}. 
    \label{eq:defocus_phase}
\end{multline} 
For instance, the phase $-k\sqrt{1-\eta^{2}}z$ in Eq.~\ref{eq:Escalarsimple}, where $\eta=\sin\theta$, is due to the out-of-focus location of the emitter. The second phase term in Eq.~\ref{eq:PSFscalar_FFT}, $\Phi\left(\theta',\phi\right)$, describes any additional phase of the pupil function (originating from optical aberrations as described in this section or PSF modulating elements described in Sec.~\ref{PSFeng_sec}), otherwise null in perfect aplanatic imaging conditions, as in Eq.~\ref{eq:Escalarsimple}. 

We start by considering monochromatic shift-invariant, \emph{i.e.}, $\left(x,y\right)$ independent, aberrations. 
In this case, aberration terms can be readily added to Eq.~\ref{eq:PSFscalar_FFT} as a phase term $\Phi\left(\theta',\phi\right)$. This phase function lives on the disk-like support $\phi \in \left\{0,2\pi\right\}$ and $\theta' \in \left\{0,\Theta'\right\}$ defined by the electric (magnetic) field Fourier amplitude distribution (see Sec.~\ref{EMField} and Fig.~\ref{fig:psfderivation}). 

It is often convenient to expand phase aberrations into a system of orthogonal basis functions, namely Zernike polynomials $Z_l^m(\xi=\sin\theta'/\sin\Theta',\phi)$ (see \emph{e.g.}, \cite{noll1976zernike,roddier1999adaptive}) 
\begin{equation}
    \Phi(\xi,\phi) = \sum_l \sum_{m=-l}^l v_{lm} Z_l^m(\xi,\phi),
\end{equation}
where $v_{lm}$ are coefficients corresponding to $Z_l^m$.
These polynomials are defined by \vspace{-3mm}
\begin{equation}
    \begin{split}
        Z_l^m(\xi,\phi) = 
        \begin{cases}
            R_l^m(\xi) \sin(m \phi),& \text{if } m>0\\
            R_l^m(\xi) \cos(m \phi),& \text{if } m\leq0
        \end{cases},
    \end{split}
    \label{eq:zernike}
\end{equation}
where the radial functions $R^{m}_l$ are given by \vspace{-2mm}
\begin{equation}
    \begin{split}
        R_l^m(\xi) = \sum_{k=0}^{(l-\vert m\vert)/2}\frac{(-1)^k(l-k)!\,\xi^{l-2k}}{k!\left[\frac{l+m}{2}-k\right]!\left[\frac{l-m}{2}-k\right]!}
    \end{split}
    \label{eq:zernike1}
\end{equation} \vspace{-1mm}
if $l-\vert m\vert$ is even, and zero otherwise; see table~\ref{tab:zernike}. 

\begin{table}[H]
\begin{tabular}{| c | c | c | c | c |} 
 \hline
 \# & $l$ & $m$ & $Z_n^m$ & name \\ [0.5ex] 
 \hline\hline
 1 & 1 & -1 & $\xi \cos\phi$ & horizontal tilt \\ 
 \hline
 2 & 1 & 1 & $\xi \sin\phi$ & vertical tilt \\ 
 \hline
 3 & 2 & 0 & $2\xi^2 -1$ & defocus \\ 
 \hline
 4 & 2 & -2 & $\xi^2 \cos 2\phi$ & vertical astigmatism \\ 
 \hline
 5 & 2 & 2 & $\xi^2 \sin 2\phi$ & oblique astigmatism \\ 
 \hline
 6 & 3 & -1 & $(3\xi^2-2)\xi \cos\phi$ & horizontal coma \\ 
 \hline
 7 & 3 & 1 & $(3\xi^2-2)\xi \sin\phi$ & vertical coma \\ 
 \hline
 8 & 4 & 0 & $6\xi^4 - 6\xi^2 + 1$ & primary spherical \\ 
 \hline
 9 & 3 & -3 & $\xi^3\cos3\phi$ & oblique trefoil \\ 
 \hline
 10 & 3 & 3 & $\xi^3\sin3\phi$ & vertical trefoil \\ 
 \hline
 11 & 4 & -2 & $(4\xi^2-3)\xi^2\cos2\phi$ & vert. secondary astigmatism \\ 
 \hline
 12 & 4 & 2 & $(4\xi^2-3)\xi^2\sin2\phi$ & obl. secondary astigmatism \\ [1ex] 
 \hline
\end{tabular}
\vspace{-1.5mm}
\caption{The first 12 Zernike polynomials.}
\label{tab:zernike}
\vspace{-5mm}
\end{table}

\begin{figure}[H]
    \centering
    \includegraphics[width=0.85\linewidth]{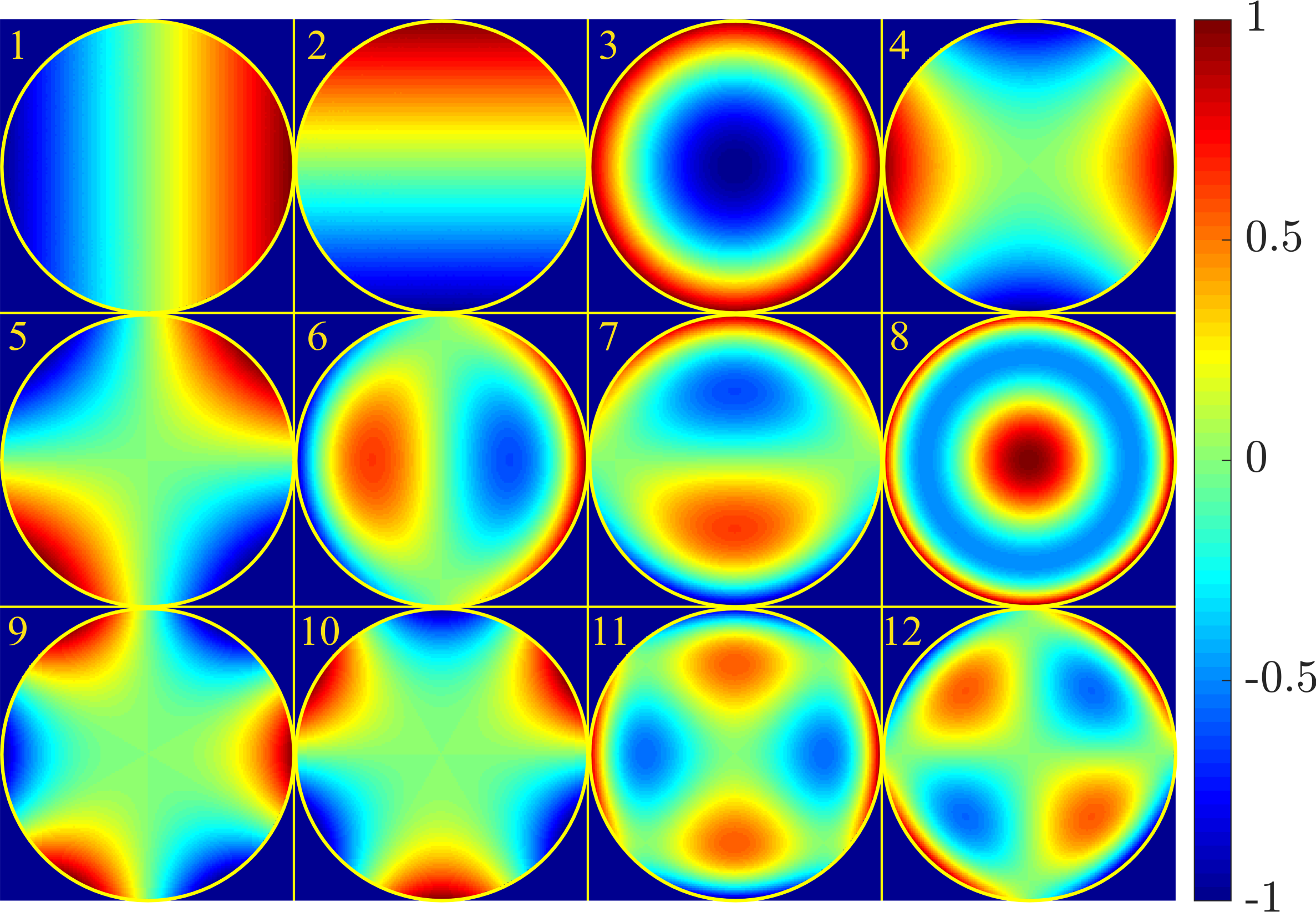}
    \caption{Density plots of the first twelve Zernike polynomials as presented in table~\ref{tab:zernike}: (1) horizontal or $x$ tilt; (2) vertical or $y$ tilt; (3) defocus; (4) vertical astigmatism; (5) oblique astigmatism; (6) horizontal coma; (7) vertical coma; (8) primary spherical aberration; (9) oblique trefoil; (10) vertical trefoil; (11) vertical secondary astigmatism; and (12) oblique secondary astigmatism.}
    \label{fig:psfwidefieldzernike}
    \vspace{-4mm}
\end{figure}

\begin{figure}[H]
    \centering
    \includegraphics[width=0.8\linewidth]{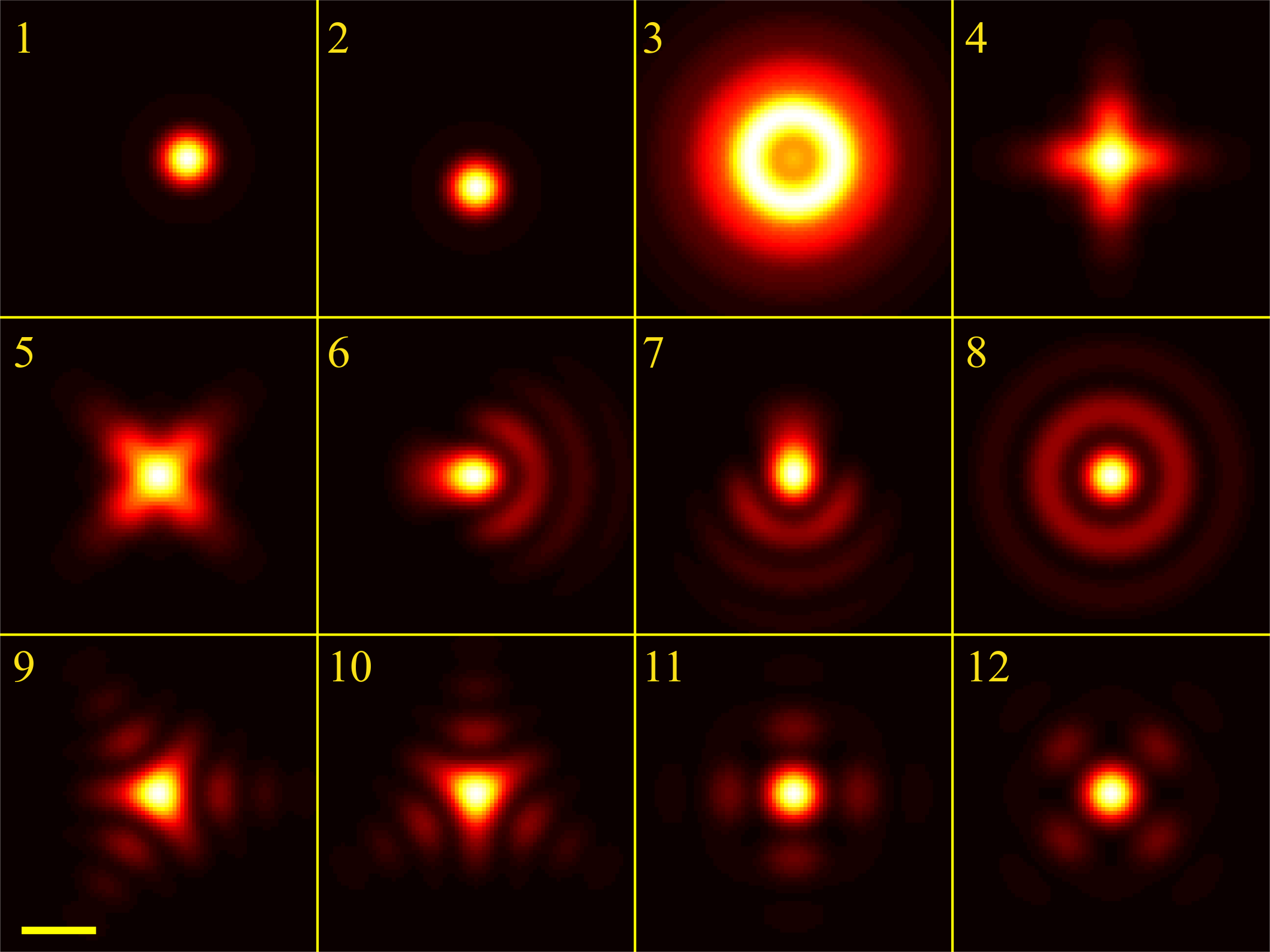}
    \caption{Model calculations of the image of an isotropic emitter (rapidly rotating dipole emitter) aberrated by a phase function given by the Zernike polynomials shown in Fig.~\ref{fig:psfwidefieldzernike}. To better visualize the effects of aberration, all Zernike polynomials were multiplied by a factor 2.5. Calculations were again done for a water immersion objective with NA~=~1.2 and for an emission wavelength of 550~nm. Yellow scale bar is 0.5~$\mu$m.}
    \label{fig:psfwidefieldaberration}
    \vspace{-4mm}
\end{figure}

Figs.~\ref{fig:psfwidefieldzernike}-\ref{fig:psfwidefieldaberration}, respectively, show density plots of the first 12 Zernike polynomials, and their impacts on the PSF for an isotropic emitter. The first three polynomials, namely horizontal tilt, vertical tilt and defocus, coincide with phases due to lateral, vertical and axial shifts in the emitter's position, respectively. All other terms describe PSF distortions due to optical aberrations. 

In some cases, aberrations may not be well described by low order Zernike polynomials. For example, when using Liquid Crystal Spatial Light Modulators (LC-SLM) \cite{moser2019model} or in some PSF engineering methods ~\cite{eliasnehme2021learning}, a sudden phase step in the pupil function may require evaluating the aberration in a pixel-wise manner~\cite{ferdman2020vipr}.

The second kind of aberration is chromatic shift-invariant. In microscopy, it is common to use achromatic objectives though dispersion from various other components inducing PSF deviations is unavoidable. These aberrations originate from the (broad, non-monochromatic) emission spectrum $S\left(\lambda\right)$ of fluorescent molecules describing the probability to emit at a wavelength $\lambda$ often with a width of a few tens of nanometers; see Sec.~\ref{Non_Chem}. In such cases, the image model follows from a super-position integral over the molecule's spectrum
\begin{equation}
    \Lambda(x,y;\mathbf{r}_0) = \int_{\lambda}S\left(\lambda\right)  U\left(x,y;\mathbf{r}_0,\lambda\right)d\lambda,
    \label{eq:psf_mc}
\end{equation}
where $U\left(x,y;\mathbf{r}_0,\lambda\right)$ is the $\lambda$-dependent PSF (as described in Sec.~\ref{ScalarApprox} as a function of $k_0 = 2\pi/\lambda$).
Such aberrations are often detrimental in 3D microscopy. For example in multi-focus microscopy, a phase mask (more details in \ref{subsec:Multiplane_mic}) with custom chromatic correction gratings are designed to correct the chromatic shifts~\cite{abrahamsson2013fast}. 

The most challenging aberrations are shift-variant, both chromatic and monochromatic, which cannot be simply described by the proposed model of Eq.~\ref{eq:PSFscalar_FFT}, as the aberration is now a function of the lateral coordinates, \emph{i.e.}, $\Phi\left(\theta',\phi,x,y\right)$. 
In microscopy, these kinds of aberrations can occur either from the sample itself or from off-axis aberrations in the optical system, namely, systematic aberrations. Sample induced aberrations occur when the sample structure has significant refractive index variations (\emph{e.g.}, imaging in deep tissue). This issue can sometimes be addressed by adaptive optics (AO)~\cite{roddier1999adaptive,booth2002adaptive,ji2010adaptive,ji2010adaptive,tao2011adaptive,gould2012adaptive,ji2017adaptive,liu2018observing}. Typically,  in AO techniques, the wavefront distortion (due to aberrations) of light from fluorescent markers embedded within the sample, called guide stars, is measured and then used for wavefront correction, using deformable mirrors to remove the aberrations and achieve a flat wavefront.
Off-axis aberrations often caused by the optical system, rather than by the sample itself, are typically easier to model as they tend to vary more smoothly. 
These aberrations can be modeled as 2D polynomial coefficients over the FOV~\cite{shajkofci2020spatially} (which multiply Zernike coefficients for example) or addressed by Nodal Aberration Theory ~\cite{shack1980influence}. 

\vspace{-4mm}
\section{Fluorescence microscopy: modalities \label{sec:modality}} \vspace{-3mm}
In the previous section, we described the fundamental optics of the wide-field microscope and derived its OTF and PSF. We also tied the lack of optical sectioning in wide-field microscopes to OTF's missing cone; see Fig.~\ref{fig:psfderivation}. Here, we turn to different fluorescence microscopy modalities achieving optical sectioning and higher resolutions, \textit{i.e.}, near-field; point scanning; SIM; light-sheet; and multi-plane. In deriving their OTFs, we show that these modalities accomplish optical sectioning by collecting more spatial frequencies along the axial direction through either modification to the illumination and/or detection arms. 

\vspace{-5mm}
\subsection{\label{Fluo_Wide}Near-field methods for enhanced axial resolution} \vspace{-3mm}

Here, we turn to fluorescence imaging methods improving \emph{axial} resolution using near-field effects. Electromagnetic near-fields are non-propagating (evanescent) fields with intensity gradients exceeding those of propagating waves.   

\vspace{-6mm}
\subsubsection{Total internal reflection fluorescence microscopy} \vspace{-2mm}

The first method discussed leverages TIR occurring when a plane wave is incident on an interface separating two media with different refractive indices. 

We begin with Fresnel's reflection and transmission coefficients $r_\perp$, $r_\parallel$, $t_\perp$, and $t_\parallel$ for $s$- and $p$-polarized plane waves reflected at an interface dividing a medium with refractive index $n_1$ (incidence medium) from a medium with refractive index $n_2$, given compactly as follows \cite{novotny2012principles}
\begin{equation}
    \begin{split}
        r_\perp &= \frac{n_{\star}^2-w_{\star}}{n_{\star}^2+w_{\star}}, \qquad r_\parallel = \frac{1-w_{\star}}{1+w_{\star}}, \\
        t_\perp &= \frac{2n_{\star}}{n_{\star}^2+w_{\star}}, \qquad t_\parallel = \frac{2}{1+w_{\star}},
    \end{split}
    \label{eq:fresnelsingleinterface}
\end{equation}
where we have used the abbreviations $n_{\star}=n_2/n_1$ and $w_{\star}=w_2/w_1 = \sqrt{(n_2^2-q^2)/(n_1^2-q^2)}$ defining $w_{1,2}$ as the wave vector's axial components in the first and second media, respectively. Moreover, $q=2\pi n_1 \sin\theta_\mathrm{inc}/\lambda$ is the length of the wave vector's lateral component with $\theta_\mathrm{inc}$ being its incidence angle upon the interface with respect to the normal to the interface within the first medium.
Here, it is convenient to work in a unit system where the length of the vacuum wave vector is unity. In this unit system, we have $q=n_1 \sin\theta_\mathrm{inc}$. 

Now, as electric field and wave vectors are perpendicular, the electric field amplitude of the transmitted wave reads \vspace{-3mm}
\begin{equation}
    \begin{split}
        \mathbf{E}_{\perp,\parallel} = E_0 t_{\perp,\parallel} \left(-\frac{w_2 \hat{\mathbf{q}} + q \hat{\mathbf{z}}}{n_2}\right) \exp\left[i w_2 z + i \mathbf{q}\cdot\boldsymbol{\rho}\right],
    \end{split}
    \label{eq:TIRplanewave}
\end{equation}
where $E_0$ is the amplitude of the incident field, with $\hat{\mathbf{q}}$ and $\hat{\mathbf{z}}$ unit vectors along the lateral wave vector component parallel to the interface and along the axial ($z$) direction perpendicular to the interface, respectively.

As can be seen from definitions of $w_{\star}$ following Eq.~\ref{eq:fresnelsingleinterface}, for $q=n_1 \sin\theta_\mathrm{inc}>n_2$, the axial component $w_2$ becomes purely imaginary and the \emph{absolute} values of the reflection coefficients in Eq.~\ref{eq:fresnelsingleinterface} both become unity. Here, TIR is possible only if $n_1>n_2$, and becomes manifest when the critical incidence angle (TIR angle) is $\theta_\mathrm{TIR} = \arcsin(n_2/n_1)$. However, as can be seen from Eq.~\ref{eq:TIRplanewave}, the electric field in medium 2 does not instantly go to zero but decays exponentially with increasing distance $z$ from the interface. This decaying field in the second medium is termed evanescent field or wave. The characteristic decay length $d_\mathrm{TIR}$ of the electric field intensity can be directly derived from Eq.~\ref{eq:TIRplanewave} and reads \vspace{-4.5mm}
\begin{equation}
    \begin{split}
        d_\mathrm{TIR} = \frac{1}{2\left\vert w_2\right\vert} = \frac{1}{2 \sqrt{n_1^2 \sin^2\theta_\mathrm{inc} - n_2^2}}.
    \end{split}
    \label{eq:TIRlength}
\end{equation}
As such, although evanescent waves do not penetrate far within medium 2, they can still be used to excite fluorophores within a distance of $d_\mathrm{TIR}$ from the surface, {\it e.g.}, in TIRF microscopy~\cite{axelrod1981JCB}. By the same token, (out-of-focus) fluorophores deeper than $d_{\mathrm{TIR}}$ are less likely to become excited decreasing undesired out-of-focus light.  

\begin{figure}[H]
    \centering
    \includegraphics[width=0.95\linewidth]{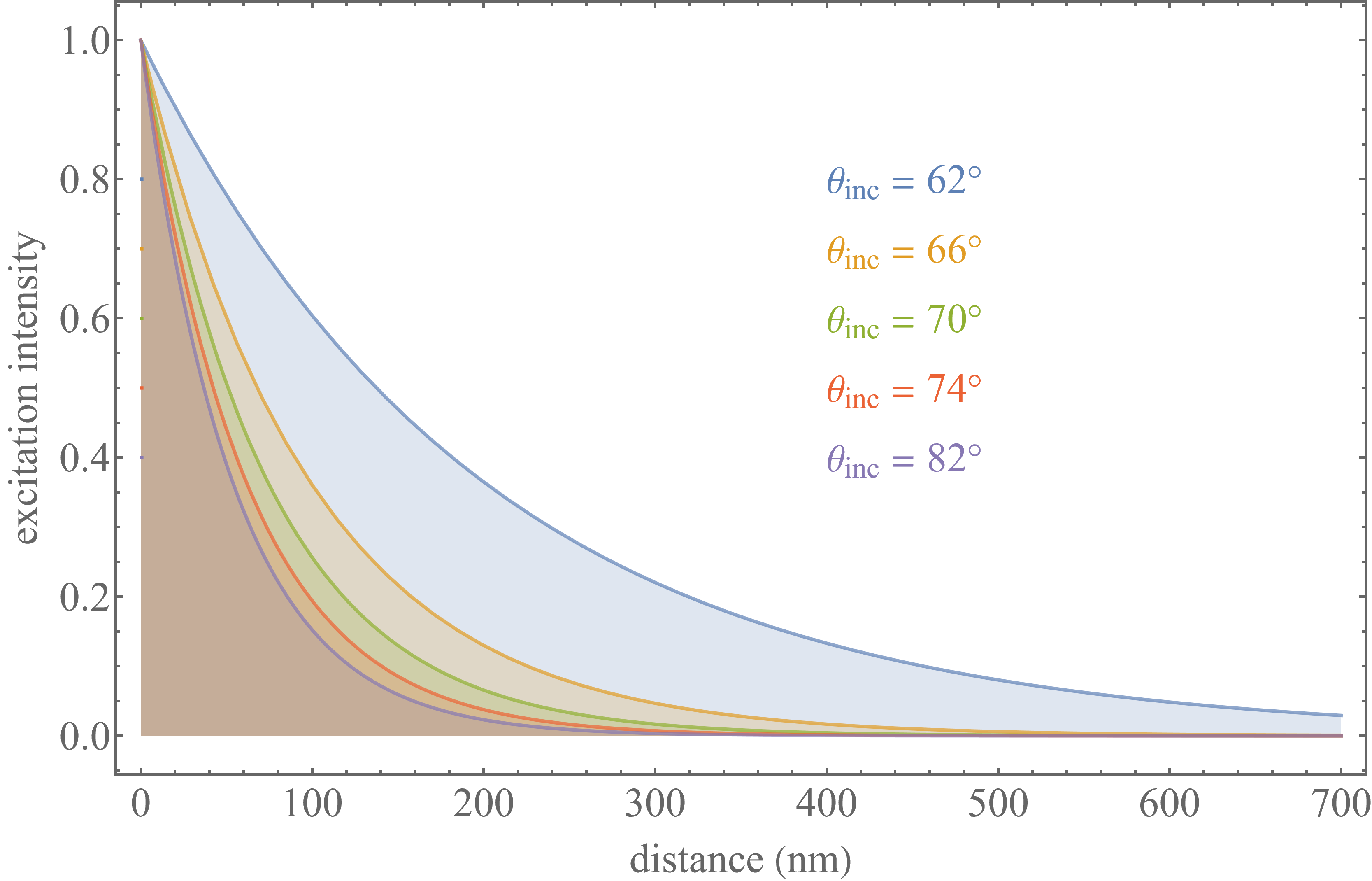}
    \vspace{-1mm}
    \caption{Total Internal Reflection Fluorescence (TIRF) microscopy. Excitation intensity above a coverslide interface with the sample medium as a function of incidence angle. The sample solution and coverslide refractive indices are, respectively, 
    1.33 (water) and 1.52, resulting in a TIR critical angle of $\approx 61^\circ$. The excitation wavelength is taken as 470~nm.}
    \label{fig:VariableAngleTIRF}
    \vspace{-4mm}
\end{figure}

To decode emitter axial location, variable angle TIRF (vaTIRF) ~\cite{stock2003variable} is used where several images are recorded at differing incidence angles of the excitation plane wave above the TIR angle. For increasing incidence angles (see Fig.~\ref{fig:VariableAngleTIRF}), the excitation intensity's decay becomes steeper. The variation in emitter brightness values across incidence angles is then used to assess its distance from the interface upon deconvolution ~\cite{saffarian2008differential,el2021advanced}
with an axial resolution in some cases down to a few nanometers, \textit{i.e.}, by \emph{ca.} 2-3 orders of magnitude better than the diffraction-limited resolution of a confocal microscope albeit within a limited range ($\approx d_{\mathrm{TIR}}$) from the interface.

\vspace{-8mm}
\subsubsection{Super-critical fluorescence microscopy \label{sec:SAF}} \vspace{-2mm}

The second near-field method discussed is SAF microscopy. This method employs the coupling of a fluorophore's near-field emission into propagating modes in the coverslide's glass to improve axial resolution ~\cite{enderlein1999highly,ruckstuhl2004supercritical,winterflood2010nanometer,deschamps20143d,oheim2020supercritical,dasgupta2021direct}. To be precise, fields due to an oscillating electric dipole have components decaying as $1/r,\,1/r^2$ and $1/r^3$ where only the first term coincides with the propagating term. The two other terms are non-propagating and represent near-field emissions decaying on short distances ($\approx \lambda$). However, when the electric dipole is located close to a coverslide's interface, non-propagating near-field dipolar components are converted into propagating modes upon coupling into the glass which can be then collected and imaged by the microscope objective. These modes can be decomposed into a super-position of plane waves traveling along directions \emph{above} the critical TIR angle for the given emission wavelength (super-critical angle fluorescence or SAF emission). The coupling of near-field modes of the fluorophore into propagating modes in the glass decrease with increasing distance from the interface. In contrast, the emission into the angle below the TIR angle (under-critical angle fluorescence or UAF emission) is due to the propagation of the emitter's far-field emission into the glass and does not depend on its distance from the surface. Thus, at its core, SAF microscopy leverages the variation in SAF to estimate the distance of an emitter from the coverslide's interface by measuring the ratio of its SAF to SAF+UAF emission intensity. 

\begin{figure}[H]
    \centering
    \includegraphics[width=0.95\linewidth]{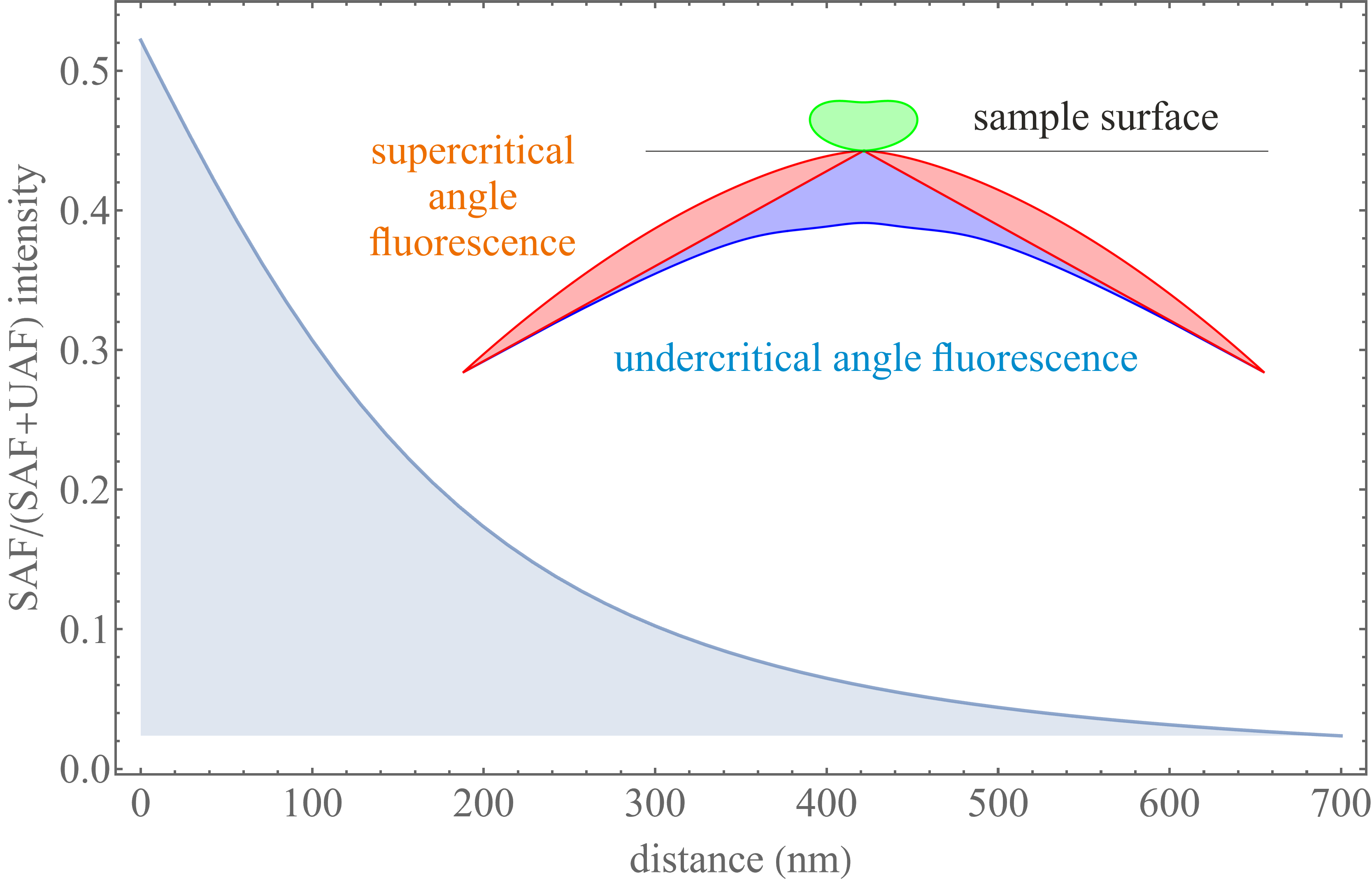}
    \vspace{-1mm}
    \caption{Super-critical Angle Fluorescence (SAF) microscopy. Ratio of super-critical to total downward fluorescence emission for a rapidly rotating molecule as a function of distance from the interface of the coverslide and the sample medium. The refractive indices of the sample solution and coverslide are, respectively, assumed to be 1.33 (water) and 1.52 (glass), with the emission wavelength of 550~nm. The inset shows the angular emission intensity distribution of an emitter directly on the interface (with the blue, red and green curves denoting UAF and SAF emissions, and emission towards sample solution, respectively). The SAF emission strongly depends on the emitter's distance to the interface, while the under-critical emission is independent of emitter axial position. By determining the ratio of SAF to SAF+UAF emission, we can find the axial position of an emitter.}
    \label{fig:SAFMicroscopy}
    \vspace{-8mm}
\end{figure}

To calculate the ratio of super-to under-critical angle emissions, we  use the theoretical framework developed in Sec.~\ref{fluorescence}. In particular, for calculating SAF emission intensity, we use Eqs.~\ref{eq:Efield} and \ref{eq:Bfield}, but with integration boundaries from $\theta' = \arcsin\left(n\sin\theta_\mathrm{TIR}/\mathcal{M}\right)$, dictated by the critical TIR angle, to $\theta' = \Theta'$, dictated by the numerical aperture. We then compute the energy flux density distribution from Eq.~\ref{eq:PSFdefinition}. The integral of the resulting energy flux density over the $xy$-plane is then proportional to the detectable SAF intensity. The UAF intensity is computed analogously but with integration boundaries from $\theta' = 0$ to $\theta' = \arcsin\left(n\theta_\mathrm{TIR}/\mathcal{M}\right)$. As an example, Fig.~\ref{fig:SAFMicroscopy} shows the SAF to SAF+UAF ratio for a glass-water interface as a function of distance, assuming an isotropic emitter with emission wavelength of 550~nm. As can be seen, the dynamic range over which one can use this ratio in determining the emitter's distance from the surface is very similar to the dynamic range over which vaTIRF is applicable; see Fig.~\ref{fig:VariableAngleTIRF}.

\subsubsection{Metal-induced energy transfer imaging}
\vspace{-5mm}

\begin{figure}[H]
    \centering
    \includegraphics[width=0.95\linewidth]{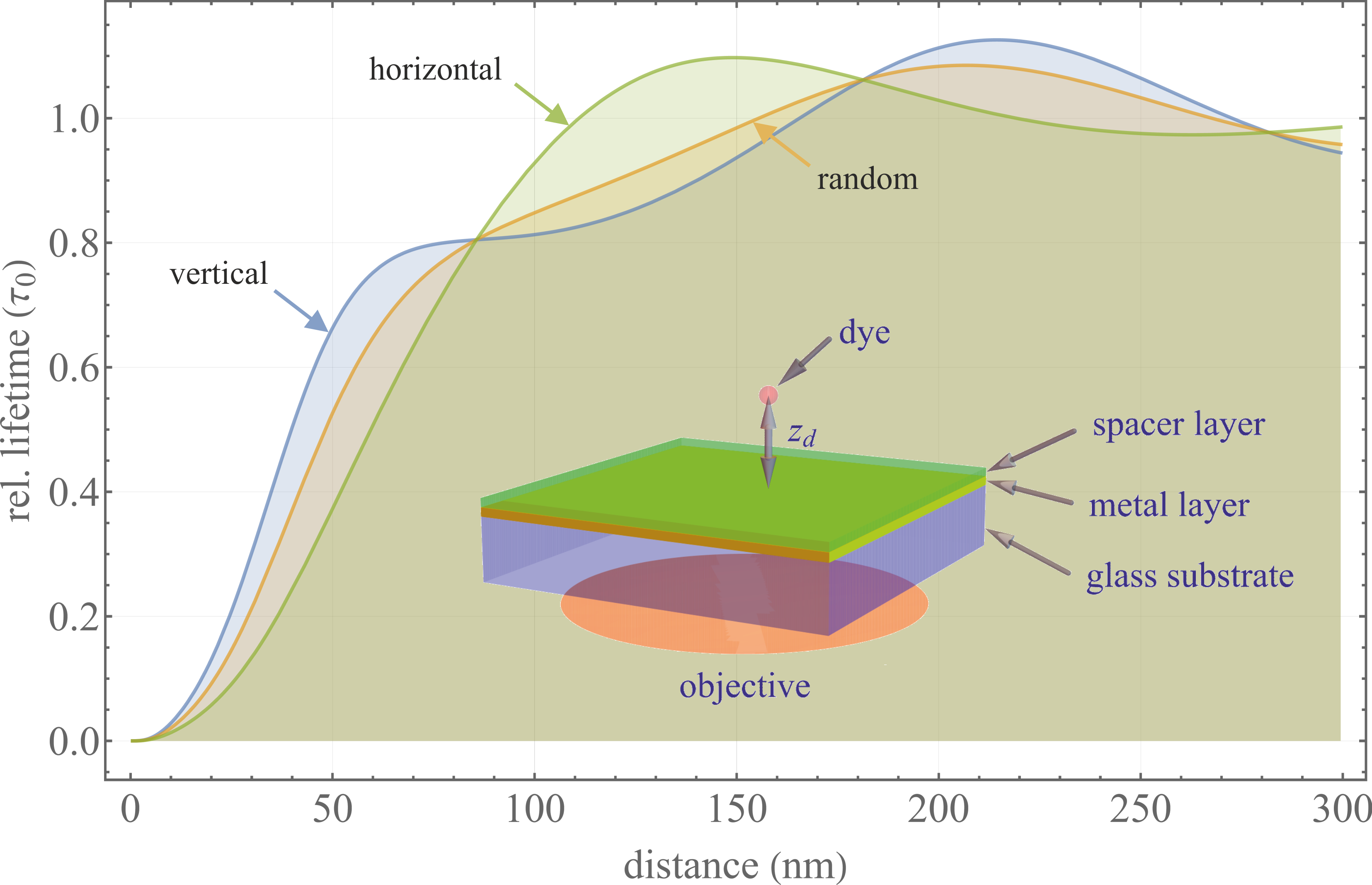}
    \vspace{-1mm}
    \caption{Metal-Induced Energy Transfer (MIET) microscopy: Dependence of the fluorescence lifetime (in terms of free space lifetime $\tau_0$) on the emitter's distance from the glass substrate (coverslide) coated with a 20~nm gold layer. Calculations were done for an emission wavelength of 550~nm, and for a unit fluorescence quantum yield. Here we show the free curves for vertical, horizontal, and random emission dipole orientations. The inset illustrates the MIET sample geometry.}
    \label{fig:MIET}
    \vspace{-3mm}
\end{figure}

MIET, another near-field method used for axial localization~\cite{chizhik2014metal}, is based on near-field coupling similar to SAF microscopy. MIET uses the fact that when a fluorescent emitter (electric dipole emitter) approaches a metal layer, its electric near-field excites surface plasmons (coherent metal electron oscillations) in the metal, accelerating de-excitation of fluorescent emitter's excited state. This is observed as a strong decrease in fluorescence lifetime with decreasing distance from the surface; see Fig.~\ref{fig:MIET} and Eq.~\ref{eq:lifetime}.

To infer distances from lifetime measurements, we use the theoretical framework developed in Sec.~\ref{fluorescence}. Briefly lifetime depends on the emission power requiring the explicit calculation of both electric and magnetic fields.

\begin{figure}[H]
    \centering
    \includegraphics[width=0.75\linewidth]{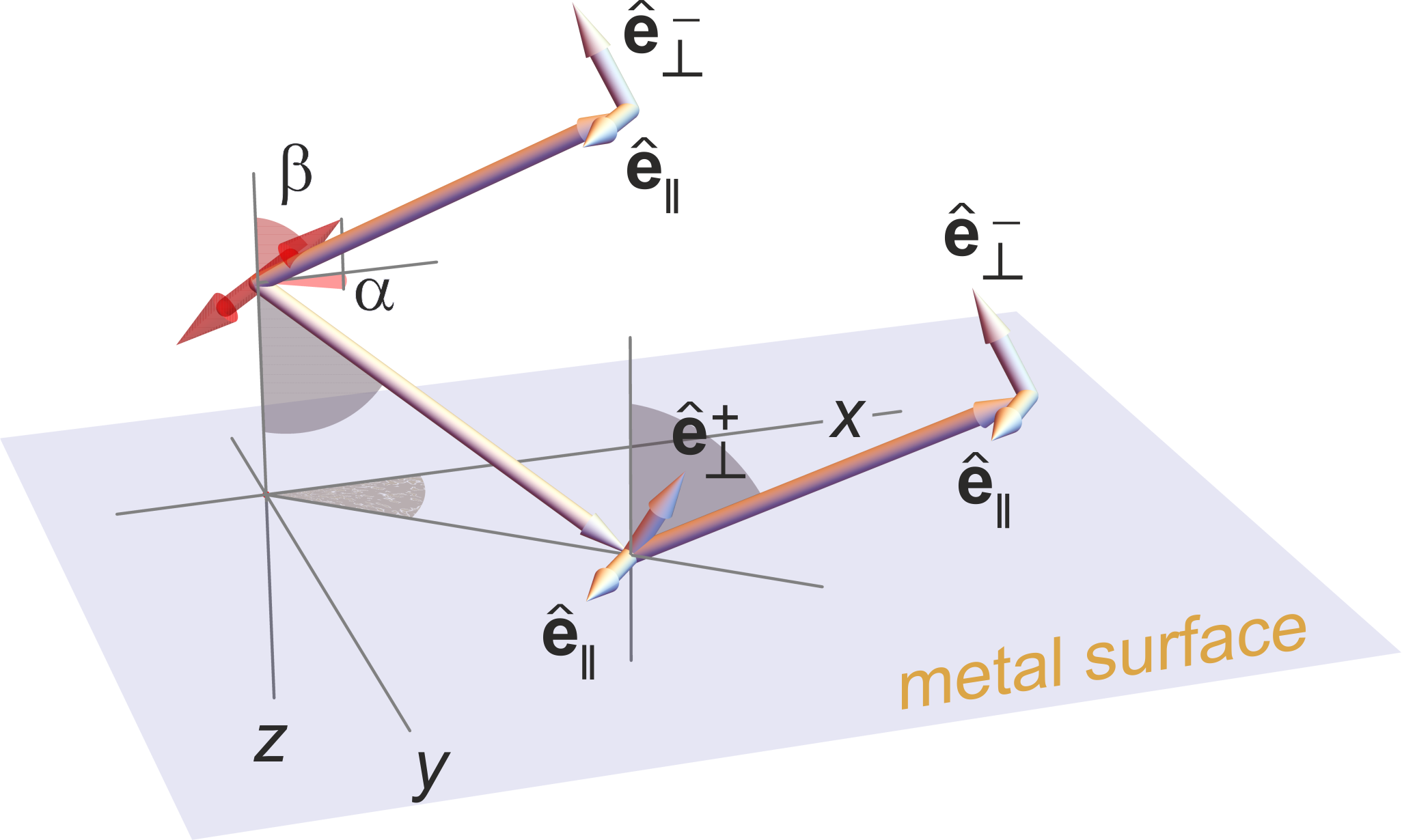}
    \caption{Geometry for deriving the electric field generated by a single dipole emitter above the MIET substrate (metal surface). The red double headed arrow shows a dipole located a distance $z_d$ above the metal surface with an orientation of $\beta$ and $\alpha$ denoting polar and inclination (azimuthal) angles, respectively. The three longer single-headed arrows show plane wave component vectors, with corresponding perpendicular polarization unit vectors $\hat{\mathbf{e}}_{\parallel}$ and $\hat{\mathbf{e}}^{\pm}_{\perp}$. Here $\hat{\mathbf{e}}^{+}_{\perp}$ is the unit vector associated with the wave vector moving toward the metal surface. Similar conventions hold for the other unit vectors.
    }
    \label{fig:DipoleMIET} \vspace{-5mm}
\end{figure}

We start from the Weyl representation of the electric field of a free dipole emitter obtained in Eq.~\ref{eq:DipoleE0} to derive the electric field distribution above a MIET substrate (denoted by a metal surface in Fig.~\ref{fig:DipoleMIET}). As shown in Fig.~\ref{fig:DipoleMIET}, two sources contribute to the electric field above this metal surface: 1) direct emission from the dipole; and 2) emission reflected from the surface (\textit{i.e.}, emission from the emitter's image) \vspace{-2mm}
\begin{equation}
    \begin{split}
        \mathbf{E}^\pm_d = &\frac{i k_0^2}{2\pi}\int \frac{d^2\mathbf{q}}{w_d} \Big[\left(\mathbf{p}\cdot\hat{\mathbf{e}}_\parallel\right)\hat{\mathbf{e}}_{\parallel} \left(1+r_\parallel e^{i w_d \left\vert z+z_d \right\vert}\right)  \\
        &+ \left(\left(\mathbf{p}\cdot\hat{\mathbf{e}}^\pm_{\perp}\right)\hat{\mathbf{e}}^\pm_{\perp}+\left(\mathbf{p}\cdot\hat{\mathbf{e}}^+_{\perp}\right)\hat{\mathbf{e}}^-_{\perp} r_\perp e^{i w_d \left\vert z+z_d \right\vert}\right) \Big]\\ &\exp\left[i\mathbf{q}\cdot\left(\boldsymbol{\rho}-\boldsymbol{\rho}_d\right) + i w_d \left\vert z-z_d \right\vert \right],
    \end{split}
    \label{eq:DipoleEmiet}
\end{equation}
where terms with the reflection coefficients $r_{\parallel,\perp}$ describe contributions from the reflected emission. Moreover, the superscripts ``+'' and ``$-$'' refer to plane waves moving towards and away from the metal surface.
The $r_{\perp,\parallel}$ are Fresnel's $q$-dependent reflection coefficients for $p$- and $s$-waves for the MIET substrate.  

For planar structures of arbitrary complexity, these coefficients are readily obtained using propagation matrix formalism in Refs. ~\cite[p.~254]{born2013principles} and ~\cite{heavens1965optical,yeh1988optical,knittel1976optics}. Here, we now have to distinguish between two $p$-wave polarization unit vectors: $\hat{\mathbf{e}}^+_\perp$ for plane waves traveling towards the substrate, and $\hat{\mathbf{e}}^-_\perp$ for plane waves traveling away from the substrate. The corresponding $s$-waves polarization unit vector $\hat{\mathbf{e}}_\parallel$ is the same for both waves. We note that the result depends on the three-dimensional orientation of the emitter (given by the Euler angles $\alpha$ and $\beta$, see Fig.~\ref{fig:DipoleMIET}) via the scalar products $\mathbf{p}\cdot\hat{\mathbf{e}}^\pm_\perp$ and $\mathbf{p}\cdot\hat{\mathbf{e}}_\parallel$. 

Analogously, we can find the magnetic field as \vspace{-1.5mm}
\begin{equation}
    \begin{split}
        \mathbf{B}^\pm_d = &\frac{i n_d k_0^2}{2\pi}\int \frac{d^2\mathbf{q}}{w_d}\\ &\Big[\left(\mathbf{p}\cdot\hat{\mathbf{e}}_\parallel\right) \left(\hat{\mathbf{e}}^\pm_{\perp}-\hat{\mathbf{e}}^+_{\perp} r_\parallel e^{i w_d \left\vert z+z_d \right\vert}\right)  \\
        &+ \left(\left(\mathbf{p}\cdot\hat{\mathbf{e}}^\pm_{\perp}\right)+\left(\mathbf{p}\cdot\hat{\mathbf{e}}^+_{\perp}\right) r_\perp e^{i w_d \left\vert z+z_d \right\vert}\right) \hat{\mathbf{e}}_{\parallel} \Big]\\ &\exp\left[i\mathbf{q}\cdot\left(\boldsymbol{\rho}-\boldsymbol{\rho}_d\right) + i w_d \left\vert z-z_d \right\vert \right].
    \end{split}
    \label{eq:DipoleBmiet}
\end{equation}

Now, given both electric and magnetic fields of Eqs.~\ref{eq:DipoleEmiet}-\ref{eq:DipoleBmiet}, the \emph{total} emission power, designated by $S(\beta)$, of the emitter follows by integrating the outwards component of the Poynting vector over two planar interfaces sandwiching the emitter 
\begin{align}
        S(\beta) = &\frac{n_d c}{8\pi}\int d^2\boldsymbol{\rho} \nonumber \\ &\hat{\mathbf{z}}\cdot\left[\left(\mathbf{E}^+\times\mathbf{B}^{+\ast}\right)_{z=0}- \left(\mathbf{E}^-\times\mathbf{B}^{-\ast}\right)_{z<z_d}\right].
    \label{eq:DipoleFlux}
\end{align}
The emission power depends only on the dipole's polar orientation angle $\beta$, and not its azimuthal angle $\alpha$. The emission power $S(\beta)$ can now be compared to the emission power $S_0$ of a ``free'' dipole within a homogeneous medium with refractive index $n_d$, given by the well-known formula in Ref.~\cite[p.~410]{jackson1999classical} (also be obtained from the above equations by neglecting the contribution from reflected emission including coefficients $r_{\perp,\parallel}$) as $S_0 = c n_d p^2 k_0^4/3$.
   
The observable enhancement of the radiative de-excitation rate $k_f$ of a fluorescence emitter due to the presence of the metal substrate with respect to the same emitter in a homogeneous environment is then given by the ratio $S(\beta)/S_0$~\cite{Chance}. 

As we recall from Sec.~\ref{Nonlinear}, there is a contribution to the excited state lifetime from non-radiative decay pathways arising by collision with surrounding molecules and thermal dissipation of the excited state energy quantified by the fluorescence quantum yield, $Q_f$. Here $Q_f$ is the probability that de-excitation proceeds radiatively with photon emission; see Eq.~\ref{eq:Q}. The observable fluorescence lifetime $\tau$ is then the inverse of the total de-excitation rate $k_f+k_{\text{non}}$ (see Eq.~\ref{eq:lifetime}), such that its change in the presence of the metal substrate is given by 
\begin{equation}
        \frac{\tau}{\tau_0} = \frac{S_0}{S(\beta) Q_f + (1-Q_f) S_0}.
    \label{eq:MIETlifetime}
\end{equation}
This is the final equation needed for calculating the dependence of fluorescence lifetime $\tau$ on emitter distance $z_d$. An example is provided in Fig.~\ref{fig:MIET} for the three cases of a vertically, horizontally, and randomly oriented emitter. In the latter case, the orientation-dependent $S(\beta)$ is substituted for its orientational average $\left<S\right> = (1/2) \int_0^\pi d\beta \sin\beta S(\beta)$.  As seen from Fig.~\ref{fig:MIET} for a randomly oriented emitter, within a range of up to 200~nm from the surface, the lifetime depends monotonically with distance and a unique distance follows from the measured lifetime. 

Further recent \r{A}ngstr\"om spatial resolution along the optical axis has been afforded by the use of materials such as Indium Tin Oxide (ITO)~\cite{moerland2016subnanometer} or single-sheet graphene (graphene induced energy transfer or GIET)~\cite{ghosh2019graphene}, leading to a distance-dependent modulation of the fluorescence lifetime on a {\it ca.} eight times smaller length scale.
\vspace{-10mm}
\subsection{\label{Fluo_Point}Point scanning microscopy} \vspace{-2mm}

Unlike wide-field imaging using multi-pixel detectors, point scanning microscopes sequentially record images by scanning samples over a set of positions and recording fluorescence signal from each position scanned. Moreover, in contrast to wide-field imaging, point scanning allows for out-of-focus light reduction thereby achieving optical sectioning. Here, we first consider image formation in the most widely used point scanning microscope: the Confocal Laser Scanning Microscope or CLSM~\cite{marvin1961USPatent,pawley2006handbook}. We then discuss enhanced-resolution achieved by ISM, 4pi, and two-photon microscopy.    

\vspace{-6mm}
\subsubsection{\label{CLSM-Confocal} Confocal laser scanning microscopy } \vspace{-2mm}

\begin{figure}[H]
    \centering
    \includegraphics[width=0.95\linewidth]{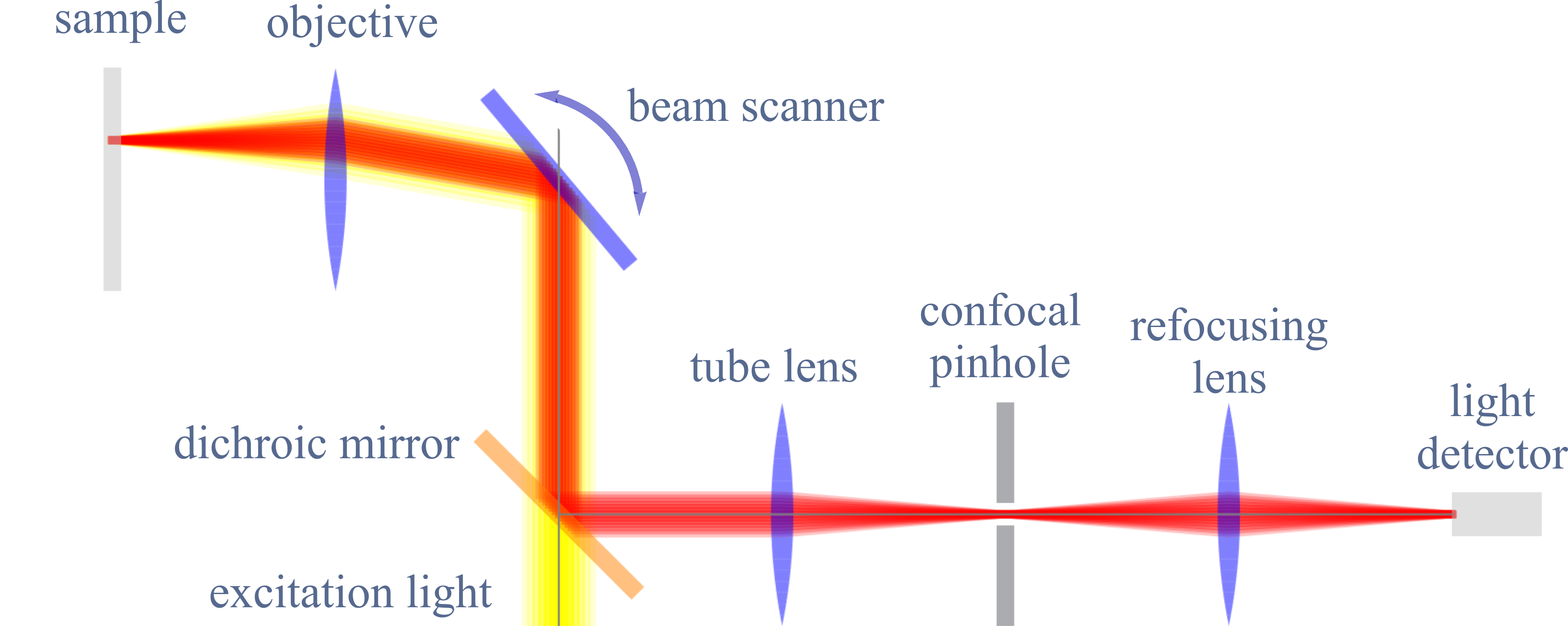}
    \caption{
    Schematic of a CLSM. 
    Yellow and red beams, respectively, show the excitation and emission light. Emission passes through a confocal pinhole suppressing out-of-focus light; see details in text.
    }
    \label{fig:CLSM} \vspace{-4mm}
\end{figure}

A schematic of a point scanning microscope is shown in Fig.~\ref{fig:CLSM}. An excitation laser beam, in yellow, is laterally deflected by a beam scanning unit along both directions perpendicular to the optical axis. Fig.~\ref{fig:CLSM} shows only one of these scanning directions where the excitation beam can be directed up and down upon reflection from the scanner by adjusting the scanner's orientation. Following deflection, the excitation light is focused by the objective into a diffraction-limited focus within the sample. The emitted fluorescence light from the illuminated spot, shown in red, is then collected by the same objective and guided back through the same beam scanner towards the dichroic mirror. This process is known as de-scanning. 

After de-scanning, fluorescence light is reflected away from the excitation beam by the dichroic mirror, which only reflects light within a range of  wavelengths. The fluorescent light is next focused by the tube lens onto the circular aperture of a confocal pinhole obstructing the undesired fluorescent light from out-of-focus fluorophores. After potentially passing additional optical filters for background suppression, the fluorescence light is refocused onto a single-pixel point detector to record the in-focus fluorescence intensity.

In what follows, we derive the confocal PSF (for a single scanning spot). To avoid notational confusion, PSFs for the wide-field and CLSM are, respectively, denoted by $U_{\mathrm{wf}}$ and $U_{\mathrm{cf}}$ for the remainder of this section.

To derive the confocal PSF for an isolated emitter sitting in an excitation focal spot in sample space, we first consider major differences with the wide-field setup (described for the most general case and its approximate analytical forms in Secs.~\ref{Fluo_PSF} and \ref{Dipole}; see Eqs.~\ref{eq:PSFscalar} and \ref{eq:PSFGaussian}). These differences include: 1) the spot illumination procedure; and 2) the existence of the confocal pinhole. 

We start from the fluorescent light from the emitter proportional to the three-dimensional excitation laser intensity at the focal spot 
$I_\mathrm{ex}(\boldsymbol{\rho},z)$ (excitation is in the sample space and thus described by non-prime coordinates). The fluorescent light is, in turn, collected by the objective and focused onto the confocal pinhole (within image space). This results in a fluorescent intensity $U_{\mathrm{wf}}I_{\mathrm{ex}}$ prior to the pinhole where $U_{\mathrm{wf}}$ is this setup's wide-field PSF in the absence of the pinhole and spot illumination. In the end, the confocal PSF (imaging PSF of a confocal microscope) is proportional to the fluorescence intensity (ignoring all constant prefactors) following the pinhole 
\begin{align}
        U_\mathrm{cf}(\boldsymbol{\rho},z) & \propto \left[A\otimes U_{\mathrm{wf}}\right] I_\mathrm{ex}(\boldsymbol{\rho},z) \nonumber \\
        & = 
        \int d\boldsymbol{\rho}' A(\boldsymbol{\rho}') U_\mathrm{wf}(\boldsymbol{\rho}'-\boldsymbol{\rho},z) I_\mathrm{ex}(\boldsymbol{\rho},z),
    \label{eq:PSFCLSM}
\end{align}
where $A$ captures the confocal pinhole, set to unity for $\rho'=\vert\boldsymbol{\rho}'\vert$ smaller than the aperture radius $a$, and zero otherwise. Here, $U_\mathrm{wf}(\boldsymbol{\rho}'-\boldsymbol{\rho},z)$ represents the wide-field PSF when imaging the fluorescence from an emitter at position $\mathbf{r}=(\boldsymbol{\rho},z)$ in sample space onto lateral position $\boldsymbol{\rho}'$ in the plane of the confocal aperture within the image space (prime coordinates). Put differently, the confocal PSF of Eq.~\ref{eq:PSFCLSM} is given as a product of: $A\otimes U_{\mathrm{wf}}$ describing the detection, sometimes termed detection PSF; and $I_{\mathrm{ex}}$ describing excitation, sometimes termed excitation PSF. 

The integral in Eq.~\ref{eq:PSFCLSM} is performed over the whole $\boldsymbol{\rho}'$-plane. The excitation PSF (excitation intensity distribution), $I_{\mathrm{ex}}$, entering the above equation is itself a function of the absorption dipole orientation $\mathbf{p}_\mathrm{ex}$ of a fluorophore via $I_\mathrm{ex}(\mathbf{r}) \propto \left\vert \mathbf{E}_\mathrm{ex}(\mathbf{r})\cdot\mathbf{p}_\mathrm{ex}\right\vert^2$, where $\mathbf{E}_\mathrm{ex}$ denotes the electric field distribution in the focal spot. 

In most cases of practical interest, one deals with rapidly rotating emitters for which the orientationally averaged excitation intensity 
reads (also see Eq.~\ref{eq:PSFarbitraryorientation})
\begin{equation}
    \begin{split}
        I_\mathrm{ex}(\mathbf{r}) \propto \left\vert E_{\mathrm{ex},x}\right\vert^2+\left\vert E_{\mathrm{ex},y}\right\vert^2 + \left\vert E_{\mathrm{ex},z}\right\vert^2.
    \end{split}
    \label{eq:Iexaverage}
\end{equation}

\begin{figure}[H]
    \centering
    \includegraphics[width=0.6\linewidth]{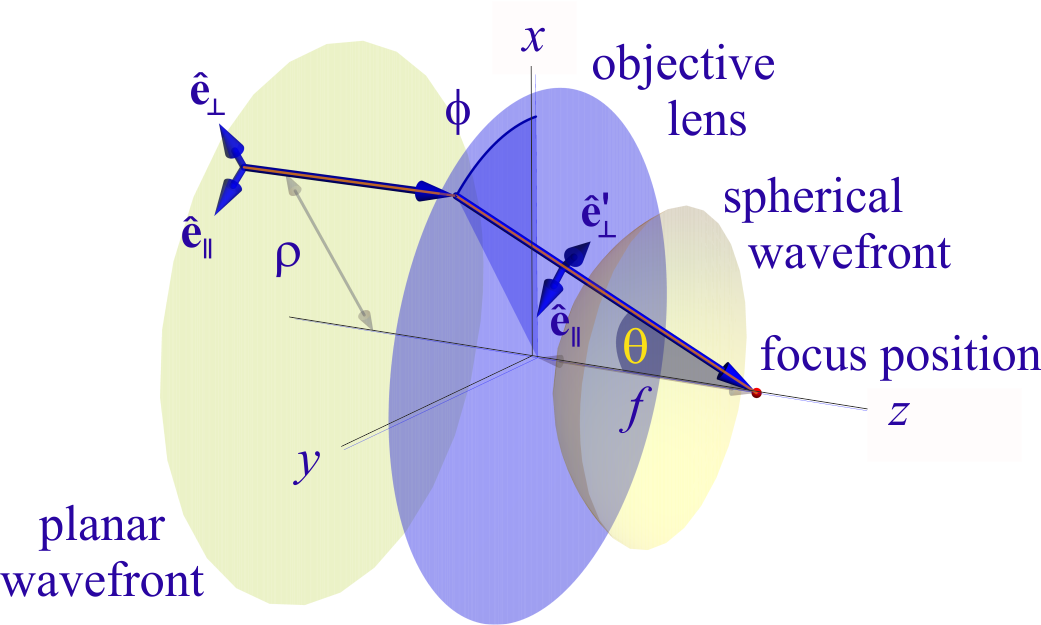}
    \caption{Schematic of the geometry of focusing a planar laser wavefront through the objective into the sample space; see Fig.~\ref{fig:CLSM}. Wavefront patches at distance $\rho$ from the optical axis in the back focal plane are converted into spherical wavefront patches traveling at angle $\theta=\arcsin\left(\rho/nf\right)$ with respect to the optical axis $z$, where $f$ is the focal length of the objective lens; see details in the main text. 
    }
    \label{fig:FocusingThroughObjective}
    \vspace{-4.5mm}
\end{figure}

To perform this calculation, we first consider the focusing of a planar wavefront through the objective into a diffraction-limited spot; see Fig.~\ref{fig:FocusingThroughObjective}. Similar to Abbe's sine condition relating propagation angles of wavefront patches in sample and image spaces, there is a similar relation between the distance $\rho$ of a patch on the planar wavefront from the optical axis, and the propagation angle $\theta$ of the corresponding patch after focusing through the objective; see Fig.~\ref{fig:FocusingThroughObjective}. This relation can be found from Abbe's sine condition when moving the focus in image space to infinity ({\it i.e.}, the focal length $f_\mathrm{tube}$ of the tube lens tends towards infinity), and remembering that the magnification $\mathcal{M}$ is given by the focal distance of the tube lens $f_\mathrm{tube}$ divided by the focal distance $f$ of the objective; see Fig.~\ref{fig:magnification}. Thus, we find $\mathcal{M}\sin\theta'=(f_\mathrm{tube}/f)\sin\theta'=n\sin\theta$. When increasing the value $f_\mathrm{tube}$ to infinity, the angle $\theta'$ tends to zero, though the product $\rho=f_\mathrm{tube}\sin\theta'$ remains finite and coincides with the distance from the optical axis in the back focal plane. Thus, one finds the relation $\rho = n f \sin\theta$ between the distance $\rho$ before the objective and the propagation angle $\theta$ in sample space. 

\begin{figure}[H]
    \centering
    \includegraphics[width=0.95\linewidth]{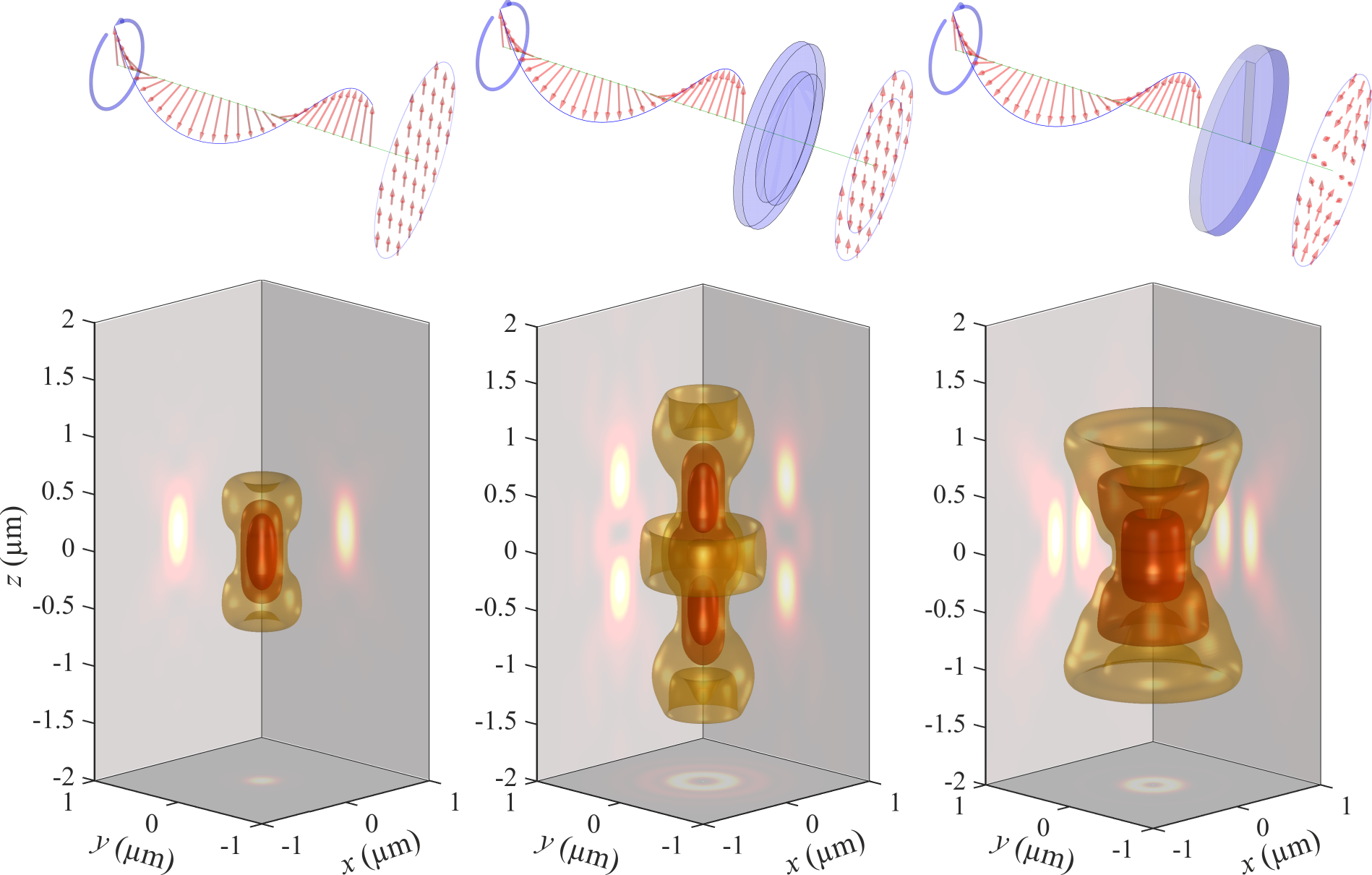}
    \caption{
    CLSM and STED intensity distributions at the focus. Comparison of intensity distribution between conventional CLSM focus (left) with $z$-STED focus (middle) and $xy$-STED focus (right). Calculations were done for water immersion objective with NA~=~1.2 at an excitation wavelength of 470~nm. On top of each column, the excitation polarization and its generating phase plate are shown. Bottom panels show 3D contour plots of the $1/e$, $1/e^2$ and $1/e^3$ intensity iso-surfaces and projections of $xy$-, $xz$-, and $yz$-cross-sections through the center.}
    \label{fig:STEDFocus}
\end{figure}

Using this relation for $\rho$, we can expand the electric field in sample space into a plane wave super-position, similar to what we did in deriving the electric field of a point emitter in image space; see Eq.~\ref{eq:Efield}. When reading Eq.~\ref{eq:Efield} in reverse, {\it i.e.}, replacing all primed by non-primed variables and vice versa (thus starting with light coming from the back side of the objective focused through the objective into sample space), and when taking into account that the angles $\theta'$ for the incoming light are all zero (plane wavefront), so that $\cos\theta'\approx 1$, we arrive at \vspace{-2mm}
\begin{equation}
    \begin{split}
        &\mathbf{E}_\mathrm{ex}(\mathbf{r}) \propto \int_0^{\Theta} d\theta \sin\theta \sqrt{\cos\theta} \int_0^{2\pi} d\phi\\
        & \left[E_{0,\parallel}(\rho,\phi) \hat{\mathbf{e}}_\parallel + E_{0,\perp}(\rho,\phi) \hat{\mathbf{e}}'_\perp\right] \exp\left(i\mathbf{k}_\mathrm{ex}\cdot\mathbf{r}\right),
    \end{split}
    \label{eq:Efocus}
\end{equation}
where $\mathbf{k}_\mathrm{ex}=2\pi n/\lambda_\mathrm{ex}\left(\cos\phi \sin\theta, \sin\phi \sin\theta, \cos\theta\right)$ is now the wave vector of a plane wave with wavelength $\lambda_\mathrm{ex}$ (excitation light wavelength), where the electric field of the incoming laser beam in the back focal plane is expanded into its radially ($E_{0,\perp}$) and azimuthally ($E_{0,\parallel}$) polarized components; see Fig.~\ref{fig:FocusingThroughObjective}. For example, for a linearly polarized laser beam with polarization direction along $x$ one has $E_{0,\perp}\propto\cos\phi$ and $E_{0,\parallel}\propto-\sin\phi$. This equation can now be used in calculating the three-dimensional excitation PSF in sample space. 
As an example, the left panel of Fig.~\ref{fig:STEDFocus} shows the CLSM PSF calculated assuming a 470~nm circularly polarized laser focused through a water immersion objective into a diffraction-limited spot (planar wavefront at the back focal plane). 

While we have focused on using Eq.~\ref{eq:Efocus} in computing the CLSM PSF, this equation is much more general. For instance, it can be used in calculating the intensity distribution of a donut excitation beam appearing in STED microscopy ~\cite{hell1994OpticsLetter}. This donut intensity distribution, with zero intensity on the crossing of the optical axis with the focal plane (focus center), can be generated in two ways. 

The first method generates a donut-shape laser intensity in the focal plane by sending a circularly polarized laser light through a ring-shaped phase plate thicker at its center. This results in retardation of the beams of light closer to the optical axis by half a wavelength with respect to the beams passing through the thinner outer part of the plate; see the central panel in Fig.~\ref{fig:STEDFocus}. A snapshot of the resulting polarization structure across the back focal plane is depicted in the top middle panel in Fig.~\ref{fig:STEDFocus}. Mathematically, this can be described by setting $E_{0,\perp}\propto\cos\phi-i\sin\phi$ and $E_{0,\parallel}\propto-\sin\phi+i\cos\phi$ for $\rho\leq\rho_\Phi$ and the same expressions but with opposite sign for $\rho_\Phi<\rho<f\sin\Theta$, where $\rho_\Phi=f\sin\Theta/\sqrt{2}$ is the radius of the thicker central part of the phase plate. This special choice of $\rho_\Phi$ assures that the total excitation intensity in the focus center is indeed zero. 

The second method sends circularly polarized light through a helical wave plate as shown at the top of the right panel in Fig.~\ref{fig:STEDFocus}. When choosing an appropriate helical pitch, this leads to an excitation beam with polarization structure $E_{0,\perp}\propto\sin 2\phi-i\cos 2\phi$ and $E_{0,\parallel}\propto\cos 2\phi+i\sin 2\phi$. Three-dimensional representations of the resulting STimulated Emission (STE) intensity distributions and  corresponding cross-sections are shown in the bottom panels of Fig.~\ref{fig:STEDFocus}. As can be seen, neither the disk phase plate (middle panel) nor the helical phase plate (right panel) lead to an ideal STE intensity distribution, \textit{i.e.}, perfect donut shape with zero intensity at the middle. Whereas the disk phase plate leads to an intensity distribution achieving excellent axial compression of the STED-PSF, it performs poorly in lateral directions. In contrast, helical wave plates lead to excellent compression of the STED-PSF laterally, but not along the optical axis. Thus, 3D-STED systems use a combination of both excitation modalities~\cite{sahl2019high}.

\begin{figure}[H]
    \centering
    \includegraphics[width=0.95\linewidth]{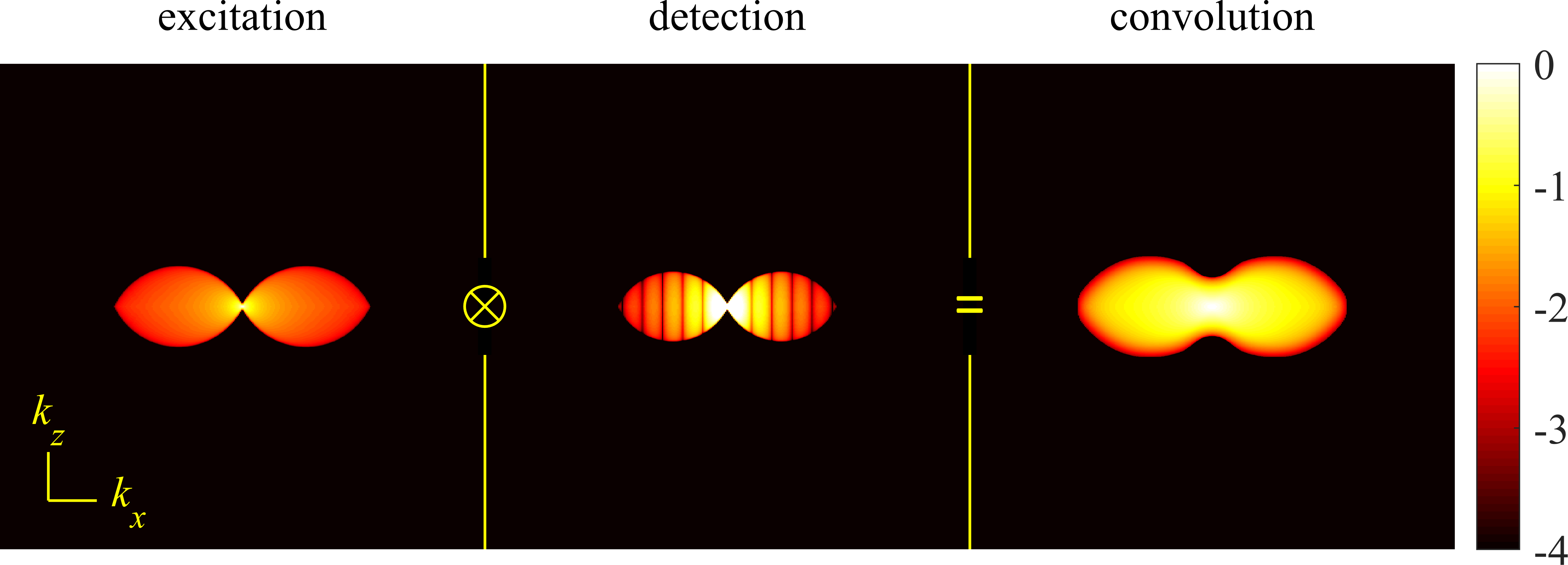}
    \caption{Anatomy of the OTF (amplitude) of a confocal microscope. The left panel shows the excitation OTF. The middle panel shows the detection OTF for a confocal pinhole with 50~$\mu$m radius and 60$\times$ magnification. The right panel shows the resulting confocal OTF obtained by a 3D convolution of the left two distributions.}
    \label{fig:OTFConfocal50}
    \vspace{-5mm}
\end{figure}

Having in place an exact description of the excitation PSF (excitation intensity distribution), we can return to the imaging PSF of a CLSM and consider its optical resolution. To do so, we consider its OTF, \textit{i.e.}, the Fourier transform of Eq.~\ref{eq:PSFCLSM}, for which we replace $I_\mathrm{ex}$ and $U_\mathrm{wf}$ of Eq.~\ref{eq:PSFCLSM} by their Fourier expansions, 
\vspace{-1mm}
\begin{equation}
\vspace{-2mm}
    \begin{split}
        U_\mathrm{wf}(\boldsymbol{\rho'}-\mathbf{r}) &= \int \frac{d\mathbf{k}}{2\pi} \tilde{U}_\mathrm{wf}(\mathbf{k}) \exp\left[i \mathbf{k}\cdot(\boldsymbol{\rho'}-\mathbf{r})\right], \\
        I_\mathrm{ex}(\mathbf{r}) &= \int \frac{d\mathbf{k}}{2\pi} \tilde{I}_\mathrm{ex}(\mathbf{k}) \exp(i \mathbf{k}\cdot\mathbf{r}),
    \end{split}
    \label{eq:UIFourier}
\end{equation}
where we recall that a tilde over a symbol denotes its Fourier amplitude. This immediately leads to \vspace{-1mm}  
\begin{equation}
    \begin{split}
        U_\mathrm{cf}(\mathbf{r}) &\propto \int d\boldsymbol{\rho}' \int d\mathbf{k} \int d\mathbf{k}' A(\boldsymbol{\rho}') \tilde{U}_\mathrm{wf}(\mathbf{k}') \\
        & \exp\left[i \mathbf{k}'\cdot(\boldsymbol{\rho}'-\mathbf{r})\right] \tilde{I}_\mathbf{ex}(\mathbf{k}) \exp\left(i \mathbf{k}\cdot \mathbf{r}\right). 
    \end{split}
    \label{eq:OTFCLSM0}
\end{equation} 
The integration over $\boldsymbol{\rho}'$ can be now be performed analytically, resulting in 
\begin{equation}
    \begin{split}
        \int d\boldsymbol{\rho}' A(\boldsymbol{\rho}') \exp\left(i \mathbf{k}'\cdot\boldsymbol{\rho}'\right)  = \frac{2\pi a}{q'}J_1(aq'),
    \end{split}
    \label{eq:FourierAperture}
\end{equation}
where $a$ is, as before, the radius of the confocal aperture, $q'=\sqrt{k_x^{\prime 2}+k_y^{\prime 2}}$ is the modulus of the radial part of the vector $\mathbf{k}'$, and $J_1$ is the first order Bessel function of the first kind. Substituting this result into Eq.~\ref{eq:OTFCLSM0}, we write \vspace{-2mm}
\begin{align}
    U_\mathrm{cf}(\mathbf{r}) \propto & \int d\mathbf{k} \int  d\mathbf{k}'  \frac{2\pi a}{q'}J_1(aq') \nonumber \\
        &\tilde{U}_\mathrm{wf}(\mathbf{k}')  \tilde{I}_\mathbf{ex}(\mathbf{k}) \exp\left[i (\mathbf{k}-\mathbf{k'})\cdot \mathbf{r}\right].
\end{align}

Following some algebra, we find for the Fourier transform of $U_\mathrm{cf}(\mathbf{r})$, \emph{i.e.}, the CLSM's OTF (up to some constant prefactor), \vspace{-2mm}
\begin{equation}
\vspace{-2mm}
        \tilde{U}_\mathrm{cf}(\mathbf{k}) \propto \int d\mathbf{k}' \frac{J_1(aq')}{q'}
        \tilde{U}_\mathrm{wf}(\mathbf{k}') \tilde{I}_\mathrm{ex}(\mathbf{k}+\mathbf{k}').
    \label{eq:OTFCLSM}
\end{equation}
Thus, the OTF of the confocal microscope is given by the three-dimensional convolution of a wide-field microscope OTF, $\tilde{U}_\mathrm{wf}(\mathbf{k})$, modulated by the aperture function, $J_1(aq')/q'$, (Fourier transform of detection PSF, Eq.~\ref{eq:PSFCLSM}, also sometimes termed detection OTF) and the Fourier transform of the excitation PSF, $\tilde{I}_\mathrm{ex}(\mathbf{k})$ (also sometimes termed excitation OTF). This is visualized in Fig.~\ref{fig:OTFConfocal50}, where the left panel shows the amplitude of the excitation OTF, $\tilde{I}_\mathrm{ex}(\mathbf{k})$, the middle panel is the detection OTF given by the absolute value of the wide-field OTF $\tilde{U}_\mathrm{wf}(\mathbf{k})$ multiplied by $J_1(aq')/q'$, and the right panel represents a cross-section of the amplitude of confocal OTF obtained by 3D convolution of the previous two panels.

\begin{figure}[H]
    \centering
    \includegraphics[width=0.85\linewidth]{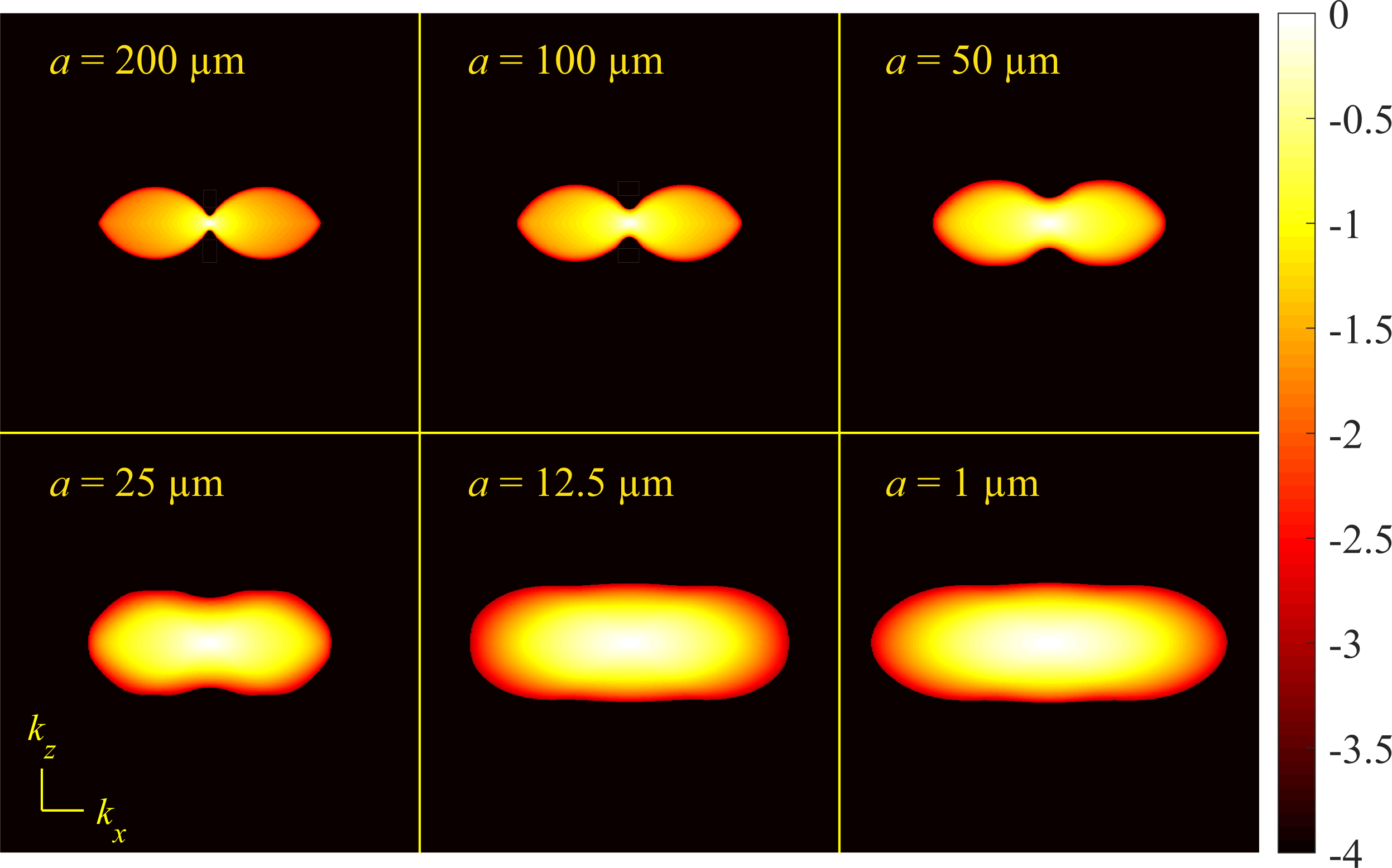}
    \caption{OTF amplitude of a confocal microscope as a function of confocal aperture size. The confocal aperture radius is given at the top of each panel. Here, we assumed an excitation wavelength of 470~nm, emission wavelength of 550~nm, and a water immersion objective of NA~=~1.2 at 60$\times$ magnification. The top most left panel shows the limit of an extremely large confocal pinhole so that the OTF approaches that of a wide-field microscope imaging at the same wavelength as the excitation wavelength of the excitation laser. The bottom right panel shows the limit of a nearly zero-size pinhole ($a=1$~$\mu$m), so that the OTF approaches that of an ISM; see Sec.~\ref{ISM}.}
    \label{fig:OTFConfocalArray}
    \vspace{-3mm}
\end{figure}

The most noticeable difference between the confocal OTF of Fig.~\ref{fig:OTFConfocal50} and the wide-field OTF of Fig.~\ref{fig:otfwidefieldscalar} is that the confocal OTF has non-zero components along the optical axis (here $k_x=0$ with the origin at the center) highlighting a confocal microscope's ability for optical sectioning. The corresponding axial resolution is given by $2\pi$ divided by the maximum frequency supported along the $k_z$-axis; see Eq.~\ref{eq:resolutionMonotonic}. 

Fig.~\ref{fig:OTFConfocalArray} shows how the confocal OTF changes with the pinhole size. As expected, for a large confocal pinhole radius of 200~$\mu$m (top left panel), the confocal OTF approaches that of a wide-field microscope at the same wavelength, as can be seen by comparing with the right panel of Fig.~\ref{fig:otfwidefieldscalar}. As the pinhole size shrinks ($a=1$~$\mu$m), optical sectioning and axial resolution are optimized; see bottom right panel of Fig.~\ref{fig:OTFConfocalArray}. In this case, the confocal aperture can be approximated by a delta function so that the integral Eq.~\ref{eq:FourierAperture} results in a constant. As such, the OTF for a very small aperture reduces to the convolution of the wide-field OTF, $\tilde{U}_\mathrm{wf}$, with the excitation OTF, $\tilde{I}_\mathrm{ex}$.
Thus the maximum frequency passed by the confocal OTF with a small aperture is given by $k_{max}=k_{max,\mathrm{ex}}+k_{max,\mathrm{em}}$ where $k_{max,\mathrm{ex}}$ and $k_{max,\mathrm{em}}$, respectively, denote the maximum extents of $\tilde{I}_\mathrm{ex}$ and $\tilde{U}_\mathrm{wf}$. 

The maximum extents of excitation and detection OTFs in the lateral direction are $k_{max,\mathrm{ex}/\mathrm{det}}=4\pi n \sin\Theta/\lambda_{\mathrm{ex}/\mathrm{em}}$, which, in turn, results in the following
lateral resolution (see Eq.~\ref{eq:resolutionMonotonic}) \vspace{-2mm}
\begin{equation}
    \begin{split}
        y_{\mathrm{min}} = \frac{1}{2 \mathrm{NA}}\left(\frac{1}{\lambda_\mathrm{ex}}+\frac{1}{\lambda_\mathrm{em}}\right)^{-1},
    \end{split}
    \label{eq:cflateralresolution}
\end{equation}
and similarly for the axial resolution \vspace{-1mm}
\begin{equation}
    \begin{split}
        z_{\mathrm{min}} = \frac{1}{2 n \left(1-\cos\Theta\right)}\left(\frac{1}{\lambda_\mathrm{ex}}+\frac{1}{\lambda_\mathrm{em}}\right)^{-1},
    \end{split}
    \label{eq:cfaxialresolution}
\end{equation}
where $\lambda_\mathrm{ex}$ and $\lambda_\mathrm{em}$ are the excitation and emission wavelengths, respectively. Thus, ignoring spectral Stokes shift between excitation and emission, \textit{i.e.}, $\lambda_\mathrm{em}\approx \lambda_\mathrm{ex}$, (see Sec.~\ref{Nonlinear}) then the confocal microscope with infinitely small pinhole has a twofold higher lateral resolution than a wide-field microscope as we can see by comparing Eq.~\ref{eq:cflateralresolution} to Eq.~\ref{eq:widefieldlateralresolution}. 
This improvement in resolution can also be explained in the spatial domain using Eq.~\ref{eq:PSFCLSM} by setting $A(\boldsymbol{\rho}')=\delta(\boldsymbol{\rho}'-\boldsymbol{\xi})$ (infinitely small aperture centered at $\boldsymbol{\xi}$) and adopting Gaussian approximations for both the wide-field PSF as in Eq.~\ref{eq:PSFGaussian} and excitation PSF $I_\mathrm{ex}$. In this case, the resulting confocal PSF would be the product of both Gaussians which is a Gaussian as well~\cite{zhang2007gaussian}
\begin{equation}
    U_{\mathrm{cf}}(\boldsymbol{\rho},z) \propto \exp\left(-\frac{(\boldsymbol{\rho}-\xi_{\rho})^2}{2\sigma_{\rho}^2}-\frac{(z-\xi_z)^2}{2\sigma^2_z}\right). 
    \label{eq:ConfocalGaussPSF}
\end{equation}
Here, the widths of the resulting Gaussian PSF, $\sigma_{\rho}$ and $\sigma_z$, are smaller than the widths of both excitation and detection PSFs leading to higher resolutions. 

\begin{figure}[H]
    \centering
    \includegraphics[width=0.94\linewidth]{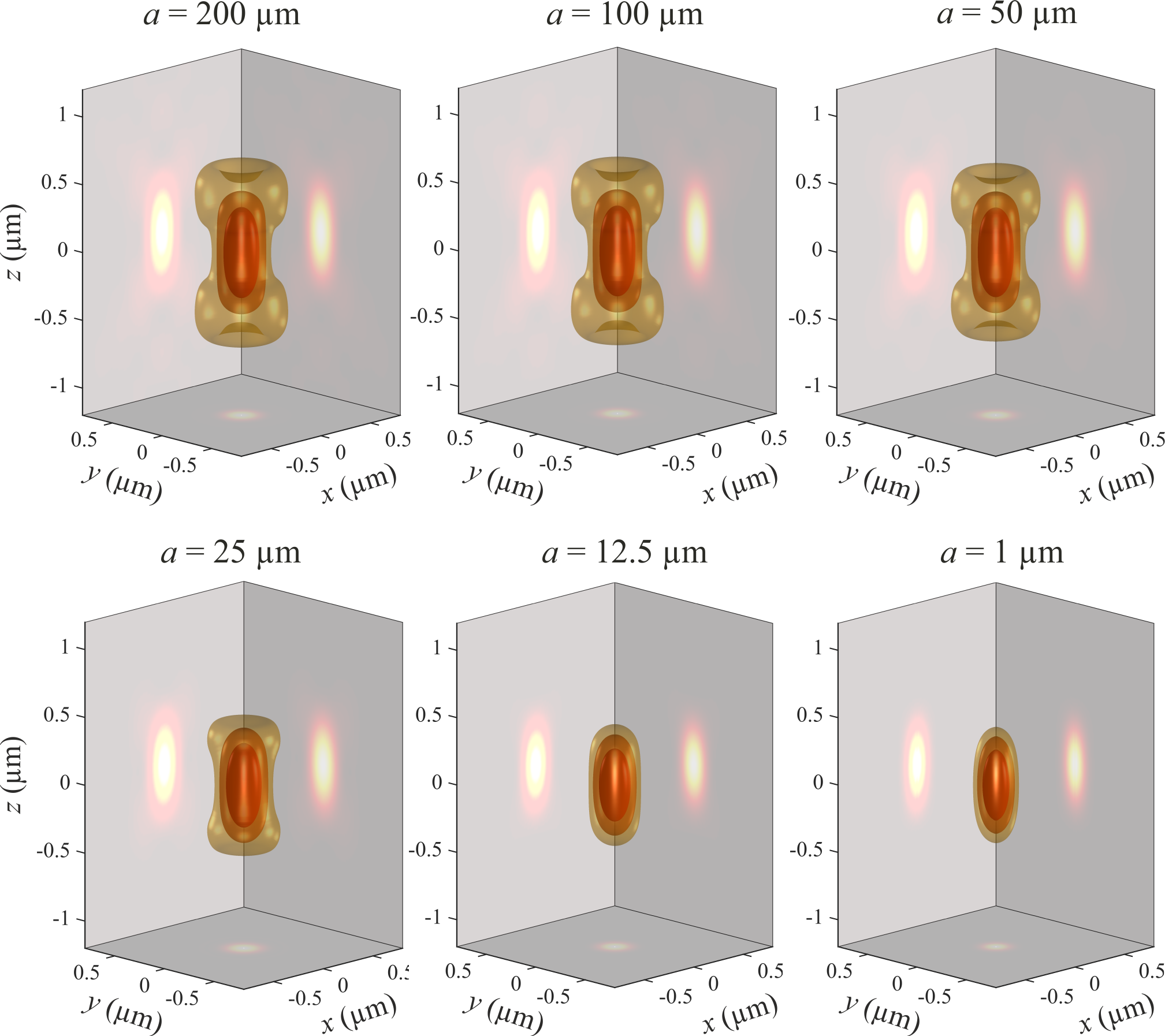}
    \caption{Confocal microscope PSF for an isotropic emitter as a function of confocal aperture size. The aperture radius is given above each panel. The parameters are similar to those in Fig.~\ref{fig:OTFConfocalArray} with 60$\times$ magnification.} 
    \label{fig:PSFConfocalArray}
    \vspace{-5mm}
\end{figure}

\begin{figure}[H]
    \centering
    \includegraphics[width=0.92\linewidth]{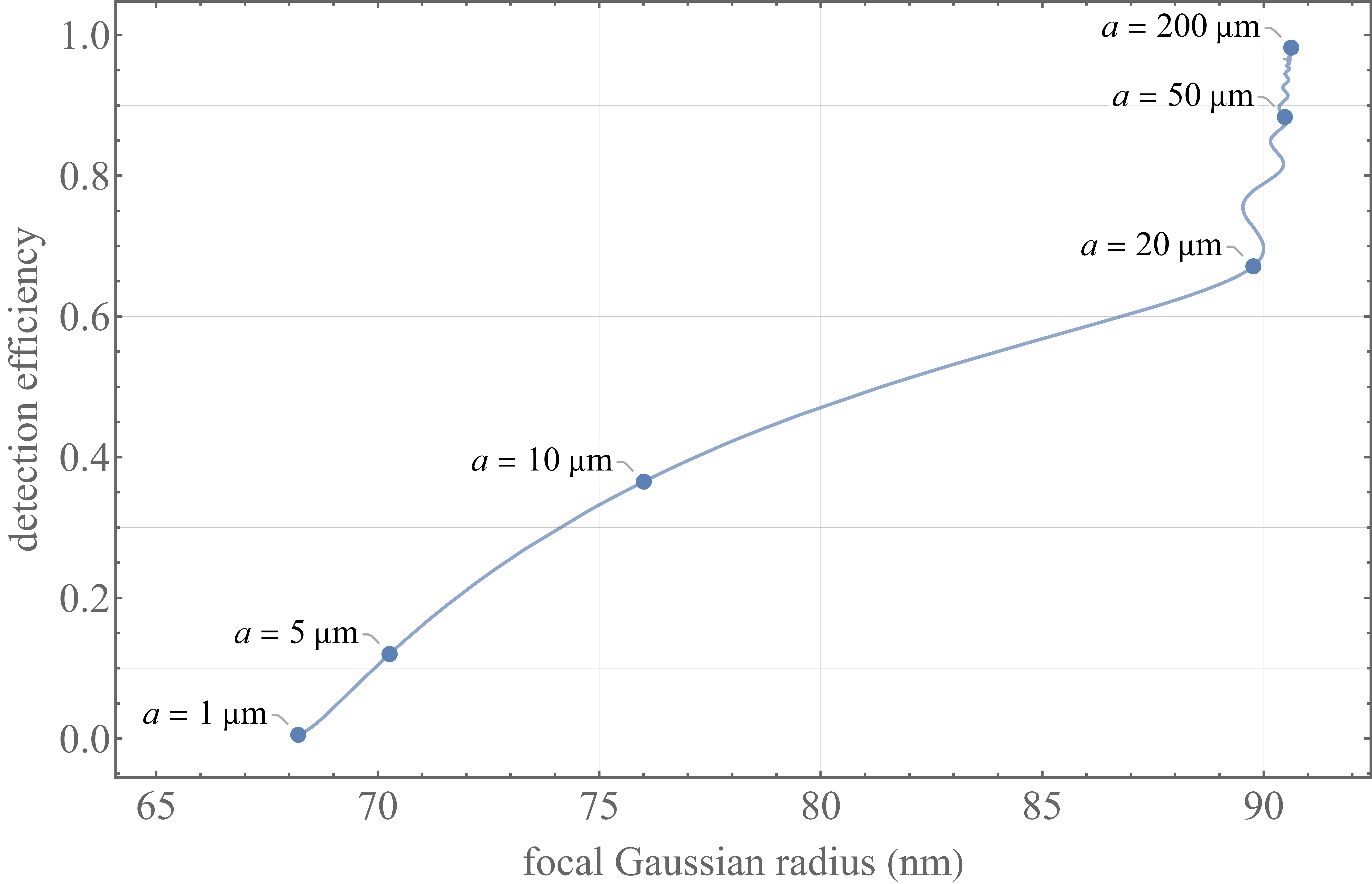} \vspace{-1mm}
    \caption{Relation between PSF size and detection efficiency in a CLSM. Here we show the light detection efficiency versus the Gaussian radius $\sigma$ of the PSF in the focal plane as a function of the confocal aperture's radius annotated $a$. 
    Calculations were done for a water immersion objective with NA~=~1.2 and image magnification of 60$\times$ (focal plane to pinhole plane). It was assumed that excitation is achieved with 470~nm circular polarized light focused into a diffraction-limited spot, and that the fluorescence emission is of 550~nm wavelength. We found the focal radius by fitting a radially symmetric Gaussian $\exp(-\rho^2/2\sigma^2)$ to the PSF in the focal plane. The curve's undulations at the upper right arise from diffraction effects of light passing through a circular pinhole.}
\label{fig:ConfocalFocusSizeVersusDetectionEfficiency}
\end{figure}

The PSFs corresponding to the OTFs shown in Fig.~\ref{fig:OTFConfocalArray} are presented in Fig.~\ref{fig:PSFConfocalArray} illustrating how the PSF's lateral width shrinks with decreasing pinhole size improving lateral resolutions albeit at a price. The smaller the confocal pinhole size, the fewer photons reach the detector thereby reducing SNR~\cite{sheppard2006signal}. This is quantified in Fig.~\ref{fig:ConfocalFocusSizeVersusDetectionEfficiency} showing the relation between PSF diameter (in the focal plane) and light detection efficiency for increasing pinhole radii (1-200~$\mu$m) assuming 470~nm excitation and 550~nm emission wavelength, and for a water immersion microscope with NA~=~1.2 objective and 60$\times$ magnification. As can be seen, light detection efficiency decreases as the confocal pinhole radius drops below 20~$\mu$m motivating the use of
ISM introduced next. \vspace{-1mm}

\vspace{-11mm}
\subsubsection{Image scanning microscopy}\label{ISM} \vspace{-2mm}

As was discussed in Sec.~\ref{CLSM-Confocal} when considering a confocal PSF, the maximum possible spatial resolution is achieved approaching an infinitely small confocal pinhole; see Eqs.~\ref{eq:cflateralresolution}-\ref{eq:ConfocalGaussPSF}. However, as this would reduce light detection efficiency to almost zero (see Fig.~\ref{fig:ConfocalFocusSizeVersusDetectionEfficiency}), such an option is often avoided in practice. To simultaneously maximize spatial resolution and light detection efficiency, now beyond three decades ago, Colin Sheppard proposed to combine scanning spot illumination of confocal microscopes and wide-field light detection of an array detector, \textit{e.g.}, EMCCD camera, without pinholes mitigating light loss~\cite{Sheppard1988}. This idea, termed Image Scanning Microscopy or ISM, was first experimentally demonstrated in 2010 by M\"uller and Enderlein~\cite{Muller2010}. The core idea of ISM is to replace the confocal pinhole and the single pixel detector of a conventional CLSM by an array detector in the image plane (pinhole plane); see Fig.~\ref{fig:CLSM}. The fluorescence light from an illumination spot at position $\mathbf{r}$ is then spread across multiple pixels of the detector array. In this setup, a pixel located at $\boldsymbol{\xi}$ records photons from the illuminated spot corresponding to a pinhole located at $\boldsymbol{\xi}$ with the same size as the pixel. The pixel size is often chosen small enough such that each pixel records an image of the illumination spot with a resolution similar to that of a CLSM with close to zero pinhole sizes; see Eqs.~\ref{eq:cflateralresolution} and \ref{eq:cfaxialresolution}. 
Moreover, as ISM builds on a CLSM, it also provides optical $z$-sectioning.

\begin{figure}[H]
    \centering
    \includegraphics[width=0.95\linewidth]{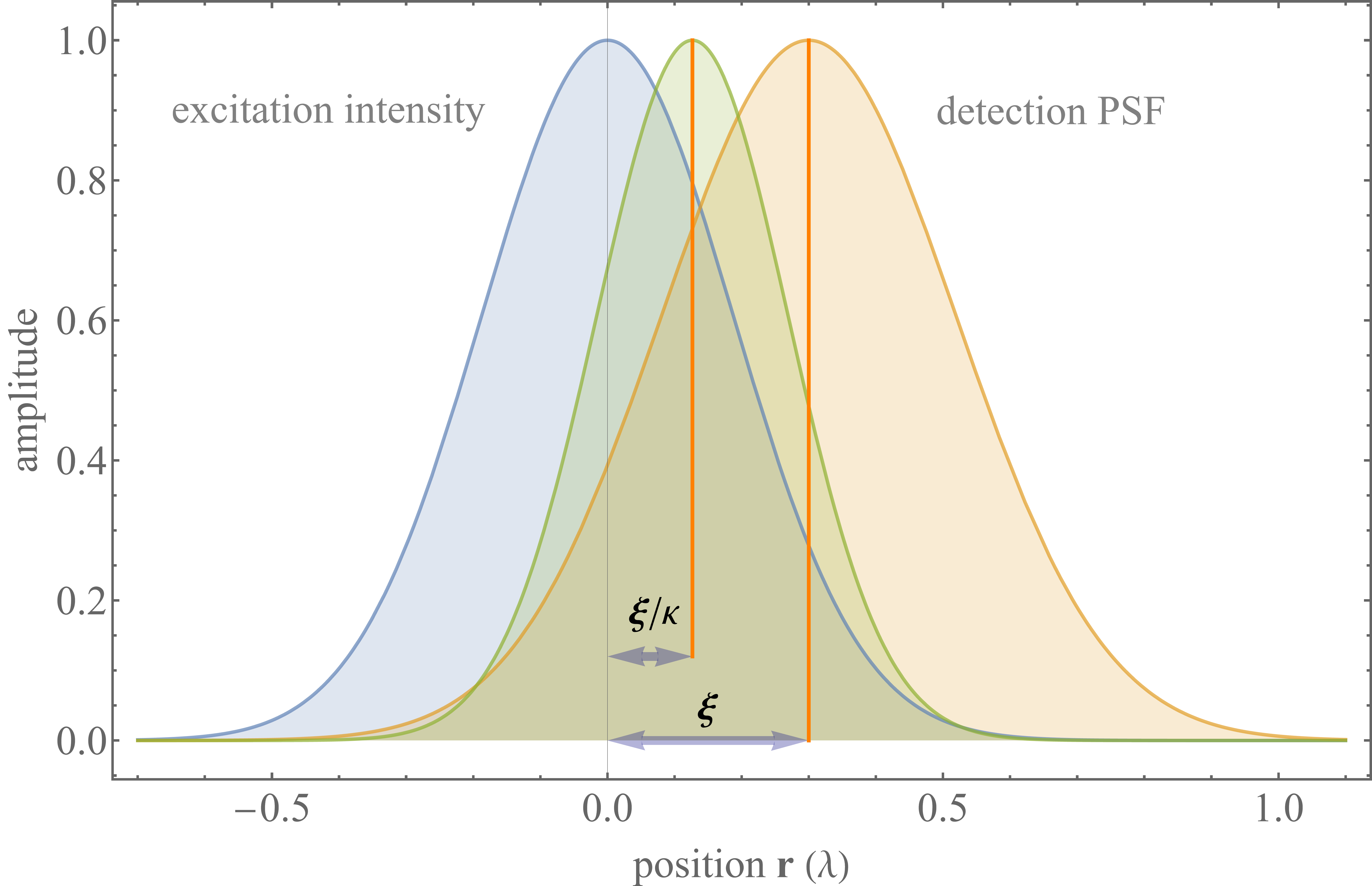}
    \caption{Image formation in ISM. The blue curve represents the excitation intensity distribution $I_\mathrm{ex}$ (excitation PSF) with its center at $\boldsymbol{\xi}=0$ (optical axis). The yellow curve shows the detection PSF ($U_\mathrm{wf}$) for a pixel located at $\boldsymbol{\xi}$ away from the optical axis. The pixel PSF ($U_\mathrm{pix}$), describing the image formation is, however, given by the product of the excitation and detection PSF, designated by the green curve and centered at $\boldsymbol{\xi}/\kappa$. Thus, a fluorophore at $\boldsymbol{\xi}=0$ (the excitation intensity's center) will 
    appear at $\boldsymbol{\xi}/\kappa$.
    }
    \label{fig:ISMReassignment}
    \vspace{-4mm}
\end{figure}

The ISM setup described here, results in $N_p$ recorded images for each illumination spot associated to all $N_p$ pixels of the detector array. As such, upon scanning the sample at $N_s$ locations, one acquires $N_p\times N_s$ images. To combine all acquired images into a single high resolution image, we first consider the scan image recorded by one pixel at a given position $\boldsymbol{\xi}$ on the array detector. The PSF of this scan image is easily found when replacing the aperture function $A(\mathbf{\boldsymbol{\rho}})$ of Eq.~\ref{eq:PSFCLSM} by the pixel area. However, as an idealization, we can consider the pixel area as a delta function $\delta(\boldsymbol{\rho}-\boldsymbol{\xi})$ as compared to the size of features we care to learn. As such, the PSF for the scan image recorded by a pixel at position $\boldsymbol{\xi}$ is \vspace{-2mm}
\begin{equation}
\vspace{-2mm}
    \begin{split}
        U_\mathrm{pix}(\mathbf{r},\boldsymbol{\xi}) \propto U_\mathrm{wf}(\boldsymbol{\xi}-\mathbf{r}) I_\mathrm{ex}(\mathbf{r}),
        \end{split}
    \label{eq:ISMraw}
\end{equation}
where, as before, $U_\mathrm{wf}$ is the wide-field imaging PSF (detection PSF), and $I_\mathrm{ex}$ is the excitation PSF. This is visualized in Fig.~\ref{fig:ISMReassignment} where a cross-section of the excitation PSF $I_\mathrm{ex}(\mathbf{r})$ is shown together with the detection PSF for a pixel at position $\boldsymbol{\xi}$ (described by $U_\mathrm{wf}(\boldsymbol{\xi}-\mathbf{r})$) and the product of both; see Eq.~\ref{eq:ISMraw}. 

\begin{figure}[H]
    \centering
    \includegraphics[width=0.81\linewidth]{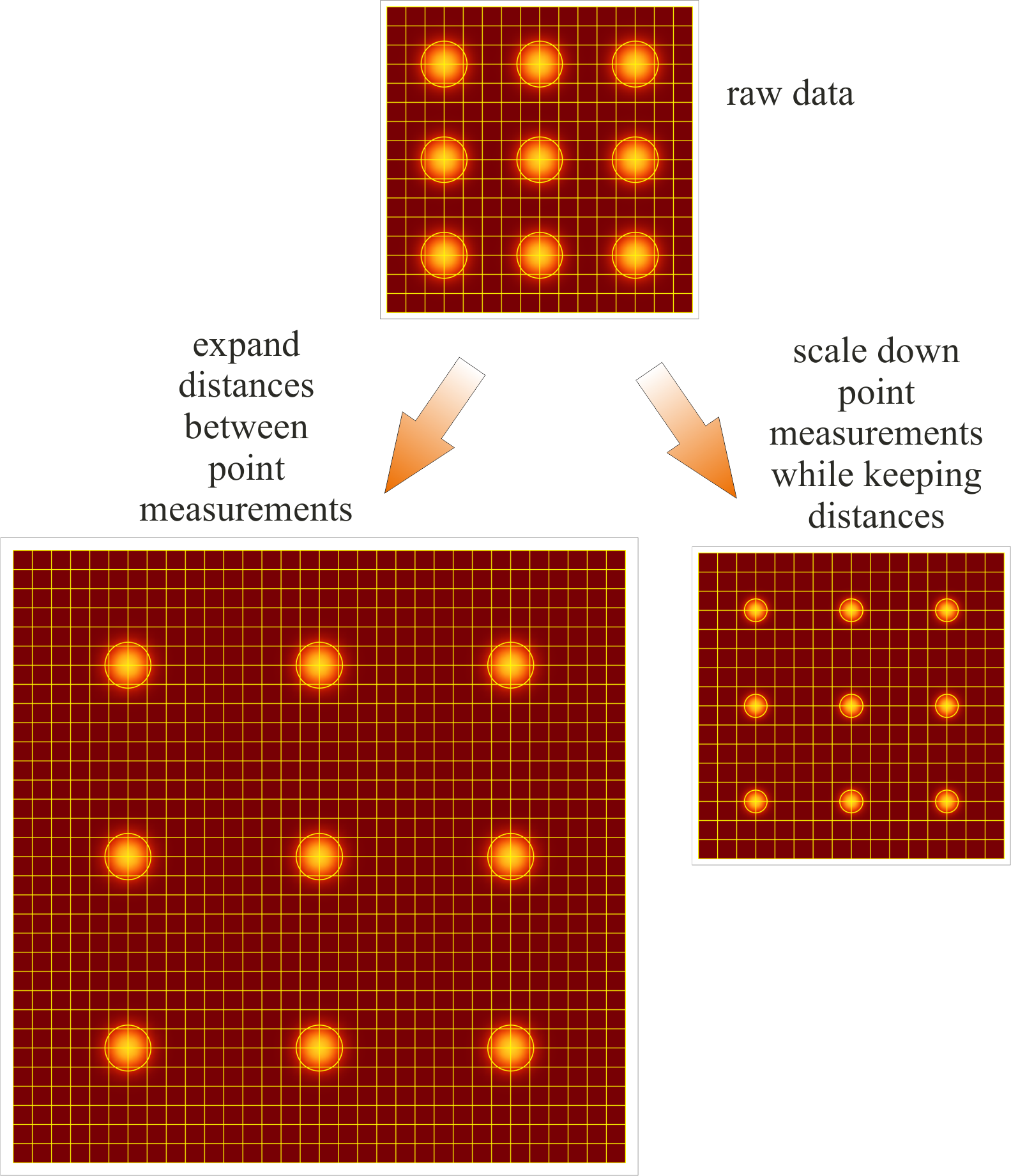} \vspace{-2mm}
    \caption{ISM image reconstruction. At each scan position, the array detector records a small image of the illuminated region (top). To reconstruct a final ISM image, we can either down-scale each recorded small image by a factor $\kappa$ (bottom right), or leave the recorded images unchanged but place them in the final ISM image by the factor $\kappa$ farther way from each other (bottom left).}
    \label{fig:ISMReassignment1} \vspace{-5mm}
\end{figure}

When approximating the excitation and detection PSFs by Gaussians with variance $\sigma^2_\mathrm{ex}$ and $\sigma^2_\mathrm{em}$, respectively, the product of both yields \vspace{-2mm}
\begin{equation}
    I_\mathrm{ex}(\mathbf{r})U_\mathrm{wf}(\mathbf{r}-\boldsymbol{\xi}) \propto \exp\left[-\frac{\left(\mathbf{r}-\boldsymbol{\xi}/\kappa\right)^2}{2 \sigma_{\mathrm{PSF}}^2}\right]
    \label{eq:ISMPSF}
\end{equation}
with $\sigma^{-2}_{\mathrm{PSF}} = \sigma^{-2}_\mathrm{ex}+\sigma^{-2}_\mathrm{em}$, and $\kappa = 1+\sigma^2_\mathrm{em}/\sigma^2_\mathrm{ex}$. Recalling that $\sigma_\mathrm{ex}$ and $\sigma_\mathrm{em}$ linearly scale with wavelength (see Eq.~\ref{eq:PSFGaussian}), we find \vspace{-3mm}
\begin{equation}
\vspace{-1mm}
    \kappa = 1+\left(\lambda_\mathrm{em}/\lambda_\mathrm{ex}\right)^2
    \label{eq:kappa}
\end{equation}
which equals 2 if one neglects the spectral Stokes shift between excitation and fluorescence emissions. Thus, the maximum of the product of excitation intensity distribution and detection PSF is located between the centers of both at position $\boldsymbol{\xi}/\kappa$, such that the scan image is shifted by the same amount with respect to an image recorded by a pixel at position $\boldsymbol{\xi}=0$; see Fig.~\ref{fig:ISMReassignment}. This insight yields a recipe for how to super-impose different scan images recorded by different pixels: an image recorded by a pixel at position $\boldsymbol{\xi}$ must be shifted by $\boldsymbol{\xi}/\kappa$ towards the optical axis before being added to the final sum image. Mathematically, this is 
expressed as 
\begin{equation}
    \begin{split}
        &U_\mathrm{ISM}(\mathbf{r}) \propto \int d\boldsymbol{\xi} U_\mathrm{pix}\left(\mathbf{r}+\frac{\boldsymbol{\xi}}{\kappa},\boldsymbol{\xi}\right) \\
        &= \int d\boldsymbol{\xi} U_\mathrm{wf}\left(\frac{\kappa-1}{\kappa}\boldsymbol{\xi}-\mathbf{r}\right) I_\mathrm{ex}\left(\mathbf{r}+\frac{\boldsymbol{\xi}}{\kappa}\right).
    \end{split}
    \label{eq:ISM}
\end{equation}
There are two ways to realize this summation in practice. As shown in Fig.~\ref{fig:ISMReassignment1}, one way is to scale down, by factor $\kappa$, all images recorded by the array detector at each scan position before adding them to the final image at the corresponding scan position (from top to bottom right in Fig.~\ref{fig:ISMReassignment1}). Alternatively, one can leave the recorded array detector images as they are, but place them a factor $\kappa$ farther away from each other when adding them to the final image (from top to bottom left in Fig.~\ref{fig:ISMReassignment1}).

Obviously, both procedures are mathematically equivalent ways to realize the algorithm described by Eq.~\ref{eq:ISM}, although the second algorithm is numerically simpler as it does not require any interpolation based down-scaling of the images recorded by the array detector. However, as first demonstrated by York and Shroff \cite{York2012} and by de Luca and Manders \cite{DeLuca2013}, both algorithms can be realized in a fully optical way. The first algorithm, scaling down the array detector images, is optically realized by inserting an extra demagnifying lens pair into the detection pathway (as realized by instant SIM~\cite{York2012,York2013}, Optical Photon Re-Assignment or OPRA~\cite{roth2013optical}, or confocal spinning disk ISM~\cite{azuma2015super}), while the second algorithm which scales up distances between recorded images is realized by a double mirror re-scan system (re-scan microscopy ~\cite{DeLuca2013}) or by re-coupling the emission into the excitation scan system (rapid two-photon excitation ISM ~\cite{gregor2017rapid}). \vspace{-2mm}

By construction, both OTF and PSF of an ISM are identical to that of a confocal microscope with an infinitely small confocal pinhole; see last panels of Fig.~\ref{fig:OTFConfocalArray} (OTF) and Fig.~\ref{fig:PSFConfocalArray} (PSF), respectively. The corresponding achievable optical lateral and axial resolutions then immediately follow from Eqs.~\ref{eq:cflateralresolution} and \ref{eq:cfaxialresolution}. One important particular property of ISM is that it also ``concentrates'' the collected fluorescence light into an area of the final image four times smaller than that of a conventional CLSM (``super-concentration of light'', ~\cite{roth2016superconcentration}, see also top and right panel of Fig. \ref{fig:ISMReassignment1}), significantly increasing image contrast. Meanwhile, multiple 
ISM variants (reviewed in Ref.~\cite{gregor2019image}), and several commercial systems are available providing CLSMs with ISM options for improved resolution and high contrast imaging.       
\vspace{-6mm}
\subsubsection{4pi microscopy}\label{4pi} 
\vspace{-3mm}

\begin{figure}[H]
    \centering
    \includegraphics[width=0.95\linewidth]{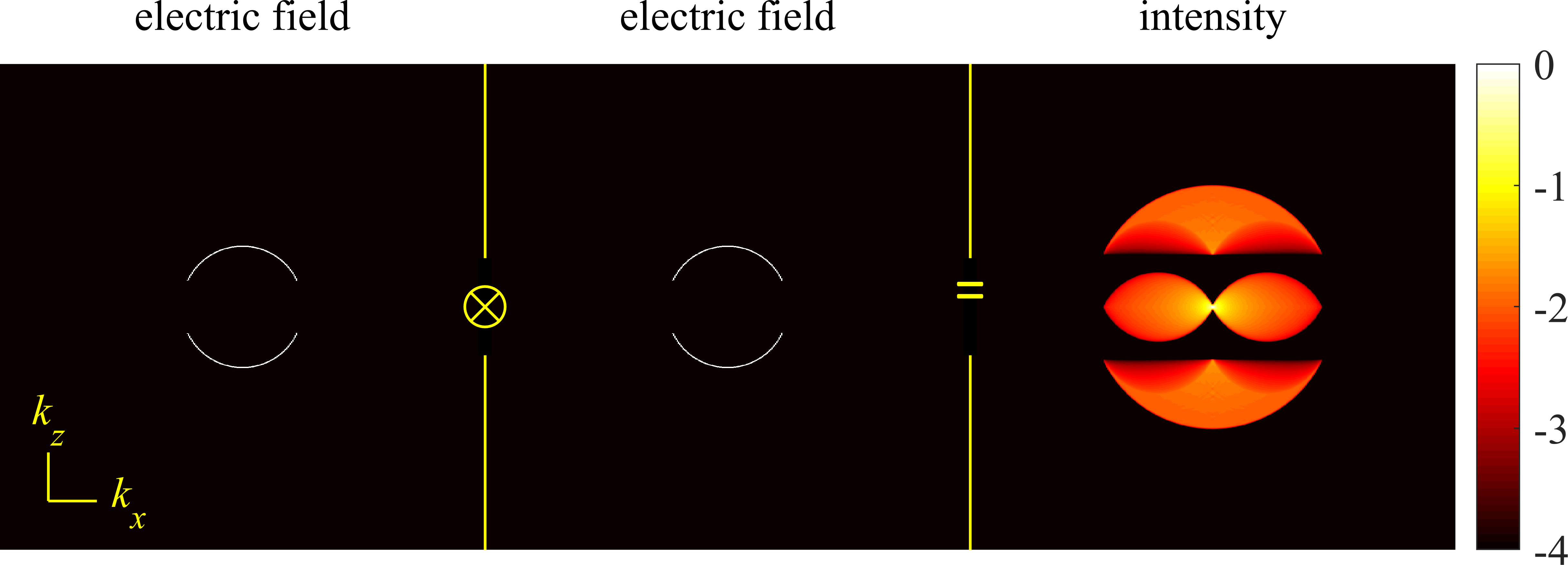}
    \caption{4pi microscope excitation OTF generated by the interference of light focused through two opposing objectives. The left and middle panel show the same Fourier transform of the excitation electric field in sample space. The resulting excitation OTF shown in the right panel is the (auto)convolution of this electric field Fourier transform and represents the Fourier transform of the excitation intensity (excitation OTF). Excitation is assumed to be done using a water immersion objective with NA~=~1.2.}
    \label{fig:OTF4pi}
    \vspace{-5mm}
\end{figure}

\begin{figure}[H]
    \centering
   \includegraphics[width=0.95\linewidth]{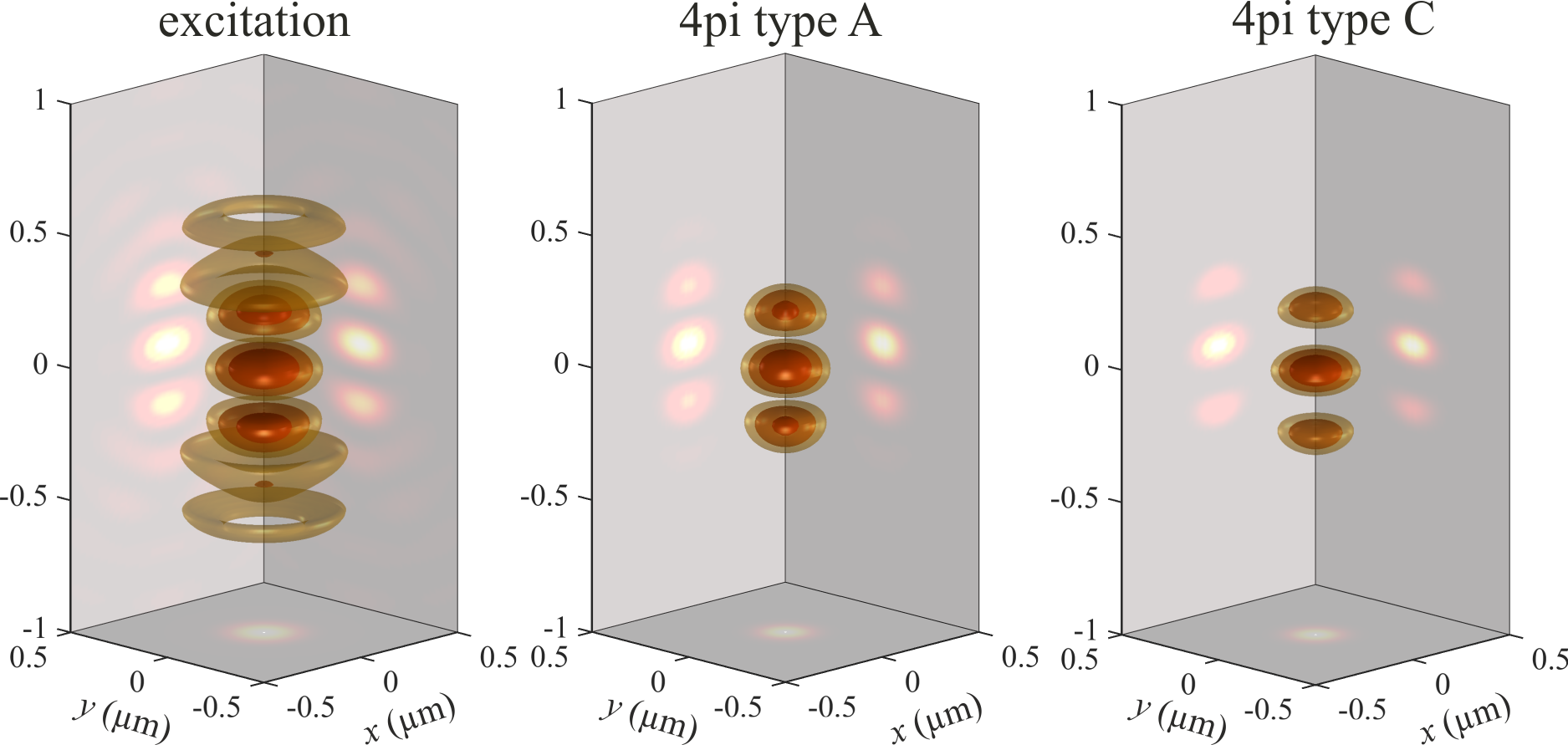}
    \caption{Excitation PSF and (imaging) PSF of 4pi microscopy for a rapidly rotating emitter. The left panel shows the excitation PSF in the focus of a 4pi microscope, the middle panel shows the (imaging) PSF of a 4pi type A microscope, and the right panel that for a 4pi type C microscope. Calculations were performed using a water immersion objective with NA~=~1.2 and 470~nm excitation wavelength and 550~nm fluorescence emission wavelength, and for a confocal detection in the limit of an infinitely small pinhole.} 
    \label{fig:PSF4pi}
    \vspace{-5mm}
\end{figure}

One peculiarity of conventional CLSM is the disparity between lateral and axial resolutions (see Eqs.~\ref{eq:cflateralresolution}-\ref{eq:cfaxialresolution}) due to the PSF's elongated shape along the optical axis yielding stretched 3D CLSM images; see Fig.~\ref{fig:PSFConfocalArray}. 
To overcome this strongly anisotropic PSF shape, Stelzer and Hell developed 4pi-microscopy using two opposing objectives to focus (and detect) light ~\cite{hell1992OpticsLetters}. When sending laser excitation light through both objectives in a \emph{coherent} manner, the resulting interference of both beams generates a multi-peaked interference pattern along the optical axis. The corresponding Fourier representations of the excitation electric fields are shown in the left and middle panels of Fig.~\ref{fig:OTF4pi}, and the convolution of both, \textit{i.e.}, the 4pi excitation OTF, is shown in the right panel of Fig.~\ref{fig:OTF4pi}. By contrast to the CLSM excitation OTF of Fig.~\ref{fig:OTFConfocal50}, its 4pi counterpart populates high frequencies along the optical axis, coinciding with a tight modulation of the excitation intensity along this axis. The corresponding excitation intensity distribution (excitation PSF) in real space is shown in the left panel of Fig.~\ref{fig:PSF4pi}. 

Detection in a 4pi microscope is done as usual in confocal detection mode, whereby two principal options are possible: 1) fluorescence is collected with both objectives and detected by two detectors resulting in two independent scan images added later to attain a single image (4pi type A microscope~\cite{lang20074pi});  2) fluorescence is collected with both objectives and coherently super-imposed onto one detector (4pi type C microscope ~\cite{bewersdorf2006comparison}). As a special case is the 4pi type B microscope, performing similarly to the type A, where excitation is done incoherently (\emph{i.e.}, with no interference pattern generation) but the collected light is super-imposed coherently~\cite{hao2022review}. 

\begin{figure}[H]
    \centering
    \includegraphics[width=0.95\linewidth]{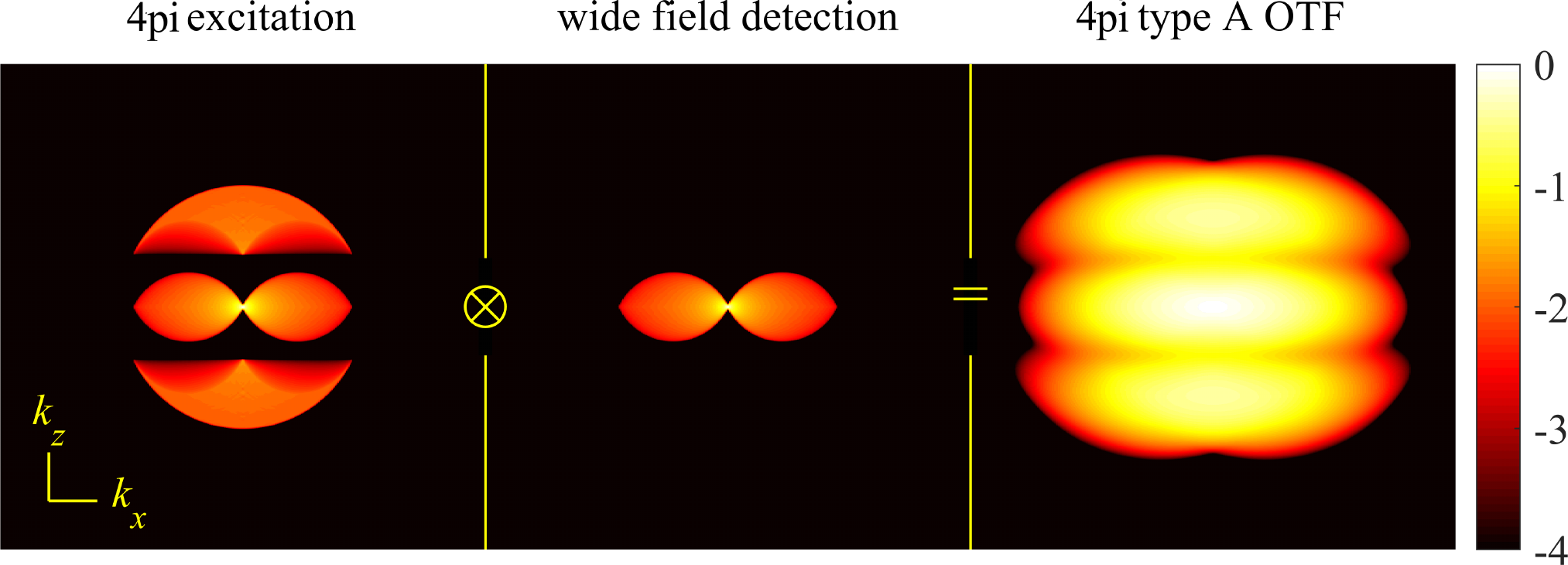}
    \caption{OTF of a type A 4pi microscope where excitation is done through two opposing objectives, and detection from one side through a confocal pinhole. For simplicity, we consider here only the limiting case of an infinitely small pinhole 
    maximizing
    spatial resolution. The left panel shows the excitation OTF, the middle panel the OTF of detection with an infinitely small pinhole, and the right panel shows the resulting 4pi OTF as a convolution of the two distributions shown on the left. Excitation and detection are achieved using a water immersion objective with NA~=~1.2, and any Stokes shift between excitation and emission light is neglected.}
    \label{fig:OTF4piA}
    \vspace{-5mm}
\end{figure}

\begin{figure}[H]
    \centering
    \includegraphics[width=0.95\linewidth]{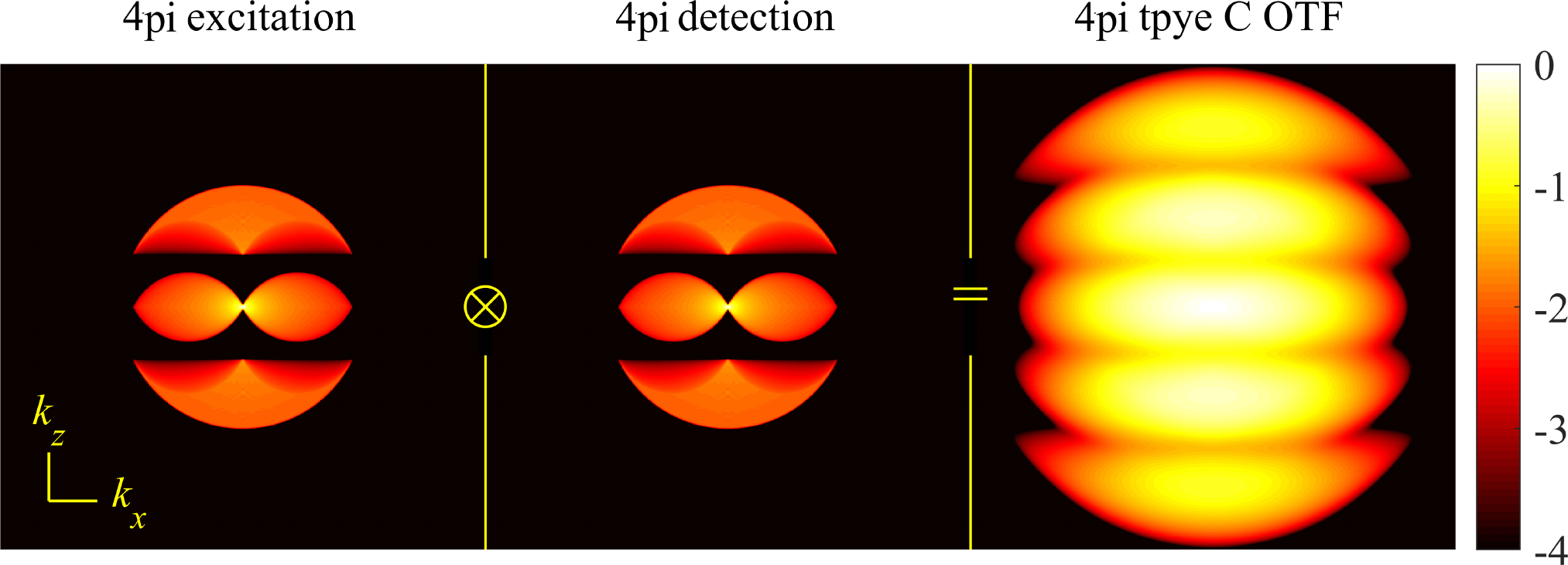}
    \caption{OTF of a type C 4pi microscope. Similar to Fig.~\ref{fig:OTF4piA}, but in this configuration, both excitation and detection occur through two opposing objectives. Again, we consider here only the limiting case of an infinitely small pinhole. 
    The left panel shows the excitation OTF, the middle panel the (identical) Fourier transform for \emph{coherent} confocal detection from both sides, and the right panel shows the resulting OTF as a convolution of the two panels shown on the left. }
    \label{fig:OTF4piC}
    \vspace{-6mm}
\end{figure}

To determine the maximal possible resolution attainable with 4pi microscopy, we show in Figs.~\ref{fig:OTF4piA} and \ref{fig:OTF4piC} the OTFs for type A and C microscopes in the limit of an infinitely small confocal pinhole (realized by combining a 4pi microscopy with ISM). Thus, the OTF of a 4pi type A microscope as shown in Fig.~\ref{fig:OTF4piA} is obtained by a convolution of the 4pi excitation OTF (see Fig.~\ref{fig:OTF4pi}) with the OTF of a simple ISM (corresponding to wide-field detection).   

As in 4pi type C microscope, detection is achieved by coherently super-posing fluorescence light from both objectives, the OTF of such detection looks similar to that of the excitation shown in Fig.~\ref{fig:OTF4pi}, except calculated for the fluorescence emission wavelength. The convolution of such a detection OTF with the excitation OTF then yields the OTF of the 4pi type C microscope; see Fig.~\ref{fig:OTF4piC}. The corresponding real space PSFs for both type A and C 4pi microscopes are shown in the middle and right panels of Fig.~\ref{fig:PSF4pi}. 

As can be seen in Figs.~\ref{fig:OTF4piA}-\ref{fig:OTF4piC}, 4pi microscopes collect more spatial frequencies than CLSMs (see Fig.~\ref{fig:OTFConfocalArray}) 
thereby improving their axial resolution. As before, we can again obtain quantitative numbers for the lateral and axial resolutions by inspecting the OTF and determining the maximum lateral and axial frequencies supported by the OTF. Concretely, the inverse of these maxima multiplied by $2\pi$ yields approximate values for the resolution; see Eq.~\ref{eq:resolutionMonotonic}. The lateral resolution of a 4pi microscope (with an infinitely small pinhole) is the same for both type A and C and equal to that of an ISM; see Eq.~\ref{eq:cflateralresolution}. However, the axial resolution of a type A 4pi microscope now reads \vspace{-3mm}
\begin{equation}
    \begin{split}
        z_{\mathrm{min}} \approx \frac{1}{2}\left[\frac{1}{\lambda_\mathrm{ex}}+\frac{1}{\lambda_\mathrm{em}}\left(1-\cos\Theta\right)\right]^{-1}
    \end{split}
    \label{eq:4piAaxialresolution}
\end{equation}
and similarly for the type C 4pi microscope \vspace{-1mm}
\begin{equation}
    \begin{split}
        z_{\mathrm{min}} \approx \frac{1}{2}\left(\frac{1}{\lambda_\mathrm{ex}}+\frac{1}{\lambda_\mathrm{em}}\right)^{-1}.
    \end{split}
    \label{eq:4piCaxialresolution}
\end{equation}
As can be seen from the PSFs of Fig.~\ref{fig:PSF4pi}, there are considerable side-lobes neighboring the central maximum along the optical axis, leading to ``ghost'' images in a recorded 3D scan image of a sample ~\cite{bewersdorf2006comparison}. These ghost images are much more pronounced for type A than C, though even for type C they must be eliminated, currently by applying deconvolution algorithms ~\cite{bewersdorf20064pi,liu2020enhanced}. Both the technical complexity of a 4pi microscope as well as image deconvolution challenges to eliminate ghost images have prevented their further distribution. 
However, the ISM lateral resolution of a 4pi type C (image scanning) microscope together with its axial resolution represent the maximum possible spatial resolutions available along the $x$ and $z$ directions using a diffraction-limited microscope. 

\subsubsection{Two-photon microscopy} 

An important variant of the point scanning microscope is the two-(or multi-)photon excitation scanning microscope~\cite{denk2007principles}. Here, a fluorophore is excited by a two-(or multi-)photon absorption process, typically with an excitation wavelength roughly twice (or multiple times) as large as that of one-photon absorption fluorescence excitation. Such two-photon excitation microscopes have several important properties~\cite{denk1997photon,williams2001multiphoton}. First, due to the longer excitation wavelength, typically in the infrared, excitation light can penetrate deeper into tissue than visible light. Thus, two-photon excitation microscopes are ideal for deep-tissue imaging in lipid and water rich tissues with high optical absorption in the visible spectrum. Second, there is a critical improvement in in-focus signal to background, \textit{i.e.}, undesired light from out-of-focus fluorophores, ratio compared to one-photon absorption fluorescence microscopy. This arises from: 1) fluorophore excitation taking place at longer wavelengths than the emission wavelength. In other words, the probability of simultaneous absorption of two or more photons is only significant at the focal spot with high photon density; 2) excitation light scattering is decreased at longer wavelengths; and 3) two-(or multi-)photon excitation does not require confocal detection for optical sectioning. This is because the two-photon excitation PSF is proportional to the \emph{square} of the excitation light intensity distribution (probability of two simultaneous photon absorption is given by \emph{square} of the one-photon excitation PSF), represented by an auto-convolution of the excitation OTF in Fourier space. A similar convolution was already considered when discussing the ISM's OTF (\textit{i.e.}, as idealized by the last panel of Fig.~\ref{fig:OTFConfocalArray}), covering higher spatial frequencies by contrast to the OTF of a wide-field microscope or a CLSM with wide pinhole shown in the first panel of Fig.~\ref{fig:OTFConfocalArray}. Thus, a two-photon excitation microscope has a similar optical sectioning capability as a confocal (one-photon excitation) microscope at the same excitation wavelength when using an infinitely small detection pinhole (neglecting the spectral Stokes shift between excitation and emission). The required peak power of the excitation pulses, orders of magnitude larger than in single-photon excitation thereby increasing photo-damage and photo-bleaching~\cite{tauer2002advantages}, is the primary downside of two-photon excitation microscopy.  \vspace{-3mm}

\begin{figure}[H]
    \centering
    \includegraphics[width=0.95\linewidth]{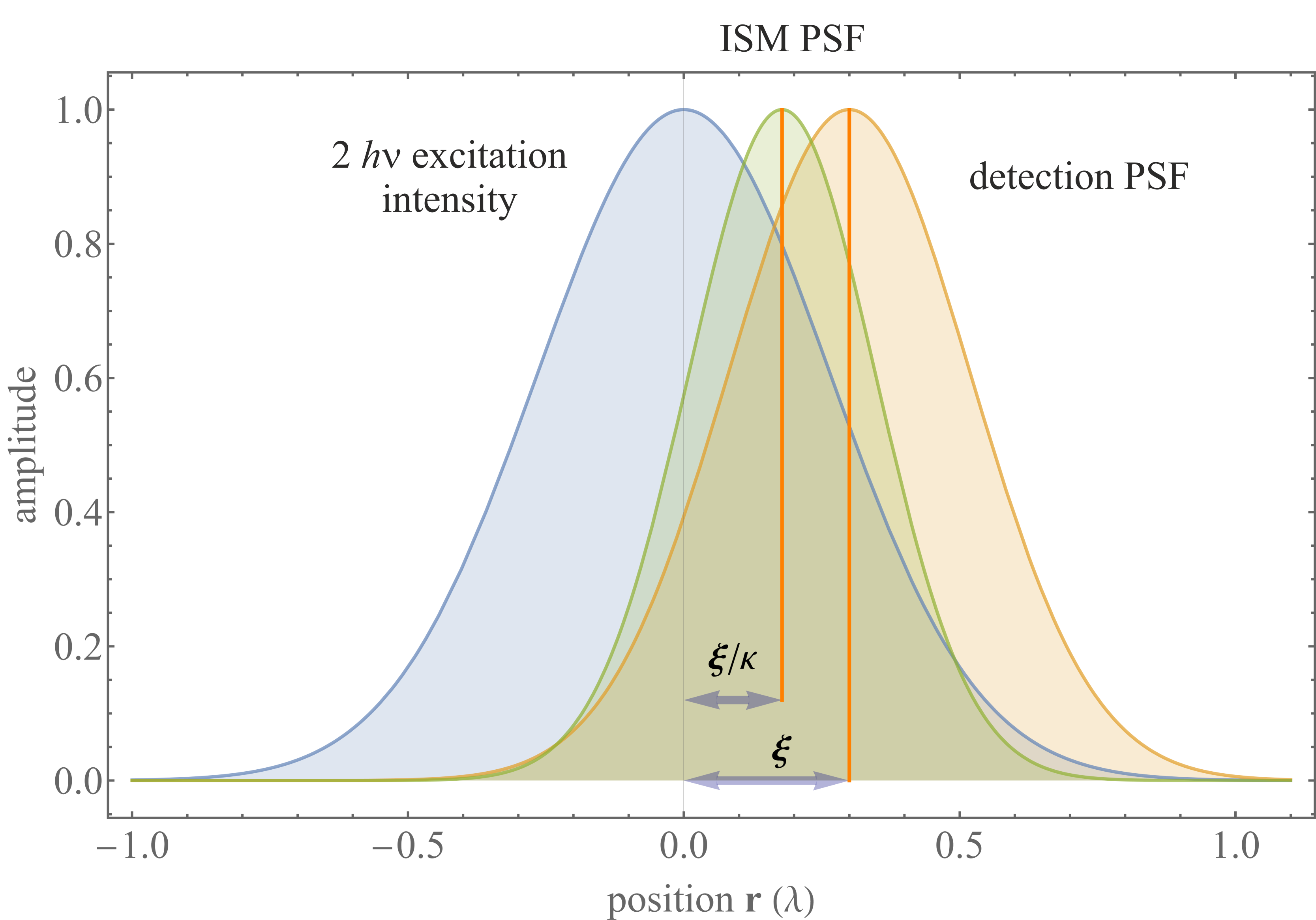}
    \caption{Pixel reassignment in two-photon excitation ISM. By contrast to the ISM in Fig.~\ref{fig:ISMReassignment}, the excitation intensity distribution (one-photon excitation PSF) in two-photon microscopy has a larger width due to the larger excitation wavelength. }
    \label{fig:TwoPhotonISMReassignment}
    \vspace{-5mm}
\end{figure}

To gain deeper insight into the best possible lateral resolution achievable by a two-photon excitation microscope, we consider two-photon excitation along with ISM detection, \textit{i.e.}, recording at each scan position a small image of the excited region and performing pixel reassignment to obtain the high resolution ISM image; see Sec.~\ref{ISM}. To do so, we approximate the one-photon excitation PSF and the single pixel detection PSF once more by Gaussians with variances $\sigma_\mathrm{ex}^2$ and $\sigma_\mathrm{em}^2$ (see Sec.~\ref{ISM}). We can visualize the PSF of the scan image recorded by one pixel at position $\boldsymbol{\xi}$ on the array detector as shown in Fig.~\ref{fig:TwoPhotonISMReassignment}; also see Eq.~\ref{eq:ISMPSF}.  

The new reassignment factor $\kappa$ (see Sec.~\ref{ISM}) is found by looking at the product of the detection PSF with the square of the one-photon excitation PSF, yielding a Gaussian function with variance $\sigma^{-2}=2\sigma_\mathrm{ex}^{-2}+\sigma_\mathrm{em}^{-2}$ and mid point position $\boldsymbol{\xi}/\kappa$ with $\kappa = 1+2\left(\lambda_\mathrm{em}/\lambda_\mathrm{ex}\right)^2$, which would yield for the case $\lambda_\mathrm{ex} = 2\lambda_\mathrm{em}$ the value $\kappa = 3/2$; also see Eq.~\ref{eq:kappa}.

\begin{figure}[H]
    \centering
    \includegraphics[width=0.95\linewidth]{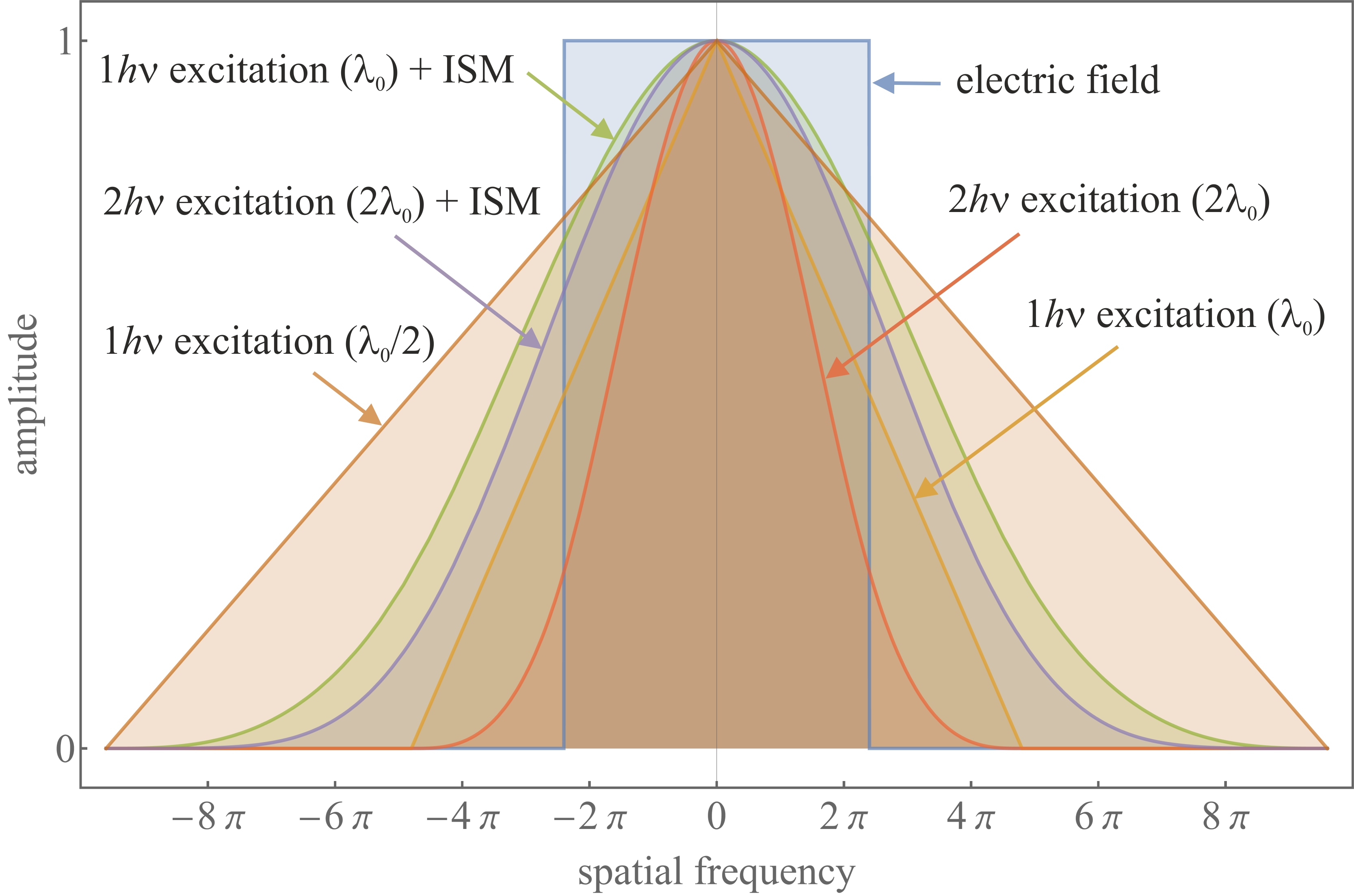}
    \vspace{-1mm}
    \caption{Comparison of one- and two-photon microscopy. For explanation see main text. }
    \label{fig:TwoPhoton}
    \vspace{-4mm}
\end{figure}

We now compare the performance of such a two-photon excitation ISM with that of a one-photon excitation CLSM and ISM at \emph{half} the wavelength. For simplicity, we consider the toy model of a one-dimensional microscope. The Fourier representation of the excitation electric field of such a one-dimensional microscope is a uniform amplitude distribution over the frequency range supported by the microscope (maximum lateral frequency transmitted is $n k_0 \sin\Theta$). This is shown in Fig.~\ref{fig:TwoPhoton} by the table-top function (electric field). The auto-convolution of this uniform amplitude distribution yields the excitation OTF and is, for the one-dimensional and one-photon case, the triangular function shown in Fig.~\ref{fig:TwoPhoton} and denoted by ``1$h\nu$ excitation ($\lambda_0$)."

The two-photon excitation PSF for an excitation with $2\lambda_0$ wavelength is given by the square of the one-photon excitation PSF. As such, its OTF corresponds to the auto-convolution of the one-photon OTF shown by ``1$h\nu$ excitation ($\lambda_0)$" in Fig.~\ref{fig:TwoPhoton}, but scaled down (along the frequency axis) by a factor of 2 (remember that we compare two-photon excitation at $2\lambda_0$ with one-photon excitation at $1\lambda_0$). The corresponding curve is denoted by ``2$h\nu$ excitation (2$\lambda_0$)". The OTFs for the extensions of one-photon and two-photon excitation fluorescence microscopy with ISM are also shown, together with the OTF of the one-photon excitation at $\lambda_0/2$ for comparison.

As can be seen, the frequency support of two-photon excitation at $2\lambda_0$ wavelength is equal to that of the one-photon excitation at $\lambda_0$, but with increased amplitudes at low frequencies and decreased amplitudes at large frequencies. In other words, a two-photon microscope transmits high lateral spatial frequencies less efficiently than a one-photon microscope operating at half the wavelength. This is also true when we compare two-photon ISM with one-photon ISM, as shown by the two curves ``1$h\nu$ excitation ($\lambda_0$) + ISM" and ``2$h\nu$ excitation ($2\lambda_0$) + ISM" in Fig.~\ref{fig:TwoPhoton}. Both modes have a frequency support equal to that of a one-photon excitation at $\lambda_0/2$, but with considerably damped amplitudes at high spatial frequencies, with one-photon ISM performing slightly better than two-photon ISM. Thus, two-photon (or multi-photon) excitation generally performs worse, in terms of resolution, than one-photon microscopes at half the wavelength, though biological tissue remains more transparent (less scattering) at long wavelengths, giving access to greater penetration depths in two- and multi-photon excitation microscopes.

\vspace{-6mm}
\subsection{\label{Im_Detect} Models for single spot confocal analysis} \vspace{-3mm}

Point scanning microscopes, including confocal and two-photon microscopes, have been used to study both dynamic~\cite{sprague2004analysis,digman2011lessons,wunderlich2013microfluidic,jazani2019method,jazani2019alternative} and static~\cite{berland1995two,rossow2010raster,kristoffersen2014testing,thiele2020confocal,karpf2020spectro} phenomena with both immobile ~\cite{berland1995two,nettels2008unfolded,wunderlich2013microfluidic,jazani2019method,jazani2019alternative} as well as scanning~\cite{digman2008phasor,rossow2010raster,digman2011lessons,fazel2022high} spots under continuous or pulsed illumination~\cite{gregor2005optical,jazani2019method}. Point scanning microscopes, particularly confocal microscopes, provide data for myriad analysis tools including fluorescence recovery after photo-bleaching (FRAP)~\cite{sprague2004analysis,loren2015fluorescence,moud2022fluorescence} used in the study of sub-cellular environments by monitoring diffusion of fluorophores into previously photo-bleached regions, FLIM~\cite{suhling2015fluorescence,datta2020fluorescence}, where photon arrival time statistics following pulsed excitation are collected and analyzed, and Fluorescence Correlation Spectroscopy (FCS) ~\cite{digman2011lessons,elson1974fluorescence,magde1974fluorescence} where photon arrival times or fluorescence intensities, often collected under constant illumination, are correlated in time to infer dynamical parameters~\cite{jazani2019method,jazani2019alternative}.

\begin{figure}[H]
    \centering
    \includegraphics[width=1\linewidth]{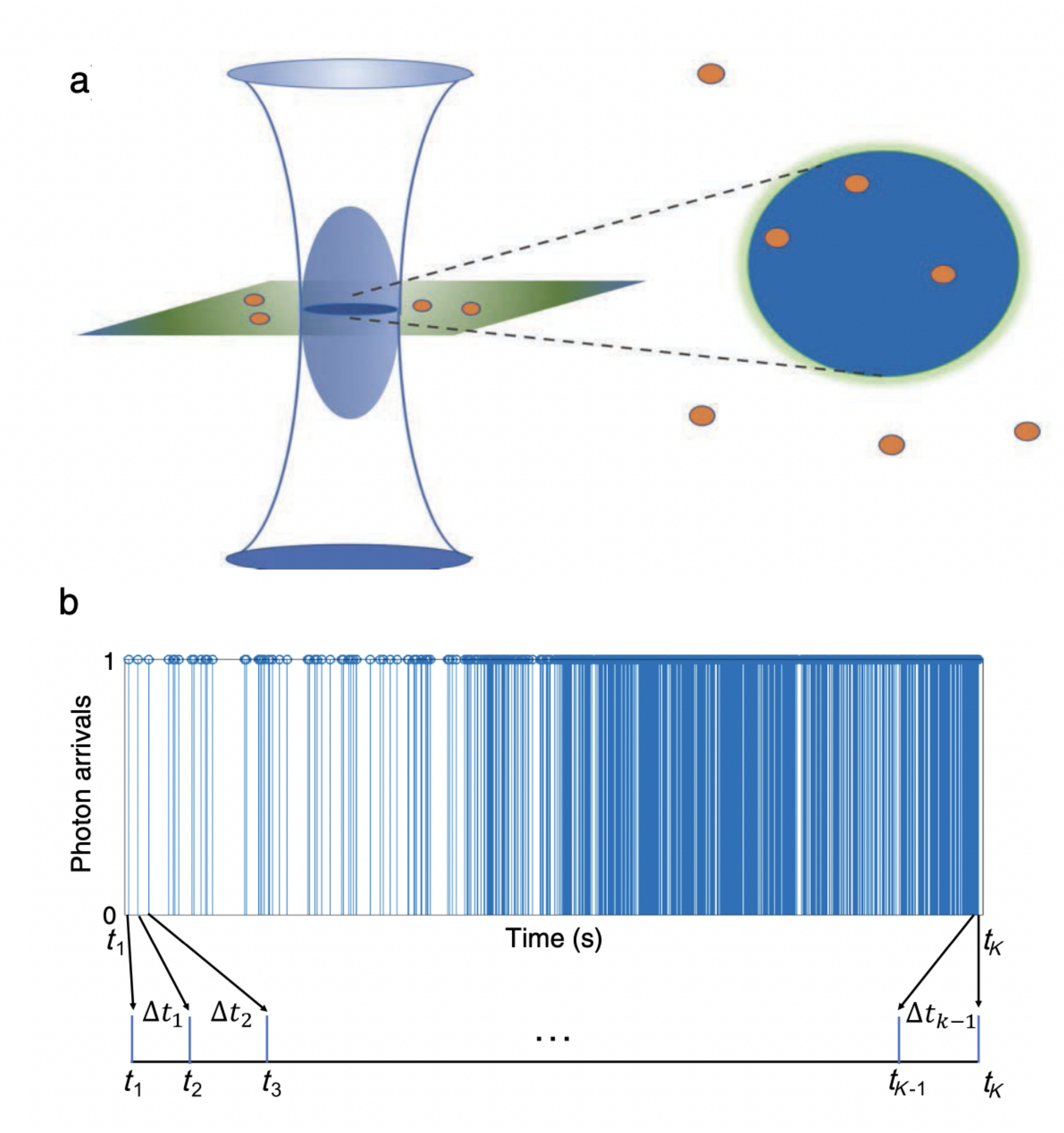}
    \caption{
    In (a) we show a schematic of confocal volume (in blue) with labeled molecules emitting photons in proportion to their degree of excitation decaying from the confocal volume center.
    In (b) we show a synthetic trace with 1500 photons generated assuming four molecules diffusing at $1\mu \mathrm{m}^2/s$ for $30 ms$ using background and molecule photon emission rates of $10^3$ photons/$s$ and $4 \times 10^4$ photons/$s$, respectively. The figure is adapted from Ref.~\cite{tavakoli2020pitching}. }
    \label{fig:Confocal-FCS} \vspace{-3mm}
\end{figure}

Here, we begin with a description of FCS where a static confocal spot is used to determine the reaction kinetics and diffusion coefficient of particles freely diffusing through the spot; see Fig.~\ref{fig:Confocal-FCS}a. In particular, this figure illustrates a scenario often analyzed using FCS
with labeled molecules freely diffusing through a static confocal spot becoming excited in proportion to the local light intensity. In traditional FCS analysis, a fraction of emitted photons are captured and dynamical properties are obtained by auto-correlating in time the emitted light intensity or photon arrival times~\cite{elson1974fluorescence,magde1974fluorescence,bright1989fluorescence,lakowicz2006principles}. While auto-correlating photon arrivals is computationally informative, it is data inefficient and eliminates single molecule information already encoded in the signal~\cite{jazani2019alternative,jazani2022computational,tavakoli2020pitching}. What is more, uncertainty is rarely propagated on derived quantities. Thus, a statistical method directly analyzing photon arrivals is warranted avoiding data post-processing including auto-correlation~\cite{jazani2019method,jazani2019alternative,tavakoli2020pitching}. Here, we begin by deriving the likelihood for the collection of $K+1$ photons whose inter-arrival intervals~\cite{tavakoli2020pitching} are designated by $\Delta t_{1:K}=\left\{\Delta t_1, ..., \Delta t_{K}\right\}$, see Fig.~\ref{fig:Confocal-FCS}b, under the assumption of continuous illumination. 

We begin by considering the confocal PSF derived earlier in this section in Eq.~\ref{eq:ConfocalGaussPSF} and, for simplicity, immediately adopt Cartesian coordinates where $\boldsymbol{r}= \left(\boldsymbol{\rho},z\right)$. For an arbitrary $M$ molecules located at $\boldsymbol{r}_k^m$ at time $t_k$, we write the following profile \vspace{-3mm}
\begin{equation}
\vspace{-2mm}
    S_k(\boldsymbol{r}) = \sum_{m=1}^M \delta\left(\boldsymbol{r}-\boldsymbol{r}^m_k\right).
    \label{FCS:density}
\end{equation}
As such, the total expected photon emission rate at time level $k$, $\mu_k$, follows from \vspace{-3mm}
\begin{equation}
    \mu_k(r) = \mu_{\mathcal{B}} + \mu_0\int d\boldsymbol{r}U_{\mathrm{cf}}(r)S_k(\boldsymbol{r}) = \mu_{\mathcal{B}} + \sum_{m=1}^M \mu_k^m(\boldsymbol{r}_k^m),
    \label{FCS:emission}
\end{equation}
where $\mu_k^m(\boldsymbol{r}_k^m)=\mu_0 U_{\mathrm{cf}}(\boldsymbol{r}_k^m)$ is the expected photon emission rate from the $m$th molecule located at $\boldsymbol{r}_k^m$, $\mu_0$ is the maximum photon emission rate associated with a molecule located at the PSF center, and $\mu_{\mathcal{B}}$ is the background photon emission rate. The photon emission rate, $\mu_k$, then dictates the photon interval time, $\Delta t_k$, 
\begin{equation}
    \Delta t_k \sim \mathrm{Exponential}\left(\mu_k(r)\right),
    \label{FCS:Dt}
\end{equation}
using notation introduced in Sec.~\ref{prob-stat}. This exponential waiting time follows from Poisson distributed photon emission per unit time implying exponentially distributed photon inter-arrival times.

Finally, under the assumption of a normal diffusion model with open boundary conditions, 
\begin{equation}
    \boldsymbol{r}_k^m|D \sim \mathrm{Normal}\left(\boldsymbol{r}_{k-1}^m,2D\Delta t_k \right),
    \label{FCS:location}
\end{equation}
where $D$ is the diffusion coefficient assumed to be constant across time and space. From Eq.~\ref{FCS:location} we see that the rate $\mu_k(r)$ inherits its stochasticity from the stochastic positions. 

\begin{figure}[H]
    \centering
    \includegraphics[width=1\linewidth]{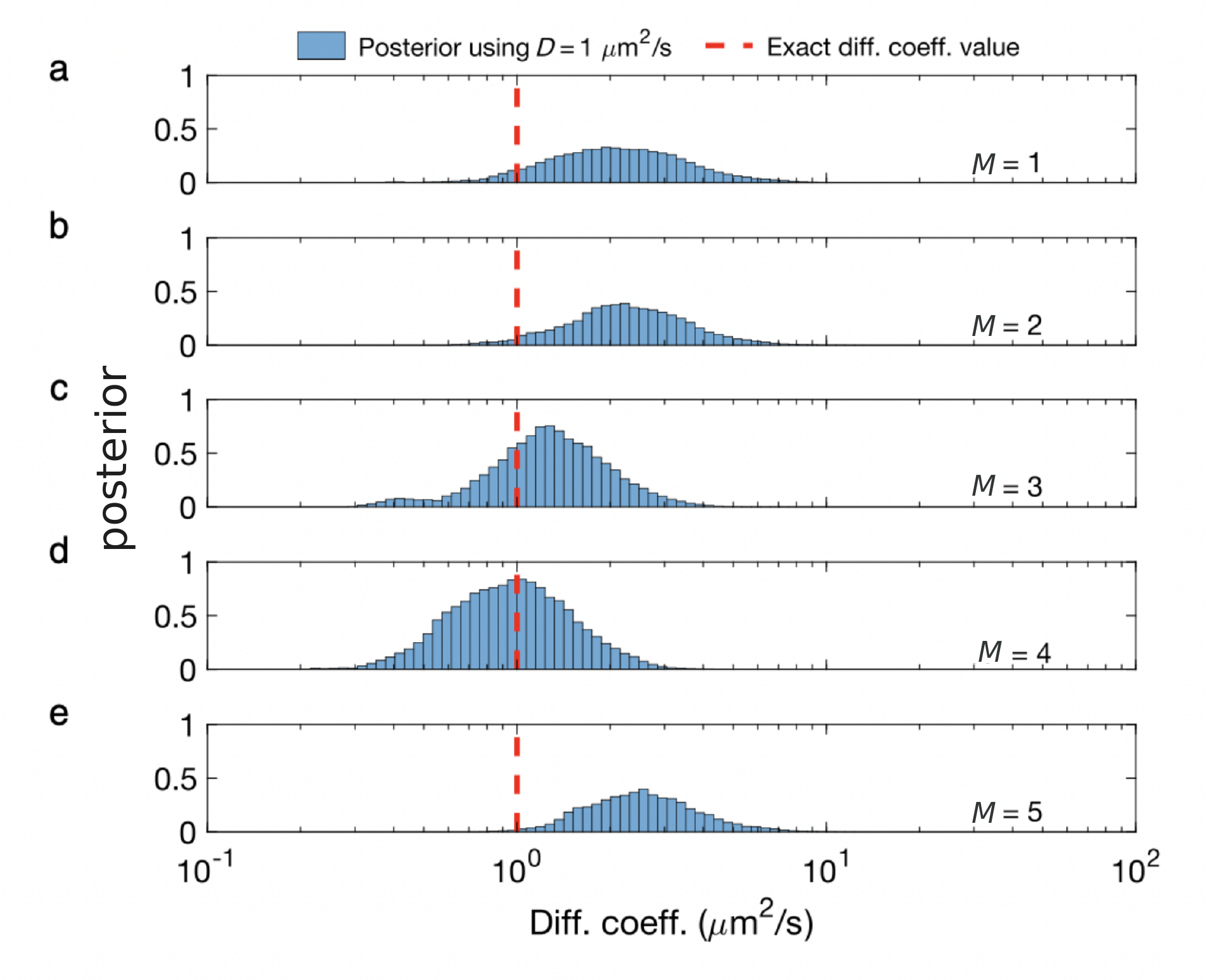}
    \vspace{-0.7cm}
    \caption{
    Posteriors over diffusion coefficients strongly depend on the pre-specified $M$ when operating within a parametric Bayesian paradigm. 
    The trace analyzed contains $\approx$1800 photons generated from 4 molecules diffusing at $D=1 \mu \mathrm{m}^2/s$ for 30 $ms$ with a background and maximum molecule photon emission rate of $10^3$ and $4\times 10^4$ photons/s, respectively. To deduce $D$ within the parametric paradigm, we assumed a fixed number of molecules: (a) $M=1$; (b) $M=2$; (c) $M=3$; (d) $M=4$; and (e) $M=5$. The correct estimate in panel d--and the mismatch in all others--highlights why we must use the available photons to simultaneously learn the number of molecules and $D$. The figure is adapted from Ref.~\cite{tavakoli2020pitching}. }
    \label{fig:Diff-FCS}
    \vspace{-3mm}
\end{figure}

Given the forward model described above, we now construct the likelihood for $K$ photon inter-arrival times, $\Delta t_{1:K}$, given by Eq.~\ref{FCS:Dt}. As $\Delta t_{1:K}$ are \textit{iid} (see Sec.~\ref{prob-stat}), the trace's likelihood is simply the product of the likelihood of every individual photon time interval
\begin{align}
    P(\Delta t_{1:K}&|M,D,\boldsymbol{\overline{\overline{r}}},\mu_0,\mu_{\mathcal{B}}) \nonumber \\
    &= \prod_k \mathrm{Exponential}\left(\Delta t_k;\mu_k(r)\right),
    \label{FCS:Likelihood}
\end{align}
where $\mu_k(r)$ is an implicit function of $M, D, \mu_0$ and $\mu_{\mathcal{B}}$; see Eqs.~\ref{FCS:emission} and \ref{FCS:location}. Moreover, double overbars represent the set of all possible values for the two associated indices, namely $m$ and $k$. 

To maximize the likelihood we would need to determine the number of molecules either in advance, \textit{i.e.}, parametric model, or work within a non-parametric paradigm and infer the number of molecules alongside other parameters. The likelihood above cannot naively be maximized to obtain parameters due to classic over-fitting problems favoring more complex models, \textit{i.e.}, larger numbers of molecules. However, in the former case, assuming a wrong parametric model with $M$ molecules~\cite{jazani2019alternative,tavakoli2020pitching} can result in incorrect estimates of other parameters, \textit{e.g}, diffusion coefficient; see Fig.~\ref{fig:Diff-FCS}. 

\begin{tcolorbox}[colback=brown!5!white,colframe=brown!75!black,title=Statistical Framework \ref{box:FCS}: Confocal under continuous illumination]
Data: photon inter-arrival times
\begin{equation}
    \Delta t_{1:K} = \left\{\Delta t_1, ..., \Delta t_{K}\right\}. \nonumber
\end{equation}
Parameters: loads, diffusion coefficient, molecule trajectories, molecule maximum emission rate, background emission
rate 
\begin{equation}
    \vartheta = \left\{\overline{b},D,\overline{\boldsymbol{\overline{r}}},\mu_0,\mu_{\mathcal{B}}\right\}. \nonumber
\end{equation}
Likelihood: 
\begin{equation}
    P\left(\Delta t_{1:K}|\vartheta\right) = \prod_k \mathrm{Exponential}\left(\Delta t_k;\mu_k(r)\right). \nonumber
\end{equation}
Priors:
\begin{align}
    q_m \sim & \, \mathrm{Beta}\left(A_q,B_q\right), \,\, m=1:\infty, \nonumber \\
    b_m \sim & \, \mathrm{Bernoulli}\left(q_m\right) \label{FCS:Bernoulli}, \nonumber \\
    D \sim & \, \mathrm{InvGamma}\left(\alpha_D,\beta_D\right), \nonumber \\
    \boldsymbol{r}_k^m \sim & \, \mathrm{Normal}\left(\boldsymbol{r}_{k-1}^m,2D\Delta t_k \right), \nonumber \\
    \mu_0 \sim & \, \mathrm{Gamma}\left(\alpha_{\mu},\beta_{\mu}\right), \nonumber \\
    \mu_{\mathcal{B}} \sim & \, \mathrm{Gamma}\left(\alpha_{b},\beta_{\mathcal{B}}\right). \nonumber
\end{align}
Posterior:
\begin{equation}
    P(\vartheta|\Delta t_{1:K}) \propto P(\Delta t_{1:K}|\vartheta)P(\vartheta). \nonumber
\end{equation}
\label{box:FCS}
\end{tcolorbox} 

As such, we abandon the parametric paradigm and start leveraging BNP tools ~\cite{ferguson1973bayesian,hjort2010bayesian,gershman2012tutorial,presse2023data}. Of particular interest within the BNP paradigm is the Beta-Bernoulli process prior (see Sec.~\ref{prob-stat}) on the number of candidate molecules, $M$, formally allowed to tend to infinity, $M\rightarrow \infty$, 
{\it a priori}. Put differently, each molecule is treated as a Bernoulli random variable (a load), $b_{m}$, learned simultaneously along with other unknowns; see Sec.~\ref{prob-stat}. The probability of the load being one, equivalently the probability of the molecule being warranted by the data, is the single parameter of the Bernoulli distribution on which we place a Beta prior.

Within this framework, 
Eq.~\ref{FCS:emission} is modified by replacing $\sum_{m=1}^M \mu_k^m(\boldsymbol{r}_k^m)$ on the right hand side with $\sum_{m=1}^{\infty}b_m\mu_k^m(\boldsymbol{r}_k^m)$ and summing over infinite molecules. The likelihood then adopts the form \vspace{-1mm}
\begin{align}
   P(\Delta t_{1:K}|\vartheta)= \prod_k \mathrm{Exponential}\left(\Delta t_k;\mu_k(r)\right),
    \label{FCS:BNP-Likelihood}
\end{align}
but where $\vartheta$ now collects all unknowns including all loads. Our non-parametric posterior is proportional to the product of this likelihood and all priors including Beta-Bernoulli process priors on each molecule; see Box \ref{box:FCS}.

\begin{figure}[H]
    \centering
    \includegraphics[width=1\linewidth]{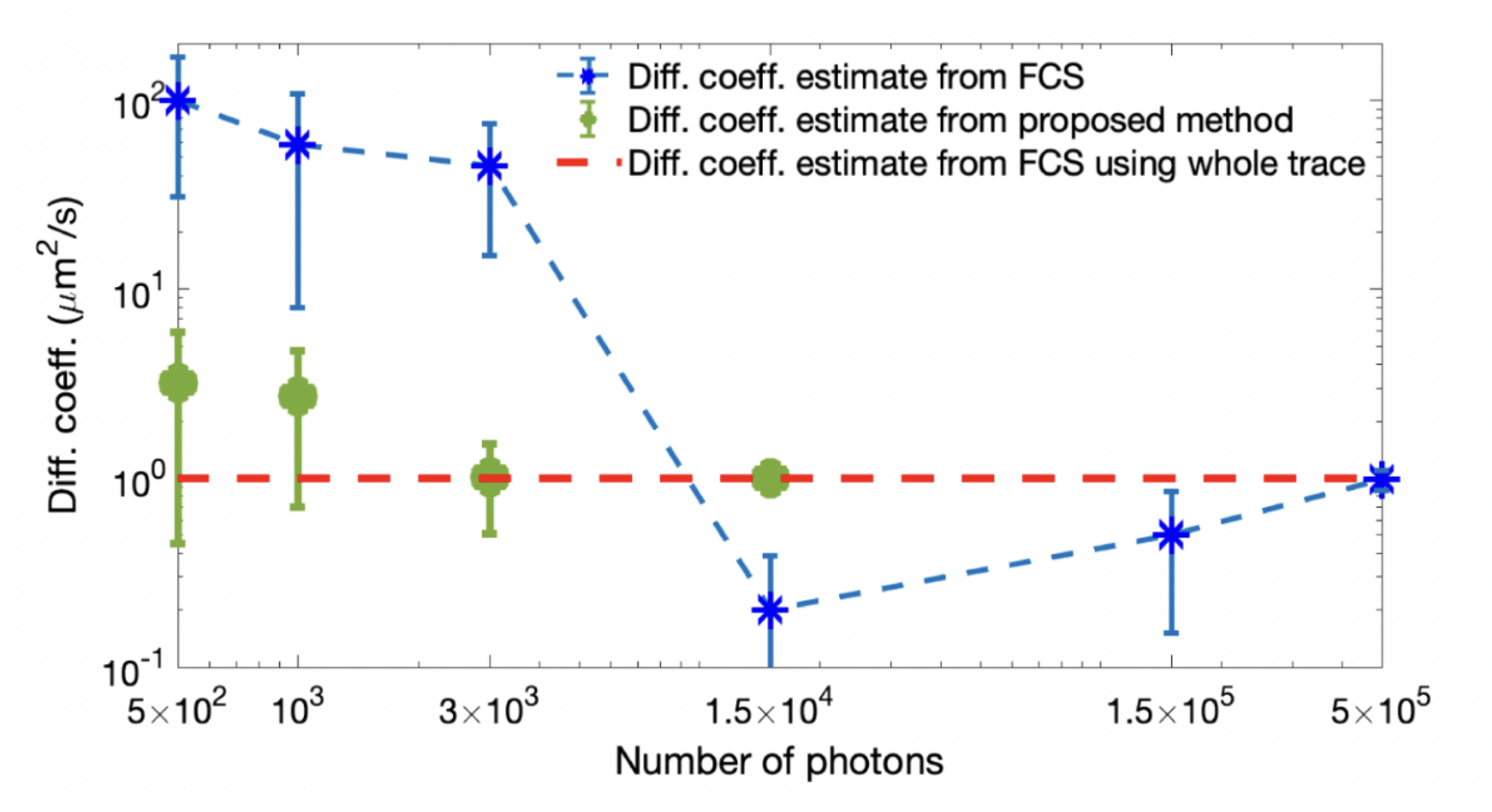}
    \caption{Comparison of diffusion coefficients, $D$, obtained from the statistical framework versus FCS plotted against photon counts used in the analysis. Photon arrival times were simulated using the parameter values in Fig.~\ref{fig:Confocal-FCS}b. The figure is adapted from Ref.~\cite{tavakoli2020pitching}.
    }
    \label{fig:Res-FCS} \vspace{-4mm}
\end{figure}

Now equipped with the posterior, we draw samples using Monte Carlo methods to learn the set of unknowns $\vartheta$. To learn the trajectories $\overline{\overline{r}}$, we use forward filtering backward sampling~\cite{scott2002bayesian,bishop2006pattern,tavakoli2020pitching,presse2023data}, while the remaining parameters are sampled either directly or using brute-force Metropolis-Hasting; see Sec.~\ref{prob-stat}. Fig. \ref{fig:Res-FCS} benchmarks the statistical framework of Box~\ref{box:FCS} versus FCS.

While the above approach returns a trajectory, due to the symmetry of the confocal PSF (see Eq.~\ref{eq:ConfocalGaussPSF}), the photon emission rate of Eq.~\ref{FCS:emission} and thus the likelihood given by Eq.~\ref{FCS:BNP-Likelihood}, are invariant under transformations leaving $\sqrt{(\boldsymbol{\rho}/\sigma_{\rho})^2+(z/\sigma_z)^2}$ unchanged. As such, equivalent positions lead to the same likelihood and thus unique positions cannot be determined using a single confocal setup.  

\begin{figure}[H]
    \centering
    \includegraphics[width=1\linewidth]{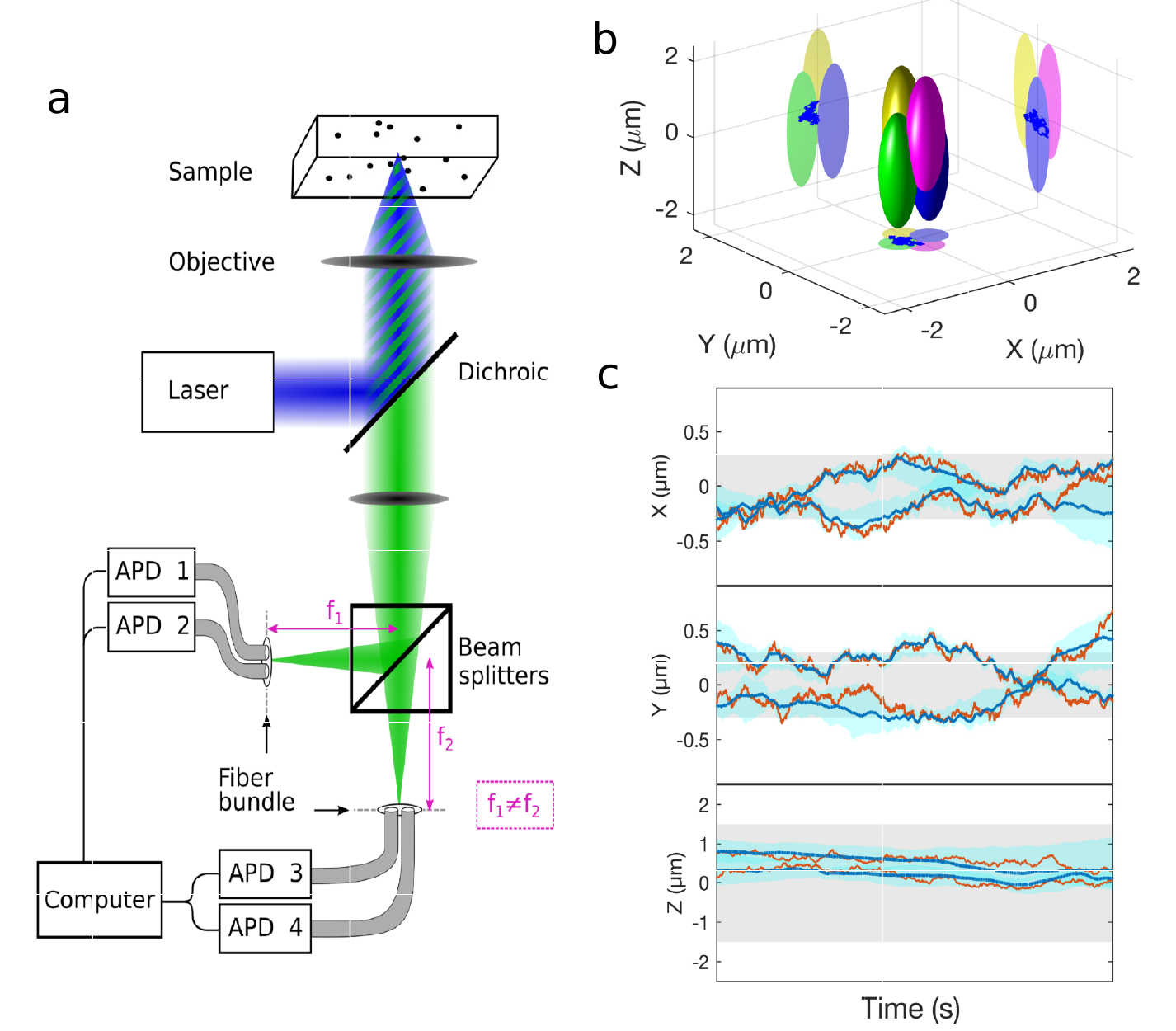}
    \caption{
    Multi-focal setup uniquely resolving many molecular trajectories simultaneously.
    (a) A beam splitter is used to divide the fluorescent emission (designated by green) into two paths later coupled into fibers and detected by 4 APDs corresponding to different focal spots. (b) PSFs associated to different light paths. (c) Trajectories for two freely diffusing molecules with $D=1~\mu\mathrm{m}^2/s, \mu_0 = 5\times 10^4$ photons/s and $\mu_{\mathcal{B}}=10^3$ photons/s. Here, the orange and blue curves represent the learned trajectories' ground truth and median, respectively. The blue and gray areas, respectively, denote the 95 percent confidence intervals and the PSF's width. The figure is adapted from Ref.~\cite{jazani2022computational}. }
    \label{fig:Multi-FCS} \vspace{-4mm}
\end{figure}

In contrast, it is possible to determine absolute molecular locations (trajectories) by breaking the spatial symmetry of the confocal spot by introducing a multi-focus confocal setup~\cite{jazani2022computational,lessard2007three,wells2010time}. Such a setup splits the confocal spot by introducing 4 detectors with axially and laterally offset detection volumes; see Fig.~\ref{fig:Multi-FCS}a-b. Photons from molecules in such a setup are detected in the $l$th detector with the following rate
\begin{equation}
    \mu_k^l(r) = \mu_{\mathcal{B}}^l + \mu_0 \sum_{m} b_m^l U_{\mathrm{cf}}^l(\mathbf{r}^m_k) 
\end{equation}
at time $k$; see Eq.~\ref{FCS:emission}. The total photon detection rate is, in turn, the sum of detection rates across all different detectors $\mu_k = \sum_l \mu_k^l$ and the likelihood is similar to the likelihood seen in Eq.~\ref{FCS:BNP-Likelihood}. From the this likelihood follows a posterior analogous to Box~\ref{box:FCS}
that when sampled yields absolute molecular trajectories; see Fig.~\ref{fig:Multi-FCS}c.

It is now conceivable to imagine generalizing the treatment above to include multiple diffusing species \cite{bacia2007practical}, species with donor and acceptor labels (FCS-FRET) \cite{torres2007measuring,schuler2018perspective}, as well as species undergoing reactions which alter their emission rate and kinetics \cite{zosel2018proline,xu2023single}. 

This brings us to the merits of statistical approaches compared to FCS. Such approaches are more data efficient, rigorously propagate error (including effects of finite data via the likelihood), can deal with any PSF shape, and optical aberrations~\cite{enderlein2004art,sarkar2019confocal}. But also, fundamentally, by avoiding data post-processing they learn more. For instance, in contrast to FCS, the statistical methods described above can learn properties of every individual molecule diffusing through the spot providing single molecule resolution albeit at computational cost.

Having dealt with continuous illumination, we now turn to pulsed illumination and, for simplicity alone, assume an immobile sample. Under pulsed illumination, the data acquired is a trace of $K$ photon arrival times, $\Delta t_{1:K}$, reported with respect to the immediate preceding pulses. These arrival times, also termed micro-times, encode the excited state lifetimes, $\tau_m$ for the $m$th species, of fluorophore species (see Sec.~\ref{Nonlinear}) present within the confocal spot. They also encode the associated photon ratios (weights) shown by $\pi_m$ for the $m$th species related to fluorophore densities as we will show later.

Although intuitive methods exist to determine excited state lifetimes~\cite{digman2008phasor}, similar to Fig.~\ref{fig:Res-FCS}, we find that lifetimes learned are sensitive to the parametric assumption on the number of lifetime species considered ~\cite{fazel2022fluorescence}. Indeed, existing techniques cannot simultaneously: 1) decode the number of fluorophore species present in a trace of photon arrival times; 2) operate on a broad range of lifetimes below the Instrument Response Function (IRF) (see Appendix~\ref{Se:Detector}) or lifetimes comparable to the laser inter-pulse times or similar lifetimes; 
3) provide uncertainties over parameter estimates; and 4) infer continuous fluorophore densities, \textit{i.e.}, lifetime maps given by $\Omega_m(\boldsymbol{r})=\mu_mS_m(\boldsymbol{r})$ where $S_m$ and $\mu_m$ are, respectively, the fluorophore densities (see Eq.~\ref{FCS:density}) and fluorophore excitation probability (for in-focus fluorophores) during a laser pulse for the $m$th species.

Here, we review statistical frameworks for FLIM analysis addressing the issues highlighted above with minimal photon budgets. In doing so, we first discuss a framework for a single confocal spot and then generalize to FLIM analysis methods using data from a scanning confocal setup to deduce lifetime maps over large FOVs.

\begin{figure}[H]
    \centering
    \includegraphics[width=1\linewidth]{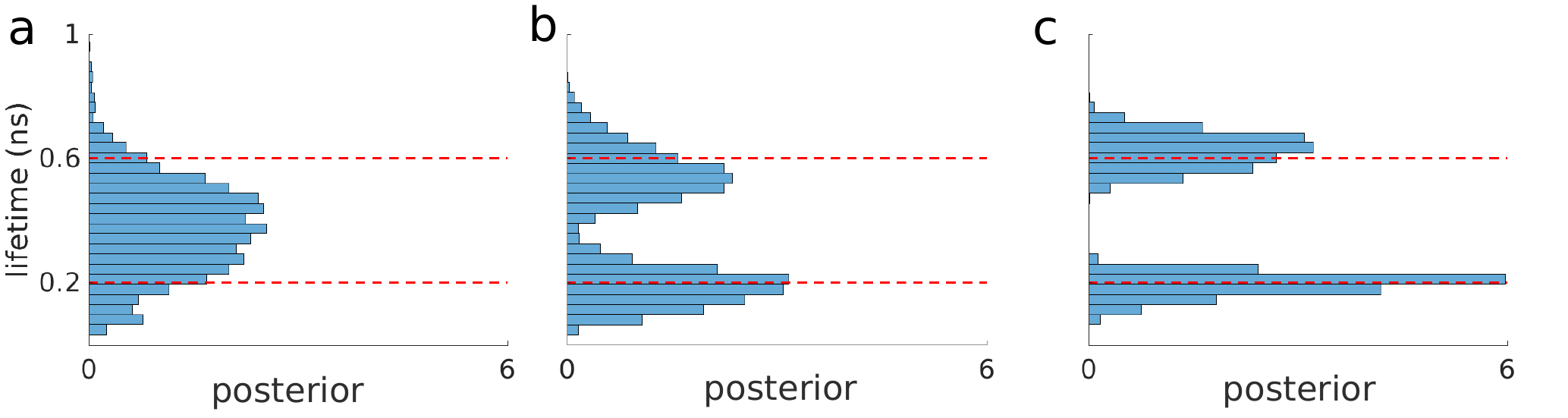}
    \caption{
    Lifetime histograms from single-pixel FLIM. Here, lifetimes are below the IRF and differ by sub-nanoseconds. Data sets used in panels (a-c) were simulated with $5\cdot10^2$, $10^3$, $2\cdot10^3$ photons, IRF width of 0.66~ns, and ground truth lifetimes of 0.2~ns and 0.6~ns denoted by dotted lines. Learning the correct number of fluorophore species here requires $>500$ photons. }
    \label{fig:FLIM_SinglePix} \vspace{-4mm}
\end{figure}

We begin by introducing the likelihood for $\Delta t_{1:K}$ collected from a single spot with $M$ species
 \begin{equation}
     P(\Delta t_{1:K}|\lambda_{1:M},\pi_{1:M}) = \prod_{k=1}^K P(\Delta t_k|\lambda_{1:M},\pi_{1:M}),
 \end{equation}
where $\lambda_m$ denotes the inverse lifetime $\tau_m=1/\lambda_m$ and $P(\Delta t_k|\lambda_{1:M},\pi_{1:M})$ denotes the likelihood of the $k$th arrival time. To derive $P(\Delta t_k|\lambda_{1:M},\pi_{1:M})$, we sum over all possibilities that could give rise to this photon including: all $M$ fluorophore species; and all $N_{pl}$ previous laser pulses. Assuming a Gaussian IRF, this leads to (see Appendix~\ref{Se:Detector} and Eq.~\ref{eq:SinglePhotonNoise}) \cite{fazel2022fluorescence,fazel2022high} 
\begin{align}
    P(\Delta t_k|&\lambda_{1:M},\pi_{1:M}) =   \bigg[\sum_{m=1}^M\pi_{m}\sum_{n=0}^{N_{pl}} \frac{\lambda_{m}}{2} \nonumber \\ 
    \times & \exp\Big(\frac{\lambda_m}{2}\left(2(\tau_{\mathrm{IRF}}-\Delta t_{k}-nT)+\lambda_{m}\sigma^2_{\mathrm{IRF}}\right)\Big) \nonumber \\
 \times & \mathrm{erfc}\left(\frac{\tau_{\mathrm{IRF}}-\Delta t_{k}-nT+\lambda_{m}\sigma^2_{\mathrm{IRF}}}{\sigma_{\mathrm{IRF}}\sqrt{2}}\right)\bigg],
\label{eq:SinglePixelLike}
\end{align}
where $\tau_{\mathrm{IRF}}, \sigma_{\mathrm{IRF}}^2$ and $T$, respectively, denote the IRF offset and variance, and the inter-pulse time; see Appendix~\ref{Se:Detector}. Ignoring excitation by previous pulses considered in Eq.~\ref{eq:SinglePixelLike}, we arrive at the likelihood obtained in Ref.~\cite{rowley2016robust}.

To summarize, parametrically, the number of fluorophore species, $M$, is pre-specified and often set to one or two for simplicity, {\it e.g.}, Refs.~\cite{rowley2016robust,kaye2017developing}. In contrast, non-parametrically, the number of fluorophore species are {\it a priori} assumed infinite~\cite{fazel2022high,fazel2022fluorescence}. 

\begin{tcolorbox}[colback=brown!5!white,colframe=brown!75!black,title=Statistical Framework \ref{box:SingleFLIM}: single spot FLIM]

Data: photon arrival times
\begin{equation}
    \Delta t_{1:K} = \left\{\Delta t_1, ..., \Delta t_{K}\right\}. \nonumber
\end{equation}
Parameters: inverse lifetimes, weights
\begin{equation}
    \vartheta = \left\{\lambda_{1:M}, \pi_{1:M}\right\}. \nonumber
\end{equation}
Likelihood:
\begin{equation}
    P\left(\Delta t_{1:K}|\vartheta\right) = \prod_k P\left(\Delta t_k|\vartheta\right). \nonumber
\end{equation}
Priors:
\begin{align}
    \lambda_m \sim & \, \mathrm{Gamma}\left(\alpha_{\lambda},\beta_{\lambda}\right), \nonumber\\
    \pi_m \sim & \mathrm{Dirichlet}_{1:M}\left(\frac{\alpha_{\pi}}{M}, ...,\frac{\alpha_{\pi}}{M}\right), \hspace{1mm} M \to \infty \nonumber 
\end{align}
Posterior:
\begin{equation}
    P(\vartheta|\Delta t_{1:K}) \propto P(\Delta t_{1:K}|\vartheta)P(\vartheta). \nonumber
\end{equation}
\label{box:SingleFLIM}
\end{tcolorbox}

Within the non-parametric paradigm, the single spot FLIM posterior is proportional to the likelihood Eq.~\ref{eq:SinglePixelLike} and priors over all unknown parameters, namely $\lambda_{1:M}$ and $\pi_{1:M}$. For $\lambda_m$, we use a Gamma prior to guarantee non-negative values. For $\pi_m$, we leverage the non-parametric Dirichlet process prior 
~\cite{neal2000markov,gelfand2005bayesian,sgouralis2017introduction} to facilitate inference over the probability in the number of species present warranted by the data, \textit{i.e.}, to address model selection; see Sec.~\ref{prob-stat}. Within this framework, as before when operating non-parametrically, we assume an {\it a priori} infinite number of species ($M\to \infty$) with associated weights $\pi_m$. As we sample these weights, the weights ascribed to species not contributing to the data attain negligible values. Fig.~\ref{fig:FLIM_SinglePix} shows lifetime histograms for two lifetimes below the IRF and with sub-nanosecond differences using 500, 1K and 2K photons.

\begin{figure}[H]
    \centering
    \includegraphics[width=1\linewidth]{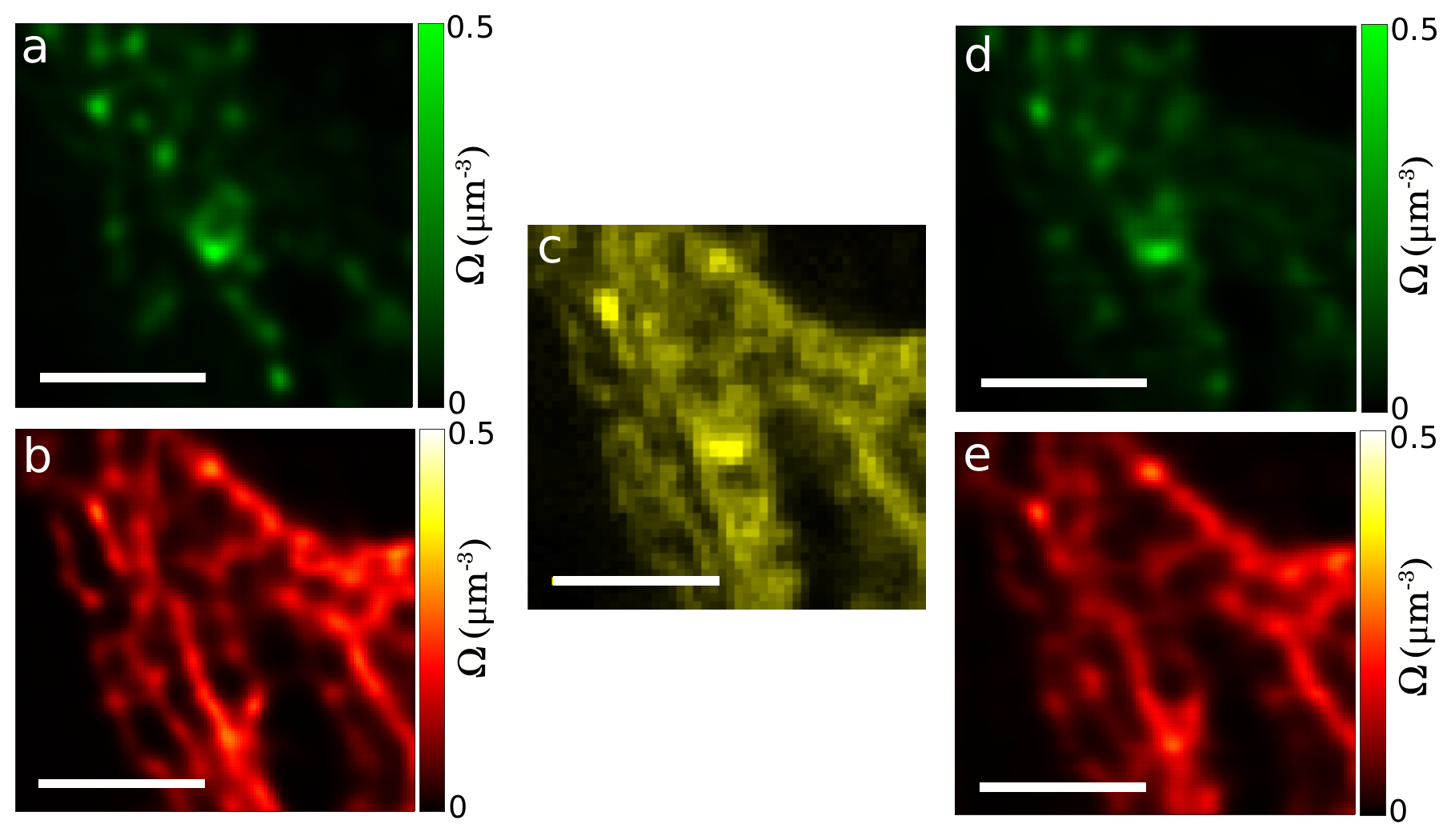}
    \caption{Experimental FLIM data from mixtures of two cellular structures (lysosome and mitochondria shown in green and red, respectively) stained with two different fluorophore species. (a-b) Ground truth lifetime maps. (c) Data acquired from mixtures of two ground truth maps. (d-e) Resulting sub-pixel interpolated lifetime maps obtained using the statistical framework of Box~\ref{box:MultiFLIM}. The average absolute difference between ground truth and learned maps is $\approx 4\%$. Scale bars are 4 $\mu$m. The figure is adapted from Ref.~\cite{fazel2022high}. }
    \label{fig:Res-FLIM} \vspace{-3mm}
\end{figure}

We now turn to FLIM over large FOVs where we show how to estimate smooth lifetime maps from confocal scanning data; see Fig.~\ref{fig:Res-FLIM}. FLIM data over large FOVs are typically collected using a CLSM to scan the sample over uniformly spaced horizontal trajectories where the spacing defines the data pixel size. The collected data is often arranged into a 2D pixel array where each pixel contains a subset of photon arrival times acquired over the pixel.

One naive way to process such data is to analyze each pixel independently using the framework of Box~\ref{box:SingleFLIM}. However, this yields pixelated lifetime maps where information from one pixel does not inform the neighboring pixels. In what follows, we review a framework for multi-pixel FLIM over large FOVs~\cite{fazel2022high,fazel2022building}
reporting lifetime maps below the data pixel size leveraging spatial correlations across pixels by invoking (non-parametric) GPs; see Sec.~\ref{prob-stat} and Fig.~\ref{fig:Res-FLIM}.

The likelihood here is now given by 
\begin{equation}    
P\left(\overline{\overline{\mathcal{W}}},\overline{\overline{\Delta t}}|\vartheta\right) = \prod_i \prod_{k_p} P\left(\mathcal{W}_{k_p}^i|\vartheta\right)P\left(\Delta t_{k_p}^i|\vartheta\right),
\end{equation}
where $\mathcal{W}_{k_p}^i$, for the $k_p$th pulse and $i$th pixel, is a binary variable designating whether a laser pulse leads to a photon detection or not. As before, $\vartheta$ collects all unknowns including inverse of lifetimes $\lambda_{1:M}$, multi-pixel lifetime maps $\Omega_{1:M}$, the loads $b_{1:M}$, and hyper-parameters $\nu_{1:M}$ over each species. Further, double overbars represent the set of all the possible values for the pair of indices associate to the corresponding parameter. Here, the likelihood associated with photon arrival times is similar to Eq.~\ref{eq:SinglePixelLike} and given by \vspace{-2mm}
\begin{align}
    P(\Delta & t_{k_p}^i |\vartheta) =   \bigg[\sum_{m=1}^M\pi_{m}\sum_{n=0}^{N_{pl}} \frac{\lambda_{m}}{2} \nonumber \\ 
    \times & \exp\Big(\frac{\lambda_m}{2}\left(2(\tau_{\mathrm{IRF}}-\Delta t_{k_p}^i-nT)+\lambda_{m}\sigma^2_{\mathrm{IRF}}\right)\Big) \nonumber \\
 \times & \mathrm{erfc}\left(\frac{\tau_{\mathrm{IRF}}-\Delta t_{k_p}^i-nT+\lambda_{m}\sigma^2_{\mathrm{IRF}}}{\sigma_{\mathrm{IRF}}\sqrt{2}}\right)\bigg]^{\mathcal{W}_{k_p}^i},
\label{eq:MultiPixelLike}
\end{align}
reducing to one for pulses that do not yield any photon detection (empty pulses with $\mathcal{W}^i_{k_p}=0$). In the above, the weights, $\pi_{1:M}$, are directly related to the lifetime maps by $\pi_m^i = (1-P_{0m}^i)\prod_{q\neq m}P_{0q}^i$~\cite{fazel2022high}, where $P_{0m}^i$ reflects the probability of no photon detection within the $i$th pixel from the $m$th species given by
\begin{equation}
    P_{0m}^i= \exp\left[-b_m\int\Omega_m(\boldsymbol{r})U_{\mathrm{cf}}(\xi^i-\boldsymbol{r})d\boldsymbol{r}\right],
    \label{FLIM:NoPhoton}
\end{equation}
where $\xi^i$ is center of the $i$th pixel. Moreover, $b_m$ denotes the loads associated to the $m$th lifetime map (see Sec.~\ref{prob-stat}) on which we place Beta-Bernoulli process priors (just as in Box~\ref{box:FCS}) to deduce the number of lifetime maps introduced by each fluorophore species present within the data. As a sanity check, we note that for species with $b_m=0$, the probability of no photon detection is one. 

After illustrating how we compute $P(\Delta t_{k_p}^i|\vartheta)$, we 
compute $P\left(\mathcal{W}_{k_p}^i|\vartheta\right)$ following the observation that $\mathcal{W}_{k_p}^i$ is Bernoulli distributed with success probability $1-\pi_0^i$
\begin{equation}
    \mathcal{W}_{k_p}^i \sim \mathrm{Bernoulli}(1-\pi_0^i).
    \label{FLIM:Detection}
\end{equation}
Here, $\pi_0^i$ is the probability of no photon detection from the $i$th pixel given by $ \pi_0^i = \prod_{m=1}^M P_{0m}^i$.

After introducing the likelihoods, we construct the posterior proportional to the product of the likelihood and priors over all unknown parameters. Our framework is doubly non-parametric as we use:
GP priors over continuous lifetime maps; and Beta-Bernoulli process priors  over the loads; see Sec.~\ref{prob-stat}. The GP priors over lifetime maps are comprised of an infinite set of correlated random variables, \textit{i.e.}, the value of the map at every point in space \vspace{-2mm}
\begin{equation}
    \Omega_m \sim \mathrm{GP}\left(\nu_m,\boldsymbol{K}\right),
\end{equation}
where $\mathbf{K}$ and $\nu_m$ denote the correlation kernel (also termed a covariance matrix) and the GP prior's mean. The remaining priors are either physically or computationally motivated; see Box~\ref{box:MultiFLIM}.

With the posterior at hand, we make inferences on $\vartheta$ once more by drawing samples from the posterior with Monte Carlo. Of note are elliptical slice samplers~\cite{murray2010elliptical} used to sample lifetime maps as the GP and likelihood do not form a conjugate pair.

\begin{tcolorbox}[colback=brown!5!white,colframe=brown!75!black,title=Statistical Framework \ref{box:MultiFLIM}: Multi-pixel FLIM]

Data: photon arrival times and pulses being empty or not
\begin{align}
    \overline{\overline{\Delta t}} &= \left\{\Delta t^{\,i=1}_{1:K_p}, ..., \Delta t^{N}_{1:K_p}\right\}, \nonumber\\
    \overline{\overline{\mathcal{W}}} &= \left\{\mathcal{W}^{\,i=1}_{1:K_p}, ..., \mathcal{W}^N_{1:K_p}\right\}. \nonumber
\end{align} 
Parameters: loads, inverse lifetimes, lifetime maps, GP prior averages (hyper-parameters)
\begin{equation}
    \vartheta = \left\{b_{1:M},\lambda_{1:M},\Omega_{1:M},\nu_{1:M}\right\}. \nonumber
\end{equation}
Likelihood:
\begin{equation}
    P\left(\overline{\overline{\mathcal{W}}},\overline{\overline{\Delta t}}\big|\vartheta\right) = \prod_k \prod_n P\left(\mathcal{W}_k^n;\vartheta\right)P\left(\Delta t_k^n;\vartheta\right). \nonumber
\end{equation}
Priors:
\begin{align}
    q_m \sim & \, \mathrm{Beta}\left(A_q,B_q\right), \,\, m=1:\infty, \nonumber \\
    b_m \sim & \, \mathrm{Bernoulli}\left(q_m\right) \label{FCS:Bernoulli}, \nonumber \\
    \lambda_m \sim & \, \mathrm{Gamma}\left(\alpha_{\lambda},\beta_{\lambda}\right), \nonumber\\
    \Omega_m \sim & \, \mathrm{GP}\left(\nu_m,\boldsymbol{K}\right), \nonumber \\
    \nu_m \sim & \, \mathrm{Normal}\left(0,\sigma^2_{\chi}\right). \nonumber
\end{align}
Posterior: 
\begin{equation}
    P\left(\vartheta\big|\overline{\overline{\mathcal{W}}},\overline{\overline{\Delta t}}\right) \propto P\left(\overline{\overline{\mathcal{W}}},\overline{\overline{\Delta t}}\big|\vartheta\right)P(\vartheta). \nonumber
\end{equation}
\label{box:MultiFLIM}
\end{tcolorbox}
\vspace{-5mm}
\subsection{\label{sec:SIM} Structured illumination microscope} \vspace{-2mm}

As discussed in Sec.~\ref{Fluo_PSF}, a major drawback of wide-field fluorescence imaging is the lack of optical sectioning arising from the OTF's missing cone; see Fig.~\ref{fig:psfderivation}. This, in turn, yields out-of-focus blur degrading the final images. Previously, we discussed near-field and point scanning methods where, for example, conventional confocal microscopes achieved optical sectioning via pinholes; see Sec.~\ref{CLSM-Confocal}. Here, we discuss how SIM achieves both optical sectioning and resolution beyond the diffraction limit~\cite{mertz2019introduction,Wu2018,strohl2016frontiers,Heintzmann1999,Gustafsson2000}.

Patterned illumination, with a high spatial stripe contrast near the focal plane~\cite{Neil1997}, was introduced in an effort to attain optical sectioning. The pattern, whose illumination contrast ideally fades away from the focal plane, was then translated twice yielding three images $\mathcal{I}_l$ with corresponding phase offsets $\phi_l$, $l=0:2$. One way to attain optical sectioning was to create three images from differences in two images, say $\Delta \mathcal{I}_{ll'}(\mathbf{r})=\mathcal{I}_l(\mathbf{r})- \mathcal{I}_{l'}(\mathbf{r})$, and combine them according to 
\begin{equation}
\begin{split}
\Lambda_{sec}(\mathbf{r}) &= \sqrt{\Delta \mathcal{I}_{01}(\mathbf{r})^2 + \Delta \mathcal{I}_{12}(\mathbf{r})^2 + \Delta \mathcal{I}_{20}(\mathbf{r})^2},\\
\phi_l &= \frac{2l\pi}{3},\, l = 0:2.
\label{eq:optical_sectioning}
\end{split}
\end{equation}
The hope was that by subtracting images, unmodulated (out-of-focus) contributions cancel as they are approximately homogeneously illuminated. 

These early efforts ultimately motivated structured illumination to achieve higher resolution~\cite{Heintzmann1999, Gustafsson2000} that we now discuss by considering the SIM image formation model. SIM images are generated from the product of the fluorophores' distribution, $S(\mathbf{r})$ (see Sec.~\ref{Im_Detect}), and the illumination intensity pattern, $I_{\text{ex}}(\mathbf{r})$, followed by convolution with the microscope's wide-field detection PSF (also see Eq.~\ref{eq:ImageFormation}) \vspace{-2mm}
\begin{equation}
\Lambda(\mathbf{r}) = \mathcal{B} + I \left[S(\mathbf{r}) I_{\text{ex}}(\mathbf{r})\right] \otimes U_{\mathrm{wf}}(\mathbf{r}),
\label{eq:structured_illumination}
\end{equation}
where $\mathcal{B}$ is the background arising from out-of-focus fluorescent features, ignored here for simplicity only, and $U_{\mathrm{wf}}(\mathbf{r})$ and $I$ are, respectively, the wide-field PSF, ({\it e.g.}, see Sec.~\ref{ScalarApprox}) and fluorophore brightness per frame. 

\begin{figure}[H]
    \centering
    \includegraphics[width=0.5\linewidth]{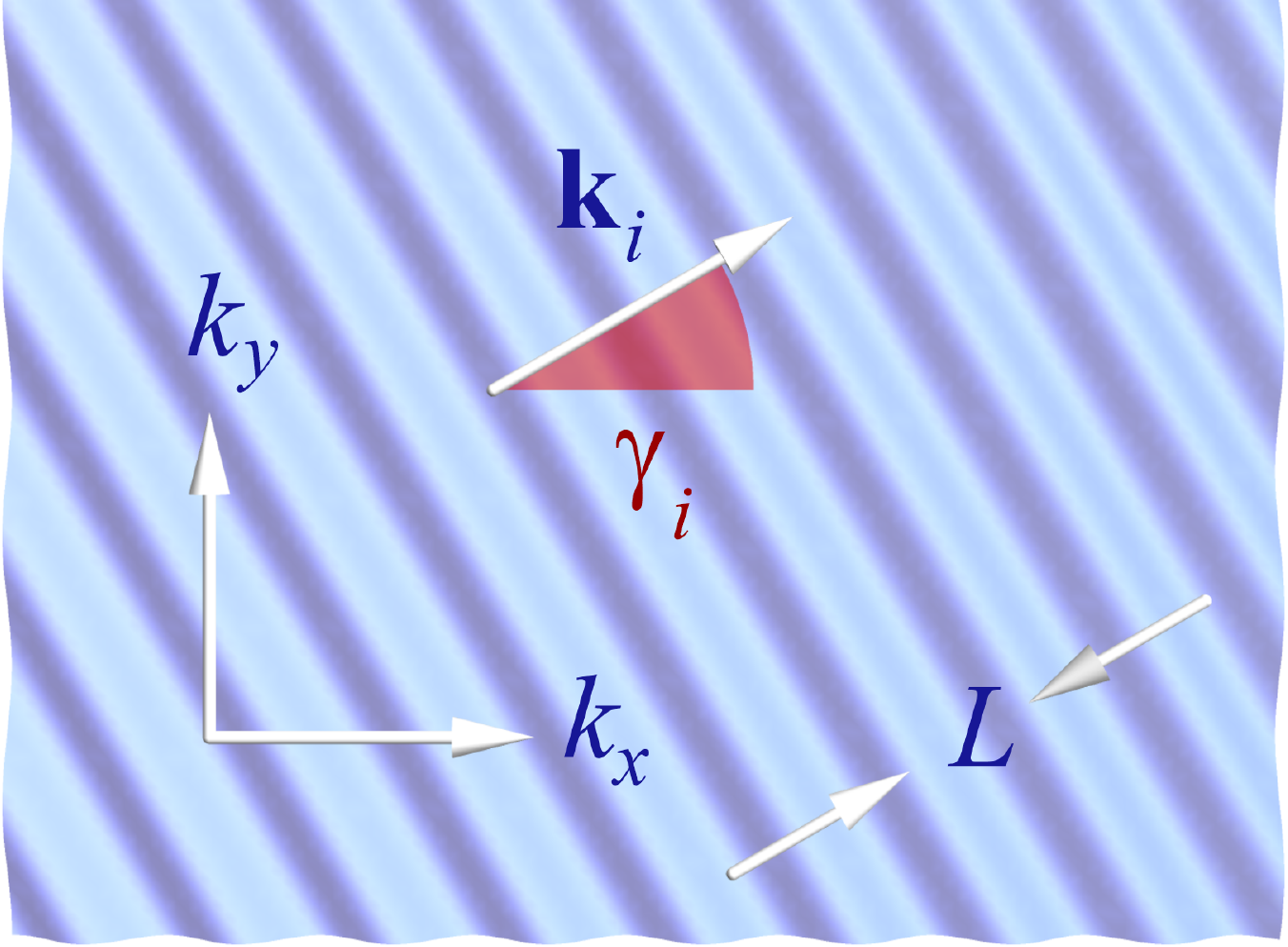}
    \caption{Sinusoidal illumination pattern for SIM microscopy. Here, $\mathbf{k}_i$ is the wave vector, $L$ is the fringe spacing, and $\gamma_i$ is the illumination's in-plane angle. The phase is related to the position of the maxima relative to the optical axis.
    }
    \label{fig:SIMPattern} \vspace{-3mm}
\end{figure}

While various modulated illumination patterns are conceivable for SIM \cite{heintzmann2003saturated,planchon2011rapid,mudry2012structured}, in practice, the sample is typically illuminated using a sinusoidal intensity, $I_{\text{ex}}(\mathbf{r})$, with different in-plane phases and angles (see Fig.~\ref{fig:SIMPattern}) achieved by interference-based~\cite{Heintzmann2017, Ma2021} methods or using laser-scanning~\cite{york2012resolution,york2013instant,Gregor2019}.

Under the former method, such intensity patterns are generated by interfering two to three laser beams, 
followed by rotation and translation of the grating embedded within the setup's illumination arm.
For two beam interference, the image formation is described by \vspace{-1mm}
\begin{align}
\Lambda^{li}(\mathbf{r}) &= I \left[S(\mathbf{r})\; \frac{1}{2}\left(1+\mathfrak{M}\cos\left(\mathbf{r}\cdot\mathbf{k}_i+\phi_l\right)\right)\right] \otimes U_{\mathrm{wf}}(\mathbf{r}), \nonumber\\ \gamma_i &= \arctan{\frac{k_x^i}{k_y^i}},\, L=2\pi/\sqrt{{k_x^i}^2+{k_y^i}^2} 
\label{eq:sinusoidal_illumination}
\end{align}
where $\mathfrak{M}$, the modulation depth, is assumed to be one in subsequent calculations for simplicity; $\mathbf{k}_i$ is the wave vector, with components $k_x^i$ and $k_y^i$, defining the oscillatory pattern's period, \textit{i.e.}, the fringe spacing denoted by $L$; see Fig.~\ref{fig:SIMPattern}. Next, $\gamma_i$ and $\phi_{l}$ are, respectively, the $l$th in-plane illumination angle and the $i$th phase offset determining the position of the maxima relative to the optical axis; see Fig.~\ref{fig:SIMPattern} and Eq.~\ref{eq:optical_sectioning}. \vspace{-2mm}

\begin{figure}[H]
    \centering
\includegraphics[width=0.95\linewidth]{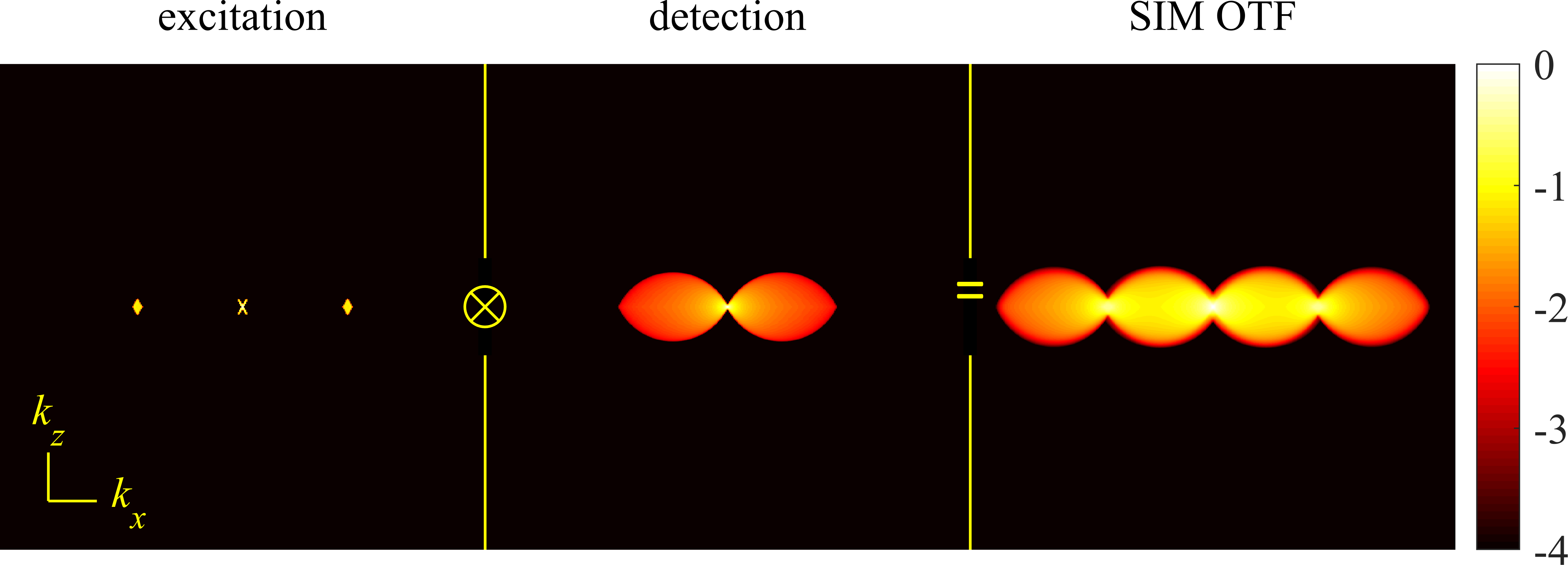}
\vspace{-1mm}
    \caption{The SIM OTF. The left and middle panels, respectively, illustrate Fourier transforms of the modulated illumination intensity (SIM excitation OTF given by the three delta-peaks) and wide-field detection. The right panel shows the SIM OTF obtained by convolution of the two other panels; see also Eq.~\ref{eq:sinusoidal_illumination_Fourier}.}
    \label{fig:OTFSIM} \vspace{-4mm}
\end{figure}

The improved resolution is achieved by exploiting the frequency mixing, \textit{i.e.}, moir{\'e} effect, between the excitation pattern and the sample's spatial frequencies. That is, previously unobservable high frequency information, beyond the wide-field OTF's support, is shifted down into the microscope's band-pass (\textit{i.e.}, frequency support of microscope's OTF); see Fig.~\ref{fig:otfwidefieldscalar} and \ref{fig:OTFSIM}. 

The effect of structured illumination is most intuitively demonstrated in Fourier space. For the sinusoidal pattern given in Eq.~\ref{eq:sinusoidal_illumination}, its Fourier representation reads \vspace{-5mm}
\begin{align}
\tilde{\Lambda}^{li}(\mathbf{k}) 
& = I\bigg[\tilde{S}(\mathbf{k}) \otimes \nonumber \tilde{I}_{ex}(\mathbf{k})\bigg] \text{OTF}_{\mathrm{wf}}(\mathbf{k}) \nonumber \\ & 
=I \bigg[\tilde{S} (\mathbf{k}) \otimes \Big( \delta(0) 
 + \frac{1}{2}e^{+ i \phi_l} \delta(\mathbf{k} + \mathbf{k}_i) \nonumber \\
& \hspace{3.5mm} + \frac{1}{2}e^{- i \phi_l} \delta(\mathbf{k} - \mathbf{k}_i)\Big)\bigg]\; \text{OTF}_{\mathrm{wf}}(\mathbf{k}) \nonumber \\ & =  I\bigg[\tilde{S}(\mathbf{k}) + \frac{1}{2}e^{+ i \phi_l} \tilde{S}(\mathbf{k + k_i}) 
\label{eq:sinusoidal_illumination_Fourier}\\
& \hspace{3.5mm} + \frac{1}{2}e^{- i \phi_l} \tilde{S}(\mathbf{k - k_i}) \bigg]\; \text{OTF}_{\mathrm{wf}}(\mathbf{k}) \nonumber\\
& = \tilde{\mathcal{I}}_0(\mathbf{k}) + \frac{1}{2} e^{+ i \phi_l} \tilde{\mathcal{I}}_+(\mathbf{k + k_i}) + \frac{1}{2} e^{- i \phi_l} \tilde{\mathcal{I}}_-(\mathbf{k - k_i}), \nonumber
\end{align}
where $\mathrm{OTF}_{\mathrm{wf}}(\mathbf{k})$ denotes the wide-field OTF (see Fig.~\ref{fig:otfwidefieldscalar} and middle panel of Fig.~\ref{fig:OTFSIM}), and the sinusoidal illumination pattern (for a given angle and phase) is described by three different frequencies in the Fourier domain (see the left panel in Fig.~\ref{fig:OTFSIM}) yielding the three SIM harmonics $\tilde{\mathcal{I}}_0$, $\tilde{\mathcal{I}}_+$, $\tilde{\mathcal{I}}_-$. \vspace{-2mm}

In Eq.~\ref{eq:sinusoidal_illumination_Fourier} the first delta function within the parenthesis coincides with the Fourier representation of the uniform (wide-field) illumination. However, the two subsequent terms arise from the illumination patterning. These additional terms are two copies of the Fourier representation of the sample $\tilde{S}(\mathbf{k})$ phase shifted by a factor $\phi_l$ and frequency shifted by $\mathbf{k}_i$, providing extra information compared to wide-field microscopy. 

Supposing the OTF cut-off frequency is $k_c$, 
the frequency shifted components contain high frequency information  otherwise absent in the central component (sum frequency $|\mathbf{k} + \mathbf{k}_i|\leq k_c$ and difference frequency $|\mathbf{k} - \mathbf{k}_i|\leq k_c$ at each sample frequency of $k$). When imaged, only frequencies inside the support of the wide-field OTF are captured. However, sample information across different (higher) frequency regions now lie within the microscope's band-pass; see Fig.~\ref{fig:OTFSIM}. 

While the three SIM harmonics $\tilde{\mathcal{I}}_0$, $\tilde{\mathcal{I}}_+$, $\tilde{\mathcal{I}}_-$ (wide-field and $\pm$ pattern wave vector) already contain frequencies beyond wide-field band-pass, no sub-diffraction resolution can yet be achieved. This is because these components overlap in frequency space. In order to unmix the overlapping parts, we need to acquire at least three images with different pattern phases $\phi_l$ designated by $\tilde{\Lambda}^{li}(\mathbf{k})$ in Fourier space. The relation between the three SIM harmonics and these images is best shown in matrix form \vspace{-1.5mm}
\begin{equation*}
\vspace{-1.5mm}
\begin{bmatrix}
    \tilde{\Lambda}^{0i}(\mathbf{k})\\
    \tilde{\Lambda}^{1i}(\mathbf{k})\\
    \tilde{\Lambda}^{2i}(\mathbf{k})\\
\end{bmatrix} = 
\begin{pmatrix}
1 & 0.5e^{i\phi_0} & 0.5e^{-i\phi_0}\\
1 & 0.5e^{i\phi_1} & 0.5e^{-i\phi_1}\\
1 & 0.5e^{i\phi_2} & 0.5e^{-i\phi_2}
\end{pmatrix}
\begin{bmatrix}
    \tilde{\mathcal{I}}_0(\mathbf{k})\\
    \tilde{\mathcal{I}}_+(\mathbf{k} + \mathbf{k}_i)\\
    \tilde{\mathcal{I}}_-(\mathbf{k} - \mathbf{k}_i)
\end{bmatrix}.
\end{equation*}
Here, we used a mixing matrix (the square matrix on the right), with different phases for the available spectra evenly spaced between $0$ and $2\pi$. This allows us to solve for $\tilde{\mathcal{I}}_0$, $\tilde{\mathcal{I}}_+$ and $\tilde{\mathcal{I}}_-$, \textit{i.e.}, unmixing the SIM harmonics. 
The unmixed components are then recombined by shifting them so that their true zero frequency is aligned with their zero frequency in Fourier space, \textit{i.e.}, $k_0$ setting. This yields an effective OTF extended to frequencies beyond the original OTF's support and thus yield high resolution SIM images, \textit{i.e.}, fluorophore densities $S(\mathbf{r})$ ~\cite{muller2016open,lal2016structured}. 

Several techniques, mostly operating within the Fourier domain, unmix the SIM harmonics to reconstruct SIM images~\cite{mudry2012structured,lukevs2014three,lal2016structured,perez2016optimal,muller2016open,huang2018fast,lai2019jointly,jin2020deep,christensen2021ml,smith2021structured,shah2021deep,cai2022total,qiao2022rationalized}. Ideally, reconstruction requires knowledge of multiple imaging system properties including the exact OTF, pattern frequency, phases, and modulation depth (\emph{e.g.}, see Eq.~\ref{eq:sinusoidal_illumination}). Inaccurately specified properties can result in imperfect SIM reconstructions typically exhibiting well-known artifacts~\cite{Demmerle2017}. For instance, refractive index mismatch (see Fig.~\ref{fig:psfwidefieldrefindex}) may lead to repeated features along the z axis known as ``ghosting". Similarly, fine hexagonal ``honeycomb" pseudo-structures can arise when neglecting background ($\mathcal{B}$ of Eq.~\ref{eq:structured_illumination}) in 2D SIM images; a false $k_0$ setting impacting the OTF leads to so-called ``hatching", \textit{i.e.}, the appearance of angle-specific stripes in one or more directions, to name only a few. 

Working in real space not only allows us to cleanly propagate uncertainty (as all data is collected in real space) but also avoids artifacts tied to Fourier domain, such as the $k_0$ setting. For this reason, we review SIM reconstruction in real space~\cite{orieux2011bayesian}.

The total likelihood over the data is the product of likelihood models corresponding to each phase $\phi_l$ and wave vector $\mathbf{k}_i$ 
\vspace{-3mm}
\begin{equation}  
\vspace{-3mm}
P(\overline{\overline{w}}_{1:N}\big|\overline{\overline{\Lambda}}_{1:N}) = \prod_{i=1}^3\prod_{l=1}^3\prod_{n=1}^N P\left(w_n^{li}|\Lambda_n^{li}\right),
\end{equation}
where double overbars represent all possible values of $i$ and $l$ (an overbar for each index) and where $P\left(w_n^{li}|\Lambda_n^{li}\right)$ is the likelihood over a single pixel. Here, $w_n^{li}$ and $\Lambda_n^{li}$, respectively, denote the observed (data) and expected photon counts over the pixel $n$ using an illumination with phase $\phi_l$ and wave vector $\mathbf{k}_i$. The expected photon count is given by (see Eqs.~\ref{eq:subset}-\ref{eq:PSFintPix}) \vspace{-2mm}
\begin{equation}
    \Lambda_n^{li} = \iint_{\mathcal{A}_n} dxdy\Lambda^{li}(\mathbf{r}),
\end{equation}
where $\Lambda^{li}(\mathbf{r})$ is given by Eq.~\ref{eq:sinusoidal_illumination} and $\mathcal{A}_n$ is the pixel area. Assuming high SNR and a Charged Coupled Devices (CCD) camera noise model of Eq.~\ref{Detector_CCDNoise}, we arrive at the following single pixel likelihood 
\begin{equation}
    P\left(w_n^{li}|\Lambda_n^{li}\right) = \mathrm{Gaussian}(w_n^{li};g\Lambda_n^{li}+o,\sigma^2_w),
\end{equation}
where $g, o$ and $\sigma^2_w$, respectively, are the camera gain, offset and read-out variance; see Appendix~\ref{Se:Detector}. 

Finally, we now present a Bayesian framework required in rigorous noise propagation from the SIM data~\cite{orieux2011bayesian}. 
Within this framework, we consider priors over unknowns including the GP priors (see Sec.~\ref{prob-stat}) over the fluorophore distributions, $S(\mathbf{r})$, and priors over the GP's covariance kernel, $\boldsymbol{K}(\nu)$. These parameters are collectively re-grouped under $\vartheta=\{S(\mathbf{r}),\nu\}$.  The complete framework is described in Box~\ref{box:SIM}.

\begin{tcolorbox}[colback=brown!5!white,colframe=brown!75!black,title=Statistical Framework \ref{box:SIM}: SIM]

Data: pixel values in ADUs
\begin{equation}
   \overline{\overline{w}}_{1:N} = \left\{\overline{\overline{w}}_1,\, ..., \overline{\overline{w}}_N \right\}. \nonumber
\end{equation}
Parameters: fluorophore distribution, GP covariance kernel parameter (hyper-parameter) 
\begin{equation}
    \vartheta = \left\{S(\mathbf{r}), \nu\right\}. \nonumber
\end{equation}
Likelihood:
\begin{equation}
   p(\overline{\overline{w}}_{1:N}\big|\vartheta) = \prod_{i=1}^3\prod_{l=1}^3\prod_{n=1}^N \mathrm{Gaussian}(w_n^{li};g\Lambda_n^{li}+o,\sigma^2_w). \nonumber
\end{equation}
Priors:
\begin{align}
    S(\mathbf{r}) \sim & \, \mathrm{GP}\left(0,\boldsymbol{K}(\nu)\right), \nonumber\\
    \nu \sim & \,\mathrm{Gamma}(\alpha_{\nu},\beta_{\nu}).
\end{align}
Posterior:
\begin{equation}
    P(\vartheta|\overline{\overline{w}}_{1:N}) \propto P(\overline{\overline{w}}_{1:N}|\vartheta)P(\vartheta). \nonumber
\end{equation}
\label{box:SIM}
\end{tcolorbox}

Finally, we numerically sample the posterior to learn the unknowns $\vartheta$. The sampling procedure is particularly straightforward for this SIM framework as the Gaussian likelihood and GP priors are conjugate to the Gaussian likelihood resulting in a closed form posterior. At low SNR, this procedure fails as it leads to negative values for the fluorophore distributions allowed by the GP prior and an alternative method is proposed.

The SIM experiment described combined with image reconstruction typically achieves resolutions up to approximately two times better than the diffraction limit.
This is because, in practice, the illumination pattern is also diffraction-limited implying that its corresponding Fourier peaks lie within the support of the system's wide-field OTF, limiting the resolution improvement to a factor of about two (not considering, \textit{e.g.}, the Stokes shift of fluorescence emission; see Sec.~\ref{Nonlinear}). 
The resolution of the SIM image is then approximately $2\pi/(k_c + k_i)$ along the direction of $\mathbf{k}_i$; see Eqs.~\ref{eq:resolutionMonotonic} and \ref{eq:widefieldlateralresolution}-\ref{eq:widefieldaxialresolution}. The process has to be repeated for at least three orientations ($\mathbf{k}_i,\,i=1:3$) to achieve near isotropic lateral resolution enhancement. 
    
Resolution improvement using structured illumination can also be combined with illumination modalities other than wide-field epi-fluorescence providing optical sectioning, such as TIRF~\cite{Chung2006}, grazing incidence illumination~\cite{Guo2018}, or light-sheet microscopy ~\cite{Chen2014,Chang2017,chen2022resolution}. 

While the above discussion was focused on 2D SIM, the principle extends to 3D by using three (or more) interfering beams generating a laterally and axially varying illumination pattern~\cite{Gustafsson2008, Shao2011, Heintzmann2017}. In three-beam interference, five phase shifts are necessary to unambiguously unmix frequencies, resulting in five SIM harmonics all three orientations ($\mathbf{k}_i$) as opposed to three for 2D SIM; see Eq.~\ref{eq:sinusoidal_illumination_Fourier}. This leads to 15 SIM harmonics requiring 15 images to unambiguously unmix the harmonics. This process has to be repeated for each z-position. Although more complicated than 2D SIM, 3D SIM achieves approximately twofold resolution improvement and optical sectioning as the OTF copies in 3D SIM overlap and fill the wide-field OTF's missing cone; see Fig.~\ref{fig:psfderivation}.

All SIM implementations mentioned thus far use linear fluorescence excitation. This has the advantage of being relatively gentle to living samples as low excitation intensities can be used compared to other super-resolution imaging methods employing non-linear response of fluorophores to excitation light; see Secs.~\ref{Nonlinear} and \ref{SR}. While the SIM resolution improvement is restricted to approximately twofold as the illumination pattern itself is limited by diffraction, higher resolution is achievable by combining SIM with non-linear fluorophore photo-physics~\cite{Heintzmann2002, Gustafsson2005}; see Secs.~\ref{Nonlinear} and \ref{SR_Target}.

For instance, resolution improvement beyond twofold was achieved by combining structured illumination with saturation of the excited state emission, \textit{i.e.}, increasing the excitation intensity above a threshold where fluorophores spend a longer time in the excited than the ground state~\cite{Gustafsson2005}, termed Saturated SIM (SSIM). In such regimes, fluorophore responses to intensities exceeding the saturation threshold remain unchanged and thus the effective intensity seen by fluorophores is the saturation intensity. As such, the effective intensity pattern seen by fluorophores beyond the saturation threshold start deviating from the sinusoidal pattern, \textit{i.e.}, flat top sinusoidal pattern. Such distorted patterns contain more than three harmonics shifting more frequencies within the band-pass of the microscope by contrast to sinusoidal patterns; see Fig.~\ref{fig:SIMPattern}. However, the frequency unmixing now provides more displaced SIM harmonics in Fourier space that require more images to be separated. When this process is repeated at multiple orientations, SSIM has achieved isotropic lateral resolution of approximately 50~nm on fluorescent beads in Ref.~\cite{Gustafsson2005}. 

Alongside higher spatial resolution comes higher computational complexity in unmixing SIM harmonics and high intensities required for saturation prevent its use for biological imaging. Instead, photo-switchable fluorescent proteins (see Sec.~\ref{Nonlinear}), cycling between dark and bright states at much lower intensities, can be used while remaining live-cell compatible. By leveraging both dye photo-switching with structured illumination patterns, resolutions similar to SSIM are achieved~\cite{Rego2012, Li2015}. 
\vspace{-6mm}
\subsection{Light-sheet microscope}\label{Sec:LSFM}
\vspace{-2mm}

Optical sectioning motivated the development of 3D microscopy such as Light-Sheet Fluorescence Microscopy (LSFM)~\cite{voie1993Microscopy}. LSFM allows optical sectioning, \textit{i.e.}, increases the OTF's $k_z$ content, by generating a thin light sheet~\cite{power2017guide,olarte2018light}. In doing so, LSFM both simultaneously minimizes out-of-focus fluorescence, otherwise present in naive wide-field microscopy (see Fig.~\ref{fig:psfderivation}), and reduces sample photo-damage~\cite{olarte2018light,stelzer2021light}. 

\begin{figure}[H]
    \centering   \includegraphics[width=0.95\linewidth]{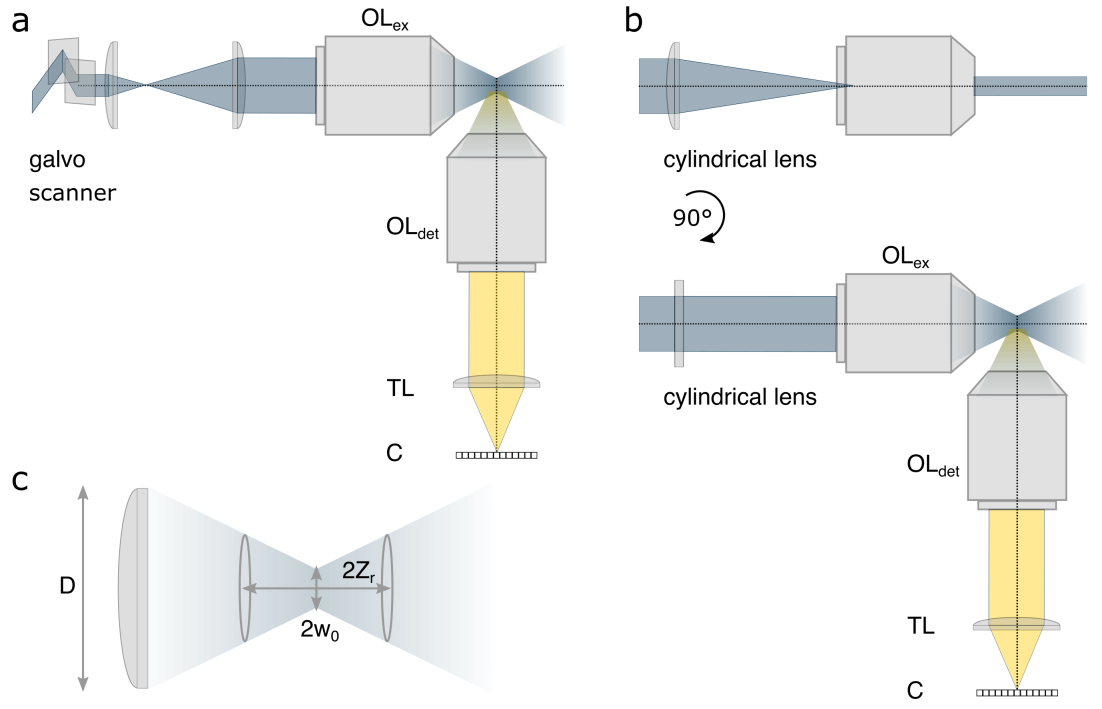}
    \caption{LSFM setups. (a) In Digitally scanned laser Light-Sheet Microscopy (DLSM) a galvanometric (galvo) scanning unit rapidly moves a Gaussian beam perpendicular to the detection axis focused in the sample through the excitation objective lens ($\text{OL}_{\text{ex}}$). Signal from the excited focal plane is collected through the detection objective lens ($\text{OL}_{\text{det}}$) and tube lens (TL) onto a camera (C). (b) In SPIM, a static light-sheet is formed by a cylindrical lens in the excitation path creating an elongated beam in one direction (above) and the same perpendicular detection optics as in panel a. 
    (c) A schematic of the Gaussian beam in panels a-b focused through a lens or objective with diameter D, beam waist $\omega_0$ and Raleigh length $z_r$. 
    }
    \label{fig:LSM} \vspace{-4mm}
\end{figure}

In LSFM, illumination and light collection paths are orthogonal providing volumetric information on the sample when axially scanning the illumination sheet; see Fig.~\ref{fig:LSM}~\cite{chakraborty2019light}.This setup facilitates faster volumetric imaging in contrast to previously discussed point-by-point scanning (see Sec.~\ref{CLSM-Confocal}). Moreover, LSFM achieves optical sectioning through illumination in contrast to other modalities, \textit{e.g.}, CLSM, where sectioning is possible only along the detection path while illuminating large portions of the specimen along the excitation path~\cite{stelzer2015light}. Indeed, while TIRF (see Sec.~\ref{Fluo_Wide}) avoids this unnecessary light dose, it is restricted to volumes neighboring the illuminated surface. 

\begin{figure}[H]
    \centering
    \includegraphics[width=0.95\linewidth]{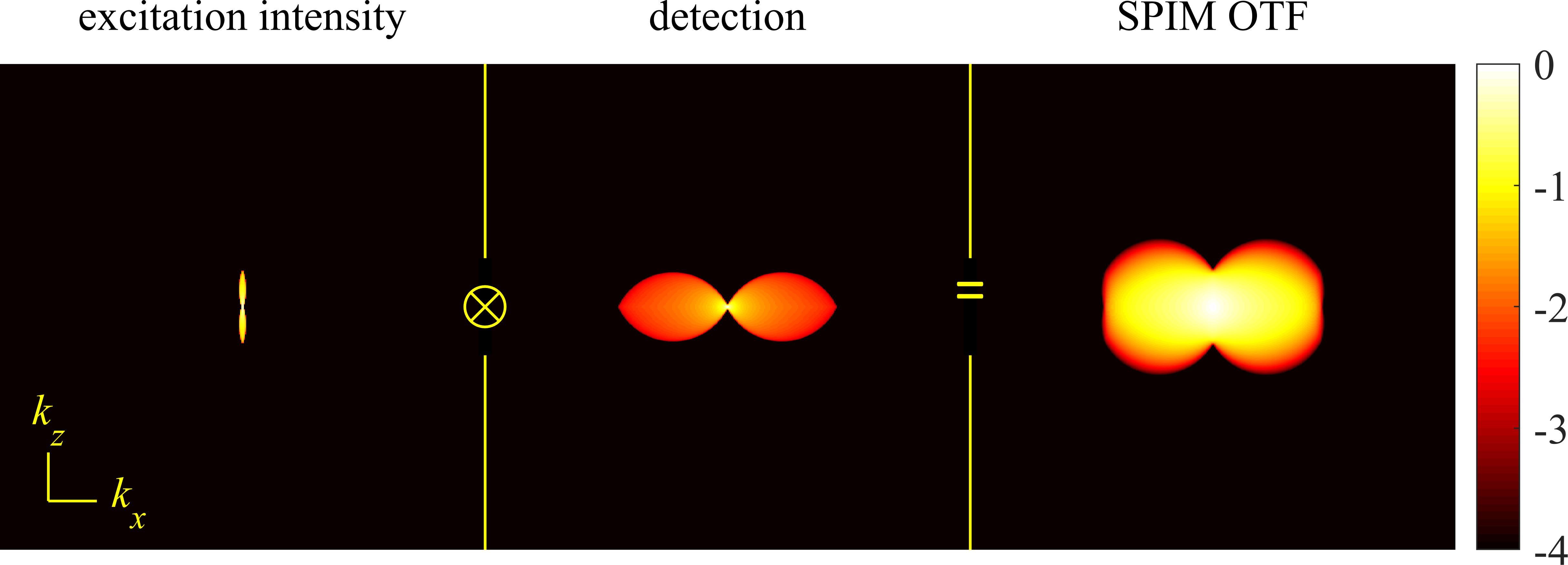}
    \caption{SPIM OTF. 
    Here, excitation is achieved by focusing a plane wave through a low-aperture lens (NA~=~0.4) from the left, resulting in a weakly diverging horizontally elongated excitation region. See further details in the main text.
    }
    \label{fig:OTFSPIM} \vspace{-4mm}
\end{figure}

In modern LSFM, there are two main approaches to generating a thin light-sheet. In the first approach, a digitally scanned laser moves rapidly along a direction perpendicular to the detection axis to achieve a thin light-sheet, termed Digitally scanned laser Light-Sheet Microscopy (DLSM)~\cite{keller2008quantitative}, see Fig.~\ref{fig:LSM}a. In the second approach, termed Selective Plane Illumination Microscopy (SPIM) \cite{huisken2004optical}, a cylindrical lens is typically used along the excitation path to form an astigmatic Gaussian beam effectively elongating the beam in one dimension to generate a thin, static light-sheet; see Fig.~\ref{fig:LSM}b. The SPIM OTF is provided on the right panel of Fig.~\ref{fig:OTFSPIM}, and obtained by convolving the SPIM light-sheet's Fourier representation (SPIM excitation OTF) on the left panel with the wide-field detection OTF in the middle panel. Compared to the wide-field OTF in Fig.~\ref{fig:otfwidefieldscalar}, the resulting SPIM OTF has a larger band-pass along the z-axis facilitating optical sectioning. 

For the Gaussian beam described above~\cite{huisken2004optical,keller2008quantitative}, LSFM's axial resolution is, as a first approximation, related to the Gaussian beam's thickness at twice the beam waist $z_{\mathrm{min}} = 2 w_0$, see Fig.~\ref{fig:LSM}c. Similarly, the FOV is related to the extent of the elongated Gaussian beam given by twice the Raleigh length $2z_r$~\cite{olarte2018light}  \vspace{-1mm}
\begin{align}    
    z_{\mathrm{min}} & \approx 2w_0 =  4\frac{\lambda f}{\pi D} = \frac{2n\lambda}{\pi \mathrm{NA}}
    \label{eq:lightsheetRes},\\
\mathrm{FOV} &= 2z_r = 2\frac{\pi w_0^2}{\lambda},
\label{eq:lightSheetFOV}
\end{align}
where $f$ and $D$ are, respectively, the focal length and lens diameter, with $\mathrm{NA}=nD/2f$ which is often smaller than 0.8 for light-sheet microscopes.

The improvement in axial resolution afforded by LSFM can be made clear when comparing to wide-field axial resolution approximately given by Eq.~\ref{eq:lightsheetRes}, and Eq.~\ref{eq:axialresolutionlimit}, as well as differently derived in Eq. ~\ref{eq:widefieldaxialresolution}, respectively. According to Eqs.~\ref{eq:lightsheetRes}, and \ref{eq:lightSheetFOV}, while thinner light-sheets (smaller $w_0$) improve axial resolution, they lead to smaller FOVs because of worsening illumination uniformity across the FOV. Such non-uniform illuminations may also result in varying PSFs and OTFs across the FOV~\cite{toader2022image}.

To soften the above trade-off and achieve simultaneous high axial resolutions and large FOVs, a few attempts have been made employing alternatives to Gaussian beams including: Bessel beams~\cite{fahrbach2010microscopy,planchon2011rapid}; Bessel beam lattices~\cite{chen2014lattice}; Airy beams~\cite{vettenburg2014light,yang2014compact}; spherically aberrated beams~\cite{fahrbach2010line}; and double beams~\cite{zhao2016multicolor}. While these beams achieve a Raleigh length typically larger than the Gaussian beam, it is unclear in practice whether high axial resolutions and contrasts are maintained~\cite{remacha2020define,chang2020systematic,shi2022quantitative,liu2023characterization}. This is because these alternative beams exhibit strong side-lobes leading to contribution of glare worsening axial resolution and contrast. Moreover, due to these side-lobes, the complex form of the resulting OTF does not lend itself to resolution estimates relying on Eq.~\ref{eq:resolutionMonotonic} or Eq.~\ref{eq:lightSheetFOV}~\cite{remacha2020define,shi2022quantitative}. \vspace{-3mm}

Further efforts at rejecting the light contribution from these side-lobes combined LSFM with CLSM, SIM, and two-photon microscopy~\cite{palero2010simple,planchon2011rapid,keller2010fast}. Moreover, the concepts of Reversible Saturable OpticaL Fluorescence Transitions (RESOLFT) (later introduced in Sec.~\ref{SR_Target}), and STED have been used in conjunction with SPIM to surpass the diffraction limit axially
~\cite{friedrich2011sted,hoyer2016breaking}. Light-sheet illumination has also been combined with non-linear fluorophore response to light (see Sec.~\ref{Nonlinear}) for SMLM~\cite{gebhardt2013single,galland20153d,meddens2016single}. 

What is more, since the lateral and axial resolutions differ, to avoid anisotropic resolutions, advanced LSFM configurations use multiple objectives generating different views of the specimen. These images are then computationally fused yielding improved isotropic resolution  ~\cite{swoger2007multi,huisken2007even,preibisch2014efficient,guo2020rapid}. Another approach involves Axial Swept Light-sheet Microscopy (ASLM)~\cite{dean2015deconvolution,chakraborty2019light,dean2022isotropic} generating isotropic images by scanning the sample laterally, \textit{i.e.}, perpendicular to the detection arm, using a tightly focused light-sheet synchronized by a moving camera shutter. This only allows fluorescence originating from the well-focused parts of the light-sheet to reach the camera.  

On the engineering front, orthogonal detection, and illumination through separate objectives (see Fig.~\ref{fig:LSM}) pose technical challenges when using two, bulky, high NA objectives, \textit{i.e.}, NA~$\approx$~0.8. As such, multiple modification to conventional LSFM have been proposed. For instance, the iSPIM (inverted SPIM) design uses two objectives (with NA~=~0.8-1.1) at 45 angle with respect to the coverslide~\cite{wu2011inverted}. More recently, different approaches have been developed achieving illumination, and fluorescent light collection using a single objective allowing use of higher NA objectives~\cite{dunsby2008optically,galland20153d,meddens2016single,sapoznik2020versatile,yang2022daxi}.
\vspace{-3mm}
\subsection{Multi-plane microscope}\label{subsec:Multiplane_mic} \vspace{-2mm}

\begin{figure}[H]
    \centering
    \includegraphics[width=0.95\linewidth]{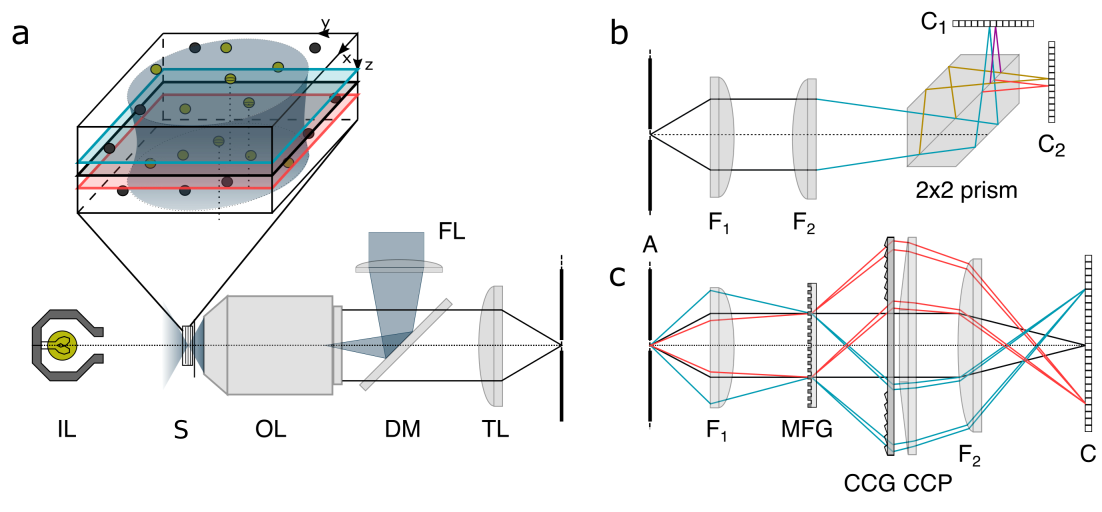} \vspace{-1mm}
    \caption{Multi-plane microscopy. (a) A conventional fluorescence microscope with epi-fluorescence (FL) and white light illumination (IL) acquire images of different focal planes across the sample by moving the objective lens (OL), and the sample with respect to each other. Here, the nominal focal plane is shown in black while the planes shown in red and blue can be also imaged by adjusting the axial positions of, for example, the sample. Shown are the sample (S), objective lens (OL), dichroic mirror (DM), and tube lens (YL). (b) A multi-plane microscope relays the optical path from the intermediate image formed in panel a via a telescope with lenses of focal lengths $\mathrm{F}_1$, and $\mathrm{F}_2$ and uses a beam-splitting prism, \textit{i.e.}, a refractive element, along the detection path to separate fluorescence emission into multiple channels (here four) with different focal planes projected next to each other on two cameras (C1, C2); see Ref.~\cite{Descloux2018}. (c) A multi-focus microscope uses a multi-focus grating (MFG), \textit{i.e.}, diffractive element, chromatic correction grating (CCG) and prism (CCP) to achieve multiple focal planes on one camera; see text for more details. 
    }
    \label{fig:multiplane} \vspace{-4mm}
\end{figure}

To improve upon wide-field microscopy's low axial resolution, we may acquire images from multiple  planes across samples. The simplest approach toward achieving this is by moving the sample and focus plane with respect to each other; see Fig.~\ref{fig:multiplane}a. However, this involves moving a large inertial object (sample, objective, camera) introducing time lags between planes and mechanical perturbation. Fast, adaptive elements or small moving components in a more complex detection path can speed this up, but do not eliminate axial scanning. Acquiring data   
across multiple focal planes simultaneously without moving the sample, or optical components, has been achieved by introducing either refractive or diffractive optical elements into the detection arm. These elements split the fluorescent emission into multiple paths leading to simultaneous acquisitions from different focal planes~\cite{Blanchard1999,Prabhat2004,Abrahamsson2013,Descloux2018,Mertz2019}. For a more in-depth review on ``snapshot" volumetric microscopy see \textit{e.g.}, Ref.~\cite{Engelhardt2022}. \vspace{-1mm}

Multi-plane, also termed multi-focus microscopy imaging, is versatile and can be combined with wide-field fluorescence, or light-sheet excitation~\cite{Ma2016} for a number of applications. These include: SPT~\cite{Ram2012, Louis2020}, super-resolution microscopy~\cite{Hajj2014, Babcock2018} (for statistical modeling of multi-plane super-resolution SMLM data see Sec.~\ref{PSFeng_sec} and Eqs.~\ref{eq:cost_PR}-\ref{eq:loglik-poiss}), Super-resolution Optical Fluctuation Imaging (SOFI)~\cite{Descloux2018, Geissbuehler2014}, structured illumination~\cite{Abrahamsson2017, Descloux2020}, as well as single cell and whole organism imaging~\cite{Abrahamsson2013, Xiao2020, Hansen2021}. Furthermore, phase imaging~\cite{Mojiri2021, Xiao2020}, polarization ~\cite{Abrahamsson2015} and dark-field~\cite{Xiao2020, Hansen2021} microscopy may also use a multi-plane setup.

In its simplest form, multi-plane microscopes use beam-splitters, \textit{i.e.}, refractive elements, in combination with optical detection paths of different lengths, or tube lenses with different foci ~\cite{Itano2016,Prabhat2004,Babcock2018,Geissbuehler2014, Hansen2021}. In such setups, the inter-plane distance, and thus axial resolution, can be independently adjusted from the pixel size (tied to lateral resolution; see Sec.~\ref{Intro_Limit}). 

However, these versatile implementations are susceptible to misalignment of the detection channels due to opto-mechanical component drift especially relevant in super-resolution microscopy; see Sec.~\ref{SR}. A more elegant solution involves a cascade of beam-splitters fused into a single piece, \textit{i.e.}, prism~\cite{Descloux2018,Xiao2020}, dividing the fluorescent light into multiple beams traveling optical paths with different lengths; see Fig.~\ref{fig:multiplane}b. Here, increased mechanical stability arises from having all beam-splitting integrated into one optical element, \textit{i.e.}, the prism, minimizing chromatic aberration. This setup can also be extended to simultaneously image several colors across planes using spectral beam-splitters \cite{Gregor2021}.

An alternative approach uses a Multi-Focus Grating (MFG), \textit{i.e.}, a diffractive element, splitting fluorescence emission into multiple paths corresponding to different diffraction orders. The grating pattern is designed to introduce diffraction order dependent de-focus phase shifts (see Sec.~\ref{Aberrations}) leading to different focal planes for each path~\cite{Blanchard1999}; see Fig.~\ref{fig:multiplane}c. However, the grating introduces chromatic dispersion, improved by introducing a Chromatic Correction Grating (CCG), and a Prism (CCP) to reverse the dispersion due to MFG~\cite{Abrahamsson2013} and separate the images laterally on the camera chip; see Fig.~\ref{fig:multiplane}c. While aberration-corrected multi-focus microscopy grating design can further improve imaging of thicker samples~\cite{Abrahamsson2013,Abrahamsson2016,Hajj2017}, gratings have lower transmission, and new gratings are required to alter inter-plane distances.

\vspace{-4mm}
\section{Super-resolution microscopy \label{SR}}
\vspace{-3mm}

Resolution across fluorescence microscopy, as  described in Sec.~\ref{sec:modality}, is fundamentally limited by the frequency band-pass given by the corresponding OTFs. This restricts the maximum achievable resolution to approximately half of the emission wavelength under optimal conditions. This limit can be surpassed by exploiting the non-linearity in fluorophore response to excitation light; see Sec.~\ref{Nonlinear}. This, in turn, has lead to the development of two main categories of super-resolution, or nanoscopy, methods to which we now turn: 1) targeted switching; and 2) stochastic switching techniques. 

\vspace{-5mm}
\subsection{\label{SR_Target}Targeted switching super-resolution microscopy}
\vspace{-3mm}
\subsubsection{Stimulated emission depletion microscopy} \label{sec:STED}
\vspace{-3mm}

Previously introduced fluorescent imaging techniques such as confocal, light-sheet, and multi-plane microscopy improve axial resolution using different optical sectioning strategies. Optical sectioning limits the collected fluorescence to an axial subset of fluorescent molecules preventing interference from fluorophores outside this axial subset. Although these techniques can significantly increase contrast, and improve axial resolution, their resolution remains limited by the diffraction of light. On the other hand, super-resolution methods such as STED microscopy~\cite{hell1994OpticsLetter,klar_breaking_2001}, and its generalization, RESOLFT~\cite{hofmann_breaking_2005,hell_far-field_2007}, are based upon a traditional point scanning microscope with confocal pinhole in the detection arm allowing higher resolution imaging while retaining the axial sectioning of confocal microscopy.

STED imaging was first achieved in the mid-nineties by Hell and Wichmann~\cite{hell1994OpticsLetter} and its popularity grew thanks to the high spatial resolution, relatively high imaging speed, and considerable imaging depth. These made possible, for instance, the visualization of biomolecular assemblies and live-cell nanoscopy~\cite{hell_far-field_2007,eggeling2015lens}. 

In terms of temporal resolution, as fast as millisecond imaging times for rapid dynamics in small fields of view was demonstrated by ultrafast STED nanoscopy~\cite{schneider_ultrafast_2015}, while spatially, the highest reported 3D isotropic resolution ($<$ 30~nm in $x,y,z$ simultaneously) was validated with the ultra-stable design of 4pi-based isoSTED~\cite{curdt_isosted_2015}.

In STED, spatial resolution improvement is achieved by adding a second de-excitation (depletion) laser quenching fluorescence around the excitation point confining fluorescence emission to a sub-diffraction limited spot. Stimulated emission is one means by which to depopulate excited states. In this process, theoretically discovered by Albert Einstein~\cite{einstein_strahlungs-emission_1916}, the incoming photon triggers the excited system to decay to its ground state, emitting a photon, with a phase, frequency, polarization, and momentum identical to the incident photon; see Sec.~\ref{Nonlinear}. 

\begin{figure}[H] \centering
\includegraphics[width=\linewidth]{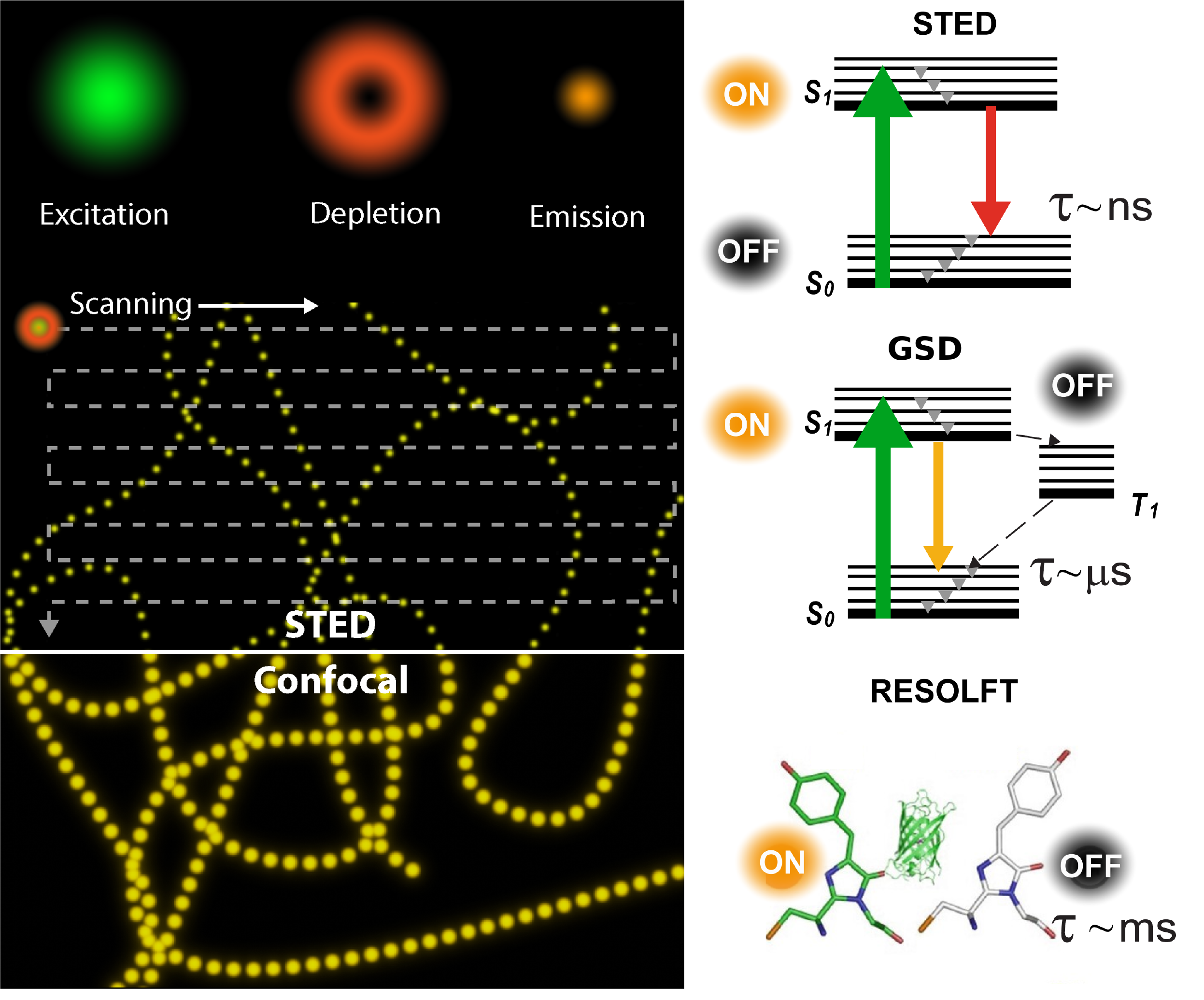} 
\caption{Schematics for STED imaging. Excitation and depletion beams are used to acquire a sub-diffraction-limited image, formed after raster scanning the full sample. The resulting image can be understood as a convolution between the effective PSF combined from the excitation, and depletion laser beams, and the fluorescent molecule distribution in the sample. The image is adapted from Refs.~\cite{hofmann_breaking_2005,hell_far-field_2007}. Schematics on the left hand side compare diffraction-limited confocal images of microtubules with the coinciding STED image. On the right panel we show the electronic transitions of excitation, and stimulated emission in STED (top), ground-state depletion GSD (middle), and RESOLFT (bottom). The figure is adapted from Ref.~\cite{sahl_fluorescence_2017}. } \vspace{-5mm}
\label{fig:STED}
\end{figure}

In STED, stimulated emission must precede spontaneous emission, requiring the excitation light to excite the sample ($\approx 200$~ps) prior to laser quenching. The whole imaging protocol is devised in two steps; see Fig.~\ref{fig:STED}. First, fluorophores are excited by a diffraction-limited laser beam with a Gaussian waist shown in green in Fig.~\ref{fig:STED}. 
If we wait until molecules spontaneously decay without stimulated emission, no gain in resolution will be achieved. Therefore, it is necessary to introduce the second step where a fraction of the fluorophores are depleted using a torus, or donut-shaped diffraction-limited beam shown in red in Fig.~\ref{fig:STED}, whose central minimum coincides with the Gaussian excitation maximum. As such, the recorded signal only originates from the ``donut hole" far narrower than the original Gaussian waist shown in orange in Fig.~\ref{fig:STED}. To understand how STED beams are generated, see Sec.~\ref{CLSM-Confocal} and Fig.~\ref{fig:STEDFocus}.

The resolution gain in STED, $y_ {\mathrm{STED}}$, given below is set by the inner donut radius \vspace{-2mm}
\begin{equation}
y_{\mathrm{STED}} = \frac{\lambda}{2\mathrm{NA}\sqrt{1+I/I_\mathrm{sat}}} = \frac{y_{\mathrm{min}}}{\sqrt{1+I/I_\mathrm{sat}}}.
\end{equation}
Here, $y_{\mathrm{min}}$ is the wide-field resolution (see Eq.~\ref{eq:abberesolutionlimit}),  $I$ is the depletion laser intensity, and $I_\mathrm{sat}$ is the depletion intensity required to outperform fluorescence emission.  

Although STED's resolution can theoretically be arbitrarily small provided high enough depletion intensity ($I\to\infty$)~\cite {pawley_handbook_2006}, in practice,
factors limiting resolution include the nature of the fluorophores used (and their absorption cross-section of the depletion beam), uncorrected aberrations (residual aberration) of the STED pattern, SNR, as well as the STED beam's relatively high power
and propensity for label photo-damage. \vspace{-2mm}

Photo-damage can be mitigated by working with solid state fluorescent nanodiamonds hosting negatively charged nitrogen-vacancy (NV) point defects. Using such photo-stable labels, resolutions of $\approx 10$~nm were demonstrated~\cite{wildanger_compact_2009,arroyo-camejo_stimulated_2013}. However, the complex functionalization of relatively large size 10-15~nm solid-state probes, including issues related to specificity and cell permeability, limit their applications especially in live-cell. 

While we have focused on 2D thus far, by using interference of two depletion beams (see implementation of 4pi microscopy introduced in Sec.~\ref{4pi}), STED super-resolution imaging has been extended to 3D  ~\cite{wildanger_solid_2012,osseforth_simultaneous_2014} though, in practice, axial resolution gain comes at the cost of lower lateral resolution.

\vspace{-5.5mm}
\subsubsection{Reversible saturable optically linear fluorescence transition microscopy} 
\vspace{-2mm}

Numerous efforts in the last two decades have been undertaken to improve upon STED's need for high power depletion beams~\cite{eggeling2015lens}. RESOLFT, a more general method encompassing STED as a special case, was one such effort proposed in the early 2000's~\cite {hell_far-field_2007}, leveraging fluorophore photo-physics. This, in turn, renders RESOLFT more appropriate for live-cell, and long-term experiments~\cite{hofmann_breaking_2005} including 3D live-cell imaging using a recent implementation of highly parallelized image acquisition with an interference pattern~\cite{boden_volumetric_2021}.

In contrast to STED, whose high laser power is required to deplete the excited state back to the ground state, RESOLFT uses donut-shape beams to transition fluorophores into any dark state, not just the ground state; Fig.~\ref{fig:STED}. Thus RESOLFT requires fluorophores controllably switchable between dark (OFF), and bright (ON) states; see Fig.~\ref{fig:STED}. For instance, such fluorophores include reversibly switchable fluorescent proteins, and dyes ~\cite{grotjohann_rsegfp2_2012,pennacchietti_fast_2018}.One such dark state is the triplet state (see Sec.~\ref{Nonlinear}) at the basis of ground state depletion (GSD) \cite{hell1995ground}, a special case of RESOLFT requiring less intense depletion laser powers; see Fig.~\ref{fig:STED}.
 
\vspace{-7mm}
\subsubsection{Minimal photon fluxes}
\vspace{-3mm}

Due to the limited photo-stability of fluorophores, \textit{e.g.}, due to photo-bleaching, first generation nanoscopy methods such as STED and RESOLFT reached practical resolution limits of 20-40~nm. This motivated the development of a second generation of fluorescence nanoscopy techniques achieving 1-10~nm resolutions~\cite{balzarotti_nanometer_2017,cnossen_localization_2020,gu_molecular_2019,jouchet_nanometric_2021,reymond2019simple,masullo2022common,masullo2022alternative} leveraging patterned illumination. 

\begin{figure}[H] \centering
\includegraphics[width=\linewidth]{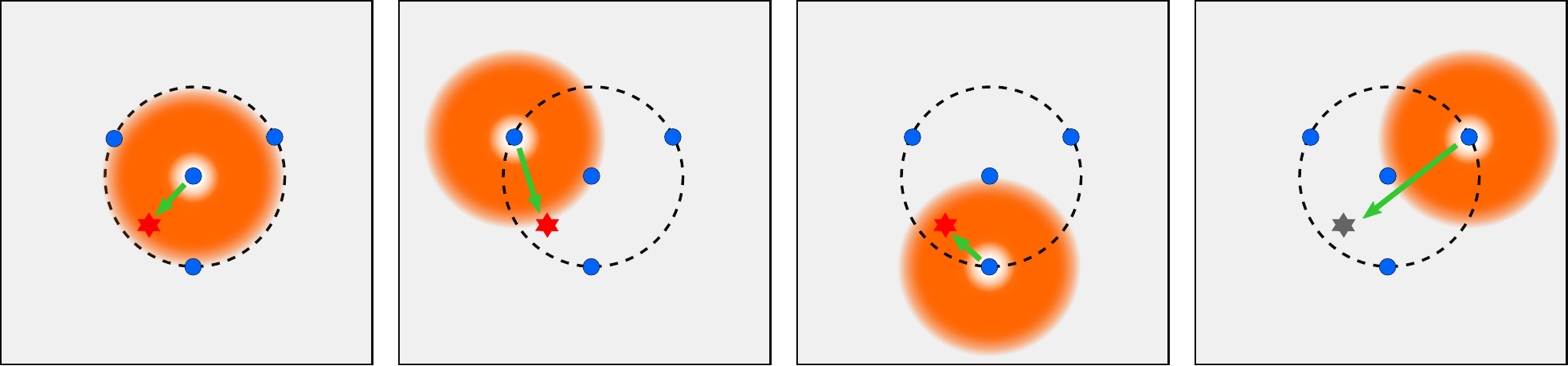} 
\caption{
MINFLUX's working principle. MINFLUX employs a donut-shape excitation beam (orange) with the donut translated to four locations (blue circles) at which fluorescence signals are measured and used to determine fluorophore's position. The red and dark stars, respectively, indicate the excited and ground state fluorophores; see details in the text.}
\label{fig:MINFLUX}
\end{figure}

The first implementations of such nanoscopy techniques includes MINimal photon FLUXes (MINFLUX) introduced in 2017~\cite{balzarotti_nanometer_2017} which extracts information from a limited photon budget and uses minimal laser intensities~\cite{balzarotti_nanometer_2017,eilers_minflux_2018,gwosch_minflux_2020}. In contrast to STED, MINFLUX uses a donut-shape beam for excitation with the intensity minimum at its center. Here, to illustrate the MINFLUX concept, we assume a single fluorophore as shown in Fig.~\ref{fig:MINFLUX}. The excitation beam is scanned across the sample and the fluorescence signal is collected by a confocal microscope. The number of collected photons depends on the excitation intensity received by the fluorophore and can be used to calculate the fluorophore's distance from the beam's center. For instance, fluorophores precisely at the donut-shape beam center, have minimal emission. However, as the exact fluorophore's location is unknown, the beam scans the area at a few locations (see Fig.~\ref{fig:MINFLUX}) and the fluorophore's distance from the beam center's locations (designated by blue dots in Fig.~\ref{fig:MINFLUX}) are calculated to pinpoint the fluorophore with nanometer precision. 

Recently, MINFLUX has been used to simultaneously perform 3D and multi-color imaging~\cite{gwosch_minflux_2020} achieving high isotropic localization precision (1–3~nm). In addition, MINFLUX has been used in SPT ~\cite{eilers_minflux_2018,pape_multicolor_2020} localizing with a precision below 20~nm within $\approx100~\mu\mathrm{s}$~\cite{schmidt_minflux_2021}.

The concept of localizing with respect to a patterned illumination has also been implemented using wide-field microscopy for faster imaging substituting donut-shaped illumination with other illumination patterns~\cite{gu_molecular_2019,cnossen_localization_2020,jouchet_nanometric_2021}. For instance, in SIMFLUX~\cite{cnossen_localization_2020} fluorophore locations are realized with respect to a sinusoidal pattern. \vspace{-3mm}

\vspace{-12mm}
\subsection{\label{SR_Stoch}Stochastic switching super-resolution microscopy} \vspace{-4mm}

Previously we described super-resolution methods based on targeted switching of fluorophores. Here, we discuss single molecule based super-resolution methods, a family of super-resolution techniques, achieving sub-diffraction resolution by imaging independent, and stochastically blinking fluorophores over time~\cite{sigal_visualizing_2018,schermelleh_super-resolution_2019,lelek_single-molecule_2021}. In these methods, the gain in spatial resolution is traded for temporal resolution as the acquisition of many camera frames is required to computationally reconstruct a single super-resolved image. In such experiments, a conventional wide-field microscope is typically used to collect fluorescent light from (photo)activatable, or switchable probes (see Sec.~\ref{Non_Chem}). Moreover, scanning image acquisitions have also been successfully used to implement super-resolution microscopy ~\cite{york2011confined}. 

\begin{figure}[H] \centering
\includegraphics[width=\linewidth]{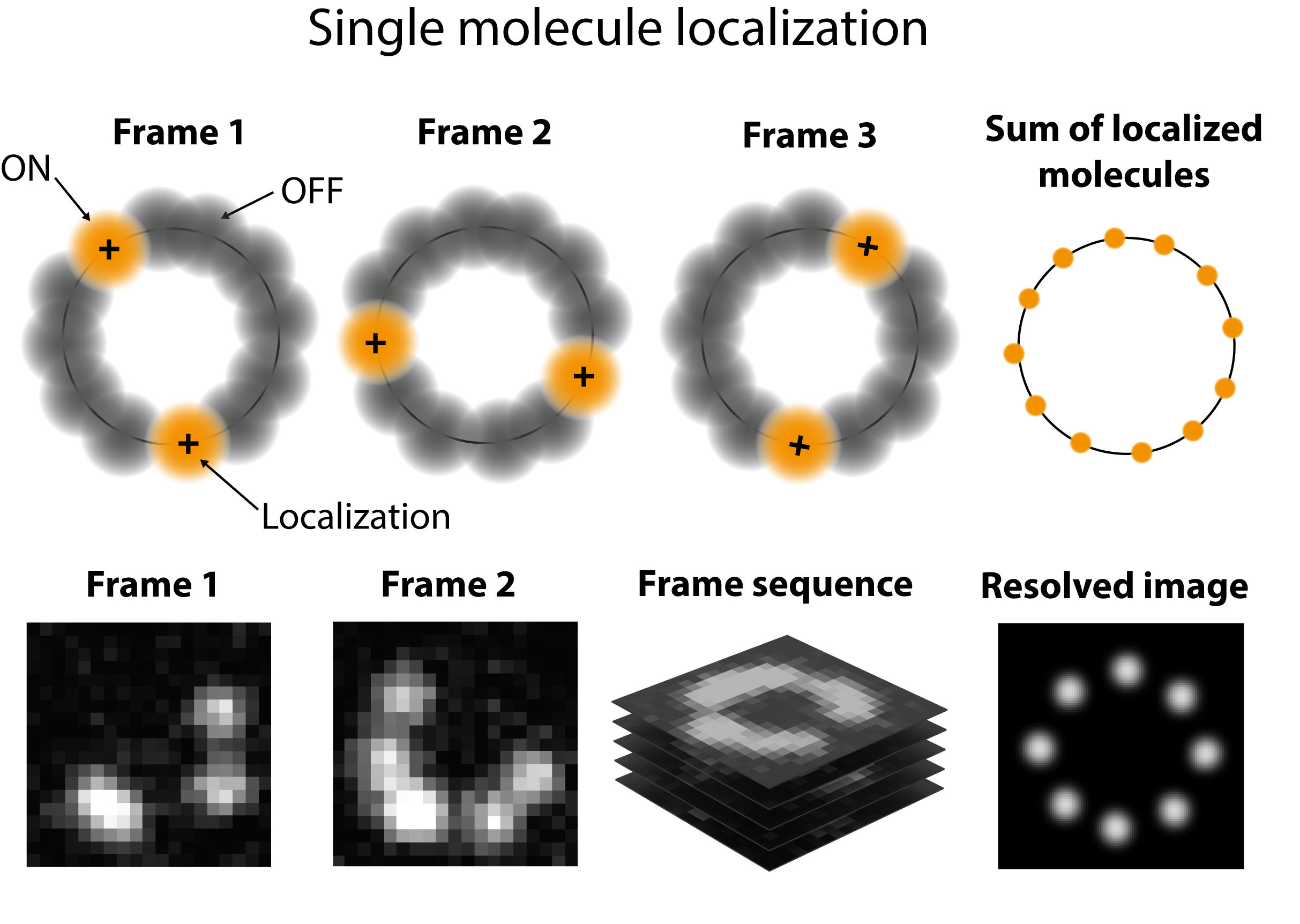} 
\caption{Single emitters are stochastically activated to become fluorescent. The activated emitters can be precisely localized provided they are spaced further apart than the Nyquist limit; see Sec.~\ref{Intro_Limit}. The process is repeated for tens of thousands of frames. In each frame, single-emitters are identified and fitted to obtain their center of mass, allowing super-resolved pointillistic image reconstruction (see bottom panel right). Repetitive activation, localization, and deactivation temporally separate spatially unresolved structures in a reconstructed image with apparent resolution gain compared to the standard diffraction-limited image; see bottom row. }
\label{fig:SMLM}
\end{figure}

The most common use of stochastic switching is applied to techniques termed Single-Molecule Localization Microscopy (SMLM)~\cite{lelek_single-molecule_2021}. In SMLM, spatially overlapping fluorophores are temporally separated by acquiring image frame sequences. As in each frame only few fluorophores switch on ($<1$\%), high precision localization is achieved by avoiding overlapping PSFs; see Fig.~\ref{fig:resolution}. The set of nanometer-resolved localizations are then used to reconstruct super-resolved structures; see Fig.~\ref{fig:SMLM}. 

The latter methods however require localizing, by chance, well-separated molecules 
thereby imposing long data acquisition times. Therefore, more recently, a range of alternative techniques have been developed to improve image resolution while avoiding identifying and localizing single molecules~\cite{Opstad2020,Pawlowska2022}. Rather, such methods analyze fluctuations in fluorescence emission over time, and tolerate a wider range of switching behavior, and imaging conditions including SOFI~\cite{dertinger_fast_2009,dertinger_superresolution_2010} (see Sec.~\ref{SOFI}), and others \textit{e.g.}, SRRF~\cite{gustafsson_fast_2016}, SPARCOM~\cite{solomon_sparsity-based_2018}, MSSR~\cite{torres2022extending}, and 3B~\cite{cox_bayesian_2012}. A common feature of fluctuation-based techniques is that they provide lower resolutions compared to SMLM methods though requiring fewer input frames, and lower laser powers as compared with SMLM, making them more live-cell compatible. \vspace{-3mm}

\vspace{-12mm}
\subsubsection{Super-resolution optical fluctuation imaging} \label{SOFI}
\vspace{-3mm}

Super-resolution Optical Fluctuation Imaging (SOFI)~\cite{Dertinger2009,Grumayer2020} is a computational post-processing tool for super-resolution single molecule data. In contrast to SMLM, SOFI is not aimed at resolving isolated molecules and is robust to the presence of overlapping PSFs. Concretely, SOFI improves resolution by exploiting correlations in the stochastic switching of the underlying fluorophores, {\it i.e.}, by leveraging the fact that a molecule's emission fluctuations only spatiotemporally correlate with itself and not with neighboring molecules. 

The data processed in SOFI consists of photon counts (intensity) $w_n^k$ at pixel $n$ in frame $k$ (time point $k$) detected on a wide-field camera \vspace{-2mm}
\begin{equation} \vspace{-2mm}
w_{n}^k=\mathcal{B} + I_0 \sum_{m=1}^{M}  U\left(\mathbf{r}_n-\mathbf{r}_{m}\right) s_{m}^k+ \varepsilon_n^k
\end{equation}
with M denoting fluorophore number, $I_0$ the molecular brightness assumed uniform across molecules,
$U$ the optical system's PSF, $s_{m}^k$ describing the state of fluorophore $m$ as
off or on-state, $\mathcal{B}$ an average background, $\mathbf{r}_n$ the location of pixel $n$, and $\varepsilon_n^k$ the additive noise. Moreover, the sample is assumed stationary over image acquisition such that the PSF's integral over the pixel area is approximated by the integrand's value at the pixel center. 

In its simplest implementation, SOFI computes cumulants, $\kappa\left(w^{1:K}_{1:N}\right)$, of the pixel intensities across frames. For instance, the second order temporal cross-cumulant coincides with the co-variance in signal intensity across frames in one pixel for different time lags. The $l$th order cumulant can be approximated as~\cite{Deschout2016} \vspace{-3mm}
\begin{equation}
\vspace{-1mm}
\kappa_l\left(w_{1:N}^{1:K}\right) \approx I^{l}_0 f_{l}\left(\rho_{\text {on }}\right) \sum_{m=1}^{M} \left(s_m^{1:K}\right)^l\, U^{l}\left(\mathbf{r}_{1:N}-\mathbf{r}_m\right),
\label{eq:cumulantL}
\end{equation}
where $f_{l}\left(\rho_{\text {on }}\right)$ denotes the $l$th order cumulant of $s_m^k$ given as an $l$th order polynomial with respect to the probability of the molecule (ratio of molecules) to be on designated by $\rho_{\mathrm{on}}$. Moreover, under assumptions of uncorrelated noise and stationary background, cumulants of the noise and background are zero. In Eq.~\ref{eq:cumulantL}, critical to SOFI analysis, the PSF is raised to the $l$th power. Thus the $l$th order cumulant, if plotted instead of the original image, yields a PSF $\sqrt{l}$ narrower than the original PSF and offers an up to $l$-fold enlarged frequency support in Fourier space. As such, the resolution can be increased up to $l$-fold with post-processing either by Fourier reweighing~\cite{Dertinger2010} or deconvolution~\cite{Dertinger2009,Geissbuehler2012} as discussed earlier, \textit{e.g.}, see confocal (Sec.~\ref{CLSM-Confocal}) and ISM (Sec.~\ref{ISM}) microscopy. This can be further generalized to spatiotemporal cross-cumulants with various time-lags across different pixel combinations to leverage spatial information albeit at higher computational cost
~\cite{Dertinger2010,Geissbuehler2012,Girsault2016}. 

One challenge with SOFI post-processing is the possibility of amplifying signal heterogeneities and potentially mask dimmer structures~\cite{Geissbuehler2012} partly addressed by a deconvolution method termed balanced SOFI (bSOFI) 
~\cite{Geissbuehler2012,Deschout2016}. Furthermore, compared to SMLM, SOFI is relatively insensitive to background, tolerates higher labeling densities, higher on-time ratios, lower SNR, and only hundreds to thousands of frames to compute cumulants allowing less photo-damaging, and faster live-cell imaging. Moreover, SOFI achieves optical sectioning and resolution improvement in the $z$-direction using simultaneously acquired multi-plane data \cite{Geissbuehler2014a,Descloux2018}.

\vspace{-6mm}
\subsubsection{Single molecule localization microscopy} \label{sec:SMLM}
\vspace{-3mm}

Almost a decade preceding its experimental realization~\cite{lidke2005superresolution,betzig_imaging_2006}, the idea underlying SMLM was theoretically proposed by Eric Betzig~\cite{betzig_proposed_1995} with experimental implementations employing photo-activatable genetically encoded proteins~\cite{lippincott-schwartz_photoactivatable_2009} and quantum dots~\cite{lidke2005superresolution}.

An initial iteration, termed (f)PALM~\cite{betzig_imaging_2006,hess_ultra-high_2006}, was followed by Stochastic Optical Reconstruction Microscopy (STORM)~\cite{rust_sub-diffraction-limit_2006} exploiting photo-switching in organic dyes. While differing only in their means to achieve temporal separation of spatially overlapping fluorophores, PALM leverages photo-activatable or photo-convertible fluorescent proteins~\cite{shroff_live-cell_2008}, allowing for genetic expression of fluorescent proteins and is compatible with live-cell imaging~\cite{shroff_live-cell_2008}, and thus stoichiometric labeling of target proteins used in counting~\cite{rollins2015stochastic,bryan2022diffraction}. On the other hand, organic fluorophore photon emission rates are typically higher compared to photo-activatable or photo-convertible fluorescent proteins, resulting in STORM's slightly better resolution. Further resolution improvements spurred the development of the more general dSTORM introducing a pallet of synthetic organic fluorophores as photo-switchable probes~\cite{heilemann_subdiffractionresolution_2008} allowing live cell imaging with site-specific tagging~\cite{wombacher_live-cell_2010}. \vspace{-3mm}

\begin{figure}[H] \centering
\includegraphics[width=\linewidth]{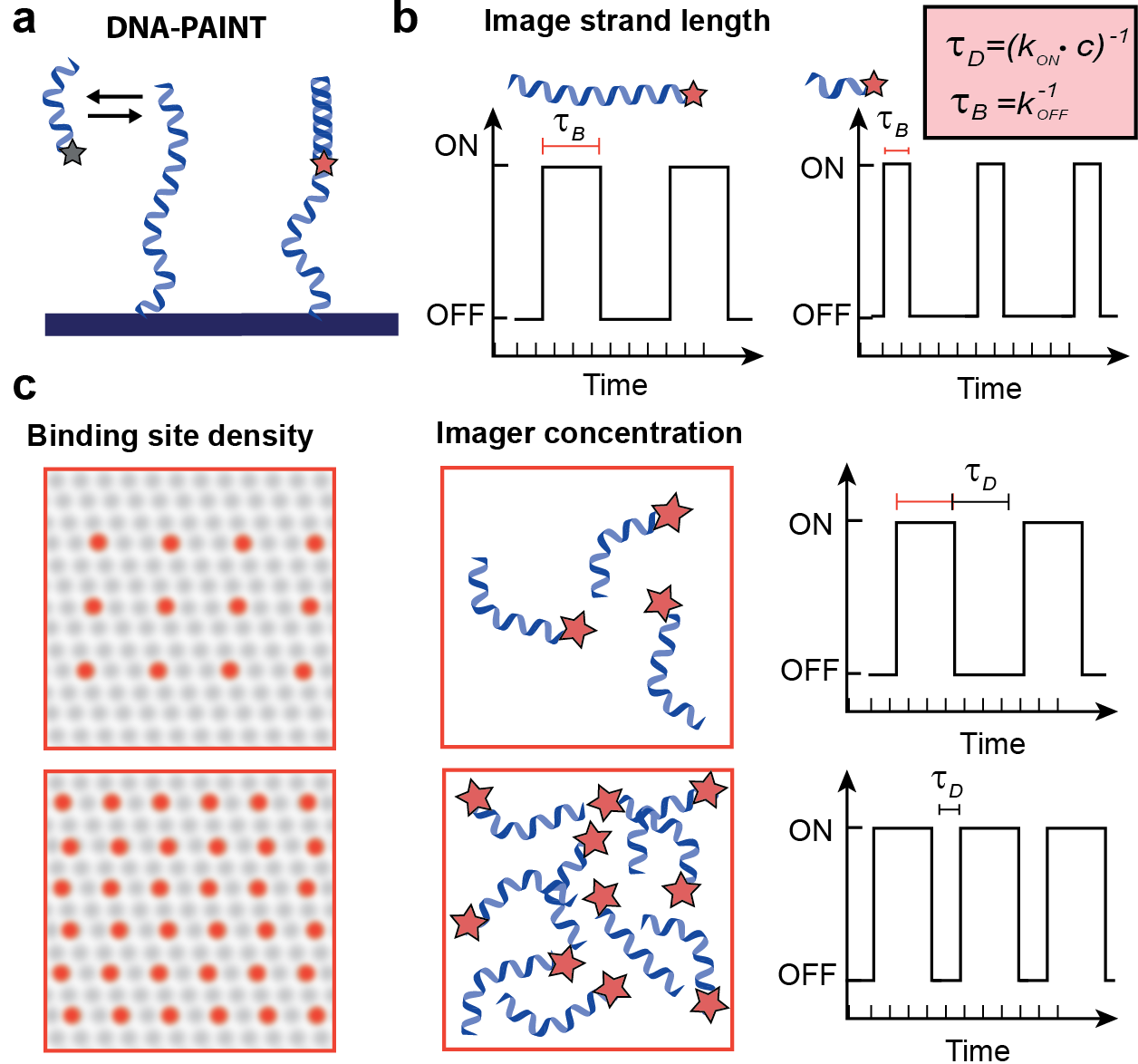} 
\caption{Imaging with DNA-PAINT. (a) Schematics illustrate DNA-PAINT where dye-conjugated oligo (imager oligo) transiently hybridizes with a complementary (docking) oligo. (b) The binding time $\tau_B$ (or the dissociation rate $1/\tau_B$) depends on imager strand length. (c) Increasing either imager strand concentration or docking site density decreases dark times, $\tau_D$ (inter-event lifetime). The figure is adapted from Ref.~\cite{schnitzbauer_super-resolution_2017}. }
\label{fig:PAINT} \vspace{-5mm}
\end{figure} 

A more recent SMLM approach termed DNA Point Accumulation for Imaging in Nanoscale Topography (DNA-PAINT) employs stochastic transient binding of diffusing dyes in solution with a complementary molecules binding to the target structure~\cite{schnitzbauer_super-resolution_2017}; see Fig.~\ref{fig:PAINT}. Upon binding, the dye molecule is temporally immobilized, and detected by the camera while the freely diffusing dyes, strongly aliased and difficult to track, are approximately treated as background. Longer imager strands, increasing binding time, typically lead to a higher photon numbers over one binding event and improved SNR alongside higher spatial resolutions; see Fig.~\ref{fig:PAINT}b. DNA-PAINT exhibits limited photo-bleaching as imaging can be continued so long as diffusing dyes are present in solution and is furthermore compatible with multiplexing using color and assortment of DNA strands' lengths~\cite{jungmann2014multiplexed,wade2019124,strauss2020up}.

\vspace{-5mm}
\subsubsection{SMLM data analysis}
\vspace{-2mm}

In SMLM, data, $w_{1:N}$, typically consist of a set of pixel values (observation) organized as 2D arrays, called image frames. Localizations are then determined, probabilistically, from pixel values, $w_n$, using a likelihood. 

To build the likelihood, we begin with the expected photon counts for the pixel $n$ given as \vspace{-2mm}
\begin{equation}
\vspace{-2mm}
    \Lambda_n =  \mathcal{B} + \sum_{m=1}^{\infty} b_m I_m \mathcal{P}_m^n.
 \label{eq:expectedPhotonsNP}
\end{equation}
where we have immediately generalized our model to the practical case with unknown emitter numbers. That is, we adopt a non-parametric framework with an infinite number of emitters ($m=1:\infty$) with load $b_m$ associated to each emitter (see Sec.~\ref{prob-stat}). The loads associated to the emitters not contributing photons are, as usual, recovered as zero. Moreover, $I_m$ and $\mathcal{B}$, respectively, represent the intensity of the $m$th emitter and uniform background. Here, $\mathcal{P}_m^n$ is the probability of a photon from 
emitter $m$ reaching pixel $n$ given by (see Eqs.~\ref{eq:subset}-\ref{eq:PSFintPix}) \vspace{-2mm}
\begin{equation} \vspace{-2mm}
    \mathcal{P}_m^n = \iint_{\mathcal{A}_n} dxdy\,U(x,y;\mathbf{r}_m),
    \label{eq:PixelProb}
\end{equation}
where $\mathcal{A}_n$ is the pixel area and $\mathbf{r}_m = (x_m, y_m, z_m)$ is the emitter position. As a simplification, the PSF is sometimes substituted for its value evaluated at the middle of the pixel~\cite{agarwal2016multiple} or the integral can be evaluated using error functions, say, for Gaussian PSFs. For more complicated cases (engineered PSFs in Sec.~\ref{PSFeng_sec}), the PSF appearing in the integral of Eq.~\ref{eq:PixelProb} can also be numerically evaluated over a sub-pixel grid. 
Further improvements are also possible by using linear or spline PSF interpolations~\cite{liu2013three,li2018real} between PSF values typically calibrated at select axial positions.

For concreteness here, we use a CCD detector noise model (see Appendix~\ref{Se:Detector})
and arrive at the following likelihood for pixel $n$
\begin{equation}
    P(w_n|\vartheta) = \mathrm{Gaussian}(w_n;g\Lambda_n(\vartheta)+o,\sigma_w^2),
    \label{eq:SMLMLikeGaussNoise}
\end{equation}
where $g,\,o$ and $\sigma_w^2$ are, respectively, the detector gain, offset, and variance. As before, we collect all  unknown parameters in $\vartheta=\{\overline{b},\mathbf{\overline{r}},\overline{I},\mathcal{B}\}$ where the overbar denotes quantities over all emitters. Finally, since pixel values are \textit{iid} (see Sec.~\ref{prob-stat}]), the likelihood of a ROI containing $N$ pixels assumes a product form \vspace{-3mm}
\begin{equation}   \vspace{-2mm}
P(w_{1:N}|\vartheta)=\prod_{n=1}^N P(w_n|\vartheta).
    \label{eq:ROILikelihood}
\end{equation}

In parametric frameworks, the likelihood from Eq.~\ref{eq:ROILikelihood} is simplified given known emitter numbers, $M$, \vspace{-3mm}
\begin{equation} \vspace{-3mm}
    \Lambda_n = \mathcal{B} + \sum_{m=1}^{M} I_m \mathcal{P}_m^n.
    \label{eq:expectedPhotons}
\end{equation}
In such frameworks, the number of emitters are typically heuristically set separately using alternate criteria, \emph{e.g.}, Bayesian Information Criteria (BIC)~\cite{quan2011high}, thresholding~\cite{babcock2012high}, or other methods~\cite{huang2011simultaneous, fazel2022analysis}. In contrast, in joint (non-parametric) optimization, the number of active emitters are treated as random variables (unknowns) on which we place priors~\cite{jazani2019alternative,jazani2022computational}. In other words, we obtain the BNP posterior from the product of the likelihood Eq.~\ref{eq:ROILikelihood}, and the priors over $\vartheta$;
see Sec.~\ref{prob-stat}. We may adopt an empirical prior for fluorophore intensity obtained by fitting isolated emitters from sparse regions of the data~\cite{fazel2019bayesian}, and adopt computationally convenient Beta-Bernoulli process priors for the loads; see Sec.~\ref{prob-stat}.  

\begin{tcolorbox}[colback=brown!5!white,colframe=brown!75!black,title=Statistical Framework \ref{box:SMLM}: SMLM]

Data: pixel values in ADUs
\begin{equation}
    w_{1:N} = \left\{w_1, ..., w_N\right\}. \nonumber
\end{equation}
Parameters: loads, fluorophore locations, intensities, background 
\begin{equation}
    \vartheta = \left\{\overline{b},\overline{\mathbf{r}},\overline{I},\mathcal{B}\right\}. \nonumber
\end{equation}
Likelihood:
\begin{equation}
    P\left(w_{1:N}|\vartheta\right) = \prod_n \mathrm{Gaussian}(w_n;g\Lambda_n(\vartheta)+o,\sigma_w^2). \nonumber
\end{equation}
Priors:
\begin{align}
    q_m \sim & \, \mathrm{Beta}(A_q,B_q), \,\, m = 1:\infty, \nonumber \\
    b_m \sim & \, \mathrm{Bernoulli}(q_m), \nonumber \\
    \mathbf{r}_m \sim & \, \mathrm{Uniform\,\, over\,\, FOV}, \nonumber \\
    I_m \sim & \, \mathrm{Empirical}, \nonumber\\
    \mathcal{B} \sim & \, \mathrm{Gamma}\left(\alpha_{\mathcal{B}},\beta_{\mathcal{B}}\right). \nonumber
\end{align}
Posterior:
\begin{equation}
    P(\vartheta|w_{1:N}) \propto P(w_{1:N}|\vartheta)P(\vartheta). \nonumber
\end{equation}
\label{box:SMLM}
\end{tcolorbox}

Here, we discussed localization of emitters using information from one frame though leveraging information across frames improves spatial resolution by increasing the photon budget available for analysis. The challenge in using multiple frames is that several low-quality putative localizations, if performed in each frame, must then be linked across frames to improve high resolution localization. This essentially becomes equivalent to the problem of single molecule tracking dealt with rigorously later in this section where molecule number determination alongside localization and linking are performed simultaneously and self-consistently. However, to avoid computational overhead, a method termed BaGoL~\cite{fazel2022BaGoL} uses frame-to-frame localization to identify which localizations belong to which emitter. Further, BaGoL efficiently accomplishes sub-nanometer precision under dense labeling conditions by removing nanometer residual drift within the input data and combining the set of identified localizations from each emitter~\cite{fazel2022BaGoL}. The idea of combining localizations to improve precision has been also employed in conjunction with orthogonal DNA-sequences to achieve Angstr{\"o}m resolutions~\cite{reinhardt2023aangstrom}. 

Having focused on static emitters thus far, we now broaden our discussion to mobile emitters, namely tracking emitters across frames. In SPT, data consists of $N$ pixel values for each frame $k=1:K$
denoted by $w_{1:N}^{1:K}=\big\{w_1^1,w_2^1,\dots,w_N^1,w_1^2,\dots,w_N^K\big\}$. The parameter set $\vartheta$ is now expanded to include particle trajectories across time, $\mathbf{r}_m(t)$ for each $m$ particle. By approximation, these may be reduced to locations across frames, $\mathbf{r}_m^{1:K}$, though, in full generality, positions can be interpolated for any inter-frame time~\cite{kilic2021generalizing,xu2023bnp,sgouralis2023dynamic}.

To obtain the SPT likelihood, similar to SMLM, we start from the expected photon count per pixel. As particles evolve over each exposure, the expected photon count for pixel $n$ in frame $k$, $\Lambda_n^k(\vartheta)$, follows from Eq.~\ref{eq:expectedPhotonsNP}~\cite{xu2023bnp,sgouralis2023dynamic} \vspace{-3mm}
\begin{equation}
    \Lambda_n^k(\vartheta)
    =\mathcal{B} + \sum_{m=1}^\infty b_m\int_{\text{exposure}_k} dt \mu(t)\mathcal{P}_m^n(t).
    \label{eq:SPT_photoncount}
\end{equation}
Here, $\mathcal{P}_m^n(t)$ is adapted from Eq.~\ref{eq:PixelProb} with time dependent location, and $\mu(t)$ is the time dependent fluorescence emission rate, \textit{e.g.}, due to blinking. The time integral of Eq.~\ref{eq:SPT_photoncount} is stochastic and numerical integration is often used in its evaluation. Under slow dynamics, for simplicity only, we may approximate the integrand as a constant 
resulting in Eq.~\ref{eq:PixelProb}~\cite{cheezum2001quantitative}. This approximation fails due to motion blurring artifacts, \textit{i.e.}, aliasing, when particles diffuse rapidly compared to the camera frame rate or exposure time~\cite{wong2010limit,michalet2012optimal}.

As an alternative, an improved approximation is afforded by the trapezoidal rule \vspace{-2mm}
\begin{align}
    \Lambda_n^k(\vartheta)
    =\mathcal{B} + \sum_{m=1}^\infty & b_m\sum_{l=1}^{L-1} \frac{\delta t}{2} 
    \Big[\mu_m\left(t_l^k\right)\mathcal{P}_m^n\left(t_l^k\right) \nonumber\\
    &+\mu_m\left(t_{l+1}^k\right)\mathcal{P}_m^n(t_{l+1}^k)\Big] 
    \label{eq:SPT_trapezoid}
\end{align}
with \vspace{-2mm}
\begin{equation}
\vspace{-1mm}
\mathcal{P}_m^n\left(t_l^k\right)=\int\int_{\mathcal{A}_n} dxdy U(x,y;\mathbf{r}_m\left(t_l^k\right)).
\end{equation}
In this equation, $t_1^k$ represents the beginning of the exposure for frame $k$ while $t_L^k$ represents its end. The entire exposure period, $\delta T$, is divided into $L-1$ equal panels of length $\delta t = \frac{\delta T}{L-1}$. A motion model, such as free diffusion or any other, can be introduced to connect positions $\left.\mathbf{r}_m\left(t_{l+1}^k\right)\middle|\mathbf{r}_m\left(t_l^k\right)\right.
    \sim\text{Normal}\left(\mathbf{r}_m\left(t_l^k\right),2D\delta t\right)$, where $D$ is the diffusion coefficient of the emitters, assuming they all satisfy the
same diffusive dynamics.

Though diffusion models are most commonly invoked, alternative models, such as anomalous diffusion, are also used~\cite{munoz2021objective}. It remains to be seen however whether alternative models can be useful in light of dramatic approximations often already made in the analysis including, but not limited to, often assuming: a number of emitters by hand \cite{tinevez2017trackmate}; a time independent integrand in Eq.~\ref{eq:SPT_photoncount}; general corrupting noise from photon count and detectors \cite{tinevez2017trackmate}, and multiple other error sources. 

The emission rates $\mu_m$ of the emitters can also be described using Markovian models \cite{rollins2015stochastic,bryan2022diffraction}; see Sec.~\ref{Nonlinear}. However, for the sake of simplicity, we
assume that all emitters maintain the same brightness throughout all frames resulting in the simplification of Eq.~\ref{eq:SPT_trapezoid} to \vspace{-2mm}
\begin{equation}
    \Lambda_n^k(\vartheta)
    =\mathcal{B} + \mu\sum_{m=1}^\infty b_m\sum_{l=1}^{L-1} \frac{\delta t}{2}
    \left[\mathcal{P}_m^n\left(t_l^k\right)+\mathcal{P}_m^n(t_{l+1}^k)\right].
    \nonumber
\end{equation}

\begin{tcolorbox}[colback=brown!5!white,colframe=brown!75!black,title=Statistical Framework \ref{box:tracking}: Tracking]
Data: pixel values in ADUs
\begin{equation*}
\vspace{-3mm}
    w_{1:N}^{1:K}=\left\{w_1^1,\dots,w_N^K\right\}.
\end{equation*}\\ 
Parameters: loads, fluorophore trajectories, emission rate, background, diffusion
coefficient
\begin{equation*}
    \vartheta = \left\{\bar{b},\bar{\mathbf{r}}\left(t_{1:L}^{1:K}\right),\mu,\mathcal{B},D\right\}.
\end{equation*}
Likelihood:
\begin{equation*}
    P\left(w_{1:N}^{1:K}\middle|\vartheta\right)
    =\prod_{n=1}^N \prod_{k=1}^K
    \mathrm{Gaussian}(w_n^k;g\Lambda_n^k(\vartheta)+o,\sigma_w^2).
\end{equation*}
Priors:
\begin{align*}
    q_m                   & \sim\text{Beta}\left(A_q,B_q\right), m=1:\infty                   \\
    b_m                   & \sim\text{Bernoulli}\left(q_m\right),                              \\
    \mathbf{r}_m\left(t_1^1\right) & \sim\text{Normal}\left(\mathbf{r}_0, \sigma_{\mathbf{r}}^2\right),\\
    \left.\mathbf{r}_m\left(t_{l+1}^k\right)\middle|\mathbf{r}_m\left(t_l^k\right)\right. &
    \sim\text{Normal}\left(\mathbf{r}_m\left(t_l^k\right),2D\delta t\right),
    \\
    \mu                   & \sim\text{Gamma}\left(\alpha_\mu,\beta_\mu\right),                 \\
    \mathcal{B}           & \sim\text{Gamma}\left(\alpha_\mathcal{B},\beta_\mathcal{B}\right), \\
    D                     & \sim\text{InvGamma}\left(\alpha_D,\beta_D\right).                  
\end{align*}
Posterior:
\begin{equation*}
    P\left(\vartheta\middle|w_{1:N}^{1:K}\right)\propto
    P\left(w_{1:N}^{1:K}\middle|\vartheta\right)P\left(\vartheta\right).
\end{equation*}
\label{box:tracking}
\end{tcolorbox}

Again assuming, for simplicity alone, a CCD camera noise model 
(see Appendix~\ref{Se:Detector}), the likelihood for pixel $n$ in frame $k$ reads \vspace{-2mm}
\begin{equation}
    P(w_n|\vartheta) = \mathrm{Gaussian}(w_n^k;g\Lambda_n^k(\vartheta)+o,\sigma_w^2).
    \label{eq:SPT_EMCCD}
\end{equation}
Now, similar to the SMLM likelihood of Eq.~\ref{eq:ROILikelihood}, the likelihood of the frame sequence is
\begin{equation}
P(\vartheta|w_{1:N}^{1:K}) = \prod_n\prod_k P(\vartheta|w_{n}^{k}).
    \label{eq:likelihoodSPT}
\end{equation}
By specifying all terms in Eq.~\ref{eq:likelihoodSPT} explicitly, we would see that $\vartheta$  now includes $\vartheta =
\left\{\bar{b},\bar{\mathbf{r}}\left(t_{1:L}^{1:K}\right),\mu,\mathcal{B},D\right\}$ where overbar denotes the set of all emitters. Sampling of the resulting posterior is outlined in the box below~\cite{xu2023bnp,sgouralis2023dynamic}.

We do highlight that, parametrically, the unknowns exclude the loads,
$\vartheta =
\left\{\bar{\mathbf{r}}\left(t_{1:L}^{1:K}\right),\mu,\mathcal{B},D\right\}$, and the number of trajectories (emitters) are individually estimated with \emph{ad hoc} metrics~\cite{tinevez2017trackmate}. In contrast, non-parametrically,
 trajectories and emitter numbers are jointly estimated alongside other parameters~\cite{jazani2019alternative,jazani2022computational,xu2023bnp,sgouralis2023dynamic}.

We note that the above tracking reveals the z-position only up to a mirror symmetry above or below the focal plane when using a single illumination plane. Thus, here, a note is warranted regarding 3D SMLM. In standard SMLM, localizing the molecule's position along the axial direction is challenging due to the limited depth-of-field and symmetry of the wide-field PSF with respect to the focal plane, \textit{i.e.}, lack of optical sectioning; see Sec.~\ref{Fluo_PSF}. To address these issues, multiple approaches have been employed including multi-plane microscopy (see Sec.~\ref{subsec:Multiplane_mic}) and PSF engineering~\cite{huang2008three,pavani2009three,lew2011corkscrew,shechtman2014optimal}, now detailed in Sec.~\ref{PSFeng_sec}.

\vspace{-7mm}
\subsection{PSF engineering}\label{PSFeng_sec}
\vspace{-2mm}

To overcome the limited optical sectioning of SMLM imposed by wide-field PSFs (see Sec.~\ref{Fluo_PSF}), engineered PSFs have been used to intentionally introduce aberrations. This typically involves inserting extra optical components into the setup ~\cite{huang2008three} or adaptive optical element, such as a deformable mirrors~\cite{izeddin_psf_2012} at the Fourier plane; see Fig.~\ref{fig:magnification}~\cite{pawley2006handbook,backer2014extending,shechtman2014optimal}. The resulting aberrations break the PSF's axial symmetry and thereby encode axial positions of molecules used in 3D localization ~\cite{siemons_comparing_2020}.
  
\begin{figure}[H]
    \centering
    \includegraphics[width=1\linewidth]{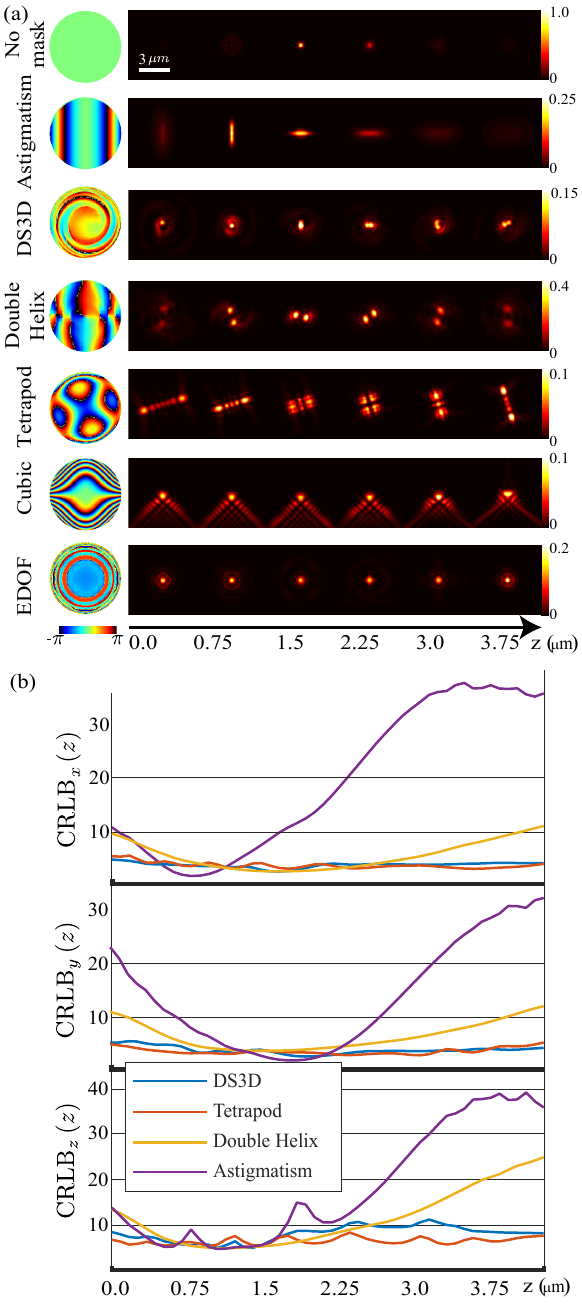} \vspace{-6mm}
    \caption{PSF engineering. (a) Frequently used engineered PSFs, simulated for an objective lens with NA~=~1.49 and pixel size of 110~nm. The top row is the wide-field PSF. Other rows present commonly used phase masks and their corresponding PSFs over a range of axial positions. (b) CRLB (see Sec.~\ref{prob-stat}) of the 3D position (each axis individually) plotted as a function of the axial position, assuming the system is laterally shift-invariant. Here, the subscripts in the axes labels indicate the coordinate for which CRLB was calculated.  
    } 
    \label{fig:PSFpanel}
\end{figure}

Most initial efforts in PSF engineering coincide with PSFs maintaining their shape throughout de-focus. One of the earliest PSF engineering applications reduced in-focus spot sizes, at the cost of larger side-lobes. This was achieved by implementing a series of amplitude and phase rings in the Fourier plane~\cite{di1952super}. As another example, toward achieving Extended Depth Of Field (EDOF), a cubic phase mask was used leading to a PSF minimally changing over a desired axial range~\cite{dowski1995extended}; see Fig.~\ref{fig:PSFpanel}a. While maintaining EDOF, other improvements were aimed at reducing the required computation and raising the SNR, \emph{e.g.}, the log-asphere lens ~\cite{chi2001electronic}, Bessel Beams~\cite{mcgloin2005bessel}, and others~\cite{ben2003all}. 

Recently, PSFs have been engineered, either heuristically, or algorithmically (more details later), to provide improved axial resolutions across different experimental conditions~\cite{von2017three} such as emitter density and wavelength. That is, at the other extreme end of design space where PSFs remain similar throughout de-focus, reside PSFs intentionally sensitive to de-focus. The purpose of such z-encoding PSFs is to encode axial information (depth) in their shape enabling 3D tracking, or imaging~\cite{von2017three}. 

An early instance of z-encoding PSF engineering is induced astigmatism,  typically implemented with a cylindrical lens, for evaluation of de-focus in compact disc players~\cite{kao1994tracking}; an idea adapted for SMLM~\cite{huang2008three}. The astigmatic PSF provides high axial resolution over an axial range of $\approx1~\mu m$. \vspace{-1mm}

Following similar ideas, larger axial ranges were attained using rotating PSFs, based on a linear combination of Laguerre-Gaussian functions~\cite{schechner1996wave}, later adapted to SMLM using the Double Helix PSF~\cite{pavani2009three}. In contrast to wide-field PSFs that spread signal over a large area resulting in low SNRs away from the focus (see first row in Fig.~\ref{fig:PSFpanel}), multiple 3D engineered PSFs have been designed including the corkscrew~\cite{lew2011corkscrew}, self-bending beams~\cite{jia2014isotropic}, tetrapods ~\cite{shechtman2014optimal}, and others \cite{baddeley2011three,prasad2013rotating}. These often attain high resolutions over wider axial ranges and maintain high SNR even at greater de-focus.

Several examples of engineered phase masks, \textit{i.e.}, phase intentionally added to the Fourier plane phase (Fourier plane phase is also sometimes termed pupil phase), 
and associated PSFs are shown in Fig.~\ref{fig:PSFpanel}. We show both PSFs maintaining their shapes over a wide axial range and those encoding axial location in their shapes.

Now, we turn to the question of how we can design phase masks to engineer a desired PSF shape, \textit{e.g.}, a PSF maintaining high axial resolution or high SNR over a wide range. This requires first finding the relation between the measured PSF and the phase mask at the Fourier plane.

To address this, we note the relation between the field at the Fourier plane, and the measured PSF intensity, as described in Eqs.~\ref{eq:Escalar} and \ref{eq:PSFscalar_FFT}. Indeed, the measured PSF intensity contains a Fourier transform of the electric field, and an absolute value operation, resulting in the loss of image plane phase information. As such, the problem of recovering the Fourier plane phase, \textit{i.e.}, pupil phase $\Phi(\theta',\phi)$ (see Eq.~\ref{eq:PSFscalar_FFT}), at the heart of PSF engineering, is known as phase retrieval~\cite{shechtman2015phase}. The phase retrieval problem in our context, involves estimating the pupil phase $\Phi(\theta',\phi)$ from the measurements $w_{1:N}$ encoding the real space PSF through, for example, detector models such as Eq.~\ref{Detector_EntireNoise}. This ill-posed non-convex optimization presents various challenges, including degenerate solutions, unstable derivatives, and more~\cite{shechtman2015phase}. As it is impossible to determine phase using data from one plane, \textit{i.e.}, a single PSF slice, we use data from several planes (z-stack) acquired, for example, by scanning the objective to capture slices of a fluorescent bead's PSF, or by using a multi-plane setup; see Sec.~\ref{subsec:Multiplane_mic}.

Following the logic presented on SMLM data analysis, to construct a likelihood, we write the expected photon count $\Lambda_n^q\left(\vartheta,\Phi\right)$, for pixel $n$ at plane $q$ of the z-stack, encoding pupil phase, $\Phi$, information. For simplicity, we consider a single fluorophore here.

Using this model, a likelihood can be constructed given data $w_n^q,\, n=1:N,\,q=1:Q$ similar to Eqs.~\ref{eq:SMLMLikeGaussNoise}-\ref{eq:ROILikelihood}. Working, for convenience, with the log-likelihood, we write the z-stack log-likelihood \vspace{-2mm}
\begin{equation}
\vspace{-2mm}
\mathcal{L}(w_{1:N}^{1:Q};\vartheta,\Phi)
    = 
    \sum_{n=1}^{N} 
    \sum_{q=1}^{Q} 
    \ell(w_n^q;\vartheta,\Phi),
    \label{eq:cost_PR}
\end{equation} 
where $\ell\left(w_n^q;\vartheta,\Phi\right)$ is the log-likelihood of pixel $n$ within plane $q$. In the most general case, detector and shot noise must both be simultaneously considered as in Eq.~\ref{Detector_EntireNoise}. However, ignoring detector noise for now, we arrive at the single pixel log-likelihood used in 
Eq.~\ref{eq:cost_PR}
 \begin{equation}
\ell\left(w_n^q;\vartheta,\Phi\right) = \Lambda_{n}^q\left(\vartheta,\Phi\right)-
    w_{n}^q \log\left(\Lambda_{n}^q\left(\vartheta,\Phi\right)\right).
\label{eq:loglik-poiss}
 \end{equation}

To maximize the likelihood in Eqs.~\ref{eq:cost_PR}-\ref{eq:loglik-poiss}, we employ iterative optimization often relying on knowledge of the likelihood's gradient with respect to the phase 
~\cite{ferdman2020vipr,smith2016simultaneous}
\begin{equation}
\frac{\partial \ell}{\partial \Phi}=\frac{\partial \ell}{\partial \Lambda_n^q}\frac{\partial \Lambda_n^q}{\partial \Phi}.
\label{eq:chainruleloglik}
\end{equation}
The first term on the right can be analytically evaluated as $\partial \ell/\partial \Lambda_{n}^q = 1 -w_{n}^q/\Lambda_{n}^q$. The next term involves the derivative of the PSF model $\Lambda_{n}^q$ with respect to the pupil phase $\Phi_l$
where, in practice, we discretize the set of spatial frequencies in the Fourier plane $l=1:L$ and write
\begin{equation}
    \frac{\partial \Lambda_{n}^q}{\partial \Phi_l} = 2 \Re{\left(\frac{\partial E_n }{\partial \Phi_l}\cdot E_n^*\right)}. 
\end{equation}
In the above, $E_n$ is the given
electric field in the image plane from Eq.~\ref{eq:Escalar}, $\Re$ indicates the real portion of the expression within the parenthesis, and $\frac{\partial E_n }{\partial \Phi_l}$ and $\frac{\partial \Lambda_{n}^q}{\partial \Phi_l}$ are, respectively, complex and real matrices of size $N\times L$. 

Finally, we must evaluate $\frac{\partial E_n }{\partial \Phi_l}$. The electric field in the image plane is obtained by a Fourier transform of the electric field in the Fourier plane (designated by $E'_{\tilde{l}}$) also containing the pupil phase $\Phi$ 
\begin{multline}
    \frac{\partial E_n}{\partial \Phi_l} = \frac{\partial }{\partial \Phi_l} \mathcal{F}_{\tilde{l}}\left[E'_{\tilde{l}}\right] =
    i \exp{\left(\frac{-i2\pi n l}{M}\right)} E'_{\tilde{l}}\,\, \delta_{l,\tilde{l}},
    \label{eq:derv_E_PM}
\end{multline}
where $\mathcal{F}_{\tilde{l}}$ is a discrete Fourier transform operation over index $ \tilde{l}$ and $ \delta_{l,\tilde{l}}$ is a Kronecker delta. Finally, if $L=N$ the summation over $n$ of Eq.~\ref{eq:cost_PR} and the exponential of Eq.~\ref{eq:derv_E_PM} can be evaluated as a compact Fourier transform providing the desired derivative 
\begin{equation}
    \frac{\partial \mathcal{L}}{\partial \Phi_l} = 2 \Re{\left[E'_l\mathcal{F}_{n}\left(i E_n^{*}\frac{\partial \mathcal{L}}{\partial\Lambda_{n}^q}\right)\right]}.
\end{equation}  
The approach described above can be used both to learn the pupil phase producing a measured PSF or, equivalently, design a PSF and learn the required pupil phase. 

In the realm of high SNR, it is also common to approximate the likelihood Eq.~\ref{eq:cost_PR} by a Gaussian distribution and use least squares minimization to determine the pupil phase. The approximate log-likelihood can then be minimized using iterative optimization, \emph{e.g.}, Gerchberg-Saxton or its variants~\cite{gerchberg1972practical,fienup1978reconstruction}, possibly estimated over a constrained Zernike polynomial set ~\cite{liu2013three,petrov2017measurement}. 

After describing the approach to derive the pupil phase for a given PSF shape, we turn to the problem of seeking an optimal PSF shape following pre-defined metrics. The engineered PSFs of Fig.~\ref{fig:PSFpanel} represent the result of various optimization metrics and numerical approaches. For instance, different PSFs exhibit different CRLBs \cite{badieirostami2010three}; CRLB optimization on the phase mask expanded in terms of Zernike polynomials yields the tetrapod PSF~\cite{shechtman2014optimal} while optimization on the phase mask expanded in terms over Laguerre-Gaussian functions yields the Double Helix PSF~\cite{pavani2008high,pavani2009three}. Similarly, in the panel on DS3D (standing for DeepSTORM3D)~\cite{nehme2020deepstorm3d}, the PSF is optimized to localize emitters within a dense environment using a neural network. Finally, for the EDOF PSF, a cost function is optimized to obtain PSFs maintaining their in-focus shape over wider axial ranges~\cite{nehme2021learning}.

As an example of optimization, to attain a PSF achieving optimal localization precision over a wide axial range, we use the Fisher information and CRLB metrics. To derive the relevant CRLB, we start from the Fisher information matrix elements $\left[\mathcal{Q}\left(\vartheta;\Phi\right)\right]_{i,j}$
of the log-likelihood given in Eq.~\ref{eq:loglik-poiss} (see Sec.~\ref{prob-stat}) \vspace{-4mm}
\begin{multline}
\left[\mathcal{Q}\left(\vartheta;\Phi\right)\right]_{i,j} = \sum\limits_{n=1}^{\text{N}} \frac{\partial}{\partial \vartheta_i}\Lambda_n\left(\vartheta,\Phi\right) \frac{\partial}{\partial \vartheta_j}\Lambda_n\left(n;\vartheta,\Phi\right) \\ \frac{1}{\Lambda_n\left(n;\vartheta,\Phi\right) + \mathcal{B}_n},
\label{eq:fisher-poiss}
\end{multline}
where $\vartheta_j$ is a parameter within the set of unknowns designated by $\vartheta$. After evaluating the Fisher information entries, we can evaluate the CRLB given by Eq.~\ref{eq:CRLB}.

In a practical implementation of an iterative optimization, the PSFs are scaled to match realistic signal counts encountered in SMLM imaging, \emph{i.e}, on the scale of a few hundred photons per emitter per frame. 
 
Heuristic and CRLB optimized PSFs, optimized for just one emitter, can drastically limit their performance at high labeling density where engineered PSFs, such as the tetrapod~\cite{shechtman2014optimal}, suffer from PSF overlaps due to their large lateral footprint. 
In such cases, fitting algorithms like MLE designed for sparse cases, exhibit a significant drop in performance with performance slightly improved for the compact DS3D PSF~\cite{nehme2020deepstorm3d}. One solution toward axial localization in dense environments is to let a neural net learn the optimal pupil phase design~\cite{nehme2020deepstorm3d}. In this case, 3D localization and the encoding pupil phase are simultaneously optimized. 

In a similar vein toward optimizing PSFs for dense localization, 
similar design strategies have been used in multi-color imaging ~\cite{shechtman2016multicolour,smith2016simultaneous} where neural networks have been used to optimize phase masks to optimally discriminate between colors ~\cite{hershko2019multicolor}.

\vspace{-4mm}
\section{Perspectives}
\vspace{-3mm}

The world of microscopy, and biology have been intertwined from the onset. As early as humankind could peer at the world beyond its visual range, it peered into life~\cite{gest2004Discovery} and we continue doing so 
from nuclear pore complexes~\cite{thevathasan2019nuclear} key in intra-cellular communication, individual synaptic spines~\cite{chakraborty2019light}, to cell adhesion~\cite{fischer2021quantitative} at the basis of tissue formation, to actin filaments~\cite{riedl2008lifeact,andrews2008actin,mazloom2021comparing} involved in cell motion and division, and may more.

Life presents events at all
spatiotemporal scales with no clear means of discriminating between object of interest and background. Discrimination from background motivated fluorescence~\cite{renz2013fluorescence}, while probing smaller and faster spatiotemporal scales continues to motivate the experimental and theoretical methodology development. 
Along these lines, major improvements in fluorescent microscopy have followed four main fronts: fluorescent probes; optical setups; detectors; and data analysis. 

Regarding fluorescent probes (see Sec.~\ref{Nonlinear}), the discovery of Green Fluorescent Proteins (GFP) was a milestone in fluorescence microscopy ~\cite{shimomura1962extraction,tsien1998green}. Next came the ability to switch  biomarkers from dark and bright states~\cite{hell1994OpticsLetter,dickson1997off} resulting in super-resolution microscopy and nanometer resolution~\cite{lelek_single-molecule_2021,huang2009super}; see Sec.~\ref{SR}. 

Concerning optical setups (see Sec.~\ref{fluorescence} and \ref{sec:modality}), the invention of the confocal microscope \cite{marvin1961USPatent} marked a milestone accomplishing optical sectioning by inserting a pinhole in the detection arm to filter out-of-focus light. Research in this area is ongoing leading to development of different microscopy modalities, \textit{e.g.}, light-sheet, SIM and others, discussed in Sec.~\ref{sec:modality}, yielding unprecedented optical sectioning as well as high lateral resolutions. 

On the detector front (see Appendix~\ref{Se:Detector}), cameras, including CCDs, EMCCDs, and CMOSs, revolutionized fluorescence microscopy and enabled rapid wide-field imaging. Indeed, the need to amplify signal lead to the development of EMCCDs capable of imaging dim fluorescent probes~\cite{madan1983experimental}. The recent advent of CMOS cameras then accelerated data acquisition up to hundreds of frames per second over large FOVs with reduced read-out noise~\cite{bigas2006review}. While we mostly focused on integrative detectors, increasingly available Single Photon Avalanche Diodes (SPAD) arrays~\cite{ulku2018512,bruschini2019single}, may herald an era of unparalleled spatiotemporal resolution.

Finally data analysis methods grounded in statistics are naturally suited to process fluorescent microscopy data while considering all sources of uncertainty; see Sec.~\ref{prob-stat}. Moreover, considering the fundamental problem of model selection inherent to fluorescence microscopy, BNP frameworks (see Sec.~\ref{prob-stat}) show promise across applications. Deep learning methods~\cite{belthangady2019applications,de2019deep,mockl2020deep,volpe2023roadmap} have also recently gained popularity and may likely be critical to the analysis of large, volumetric, fluorescence data sets~\cite{wang2021real,patel2022high} though these tools require continued model training for different applications. A concrete future avenue for data analysis might very well merge the ideas from both Bayesian and deep learning~\cite{winter2023machine}.

Despite continued progress in fluorescence microscopy~\cite{pawley2006handbook,sahl2017fluorescence,stockert2017fluorescence}, multiple challenges remain including potentially perturbative effects of fluorescent probes on the labeled systems; uncontrolled probe interaction with themselves and their environment; photo-toxic effects naturally arising from any form of illumination; labeling and detection challenges in thicker samples and complex environments; rapid volumetric imaging; manipulating large data set sizes; and many others.

Indeed, as we move to complex environments complementary read-outs beyond fluorescence are often desired and, along these fronts, a number of other methods continue to be developed. These include refractive index tomography~\cite{lee2013quantitative,kim2016optical}, Raman imaging~\cite{smith2019modern,camp2014high}, phase imaging~\cite{popescu2011quantitative,park2018quantitative}, lens-free imaging \cite{bishara2010lensfree}, ghost imaging~\cite{gatti2004ghost,shapiro2008computational}, rotating coherent scattering microscopy~\cite{ruh2018superior,junger2022100}, expansion microscopy~\cite{chen2015expansion,gambarotto2019imaging} and others proven useful at the nanoscale. 

Together, these approaches, alongside the development of theoretical and numerical tools, may help us visualize life's events otherwise unfolding in environments that remain impenetrable and at scales still beyond our reach.   

\begin{acknowledgments}
We thank the Quantitative BioImaging Society (QBI) for providing a venue in which many of the authors first met and discussed the topics presented herein. We also deeply thank Weiqing Xu, Peter T. Brown, Ayush Saurabh, Shep Bryan IV, Ioannis Sgouralis, Alex Rojewski, Tristan Manha, Douglas Shepherd, Thorsten Wohland, Sheng Liu, Kunihiko Ishii, and Tahei Tahara for carefully reading portions of this review and providing detailed feedback. SP also acknowledges NIH NIGMS (R01GM130745), NIH NIGMS (R01GM134426) and NIH MIRA R35GM148237. BF and YS acknowledge funding from the European Research Council (ERC) under the European Union's Horizon 2020 research and innovation program (Grant agreement No. 802567). AR acknowledges support from the Swiss National Science Foundation through the National Centre of Competence in Research Bio-Inspired Materials. JE thanks the European Research Council (ERC) for financial support via project ``smMIET'' (grant agreement No. 884488) under the European Union's Horizon 2020 research and innovation program. KS thanks the Department of Bionanoscience, TU Delft for support and the Kavli Institute for Nanosciences Delft for KIND Synergy seed funding.

\end{acknowledgments}

\appendix
\section{\label{Se:Detector} Detectors physics}

Every photon carries with it information that can be recorded by detectors and later employed to draw inferences. These detectors are comprised of one or many pixels arranged as 2D arrays. The former is mostly employed in point scanning microscopy to record single photon arrival times, {\it e.g.}, in FLIM and FRET. The latter is suitable to wide-field fluorescence. 

\begin{figure}[H]
    \centering
    \includegraphics[scale=0.25]{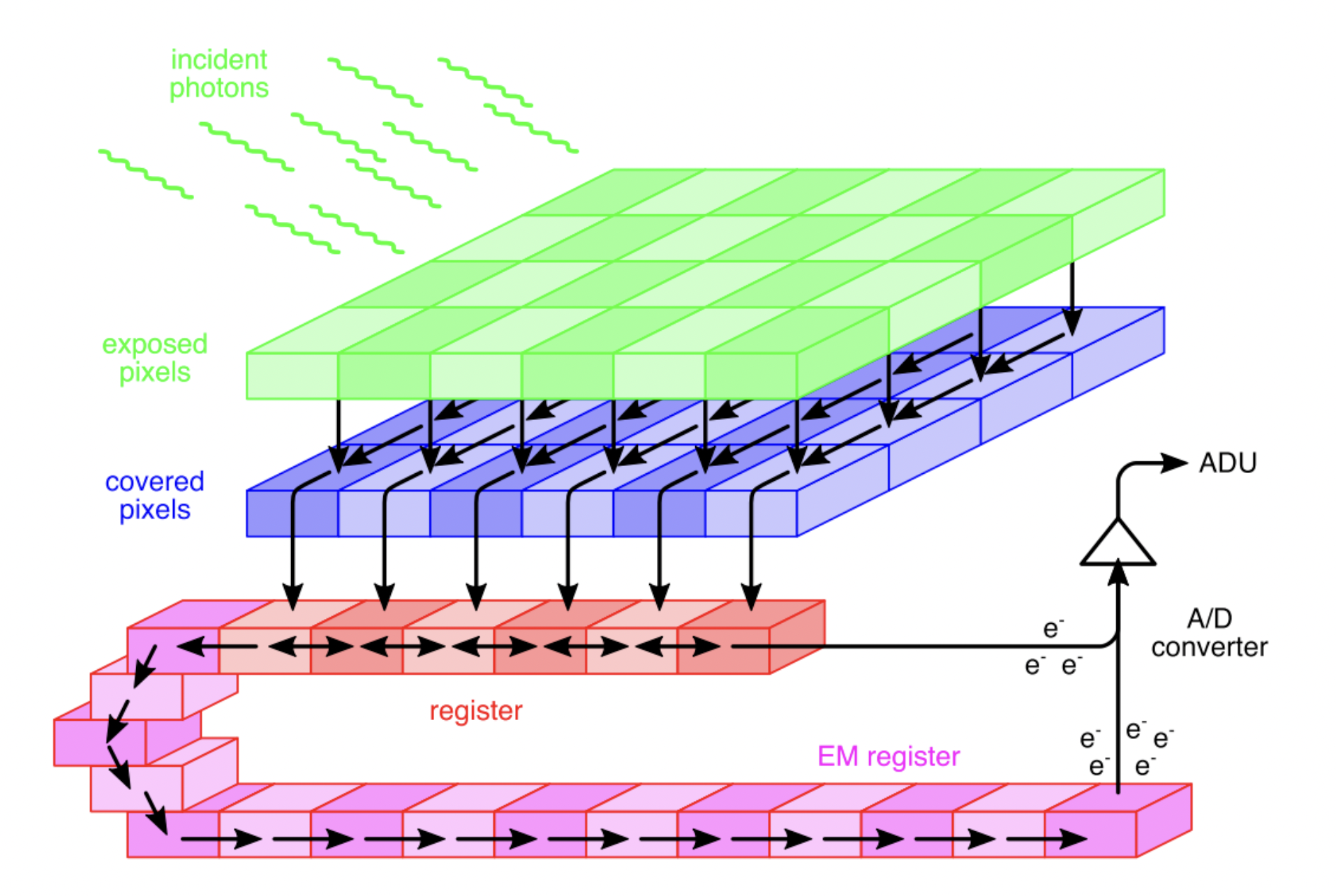}
    \caption{A cartoon illustration of the CCD/EMCCD detector design detailed in the text.}
    \label{EMCCD}
\end{figure}

Ideally, in wide-field detectors, pixel values would histogram the photon counts incident on a particular pixel over the course of an exposure. Similarly, single photon detectors would record precise photon arrival times. However, due to the stochastic noise inherent to detectors, pixel values and recorded arrival times are only probabilistically related to photon counts and direct photon emission times, respectively~\cite{tavakoli2020direct,fazel2022high,michalet2007detectors}.  This section lays out noise models for reported values by different detectors motivated by the detector physics. Once the model is formulated, its parameters are estimated for specific detectors using data from calibration experiments ~\cite{magde1974fluorescence,elson1974fluorescence,weiss1999fluorescence,huang2013video,afanasyev2015posteriori,heintzmann2016calibrating}. In what follows, we first describe wide-field detectors (integrative detectors) and next turn to single photon detectors.

There are three common types of wide-field detectors used in fluorescence microscopy: CCDs~\cite{boyle1970charge,amelio1970experimental,fossum2014review}; EMCCDs~\cite{madan1983experimental,jerram2001llccd,basden2003photon}; and CMOS~\cite{tian2000noise,fossum2014review}. In what follows, we describe the architecture and physics of each detector and, in turn, derive the appropriate noise model.

We begin with a cartoon of detector devices. Fig.~\ref{EMCCD} depicts the main components of CCDs/EMCCDs. The green pixel grid represents photo-active capacitors accumulating photo-electrons proportional to the incident photon counts. The blue grid is a set of capacitors that temporarily hold the resulting photo-electrons. The blue grid then transfers its electrons to the red register, one row at a time. In CCD cameras there is no Electron Multiplier (EM) stage and the transferred electrons follow the arrows to the right in Fig.~\ref{EMCCD} and go to the charge-to-voltage converter. The voltages are then converted into ADUs and recorded as pixel values.

\begin{figure}[H]
    \centering
    \includegraphics[scale=0.25]{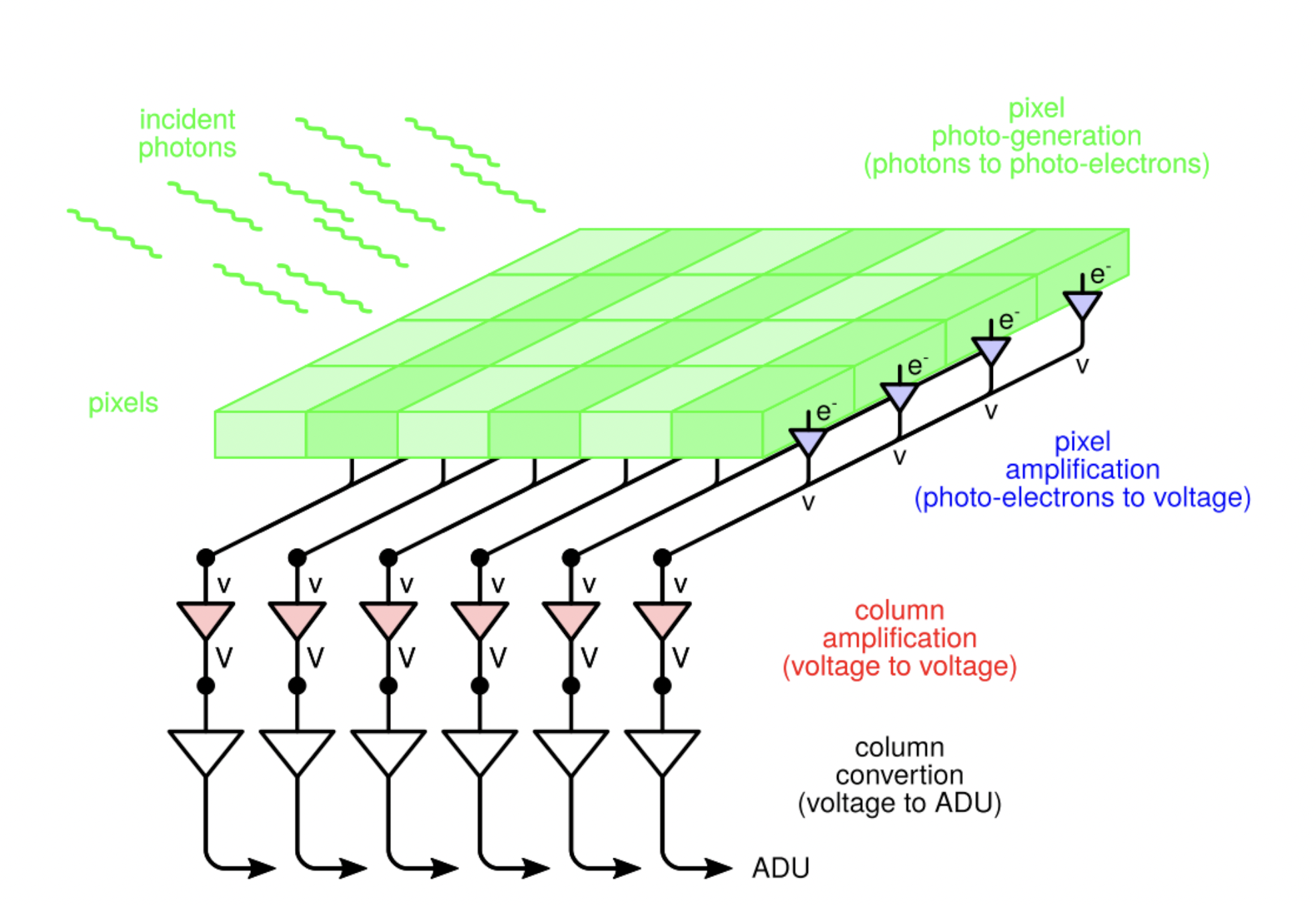}
    \caption{A cartoon illustration of CMOS detector design detailed in the text. }
    \label{sCMOS} \vspace{-4mm}
\end{figure}

By contrast, in EMCCD detectors, the  electrons transferred follow the arrows to the left in the red register and undergo an amplification stage shown in pink before going to the charge-to-voltage converter; see Fig.~\ref{EMCCD}. 
In the EM stage, electrons are fed through a chain of avalanche EMs where an electric field is applied to the electrons, giving them sufficient kinetic energy to knock other electrons into the material's conducting band. This creates new electron-hole pairs thereby amplifying the current. Each stage of the EM process has a small expected gain ($\approx 1\%$) but the device has many stages dramatically amplifying the current prior to reaching the charge-to-voltage converter. 

While CMOS detectors have similar architectures, they use transistors instead of capacitors and every CMOS pixel has its own amplifier; see Fig.~\ref{sCMOS}. This allows for faster data acquisition, a larger FOV, lower power consummation, and larger quantum efficiency. However, such architecture imposes pixel-dependent noise requiring maps of pixel 
gain, variance and offset~\cite{huang2013video,liu2017scmos,mandracchia2020fast}. \vspace{1mm}

\noindent
\textbf{Noise models}\\ \vspace{1mm}
Every stage involved in detecting and converting incident photons to ADUs in detectors introduces 
noise to the final reported pixel values. Here, we discuss main noises introduced at every stage: \\
1) The first source of noise stems from the discrete nature of the photons. Given the expected photon count for pixel $n$, $\Lambda_n$, over a fixed exposure time (\textit{e.g.}, see Eq.~\ref{eq:expectedPhotons}), the measured photon count, $N_{\mathrm{Ph},n}$, is Poisson distributed, \textit{i.e.}, shot noise limited ~\cite{ober2004localization,zhang2009signal,harpsoe2012bayesian,huang2013video}
\begin{equation}
    N_{\mathrm{Ph},n} \sim \mathrm{Poisson}\left(\Lambda_n\right),
    \label{Detector_Poisson}
\end{equation}
where we use notation introduced in Sec.~\ref{prob-stat}.\\
2) Only a fraction of the photons incident on the detector generate photo-electrons where the expected number of photo-electrons per incident photon is called the quantum efficiency $\beta$ ~\cite{zhang2009signal,hirsch2013stochastic,quan2010localization}. The number of generated photo-electrons, $N_{\mathrm{pe},n}$, therefore follows a Binomial distribution 
~\cite{hirsch2013stochastic} 
\begin{equation}
    N_{\mathrm{pe},n} \sim \mathrm{Binomial}\left(N_{\mathrm{Ph},n},\beta\right).
    \label{Detector_QuantumEffic}
\end{equation}
The distribution over the number of photo-electrons given the expected number of photons, $\Lambda_n$, can then be obtained by 
marginalizing over the incident number of photons (see Sec.~\ref{prob-stat}), as follows
\begin{align}
    \mathrm{Poisson}(N_{\mathrm{pe},n};\beta\Lambda_n & ) =  \sum_{N_{\mathrm{Ph},n}=N_{\mathrm{pe},n}}^{\infty} \mathrm{Poisson}\left(N_{\mathrm{Ph},n};\Lambda_i\right) \nonumber \\ &
    \times \mathrm{Binomial}\left(N_{\mathrm{pe},n};N_{\mathrm{Ph},n},\beta\right).
    \label{Detector_PhotoE}
\end{align} \\
where, to be clear, we have distinguished between the Binomial distribution of Eq. A1 and the Binomial density of Eq.~\ref{Detector_PhotoE} as detailed in Sec.~\ref{prob-stat}.\\
3) The third source of noise is due to spurious charge consisting of unwanted electrons generated during the transfer process, termed Clock Induced Charge (CIC) ~\cite{daigle2009extreme,hirsch2013stochastic}. The CIC noise follows a Poisson distribution and gives rise to additional electrons while transferring to the register 
\begin{equation}
    N_{\mathrm{te},n}\sim \mathrm{Poisson}\left(\beta\Lambda_n+C\right),
    \label{Detector_CIC}
\end{equation}
where $N_{\mathrm{te},n}$ and $C$ are the number of electrons after the transferring stage and the mean value of the CIC. This noise is small but can be greatly amplified during the electron multiplier step of EMCCD detectors. \\
4) The EM process consists of many stages in which new electrons are excited through impact ionization, which can be considered a cascade of stochastic events. These steps are approximately identical thus the EM process can be modeled as a cascade of Poisson processes, or Bernoulli trials ~\cite{tubbs2003lucky,hirsch2013stochastic}, or a geometric model of multiplication ~\cite{chao2012fisher,chao2013two}. 

The number of electrons after the EM stage, $N_{\mathrm{ae},n}$, is a random variable that approximately follows a Gamma distribution   ~\cite{basden2003photon,harpsoe2012bayesian,hirsch2013stochastic}
\begin{equation}
    N_{\mathrm{ae},n} \sim \mathrm{Gamma}\left(N_{\mathrm{te},n},\hat{g}\right),
    \label{Detector_EMgamma}
\end{equation}
where $\hat{g}$ denotes the amplification gain given by the ratio of the output and input electrons to the EM stage. For large values of $N_{\mathrm{te},n}$, this process is further approximated by a Gaussian, which is computationally more efficient ~\cite{basden2003photon,hirsch2013stochastic}
\begin{equation}
    N_{\mathrm{ae},n} \sim \mathrm{Gaussian}\left(\hat{g}N_{\mathrm{te},n},\hat{g}^2N_{\mathrm{te},n}\right).
    \label{Detector_EMgauss}
\end{equation}
5) The last stage termed ``read-out" takes the input electrons following the EM stage,
$N_{\mathrm{ae},n}$, and converts these (with noise) into discrete pixel values reported as the data $w_n$ in ADUs. This stage introduces another gain $\gamma$ (ADUs per electron, also sometimes referred to as the sensitivity and typically smaller than one) and offset $\mu$ (output ADU at zero input electron often added to avoid negative pixel values) modeled by a Gaussian distribution and termed ``read-out noise"
\begin{equation}
    w_n \sim \mathrm{Gaussian}\left(\gamma N_{\mathrm{ae},n}+\mu,\sigma_{\mathrm{ro}}^2\right),
    \label{Detector_ADUnoise}
\end{equation}
where $\sigma_{\mathrm{ro}}^2$ is the read-out noise variance.

The combination of the noises introduced via the amplification and read-out stages is obtained by marginalizing the intermediate parameter $N_{\mathrm{ae},n}$ (namely the number of electrons after the EM stage) between Eqs.~\ref{Detector_EMgauss} and \ref{Detector_ADUnoise}, resulting in 
\begin{equation}
    w_n \sim \mathrm{Gaussian}(\tilde{g}N_{\mathrm{te},n}+\mu,\sigma_w^2),
    \label{Detector_CombinedNoise}
\end{equation}
where $\tilde{g}=\gamma\hat{g}$ and $\sigma^2_w = \gamma^2\hat{g}^2N_{\mathrm{te},n}+\sigma_{\mathrm{ro}}^2$ denote total gain and variance, respectively. Finally, the entire detector model, which relates the expected photon count ($\Lambda_n$) to the reported pixel value ($w_n$) is obtained by marginalizing the other intermediate parameter $N_{\mathrm{te},n}$ (namely the number of electrons after the transferring stage) between Eqs.~\ref{Detector_CIC} and \ref{Detector_CombinedNoise}
\begin{align}
    P\left(w_n|\Lambda_n\right) & =  \sum_{N_{\mathrm{te},n}=0}^{\infty} \mathrm{Poisson}(N_{\mathrm{te},n};\beta\Lambda_n+C) \nonumber \\ & \times \mathrm{Gaussian}(w_n;\tilde{g}N_{\mathrm{te},n}+\mu,\sigma_w^2).
    \label{Detector_EntireNoise}
\end{align}
Since we did not make any assumptions about the gain, offset and other parameters to derive the noise model above, it is valid for both CCD and EMCCD detectors. Moreover, if we assume pixel-dependent parameters, \textit{e.g.}, gain, offset and others, this model would be valid for CMOS detectors as well. As  Eq.~\ref{Detector_EntireNoise} remains complex, 
we make appropriate approximations for the sake of computational efficiency to derive simpler noise models specialized to each detector.

We start with CCD detectors lacking an EM amplification stage ($\hat{g}\approx 1$ and $\sigma_w^2 \approx \sigma_{ro}^2$). These are therefore suitable in detecting large input signals compared to the read-out noise variance. This can be quantitatively expressed as
\begin{equation}
    \mathrm{SNR} = \frac{\Lambda_n}{\sigma_{\mathrm{ro}}} \gg 1,
    \label{Detector_SNR}
\end{equation}
implying the signal is not buried by read-out noise. Under the large signal ($\Lambda_n$) assumption, the Poisson distribution Eq.~\ref{Detector_CIC} is approximated by a Gaussian where both mean and variance are given by the Poisson's mean
\begin{align}
    P\left(w_n|\Lambda_n\right) & \approx  \sum_{N_{\mathrm{te},n}=0}^{\infty} \mathrm{Gaussian}(N_{\mathrm{te},n};\beta\Lambda_n,\beta\Lambda_n) \nonumber \\ & \times 
    \mathrm{Gaussian}(w_n;\gamma N_{\mathrm{te},n}+\mu,\sigma_w^2),
    \label{Detector_tCCDNoise}
\end{align}
where we assumed $\tilde{g}=\gamma$ and $\sigma_w^2 = \sigma_{\mathrm{ro}}^2$ and further
neglected the spurious charge $C$ in the absence of amplification in CCD cameras. Therefore, Eq.~\ref{Detector_tCCDNoise} leads to 
\begin{equation}
    w_n|\Lambda_n \sim \mathrm{Gaussian}(g\Lambda_n+o,\sigma^2_w),
    \label{Detector_CCDNoise}
\end{equation}
where $g=\gamma\beta, o=\mu$ and $\sigma^2_w=\sigma_{\mathrm{ro}}^2$, respectively, denote gain, offset, and variance for CCD detectors. It is also common to apply the offset and gain to the pixel values (data) and write the above equation for gain and offset corrected pixel values
\begin{equation}
    (w_n-o)/g\big|\Lambda_n \sim \mathrm{Gaussian}(\Lambda_n,\sigma^2_w/g^2).
    \label{Detector_CCDNoise_Mod}
\end{equation}

Next, we consider EMCCD detectors. These detectors are suitable for low SNR. The EM stage of these detectors amplify signal above the read-out noise ($\hat{g} \gg \sigma_w$).
In an effort to simplify Eq.~\ref{Detector_EntireNoise} for EMCCDs, we write Eq.~\ref{Detector_EntireNoise} in explicit form 
\begin{align}
    P\left(w_n|\Lambda_n\right) & =  \sum_{N_{\mathrm{te},n}=0}^{\infty}  \frac{(\beta\Lambda_n)^{N_{\mathrm{te},n}}e^{-\beta\Lambda_n}}{N_{\mathrm{te},n}!} \nonumber \\  & \times \frac{1}{\sqrt{2\pi\sigma^2_D}}\exp\left[{-\frac{\left(N_{\mathrm{te},n}-\frac{w_n-\mu}{\tilde{g}}\right)^2}{2\sigma^2_w/\tilde{g}^2}}\right]
    \label{Detector_tEMCCDNoise}
\end{align}
where we have factorized $\tilde{g}$ in the exponent. For large amplifications the standard deviation $\sigma^2_w/\tilde{g}^2$ becomes small and results in a narrow Gaussian approximated by a delta function. Therefore, upon marginalization and some algebra we recover~\cite{huang2013video}
\begin{equation}
    (w_n-o)/g|\Lambda_n \sim \frac{1}{\Gamma(1+\frac{w_n-o}{g})}e^{-\beta\Lambda_n}(\beta\Lambda_n)^{\frac{w_n-o}{g}}
\end{equation}
where $o=\mu$ and $g=\tilde{g}$ denote offset and gain, respectively.
The above EMCCD model for the corrected pixel values is similar to a Poisson noise model where the corrected pixel values do not need to be integers~\cite{huang2013video}. An alternative EMCCD camera noise model could be obtained by convolution of the Poisson distribution Eq.~\ref{Detector_CIC} and the Gamma noise model for EM amplification Eq.~\ref{Detector_EMgamma} resulting in an approximate Gamma noise model for EMCCD detectors  ~\cite{basden2003photon,hirsch2013stochastic,bryan2022diffraction}.
Both noise models asymptotically converge to the same model for large gain $g$.

After deriving noise models for CCD and EMCCD detectors with pixel-independent gain, $g$, offset, $o$, and variance, $\sigma_w^2$, we continue in deriving the noise model for CMOSs where the gain, variance and offset are pixel-dependent~\cite{huang2013video,liu2017scmos,mandracchia2020fast}. Therefore, every pixel follows a different noise model similar to Eq.~\ref{Detector_EntireNoise}
\begin{align}
    P\left(w_n|\Lambda_n\right) & =  \sum_{N_{\mathrm{te},n}=0}^{\infty} \mathrm{Poisson}(N_{\mathrm{te},n};\beta\Lambda_n) \nonumber \\ & \times \mathrm{Gaussian}(w_n;\tilde{g}_nN_{\mathrm{te},n}+\mu_n,\sigma_{w,n}^2),
\end{align}
where $n$ indexes pixels.
Provided small gain $\tilde{g}_n$ for CMOSs, the above convolution can be approximated by a Poisson distribution assuming an extra source of photon for each $n$ pixel  contributing $\sigma^2_{w,n}/g^2_n$ photons~\cite{huang2013video}
\begin{equation}
    \hat{w}_n|\Lambda_n \sim \mathrm{Poisson}(\Lambda_n+\sigma^2_{w,n}/g^2_n),
    \label{Detector_CMOSNoise}
\end{equation}
where $\hat{w}_n = (w_n-o_n)/g_n+\sigma^2_{w,n}/g^2_n$, $g_n$ is the gain, characterized from calibration experiments, and $o_n=\mu_n$ is the pixel-dependent offset.

\begin{figure}[H]
    \centering
    \includegraphics[scale=0.45]{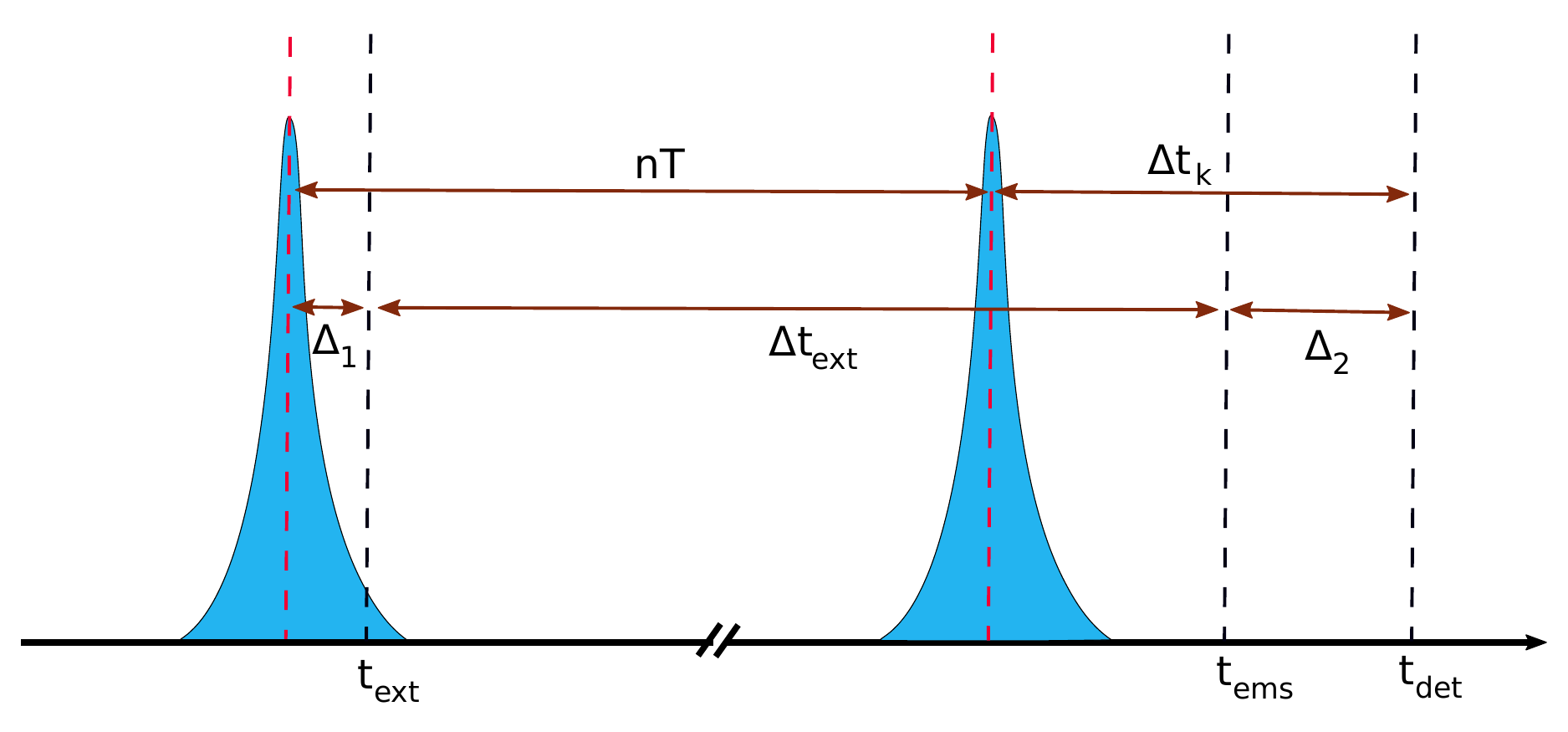}
    \caption{Single photon detector. Laser pulses and their centers are, respectively, shown by blues spikes and red dashed lines with inter-pulse window $T$. The fluorophore excitation, photon emission and photon detection events take place, respectively, at $t_{\mathrm{ext}}, t_{\mathrm{ems}}$ and $t_{\mathrm{det}}$ designated by black dashed lines. The fluorophore spends time $\Delta t_{\text{ext}}$ in the excited state and emits a photon after $n$ pulses. The reported photon arrival time, $\Delta t_k$, is measured with respect to the immediate previous pulse center. Moreover, $\Delta_1$ and $\Delta_2$ denote the difference of the excitation pulse center and the detector delay in reporting the photon arrival time.}
    \label{SingleDetector}
\end{figure}

After considering wide-field detectors, we proceed to describe noise models for single photon detector. We do so by assuming fluorophore excitation using a pulsed laser. This is illustrated in Fig.~\ref{SingleDetector} where the laser pulses are designated by blue spikes with inter-pulse window $T$. Here, a fluorophore gets excited during a pulse at time $t_{\mathrm{ext}}$, spends $\Delta t_{\mathrm{ext}}$ time in the excited state and emits a photon, in the most general case, after $n$ pulses at $t_{\mathrm{ems}}$. However, the photon arrival time is recorded as $t_{\mathrm{det}}$ by a $\Delta_2$ delay in the detector and is reported with respect to the immediate previous pulse given by $\Delta t_k$ for the $k$th photon. 

From Fig.~\ref{SingleDetector}, we can write the following relation for the reported photon arrival time 
\begin{equation}
    \Delta t_k = \Delta t_{\mathrm{ext}} + \Delta t_{\mathrm{IRF}} - nT,
\end{equation}
where $\Delta t_{\mathrm{IRF}}=\Delta_1 + \Delta_2$ is the noise introduced by the IRF due to the laser pulses' finite width and stochastic delay of the detector. Here, the reported arrival time is the sum of three random variables. As such, the noise model is given by the convolution of three probability distributions 
\begin{align}
    P(\Delta t_k|\lambda) = & P(n|N) \\
    \otimes & \left[P(\Delta t_{\mathrm{ext}}|\lambda)\otimes P(\Delta t_{\mathrm{IRF}}|\tau_{\mathrm{IRF}}, \sigma^2_{\mathrm{IRF}})\right], \nonumber
    \label{eq:ConvTimes}
\end{align}
where $\lambda, \tau_{\mathrm{IRF}}, \sigma^2_{\mathrm{IRF}}$ and $N$ are, respectively, the rate of excited state decay (inverse of the excited state lifetime; see Eq.~\ref{eq:lifetime}), the IRF offset, the IRF variance, and the maximum possible number of pulses after which the fluorophore emits. These distributions are given by 
\begin{align}
    \Delta t_{\mathrm{ext}}|\lambda \sim & \mathrm{Exponential}\left(\Delta t_{\mathrm{ext}};\lambda \right), \\
    \Delta t_{\mathrm{IRF}}|\tau_{\mathrm{IRF}}, \sigma^2_{\mathrm{IRF}} \sim & \mathrm{Gaussian}(\Delta t_{\mathrm{IRF}};\tau_{\mathrm{IRF}}, \sigma^2_{\mathrm{IRF}}), \\
    n|N \sim & \mathrm{Categorical}(A_0, ..., A_N),
\end{align}
where the time spent in the excited state and the IRF time are sampled from Exponential and Gaussian distributions. The pulse at which the fluorophore emits is sampled from a Categorical distribution where $A_n$ is given by the integral of the term inside brackets in Eq.~\ref{eq:ConvTimes} over pulse $n$~\cite{fazel2022high}. Finally, calculating the convolutions in Eq.~\ref{eq:ConvTimes}, we obtain the following noise model for single photon detectors under pulsed illumination~\cite{fazel2022high}
\begin{align}
P(&\Delta t_{k}|\lambda) = 
\Bigg[ \sum_{n=0}^N \frac{\lambda}{2} 
\mathrm{erfc}\left(\frac{\tau_{\mathrm{IRF}}-\Delta t_k - nT+\lambda \sigma^2_{\mathrm{IRF}}}{\sigma_{\mathrm{IRF}}\sqrt{2}}\right)\nonumber\\
\times & \exp\left(\frac{\lambda}{2}\left(2(\tau_{\mathrm{IRF}}-\Delta t_k - nT )+\lambda \sigma^2_{\mathrm{IRF}}\right)\right) \Bigg],
\label{eq:SinglePhotonNoise}
\end{align}
where $\mathrm{erfc}(.)$ denotes the complementary error function ~\cite{olver2010nist}. In many practical cases, the inter-pulse time is much larger than the fluorophore lifetime (inverse of the fluorophore radiative decay, $T \gg 1/\lambda$) where the fluorophore emits before the next pulse. In such case, the noise model can be simplified by setting $N=0$ ~\cite{tavakoli2020direct}.


\bibliography{Ref}

\end{document}